\documentclass[longauth,traditabstract]{aa}
\pdfoutput=1
\usepackage{graphicx}
\usepackage{amsmath,amsfonts,amssymb}
\usepackage{txfonts}
\usepackage[breaklinks,colorlinks,citecolor=blue]{hyperref}
\usepackage{color}
\usepackage{fixltx2e}
\usepackage{natbib}
\usepackage{float}
\usepackage[usenames,dvipsnames,svgnames,table]{xcolor}
\bibpunct{(}{)}{;}{a}{}{,}
\usepackage[switch]{lineno}
\usepackage{ifthen}
\usepackage[T1]{fontenc}
\usepackage{lmodern}
\usepackage{ifxetex,ifluatex}
\usepackage{dblfloatfix}
\usepackage{morefloats}

\newcommand{\cruler}{{\tt COMMANDER}}

\newcommand{\nilc}{{\tt NILC}}
\newcommand{\sevem}{{\tt SEVEM}}
\newcommand{\smica}{{\tt SMICA}}

\newcommand{\wmap}{\textit{WMAP}}
\newcommand{\lrg}{SDSS-CMASS/LOWZ}
\newcommand{\mg}{SDSS-MphG}
\newcommand{\tmpz}{2MPZ}
\newcommand{\nvss}{NVSS}
\newcommand{\wg}{WISE-GAL}
\newcommand{\wagn}{WISE-AGN}
\newcommand{\wise}{WISE}
\newcommand{\sdss}{SDSS}
\newcommand{\kap}{Kappa}
\newcommand{\gr}{GR08}
\newcommand{\nside}{$N_\textrm{side}$}
\newcommand{\tce}{$T_{E-\mathrm{c}}$}
\newcommand{\tue}{$T_{E-\mathrm{u}}$}
\newcommand{\tcem}{T_{E-\mathrm{c}}}
\newcommand{\tuem}{T_{E-\mathrm{u}}}
\newcommand{\vn}{\vec{\hat{n}}}
\newcommand{\tp}{$T$}
\newcommand{\qp}{$Q$}
\newcommand{\up}{$U$}
\newcommand{\ep}{$E$}

\newcommand{\stn}{signal-to-noise}
\newcommand{\dd}{\mathrm{d}}
\newcommand{\healpix}{{\tt HEALPix}}

\usepackage{multirow}

\def\setsymbol#1#2{\expandafter\def\csname #1\endcsname{#2}}
\def\getsymbol#1{\csname #1\endcsname}

\def\Planck{\textit{Planck}}

\def\alltwentyfifteenresultspapers{\nocite{planck2014-a01, planck2014-a03, planck2014-a04, planck2014-a05, planck2014-a06, planck2014-a07, planck2014-a08, planck2014-a09, planck2014-a11, planck2014-a12, planck2014-a13, planck2014-a14, planck2014-a15, planck2014-a16, planck2014-a17, planck2014-a18, planck2014-a19, planck2014-a20, planck2014-a22, planck2014-a24, planck2014-a26, planck2014-a28, planck2014-a29, planck2014-a30, planck2014-a31, planck2014-a35, planck2014-a36, planck2014-a37, planck2014-ES}}

\newbox\tablebox    \newdimen\tablewidth
\def\leaderfil{\leaders\hbox to 5pt{\hss.\hss}\hfil}
\def\endPlancktable{\tablewidth=\columnwidth 
    $$\hss\copy\tablebox\hss$$
    \vskip-\lastskip\vskip -2pt}
\def\endPlancktablewide{\tablewidth=\textwidth 
    $$\hss\copy\tablebox\hss$$
    \vskip-\lastskip\vskip -2pt}
\def\tablenote#1 #2\par{\begingroup \parindent=0.8em
    \abovedisplayshortskip=0pt\belowdisplayshortskip=0pt
    \noindent
    $$\hss\vbox{\hsize\tablewidth \hangindent=\parindent \hangafter=1 \noindent
    \hbox to \parindent{$^#1$\hss}\strut#2\strut\par}\hss$$
    \endgroup}
\def\doubleline{\vskip 3pt\hrule \vskip 1.5pt \hrule \vskip 5pt}

\def\L2{\ifmmode L_2\else $L_2$\fi}

\def\DeltaT{\ifmmode \Delta T\else $\Delta T$\fi}
\def\deltat{\ifmmode \Delta t\else $\Delta t$\fi}
\def\fknee{\ifmmode f_{\rm knee}\else $f_{\rm knee}$\fi}
\def\Fmax{\ifmmode F_{\rm max}\else $F_{\rm max}$\fi}
\def\solar{\ifmmode{\rm M}_{\mathord\odot}\else${\rm M}_{\mathord\odot}$\fi}
\def\Msolar{\ifmmode{\rm M}_{\mathord\odot}\else${\rm M}_{\mathord\odot}$\fi}
\def\Lsolar{\ifmmode{\rm L}_{\mathord\odot}\else${\rm L}_{\mathord\odot}$\fi}
\def\inv{\ifmmode^{-1}\else$^{-1}$\fi}
\def\mo{\ifmmode^{-1}\else$^{-1}$\fi}
\def\sup#1{\ifmmode ^{\rm #1}\else $^{\rm #1}$\fi}
\def\expo#1{\ifmmode \times 10^{#1}\else $\times 10^{#1}$\fi}
\def\,{\thinspace}
\def\lsim{\mathrel{\raise .4ex\hbox{\rlap{$<$}\lower 1.2ex\hbox{$\sim$}}}}
\def\gsim{\mathrel{\raise .4ex\hbox{\rlap{$>$}\lower 1.2ex\hbox{$\sim$}}}}

\def\simprop{\mathrel{\raise .4ex\hbox{\rlap{$\propto$}\lower 1.2ex\hbox{$\sim$}}}}
\def\deg{\ifmmode^\circ\else$^\circ$\fi}
\def\pdeg{\ifmmode $\setbox0=\hbox{$^{\circ}$}\rlap{\hskip.11\wd0 .}$^{\circ}
          \else \setbox0=\hbox{$^{\circ}$}\rlap{\hskip.11\wd0 .}$^{\circ}$\fi}
\def\arcs{\ifmmode {^{\scriptstyle\prime\prime}}
          \else $^{\scriptstyle\prime\prime}$\fi}
\def\arcm{\ifmmode {^{\scriptstyle\prime}}
          \else $^{\scriptstyle\prime}$\fi}
\newdimen\sa  \newdimen\sb
\def\parcs{\sa=.07em \sb=.03em
     \ifmmode \hbox{\rlap{.}}^{\scriptstyle\prime\kern -\sb\prime}\hbox{\kern -\sa}
     \else \rlap{.}$^{\scriptstyle\prime\kern -\sb\prime}$\kern -\sa\fi}
\def\parcm{\sa=.08em \sb=.03em
     \ifmmode \hbox{\rlap{.}\kern\sa}^{\scriptstyle\prime}\hbox{\kern-\sb}
     \else \rlap{.}\kern\sa$^{\scriptstyle\prime}$\kern-\sb\fi}
\def\ra[#1 #2 #3.#4]{#1\sup{h}#2\sup{m}#3\sup{s}\llap.#4}
\def\dec[#1 #2 #3.#4]{#1\deg#2\arcm#3\arcs\llap.#4}
\def\deco[#1 #2 #3]{#1\deg#2\arcm#3\arcs}
\def\rra[#1 #2]{#1\sup{h}#2\sup{m}}

\def\dots{\relax\ifmmode \ldots\else $\ldots$\fi}
\def\WHzsr{\ifmmode $W\,Hz\mo\,sr\mo$\else W\,Hz\mo\,sr\mo\fi}
\def\mHz{\ifmmode $\,mHz$\else \,mHz\fi}
\def\GHz{\ifmmode $\,GHz$\else \,GHz\fi}
\def\mKs{\ifmmode $\,mK\,s$^{1/2}\else \,mK\,s$^{1/2}$\fi}
\def\muKs{\ifmmode \,\mu$K\,s$^{1/2}\else \,$\mu$K\,s$^{1/2}$\fi}
\def\muKRJs{\ifmmode \,\mu$K$_{\rm RJ}$\,s$^{1/2}\else \,$\mu$K$_{\rm RJ}$\,s$^{1/2}$\fi}
\def\muKHz{\ifmmode \,\mu$K\,Hz$^{-1/2}\else \,$\mu$K\,Hz$^{-1/2}$\fi}
\def\MJysr{\ifmmode \,$MJy\,sr\mo$\else \,MJy\,sr\mo\fi}
\def\MJysrmK{\ifmmode \,$MJy\,sr\mo$\,mK$_{\rm CMB}\mo\else \,MJy\,sr\mo\,mK$_{\rm CMB}\mo$\fi}
\def\microns{\ifmmode \,\mu$m$\else \,$\mu$m\fi}

\def\muK{\ifmmode \,\mu$K$\else \,$\mu$\hbox{K}\fi}
\def\microK{\ifmmode \,\mu$K$\else \,$\mu$\hbox{K}\fi}
\def\muW{\ifmmode \,\mu$W$\else \,$\mu$\hbox{W}\fi}
\def\kms{\ifmmode $\,km\,s$^{-1}\else \,km\,s$^{-1}$\fi}
\def\kmsMpc{\ifmmode $\,\kms\,Mpc\mo$\else \,\kms\,Mpc\mo\fi}
\providecommand{\sorthelp}[1]{}

\title{\Planck\ 2015 results. XXI. The integrated Sachs-Wolfe effect}

\begin{document}
\author{\small
Planck Collaboration: P.~A.~R.~Ade\inst{97}
\and
N.~Aghanim\inst{64}
\and
M.~Arnaud\inst{80}
\and
M.~Ashdown\inst{76, 7}
\and
J.~Aumont\inst{64}
\and
C.~Baccigalupi\inst{94}
\and
A.~J.~Banday\inst{109, 12}
\and
R.~B.~Barreiro\inst{71}
\and
N.~Bartolo\inst{34, 72}
\and
S.~Basak\inst{94}
% ...
}
\institute{\small
APC, AstroParticule et Cosmologie, Universit\'{e} Paris Diderot, CNRS/IN2P3, CEA/lrfu, Observatoire de Paris, Sorbonne Paris Cit\'{e}, 10, rue Alice Domon et L\'{e}onie Duquet, 75205 Paris Cedex 13, France\goodbreak
\and
Aalto University Mets\"{a}hovi Radio Observatory and Dept of Radio Science and Engineering, P.O. Box 13000, FI-00076 AALTO, Finland\goodbreak
\and
African Institute for Mathematical Sciences, 6-8 Melrose Road, Muizenberg, Cape Town, South Africa\goodbreak
\and
Agenzia Spaziale Italiana Science Data Center, Via del Politecnico snc, 00133, Roma, Italy\goodbreak
\and
Aix Marseille Universit\'{e}, CNRS, LAM (Laboratoire d'Astrophysique de Marseille) UMR 7326, 13388, Marseille, France\goodbreak
\and
Aix Marseille Universit\'{e}, Centre de Physique Th\'{e}orique, 163 Avenue de Luminy, 13288, Marseille, France\goodbreak
\and
Astrophysics Group, Cavendish Laboratory, University of Cambridge, J J Thomson Avenue, Cambridge CB3 0HE, U.K.\goodbreak
\and
Astrophysics \& Cosmology Research Unit, School of Mathematics, Statistics \& Computer Science, University of KwaZulu-Natal, Westville Campus, Private Bag X54001, Durban 4000, South Africa\goodbreak
\and
Atacama Large Millimeter/submillimeter Array, ALMA Santiago Central Offices, Alonso de Cordova 3107, Vitacura, Casilla 763 0355, Santiago, Chile\goodbreak
\and
CGEE, SCS Qd 9, Lote C, Torre C, 4$^{\circ}$ andar, Ed. Parque Cidade Corporate, CEP 70308-200, Bras\'{i}lia, DF, Brazil\goodbreak
\and
CITA, University of Toronto, 60 St. George St., Toronto, ON M5S 3H8, Canada\goodbreak
\and
CNRS, IRAP, 9 Av. colonel Roche, BP 44346, F-31028 Toulouse cedex 4, France\goodbreak
\and
CRANN, Trinity College, Dublin, Ireland\goodbreak
\and
California Institute of Technology, Pasadena, California, U.S.A.\goodbreak
\and
Centre for Theoretical Cosmology, DAMTP, University of Cambridge, Wilberforce Road, Cambridge CB3 0WA, U.K.\goodbreak
\and
Centro de Estudios de F\'{i}sica del Cosmos de Arag\'{o}n (CEFCA), Plaza San Juan, 1, planta 2, E-44001, Teruel, Spain\goodbreak
\and
Computational Cosmology Center, Lawrence Berkeley National Laboratory, Berkeley, California, U.S.A.\goodbreak
\and
Consejo Superior de Investigaciones Cient\'{\i}ficas (CSIC), Madrid, Spain\goodbreak
\and
DSM/Irfu/SPP, CEA-Saclay, F-91191 Gif-sur-Yvette Cedex, France\goodbreak
\and
DTU Space, National Space Institute, Technical University of Denmark, Elektrovej 327, DK-2800 Kgs. Lyngby, Denmark\goodbreak
\and
D\'{e}partement de Physique Th\'{e}orique, Universit\'{e} de Gen\`{e}ve, 24, Quai E. Ansermet,1211 Gen\`{e}ve 4, Switzerland\goodbreak
\and
Departamento de Astrof\'{i}sica, Universidad de La Laguna (ULL), E-38206 La Laguna, Tenerife, Spain\goodbreak
\and
Departamento de F\'{\i}sica, Universidad de Oviedo, Avda. Calvo Sotelo s/n, Oviedo, Spain\goodbreak
\and
Department of Astronomy and Astrophysics, University of Toronto, 50 Saint George Street, Toronto, Ontario, Canada\goodbreak
\and
Department of Astrophysics/IMAPP, Radboud University Nijmegen, P.O. Box 9010, 6500 GL Nijmegen, The Netherlands\goodbreak
\and
Department of Physics \& Astronomy, University of British Columbia, 6224 Agricultural Road, Vancouver, British Columbia, Canada\goodbreak
\and
Department of Physics and Astronomy, Dana and David Dornsife College of Letter, Arts and Sciences, University of Southern California, Los Angeles, CA 90089, U.S.A.\goodbreak
\and
Department of Physics and Astronomy, University College London, London WC1E 6BT, U.K.\goodbreak
\and
Department of Physics, Florida State University, Keen Physics Building, 77 Chieftan Way, Tallahassee, Florida, U.S.A.\goodbreak
\and
Department of Physics, Gustaf H\"{a}llstr\"{o}min katu 2a, University of Helsinki, Helsinki, Finland\goodbreak
\and
Department of Physics, Princeton University, Princeton, New Jersey, U.S.A.\goodbreak
\and
Department of Physics, University of California, Santa Barbara, California, U.S.A.\goodbreak
\and
Department of Physics, University of Illinois at Urbana-Champaign, 1110 West Green Street, Urbana, Illinois, U.S.A.\goodbreak
\and
Dipartimento di Fisica e Astronomia G. Galilei, Universit\`{a} degli Studi di Padova, via Marzolo 8, 35131 Padova, Italy\goodbreak
\and
Dipartimento di Fisica e Scienze della Terra, Universit\`{a} di Ferrara, Via Saragat 1, 44122 Ferrara, Italy\goodbreak
\and
Dipartimento di Fisica, Universit\`{a} La Sapienza, P. le A. Moro 2, Roma, Italy\goodbreak
\and
Dipartimento di Fisica, Universit\`{a} degli Studi di Milano, Via Celoria, 16, Milano, Italy\goodbreak
\and
Dipartimento di Fisica, Universit\`{a} degli Studi di Trieste, via A. Valerio 2, Trieste, Italy\goodbreak
\and
Dipartimento di Matematica, Universit\`{a} di Roma Tor Vergata, Via della Ricerca Scientifica, 1, Roma, Italy\goodbreak
\and
Discovery Center, Niels Bohr Institute, Blegdamsvej 17, Copenhagen, Denmark\goodbreak
\and
Discovery Center, Niels Bohr Institute, Copenhagen University, Blegdamsvej 17, Copenhagen, Denmark\goodbreak
\and
European Southern Observatory, ESO Vitacura, Alonso de Cordova 3107, Vitacura, Casilla 19001, Santiago, Chile\goodbreak
\and
European Space Agency, ESAC, Planck Science Office, Camino bajo del Castillo, s/n, Urbanizaci\'{o}n Villafranca del Castillo, Villanueva de la Ca\~{n}ada, Madrid, Spain\goodbreak
\and
European Space Agency, ESTEC, Keplerlaan 1, 2201 AZ Noordwijk, The Netherlands\goodbreak
\and
Gran Sasso Science Institute, INFN, viale F. Crispi 7, 67100 L'Aquila, Italy\goodbreak
\and
HGSFP and University of Heidelberg, Theoretical Physics Department, Philosophenweg 16, 69120, Heidelberg, Germany\goodbreak
\and
Helsinki Institute of Physics, Gustaf H\"{a}llstr\"{o}min katu 2, University of Helsinki, Helsinki, Finland\goodbreak
\and
INAF - Osservatorio Astronomico di Padova, Vicolo dell'Osservatorio 5, Padova, Italy\goodbreak
\and
INAF - Osservatorio Astronomico di Roma, via di Frascati 33, Monte Porzio Catone, Italy\goodbreak
\and
INAF - Osservatorio Astronomico di Trieste, Via G.B. Tiepolo 11, Trieste, Italy\goodbreak
\and
INAF/IASF Bologna, Via Gobetti 101, Bologna, Italy\goodbreak
\and
INAF/IASF Milano, Via E. Bassini 15, Milano, Italy\goodbreak
\and
INFN, Sezione di Bologna, viale Berti Pichat 6/2, 40127 Bologna, Italy\goodbreak
\and
INFN, Sezione di Ferrara, Via Saragat 1, 44122 Ferrara, Italy\goodbreak
\and
INFN, Sezione di Roma 1, Universit\`{a} di Roma Sapienza, Piazzale Aldo Moro 2, 00185, Roma, Italy\goodbreak
\and
INFN, Sezione di Roma 2, Universit\`{a} di Roma Tor Vergata, Via della Ricerca Scientifica, 1, Roma, Italy\goodbreak
\and
INFN/National Institute for Nuclear Physics, Via Valerio 2, I-34127 Trieste, Italy\goodbreak
\and
IPAG: Institut de Plan\'{e}tologie et d'Astrophysique de Grenoble, Universit\'{e} Grenoble Alpes, IPAG, F-38000 Grenoble, France, CNRS, IPAG, F-38000 Grenoble, France\goodbreak
\and
IUCAA, Post Bag 4, Ganeshkhind, Pune University Campus, Pune 411 007, India\goodbreak
\and
Imperial College London, Astrophysics group, Blackett Laboratory, Prince Consort Road, London, SW7 2AZ, U.K.\goodbreak
\and
Infrared Processing and Analysis Center, California Institute of Technology, Pasadena, CA 91125, U.S.A.\goodbreak
\and
Institut N\'{e}el, CNRS, Universit\'{e} Joseph Fourier Grenoble I, 25 rue des Martyrs, Grenoble, France\goodbreak
\and
Institut Universitaire de France, 103, bd Saint-Michel, 75005, Paris, France\goodbreak
\and
Institut d'Astrophysique Spatiale, CNRS, Univ. Paris-Sud, Universit\'{e} Paris-Saclay, B\^{a}t. 121, 91405 Orsay cedex, France\goodbreak
\and
Institut d'Astrophysique de Paris, CNRS (UMR7095), 98 bis Boulevard Arago, F-75014, Paris, France\goodbreak
\and
Institut f\"ur Theoretische Teilchenphysik und Kosmologie, RWTH Aachen University, D-52056 Aachen, Germany\goodbreak
\and
Institute for Space Sciences, Bucharest-Magurale, Romania\goodbreak
\and
Institute of Astronomy, University of Cambridge, Madingley Road, Cambridge CB3 0HA, U.K.\goodbreak
\and
Institute of Theoretical Astrophysics, University of Oslo, Blindern, Oslo, Norway\goodbreak
\and
Instituto de Astrof\'{\i}sica de Canarias, C/V\'{\i}a L\'{a}ctea s/n, La Laguna, Tenerife, Spain\goodbreak
\and
Instituto de F\'{\i}sica de Cantabria (CSIC-Universidad de Cantabria), Avda. de los Castros s/n, Santander, Spain\goodbreak
\and
Istituto Nazionale di Fisica Nucleare, Sezione di Padova, via Marzolo 8, I-35131 Padova, Italy\goodbreak
\and
Jet Propulsion Laboratory, California Institute of Technology, 4800 Oak Grove Drive, Pasadena, California, U.S.A.\goodbreak
\and
Jodrell Bank Centre for Astrophysics, Alan Turing Building, School of Physics and Astronomy, The University of Manchester, Oxford Road, Manchester, M13 9PL, U.K.\goodbreak
\and
Kavli Institute for Cosmological Physics, University of Chicago, Chicago, IL 60637, USA\goodbreak
\and
Kavli Institute for Cosmology Cambridge, Madingley Road, Cambridge, CB3 0HA, U.K.\goodbreak
\and
Kazan Federal University, 18 Kremlyovskaya St., Kazan, 420008, Russia\goodbreak
\and
LAL, Universit\'{e} Paris-Sud, CNRS/IN2P3, Orsay, France\goodbreak
\and
LERMA, CNRS, Observatoire de Paris, 61 Avenue de l'Observatoire, Paris, France\goodbreak
\and
Laboratoire AIM, IRFU/Service d'Astrophysique - CEA/DSM - CNRS - Universit\'{e} Paris Diderot, B\^{a}t. 709, CEA-Saclay, F-91191 Gif-sur-Yvette Cedex, France\goodbreak
\and
Laboratoire Traitement et Communication de l'Information, CNRS (UMR 5141) and T\'{e}l\'{e}com ParisTech, 46 rue Barrault F-75634 Paris Cedex 13, France\goodbreak
\and
Laboratoire de Physique Subatomique et Cosmologie, Universit\'{e} Grenoble-Alpes, CNRS/IN2P3, 53, rue des Martyrs, 38026 Grenoble Cedex, France\goodbreak
\and
Laboratoire de Physique Th\'{e}orique, Universit\'{e} Paris-Sud 11 \& CNRS, B\^{a}timent 210, 91405 Orsay, France\goodbreak
\and
Lawrence Berkeley National Laboratory, Berkeley, California, U.S.A.\goodbreak
\and
Lebedev Physical Institute of the Russian Academy of Sciences, Astro Space Centre, 84/32 Profsoyuznaya st., Moscow, GSP-7, 117997, Russia\goodbreak
\and
Max-Planck-Institut f\"{u}r Astrophysik, Karl-Schwarzschild-Str. 1, 85741 Garching, Germany\goodbreak
\and
McGill Physics, Ernest Rutherford Physics Building, McGill University, 3600 rue University, Montr\'{e}al, QC, H3A 2T8, Canada\goodbreak
\and
National University of Ireland, Department of Experimental Physics, Maynooth, Co. Kildare, Ireland\goodbreak
\and
Nicolaus Copernicus Astronomical Center, Bartycka 18, 00-716 Warsaw, Poland\goodbreak
\and
Niels Bohr Institute, Blegdamsvej 17, Copenhagen, Denmark\goodbreak
\and
Niels Bohr Institute, Copenhagen University, Blegdamsvej 17, Copenhagen, Denmark\goodbreak
\and
Nordita (Nordic Institute for Theoretical Physics), Roslagstullsbacken 23, SE-106 91 Stockholm, Sweden\goodbreak
\and
Optical Science Laboratory, University College London, Gower Street, London, U.K.\goodbreak
\and
SISSA, Astrophysics Sector, via Bonomea 265, 34136, Trieste, Italy\goodbreak
\and
SMARTEST Research Centre, Universit\`{a} degli Studi e-Campus, Via Isimbardi 10, Novedrate (CO), 22060, Italy\goodbreak
\and
School of Chemistry and Physics, University of KwaZulu-Natal, Westville Campus, Private Bag X54001, Durban, 4000, South Africa\goodbreak
\and
School of Physics and Astronomy, Cardiff University, Queens Buildings, The Parade, Cardiff, CF24 3AA, U.K.\goodbreak
\and
School of Physics and Astronomy, University of Nottingham, Nottingham NG7 2RD, U.K.\goodbreak
\and
Sorbonne Universit\'{e}-UPMC, UMR7095, Institut d'Astrophysique de Paris, 98 bis Boulevard Arago, F-75014, Paris, France\goodbreak
\and
Space Research Institute (IKI), Russian Academy of Sciences, Profsoyuznaya Str, 84/32, Moscow, 117997, Russia\goodbreak
\and
Space Sciences Laboratory, University of California, Berkeley, California, U.S.A.\goodbreak
\and
Special Astrophysical Observatory, Russian Academy of Sciences, Nizhnij Arkhyz, Zelenchukskiy region, Karachai-Cherkessian Republic, 369167, Russia\goodbreak
\and
Stanford University, Dept of Physics, Varian Physics Bldg, 382 Via Pueblo Mall, Stanford, California, U.S.A.\goodbreak
\and
Sub-Department of Astrophysics, University of Oxford, Keble Road, Oxford OX1 3RH, U.K.\goodbreak
\and
The Oskar Klein Centre for Cosmoparticle Physics, Department of Physics,Stockholm University, AlbaNova, SE-106 91 Stockholm, Sweden\goodbreak
\and
Theory Division, PH-TH, CERN, CH-1211, Geneva 23, Switzerland\goodbreak
\and
UPMC Univ Paris 06, UMR7095, 98 bis Boulevard Arago, F-75014, Paris, France\goodbreak
\and
Universit\"{a}t Heidelberg, Institut f\"{u}r Theoretische Astrophysik, Philosophenweg 12, 69120 Heidelberg, Germany\goodbreak
\and
Universit\'{e} de Toulouse, UPS-OMP, IRAP, F-31028 Toulouse cedex 4, France\goodbreak
% ...
}

\modulolinenumbers[5]

\authorrunning{Planck Collaboration}  
\titlerunning{The ISW effect with \Planck}

\abstract{This paper presents a study of the integrated Sachs-Wolfe (ISW) effect from the \Planck\ 2015 temperature and polarization data release.
This secondary cosmic microwave background (CMB) anisotropy caused by the large-scale time-evolving gravitational potential is probed from different perspectives.
The CMB is cross-correlated with different large-scale structure (LSS) tracers: radio sources from the NVSS catalogue; galaxies from the optical SDSS and the infrared WISE surveys; and the \Planck\ 2015 convergence lensing map. The joint cross-correlation of the CMB with the tracers yields a detection at $4\,\sigma$ where most of the signal-to-noise is due to the \Planck\ lensing and the NVSS radio catalogue.
In fact, the ISW effect is detected from the \Planck\ data only at $\approx 3\,\sigma$ (through the ISW-lensing bispectrum), which is similar to the detection level achieved by combining the cross-correlation signal coming from all the galaxy catalogues mentioned above. 
We study the ability of the ISW effect to place constraints on the dark-energy parameters; in particular, we show that $\Omega_\Lambda$ is detected at more than $3\,\sigma$. 
This cross-correlation analysis is performed only with the  \Planck\ temperature data, since the polarization scales available in the 2015 release do not permit significant improvement of the CMB-LSS cross-correlation detectability. Nevertheless, the \Planck\ polarization data are used to study the anomalously large ISW signal previously reported through the aperture photometry on stacked CMB features at the locations of known superclusters and supervoids, which is in conflict with $\Lambda$CDM expectations. We find that the current \Planck\ polarization data do not exclude that this signal could be caused by the ISW effect. In addition, the stacking of the \Planck\ lensing map on the  locations of superstructures exhibits a positive cross-correlation with these large-scale structures. Finally, we have improved our previous reconstruction of the ISW temperature fluctuations by combining the information encoded in all the previously mentioned LSS tracers. In particular, we construct a map of the ISW secondary anisotropies and the corresponding uncertainties map, obtained from simulations. We also explore the reconstruction of the ISW anisotropies caused by the large-scale structure traced by the 2MASS Photometric redshift survey (2MPZ) by directly inverting the density field into the gravitational potential field.}


\maketitle


\section{Introduction}
\label{sec:intro}
This paper, one of a set associated with the 2015 release of data from the \Planck\footnote{\Planck\ (\url{http://www.esa.int/Planck}) is a project of the European Space Agency  (ESA) with instruments provided by two scientific consortia funded by ESA member states and led by Principal Investigators from France and Italy, telescope reflectors provided through a collaboration between ESA and a scientific consortium led and funded by Denmark, and additional contributions from NASA (USA).} mission, describes the detection and characterization of the integrated Sachs-Wolfe (ISW) effect using external (galaxy-survey catalogues) and internal (\Planck\ lensing map) large-scale tracers.
The 2015 \Planck\ data release \alltwentyfifteenresultspapers offers polarization information on the cosmic microwave background (CMB) for angular scales smaller than $5^\circ$. Whenever possible, this polarization information is used to improve our characterization of the ISW signal.

The ISW effect~\citep{Sachs1967, Rees1968, Martinez1990b, Sugiyama1995} is a secondary anisotropy in the CMB, which is caused by gravitational interaction of CMB photons with the growing cosmic large-scale structure (LSS):
\begin{linenomath*}
\begin{equation}
\Theta = \frac{\Delta T}{T_\mathrm{CMB}} = -\frac{2}{c^3}\int_{0}^{\chi_\mathrm{CMB}}\dd\chi\frac{\partial\Phi}{\partial\chi}.
\end{equation}
\end{linenomath*}
Here, the fractional temperature perturbation $\Theta$ is given as a line of sight integral over the time-evolving potentials $\Phi$ in the LSS. The integral is expressed in terms of comoving distance $\chi$, which is related to the scale factor $a$ according to $\dd a/\dd\chi = a^2H(a)/c$, with the Hubble function $H(a)$ and the speed of light $c$. The integration is extended to the surface of last scattering $\chi_\mathrm{CMB}\simeq10~\mathrm{Gpc}/h$ corresponding to a redshift of $z\simeq1100$ in a $\Lambda$CDM cosmology.

The ISW effect measures the rate of growth of gravitational potentials relative to universes with a critical density of matter through frequency shifts in the photon distribution. It is measured by cross-correlating with a tracer of the LSS, such as a galaxy catalogue or a reconstructed weak gravitational lensing map, in order to distinguish it from primary CMB anisotropies; this is because gravitational interaction conserves the Planckian shape of the photon spectrum. The ISW effect is generated at late times when the growth of structure is influenced by a cosmological constant, dark energy \citep{Crittenden1996}, modified gravity \citep{Hu2002a}, or spatial curvature \citep{Kamionkowski1996}.

The most direct way of detecting the ISW effect is the determination of the cross-correlation or the cross-angular power spectrum between the CMB temperature and the density of tracer objects such as galaxies. In this way, the first detection was reported by \cite{Boughn2004} which was subsequently  refined by many groups on the basis of \wmap\ data, yielding values for the detection significance in excess of $4\,\sigma$ \citep[e.g.,][]{Fosalba2003, Nolta2004, Corasaniti2005, Padmanabhan2005, Vielva2006, Giannantonio2006b, Cabre2007, Rassat2007, McEwen2007, Giannantonio2012}. Corresponding constraints on cosmological parameters were derived for standard models with a cosmological constant and for dark energy models \citep[e.g.,][]{Pietrobon2006a, McEwen2007, Vielva2006, Giannantonio2008a, Ho2008, Xia2009b}, as well as for models with modified gravity \citep[e.g.,][]{Zhao2010}. A Bayesian ISW detection method, which estimates the ISW amplitude conditionally to the observed LSS, can be expected to provide 10\,\% better signal-to-noise ratio compared to a direct CMB-LSS cross-correlation study~\citep{Frommert2008}, as used traditionally and in this psper beacuse of its lower computational complexity.

In fact, using the ISW signal alone (but fixing the remaining cosmological parameters), the dark energy density parameter $\Omega_\Lambda$ was estimated to be $\approx 0.75$ with an error of about 20\,\%~\citep[e.g.,][]{Nolta2004,Vielva2006,Giannantonio2006b}, the dark energy equation of state parameter was found to be close to $w=-1$~\citep[e.g.,][]{Vielva2006,Giannantonio2006b,Ho2008}, and tests on spatial flatness yielded upper limits of a few percent for $\Omega_K$~\citep[e.g.,][]{Ho2008, Li2010}, thus confirming the concordance cosmological model.

The presence of systematics at large angular scales in LSS surveys and their possible impact on ISW studies was first emphasized in \citet{Hernandez2010} and formally addressed in \citet{Giannantonio2012} and \citet{Hernandez2014}.
The ISW analysis with the \Planck\ data release in 2013 \citep{planck2013-p14} was consistent with \wmap\ results using the NVSS radio catalogue and catalogues of tracer objects derived with optical \sdss\ data, while lowering the claimed detection levels to smaller numbers (from $>4\,\sigma$ down to around $2.5\,\sigma$). In addition, a non-zero correlation between the reconstructed CMB-lensing map as an LSS tracer and the microwave background was reported for the first time, using the non-vanishing bispectrum of the CMB anisotropies on the relevant scales. The strength of this correlation was measured to be $3\,\sigma$, and provides further evidence for a late-time accelerated expansion of the Universe, as theoretically shown by \cite{Hu2002b} and \citet{Okamoto2003}.

An alternative method for detecting the ISW effect is the stacking of CMB fields at the positions of known superstructures; if the ISW effect is associated with regions of large density, it should be possible to reduce the noise due to primary, uncorrelated CMB anisotropies by superposition and to reach a reduction inversely proportional to the square root of the number of stacked fields. Detections using this method range between $2\,\sigma$ and $4\,\sigma$, based on \wmap\ data 
\citep[e.g.,][]{Granett2008a, Papai2010a} and on \Planck\ data \citep{planck2013-p14}.

A third application of the ISW effect is the reconstruction of a large-scale map of projected gravitational potentials \citep{Barreiro2008}. Using the correlation between temperature anisotropies and a map of the tracer density, it is possible to estimate these secondary temperature anisotropies directly. 

The purpose of this paper is the measurement of the ISW effect with the full \Planck\ 2015 data set and to establish the corresponding constraints on cosmological parameters. In principle, including polarization data allows us to reduce the error bars in estimating angular cross-power spectra~\citep{Frommert2009}, and it provides a separation of the temperature anisotropies into those correlated and uncorrelated with polarization, through which the secondary nature of the ISW effect can be better investigated. Furthermore, the reconstruction of the weak lensing potential is improved, and a better template for cross-correlation is provided. 
However, as mentioned above, the current polarization information provided in the CMB maps of the 2015 \Planck\ data release is limited to angular scales smaller than $5^\circ$ (more precisely, only multipoles $\ell \geq 20$ are kept, with a cosine transition in the range $20 < \ell <40$). This limits the amount of information on the ISW effect that can be obtained from the polarization data, since this secondary anisotropy is mostly significant on the largest angular scales. Therefore, in this paper, polarization is not used for the CMB cross-correlation with LSS tracers, although it is considered in the analysis of the CMB anisotropies stacked on the positions of known superstructures. 

The paper is organized as follows. In Sect.~\ref{sec:data} we present the data used in this work (both for the CMB and the LSS tracers). The cross-correlations of these tracers are investigated in Sect.~\ref{sec:xcorr}. In Sect.~\ref{sec:stack} we present the results of the stacking analysis using temperature and polarization data. The recovery of the ISW anisotropy map is described in Sect.~\ref{sec:recov}. Finally, we discuss our main results and their cosmological implications in Sect.~\ref{sec:discussion}.

\section{Data sets}

\label{sec:data}
In this section we describe the data sets and the simulations used throughout the paper. 
In Sect.~\ref{sec:data_cmb} we describe the CMB related data (temperature and
polarization anisotropies), whereas the LSS data sets are discussed in
Sect.~\ref{sec:data_lss}, including galaxy, cluster and void catalogues from redshift and photometric
surveys, and the \Planck\ lensing map. In Sect.~\ref{sec:data_sims} we explain the specific simulations performed to study the CMB-LSS cross-correlation.

\subsection{CMB data}
\label{sec:data_cmb}
%

\begin{figure*}[ht]
\centering
\includegraphics[width=0.495\textwidth]{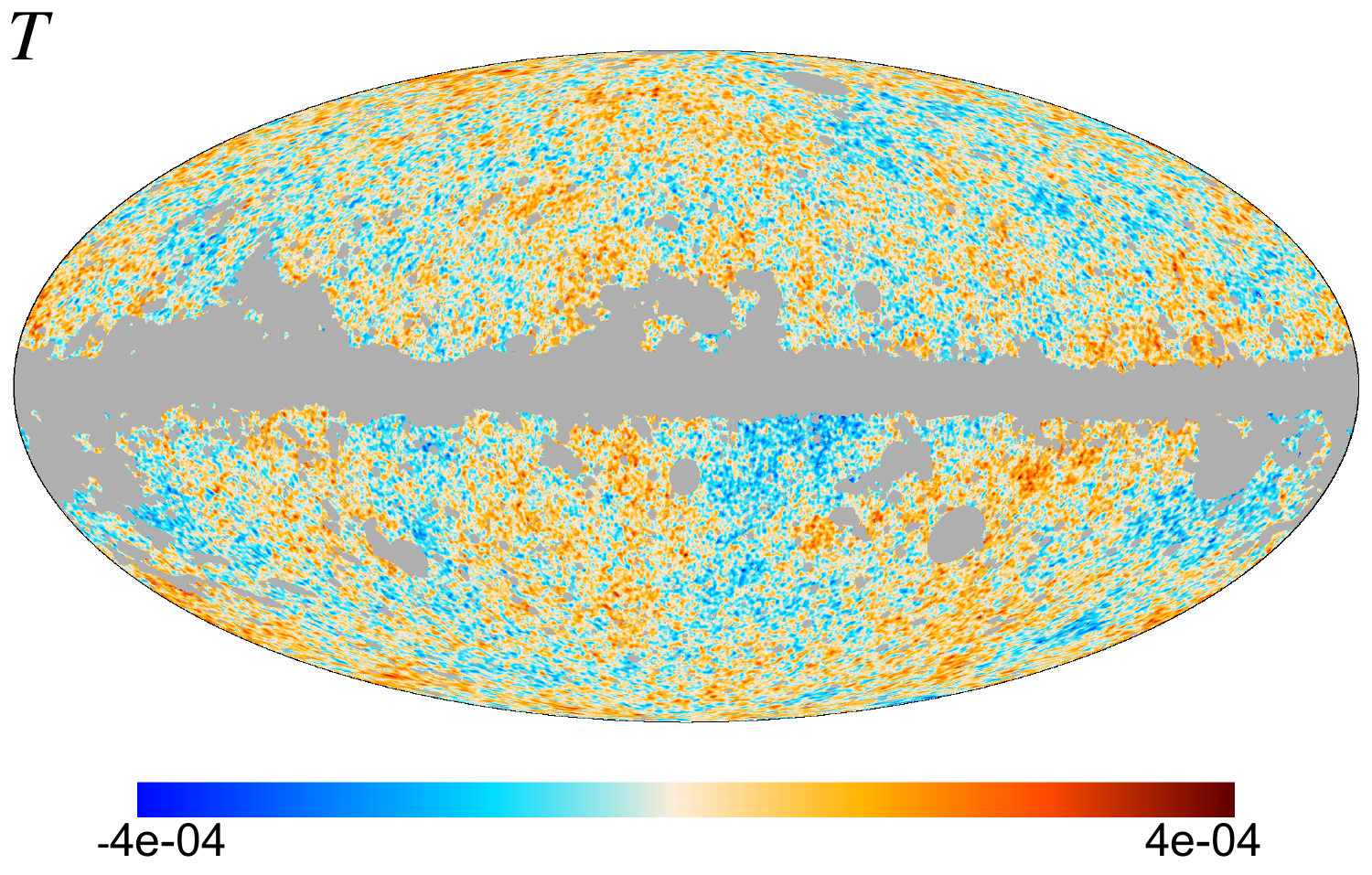}
\includegraphics[width=0.495\textwidth]{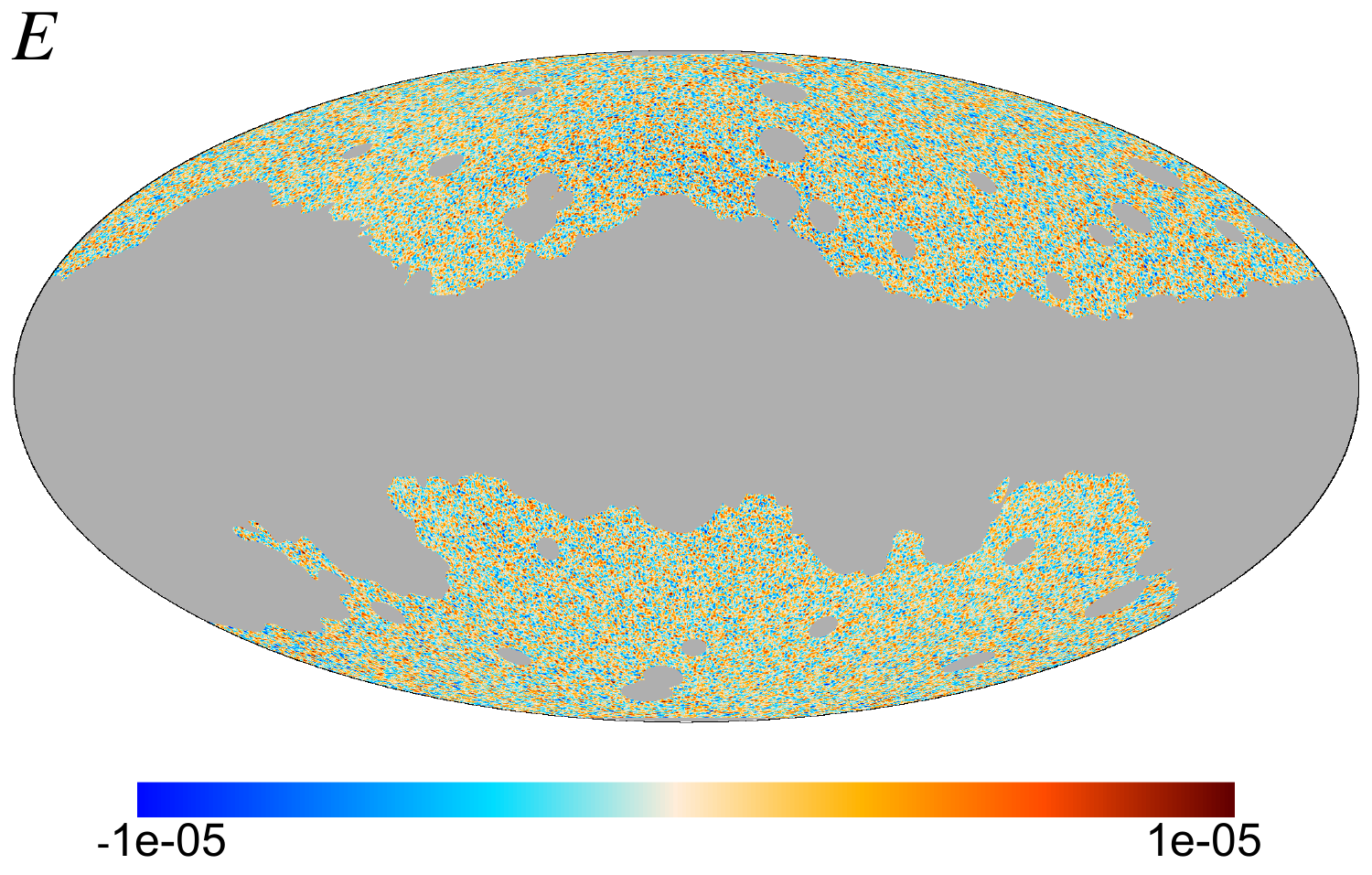}
\includegraphics[width=0.495\textwidth]{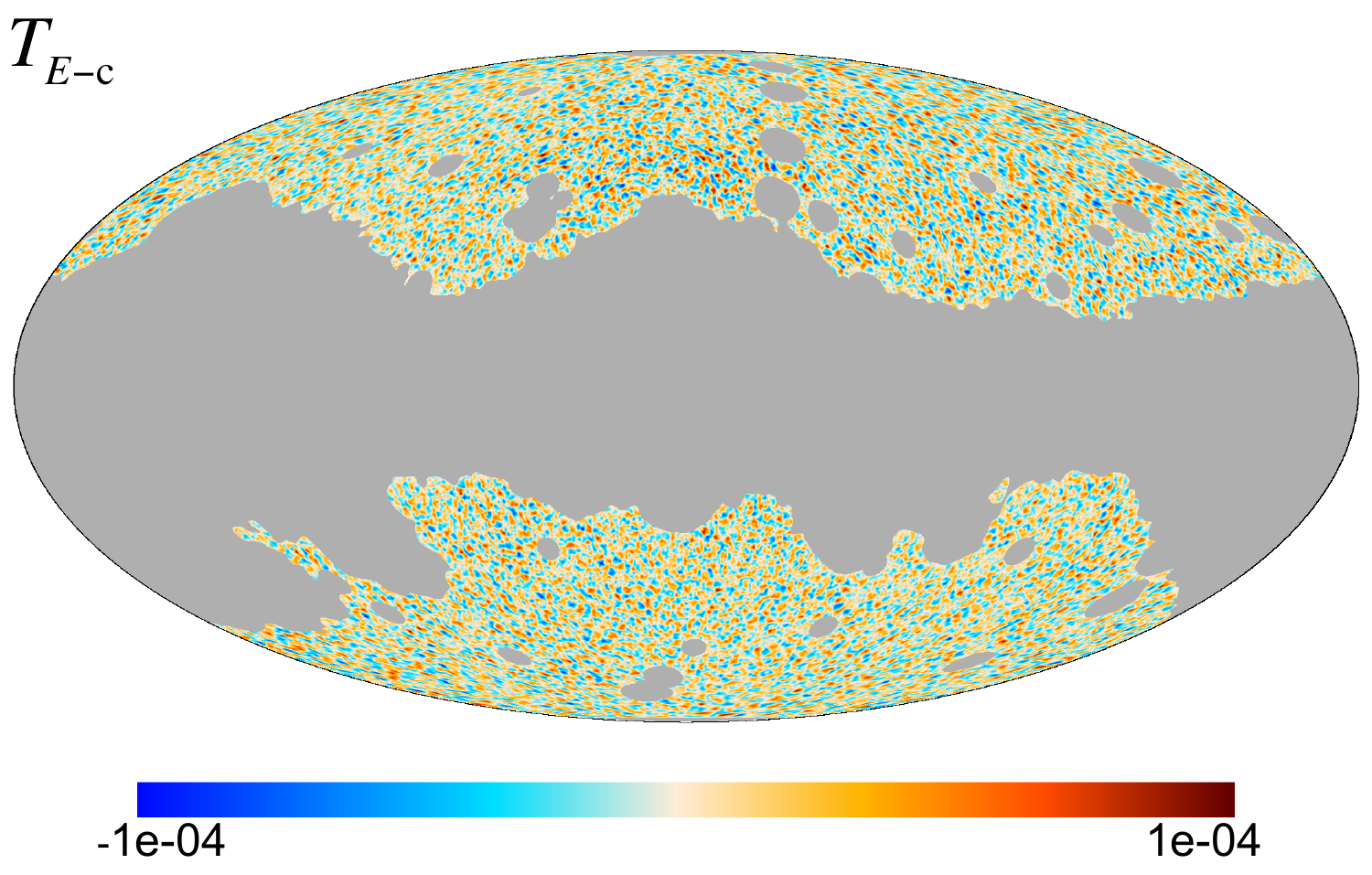}
\includegraphics[width=0.495\textwidth]{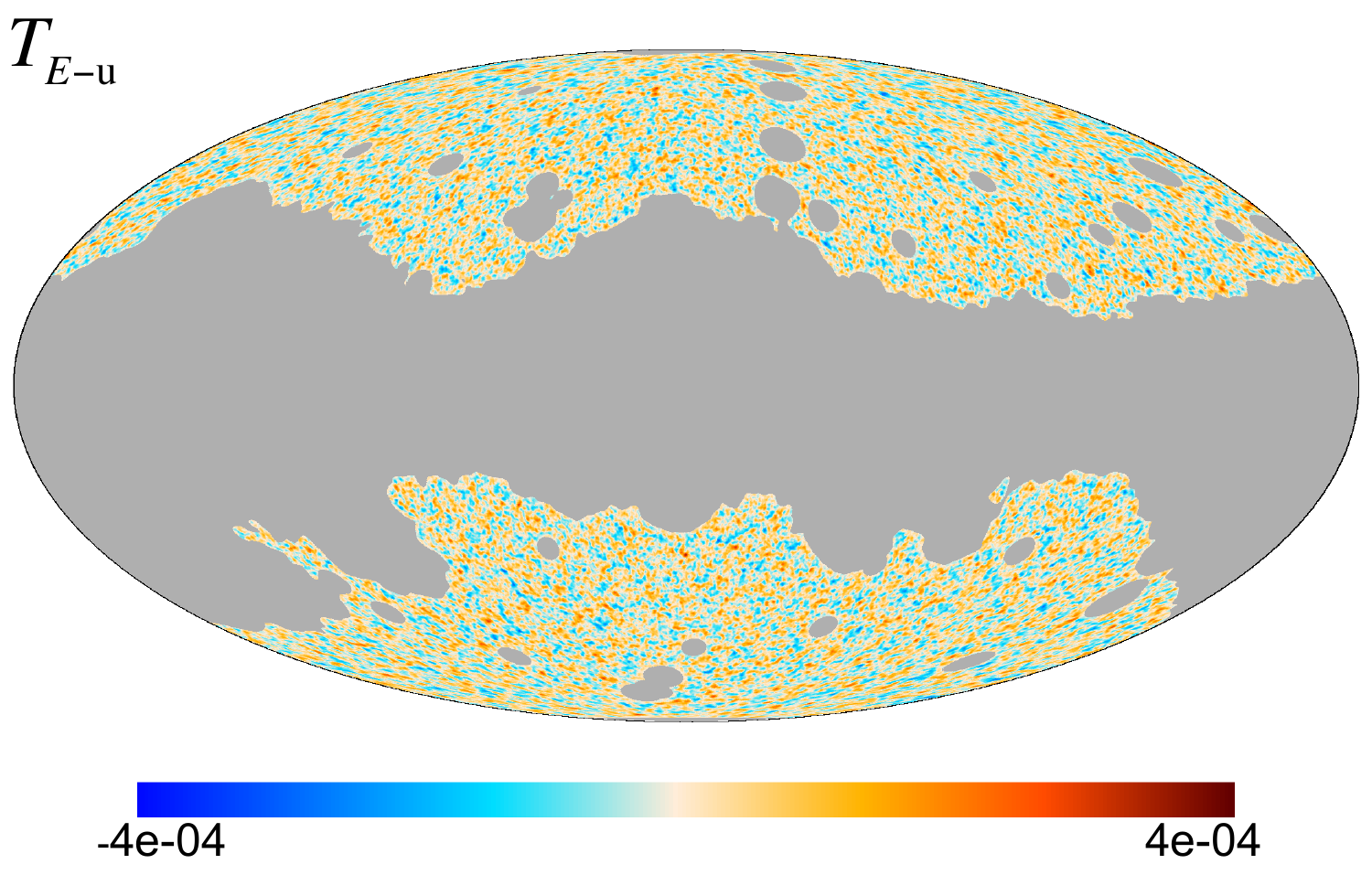}
\caption{\Planck\ CMB temperature and polarization anisotropies as provided by the \sevem\ component separation method at a resolution of \nside=512. From left to right and from top to bottom, the panels show the maps of temperature,the \ep -mode, and the \ep -correlated (\tce) and \ep -uncorrelated (\tue) temperature maps. The units are Kelvin.} 
\label{fig:cmb_data}
\end{figure*}

There are four major \Planck\ foreground-cleaned CMB temperature and polarization maps, namely, the \cruler, \nilc, \sevem, and \smica\ maps, named after their respectively generating component separation methods~\citep[see][for details]{planck2014-a11}. All these maps are used here, in comparison, in order to test the robustness of our results. Together with the common \qp\ and \up\ Stokes parameter polarization maps, the \Planck\ 2015 data release also provides \ep-mode maps based on the four component separation methods. In addition, the \sevem\ method also provides foreground-cleaned CMB maps at specific frequencies, in temperature at 100, 143, and 217\,GHz, and in polarization at 70, 100, and 143\,GHz.

The \Planck\ 2015 CMB maps are provided at different resolutions~\citep{planck2014-a11}. In this paper we consider two different resolutions, depending on the application. First, maps with a \healpix~\citep{gorski2005} resolution parameter \nside = 64 (FWHM = $160$ arcmin) are adopted for studying the CMB-LSS cross-correlation (Sect.~\ref{sec:xcorr}) and for recovering the ISW anisotropies (Sect.~\ref{sec:recov}). Second, \nside = 512 
(FWHM = $20$ arcmin) maps are used to study the ISW effect through the stacking of CMB maps on the positions of known superstructures (Sect.~\ref{sec:stack}). 
Each resolution has an associated set of masks, one for temperature (called UT78, $f_{\rm sky}$ = 74\,\% at \nside\,= 512), another for \qp\ and \up\ Stokes parameters (called UPB77, $f_{\rm sky}$ = 76\,\% at \nside\,= 512), and a final one for the \ep -mode ($f_{\rm sky}$ = 45\,\% at \nside\,= 512). The $f_{\rm sky}$ parameter indicates the fraction of the sky that is retained after masking.

In addition, there are 1000 simulations associated with each delivered map, which allow us to characterize the instrumental properties of \Planck\ CMB maps. In the context of this work, these simulations are used for the stacking analyses in Sect.~\ref{sec:stack}. The other ISW studies require specific coherent simulations between the CMB and the LSS tracers. These simulations are described in Sect.~\ref{sec:data_sims}.

As mentioned in Sect.~\ref{sec:intro}, the polarized CMB maps of the 2015 release have been high-pass filtered~\citep[see][for details]{planck2014-a08,planck2014-a11}. In particular, all the multipoles with $\ell \leq 20$ were removed, and a cosine transition between $20 < \ell <40$ was imposed.
Obviously, this high-pass filtering very much limits the usefulness of the polarization information for the ISW analyses. More precisely, the expected 16\,\% increase of the ISW detection significance by exploiting polarization information in the CMB-LSS cross-correlation~\citep{Frommert2009} depends,  mainly, on the filtered-out scales (up to $\approx 80\%$ for $\ell \lesssim 20$, and $\approx 90\%$ for $\ell \lesssim 40$). In addition, the approach to derive the \ep-correlated (\tce) and the \ep-uncorrelated (\tue) maps (see below), are based on an $E$-mode map, with a corresponding mask that, as mentioned above, covers significantly less sky (45\%) than the temperature one
(74\%). Therefore, in practice, there is no real gain in the signal-to-noise level.
Nevertheless, some of the information kept at smaller scales may still be useful for particular analyses such as the stacking of the CMB anisotropies on the positions of known superstructures.
First, because these structures are within the part of the sky covered by the \tce\ and \tue\ maps and, second, because the multipole range that mainly contributes to the angular scales of the stacked profiles corresponds to smaller scales than those for the CMB-LSS cross-correlation. 
For that reason, the polarization information is not used in the ISW study through the correlation of the CMB and the LSS tracers (Sect.~\ref{sec:xcorr}), but it was considered in the stacking analyses (Sect.~\ref{sec:stack}). A final study of the ISW effect using full polarization information is expected to be done with the next \Planck\ data release.
 
The primary CMB temperature anisotropies act effectively as a noise source for the measurement of secondary CMB anisotropies by increasing its cosmic variance. This is true for the ISW effect, which does not produce a notable \ep -mode polarization. Hence, polarization data permit us to identify the part of the primary temperature anisotropies that is correlated with the \ep -mode polarization, and to remove it from the maps. The resulting CMB temperature map, partly cleaned form primary anisotropies, provides up to a 16\,\% better signal-to-noise ratio for secondary fluctuations~\citep{Frommert2009}. To this end, we separate the temperature map in two components: an \ep-correlated (\tce) and an \ep-uncorrelated (\tue) part.
Following the approach of~\cite{Frommert2009}, we have produced these maps from the delivered CMB inputs described above. An estimation of the \ep-correlated temperature anisotropies (\tce) is given, in terms of its spherical harmonic coefficients $a_{\ell m}^{T_{E-\mathrm{c}}}$, by
\begin{linenomath*}
\begin{equation}
a_{\ell m}^{\tcem} = a_{\ell m}^E w_\ell,
\end{equation}
\end{linenomath*}
where the filter $w_\ell$ is defined by the $TE$ and the $EE$ angular power spectra:
\begin{linenomath*}
\begin{equation}
w_\ell = \frac{C_\ell^{TE} + F_\ell^{TE}}{C_\ell^{EE} + F_\ell^{EE} + N_\ell^{EE}},
\label{eq:filter}
\end{equation} 
\end{linenomath*}
with $C_\ell$, $F_\ell$, and $N_\ell$ representing the angular power spectra of the CMB, residual foregrounds, and noise, respectively. Hence, the \tce\ map is given by:
\begin{linenomath*}
\begin{equation}
\tcem \left(\vn\right) = \sum_{\ell = 0}^{\ell_\mathrm{max}} \sum_{m = -\ell}^{\ell} a_{\ell m}^{\tcem} \textrm{Y}_{\ell m}\left(\vn\right),
\end{equation}
\end{linenomath*}
with $\textrm{Y}_{\ell m}\left(\vn\right)$ the spherical harmonic functions. The \tue\ map is build by subtraction: $\tuem\left(\vn\right) = T\left(\vn\right) - \tcem\left(\vn\right)$. The above procedure is performed by applying an apodized version of the corresponding masks.
In Fig.~\ref{fig:cmb_data} we show the \tp, \tce, and \tue\ maps for \sevem.
In practice, the determination of the filter $w_{\ell}$ is not straightforward; although the CMB and noise contributions can be obtained directly from the \Planck\ best-fit cosmological model~\citep{planck2014-a15} and the FFP8 simulations~\citep{planck2014-a14,planck2014-a11}, information about the residual foregrounds ($F_{\ell}$) present in the CMB temperature and polarization is also needed. We verified that the expected CMB and noise power spectra account well for the observed $TE$ and the $EE$ angular power spectra at $\ell < 200$. Although the foreground spectra are not fully known, their impact is minor on these scales due to the large mask imposed on the \ep-mode map and the high-pass filtering applied to the polarization data. However, at smaller angular scales some foreground residuals exist.

\begin{figure}
\centering
\includegraphics[width=0.495\textwidth]{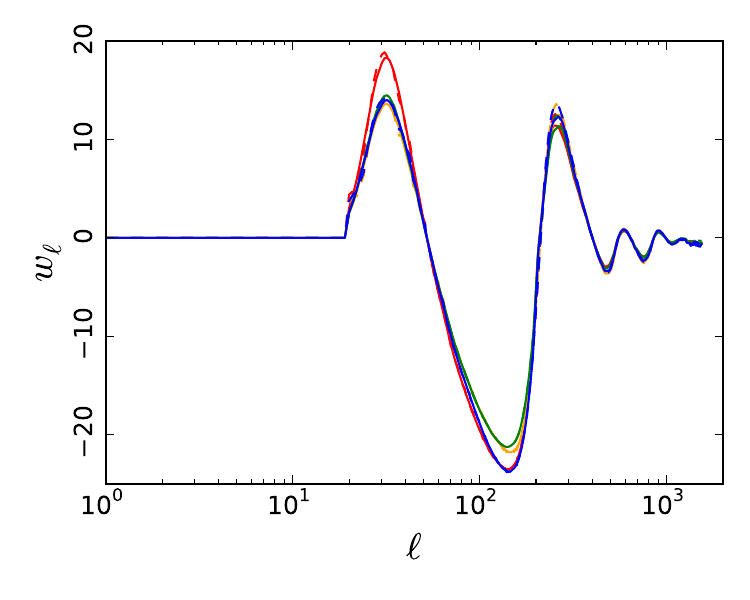}
\caption{Filter used to construct the \tce correlated maps for all component separation methods: \cruler\ in red; \nilc\ in orange; \sevem\ in green; and \smica\ in blue. The solid lines are obtained directly from the data, whereas the dashed ones represent the theoretical shape of the filters, only considering the instrumental noise characteristics of the data and the fiducial \Planck\ angular power spectra.}
\label{fig:filter}
\end{figure}

An alternative way to construct such a filter to reduce primary anisotropies is to extract the relevant correlation functions directly from the data. In particular, we have constructed filters $w_{\ell}$ using a smooth fit of the filter constructed as the ratio of the  $TE$ and the $EE$ angular power spectra of the different CMB component separation maps. The procedure followed to build the filter distinguishes between high- and low-$\ell$ regimes. For small scales ($\ell>200$), we compute the ratio of $C_{\ell}^{TE}$ and $C_{\ell}^{EE}$ obtained from the data using an apodized mask, which is afterwards smoothed following the Savitzky-Golay procedure \citep{Savitzky1964}. In the low-$\ell$ regime ($\ell < 200$) the filter is constructed using the average value obtained from 1000 simulations of CMB plus noise, using the same apodized mask. 
The resulting filters (solid lines) are shown in Fig.~\ref{fig:filter}; for comparison, the corresponding theoretical filters, computed only from the instrumental properties and the \Planck\ fiducial angular power spectra, are also plotted (dashed lines).

\subsection{LSS tracers}
\label{sec:data_lss}

\begin{figure}
\centering
\includegraphics[width=\hsize]{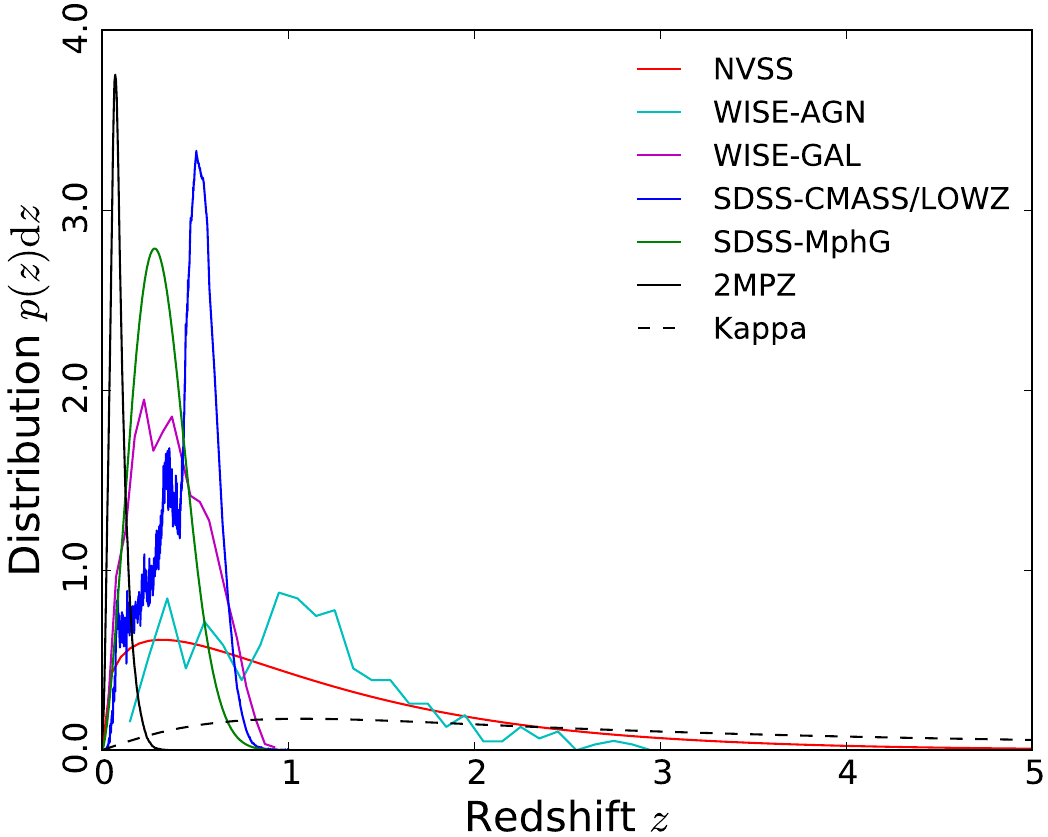}
\caption{Redshift distributions of the different surveys used as LSS tracers. To facilitate comparison, the distributions of the external tracers have been normalized to unity (and multiplied by a factor 10 for the \tmpz\ catalogues). For completeness, we also include the contribution of the gravitational potential to the lensing convergence map, as a function of redshift (without any additional normalization).}
\label{fig:surveys_dndz}
\end{figure}

As mentioned in Sect.~\ref{sec:intro}, tracers of the gravitational potential of the LSS are required to extract the secondary ISW anisotropies from the dominant primary CMB anisotropies. These tracers are used to perform the CMB-LSS cross-correlation, but also for studying the ISW effect through the stacking of the CMB anisotropies on the position of known superstructures (such as clusters or voids), and for producing a map of the ISW anisotropies.

We have included three additional galaxy catalogues with respect to to the ones used in~\cite{planck2013-p14}, which were the radio \nvss\ catalogue and the optical luminous galaxies (\lrg) as well as the main photometric galaxy sample (\mg) catalogues from the {\it Sloan Digital Sky Survey} (\sdss).
These additional catalogues consist of star forming galaxies (\wg), of active galactic nuclei (AGN; \wagn), both sets taken from the catalogue of extragalactic sources detected by the 
\textit{Wide-Field Infrared Survey Explorer}~\citep[\wise, see][]{Wright2010}, and of photometric redshifts (\tmpz) obtained from the \textit{Two Micron All Sky Survey} Extended Source Catalogue (2MASS-XSC), WISE and SuperCOSMOS data sets. This last catalogue is only used to build an estimation of the ISW anisotropies based on a reconstruction of the gravitational potential from the 3D distribution of the galaxies (see Appendix.~\ref{sec:appen}). The reason is that the expected CMB-LSS cross-correlation signal is very low to be used in this cross-correlation study but, however, the galaxy redshift estimation error is sufficiently low to attempt the gravitational potential reconstruction. Finally, we also cross-correlate the \Planck\ lensing map as a LSS tracer with the CMB. In particular, we use the lensing convergence map (\kap) obtained in~\cite{planck2014-a17}.

The redshift distributions of these catalogues are shown in Fig.~\ref{fig:surveys_dndz}. We note that lensing, \nvss, and \wagn\ offer the widest redshift coverage. Some basic properties of the galaxy catalogues used (\nvss, \wagn, \wg, \lrg, \mg, and \tmpz) are summarized in Table~\ref{tab:surveys}.

\begin{figure*}
\centering
\includegraphics[width=0.495\textwidth]{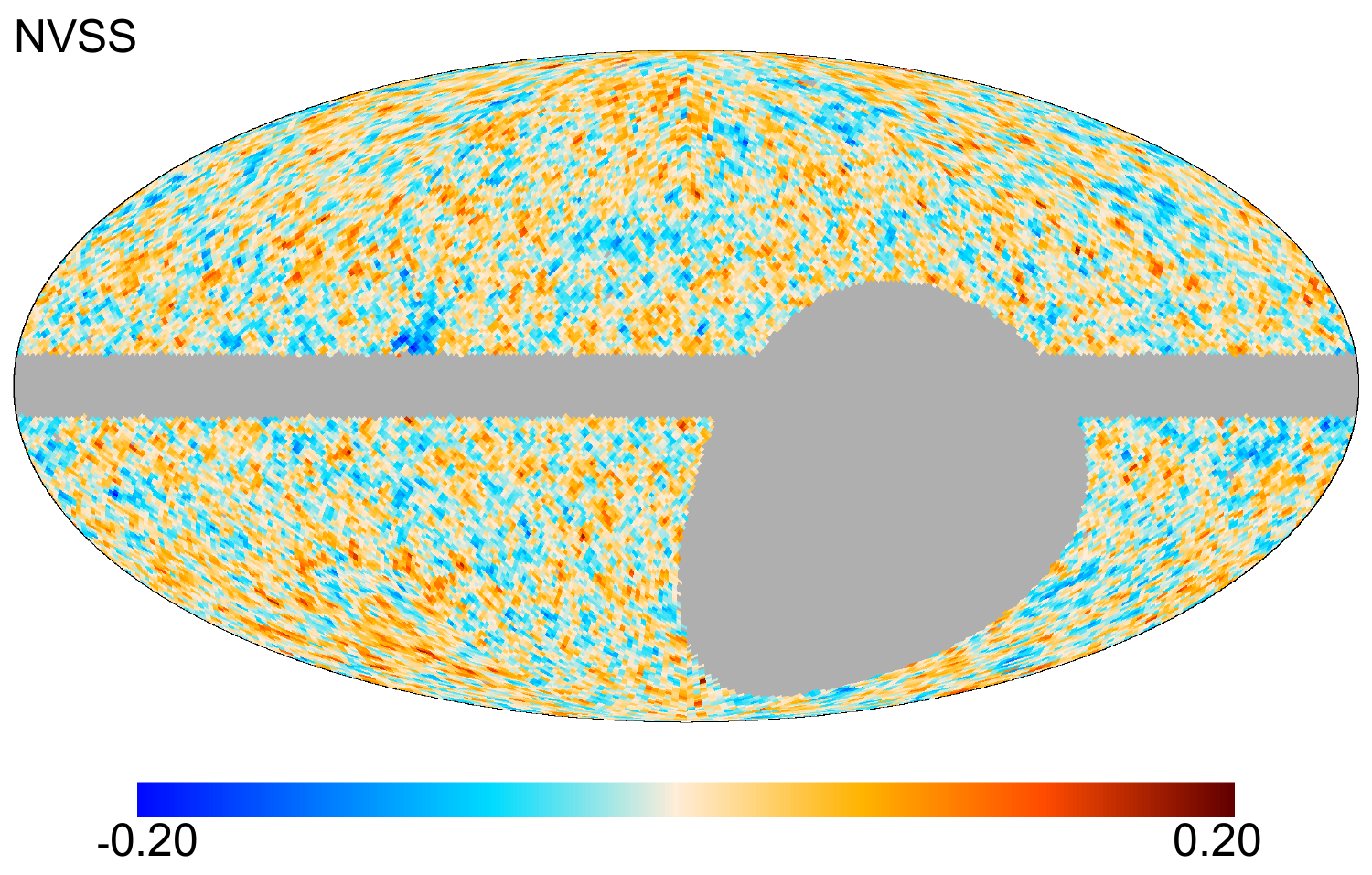}
\includegraphics[width=0.495\textwidth]{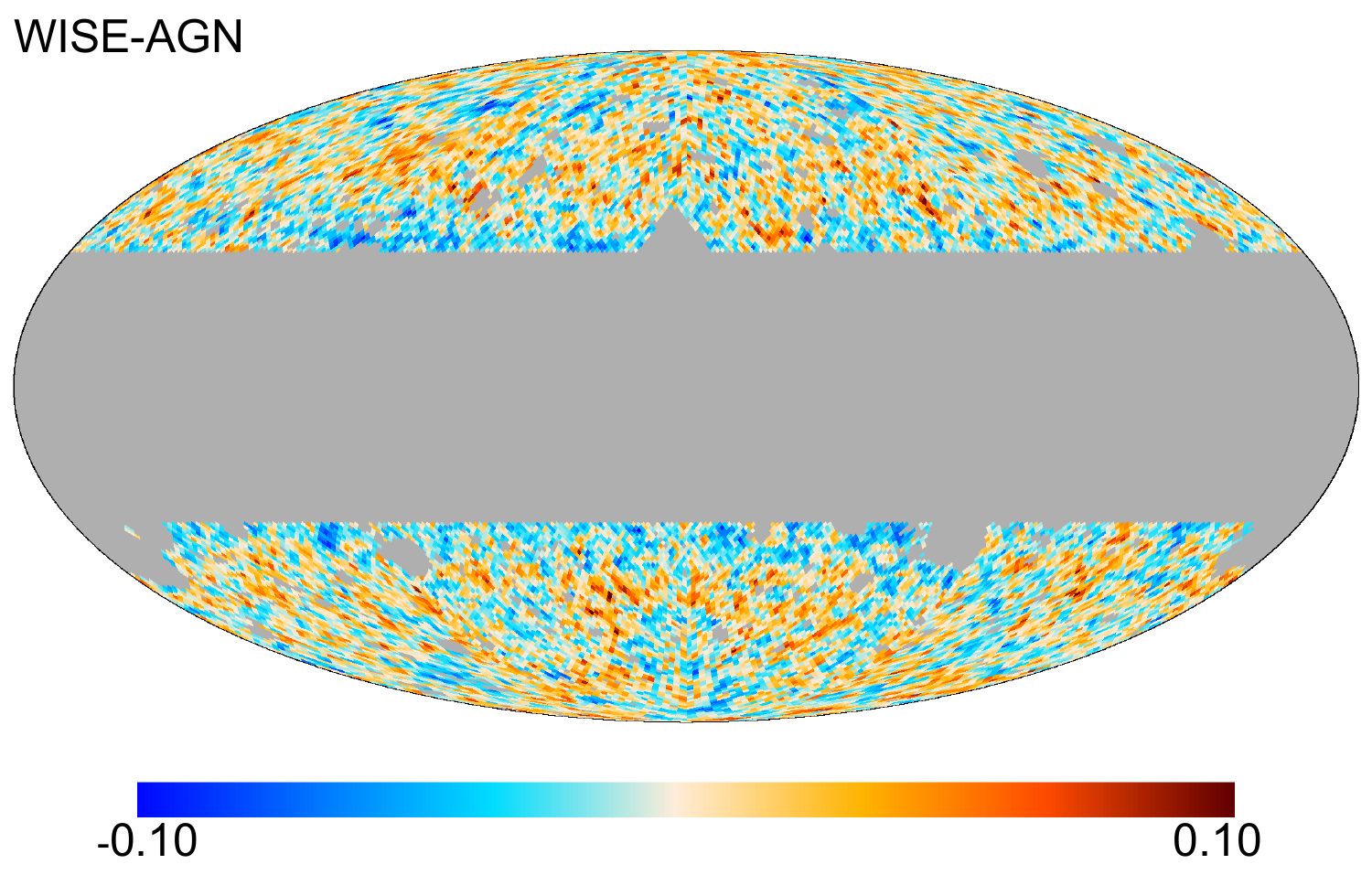}
\includegraphics[width=0.495\textwidth]{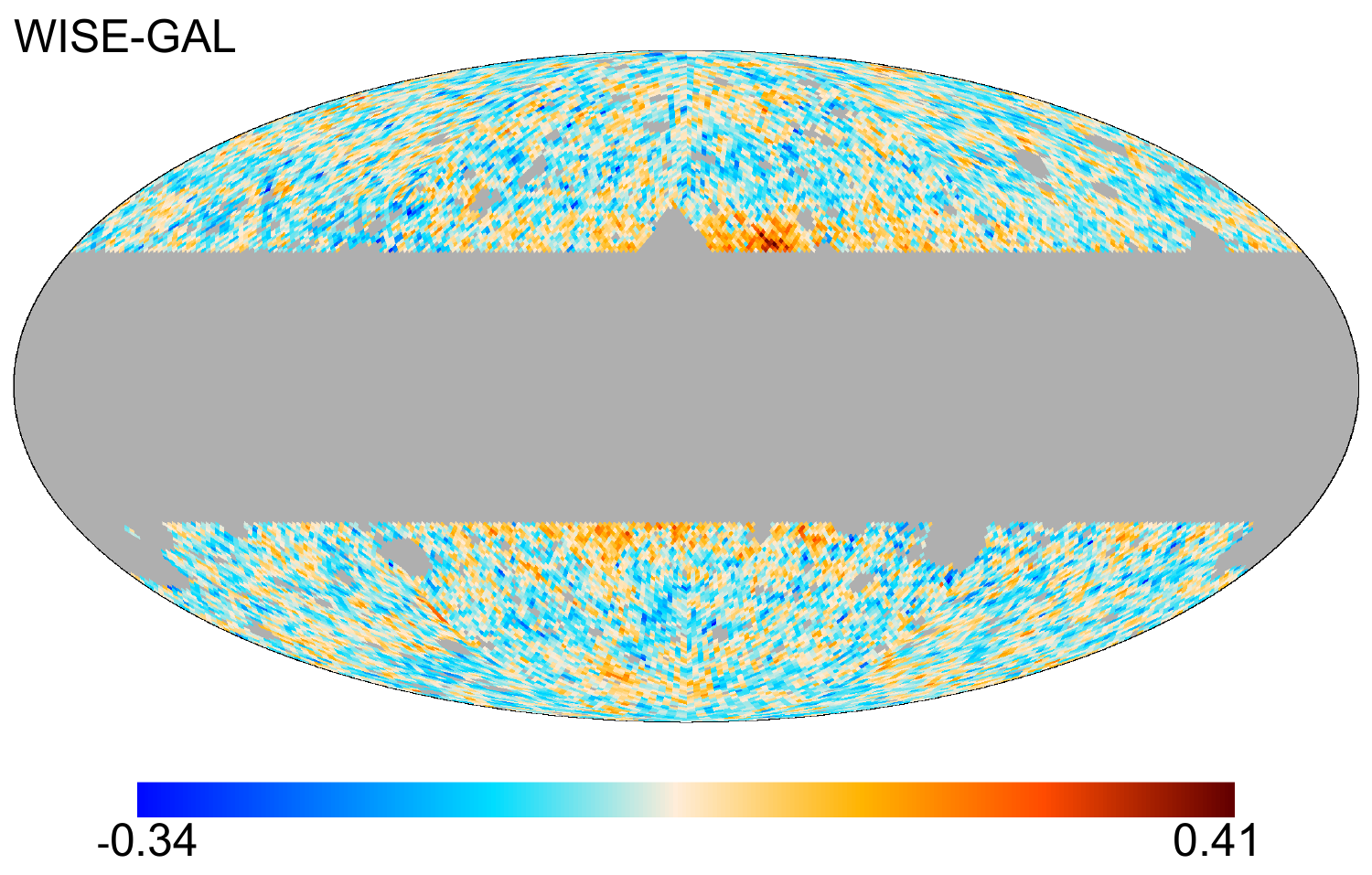}
\includegraphics[width=0.495\textwidth]{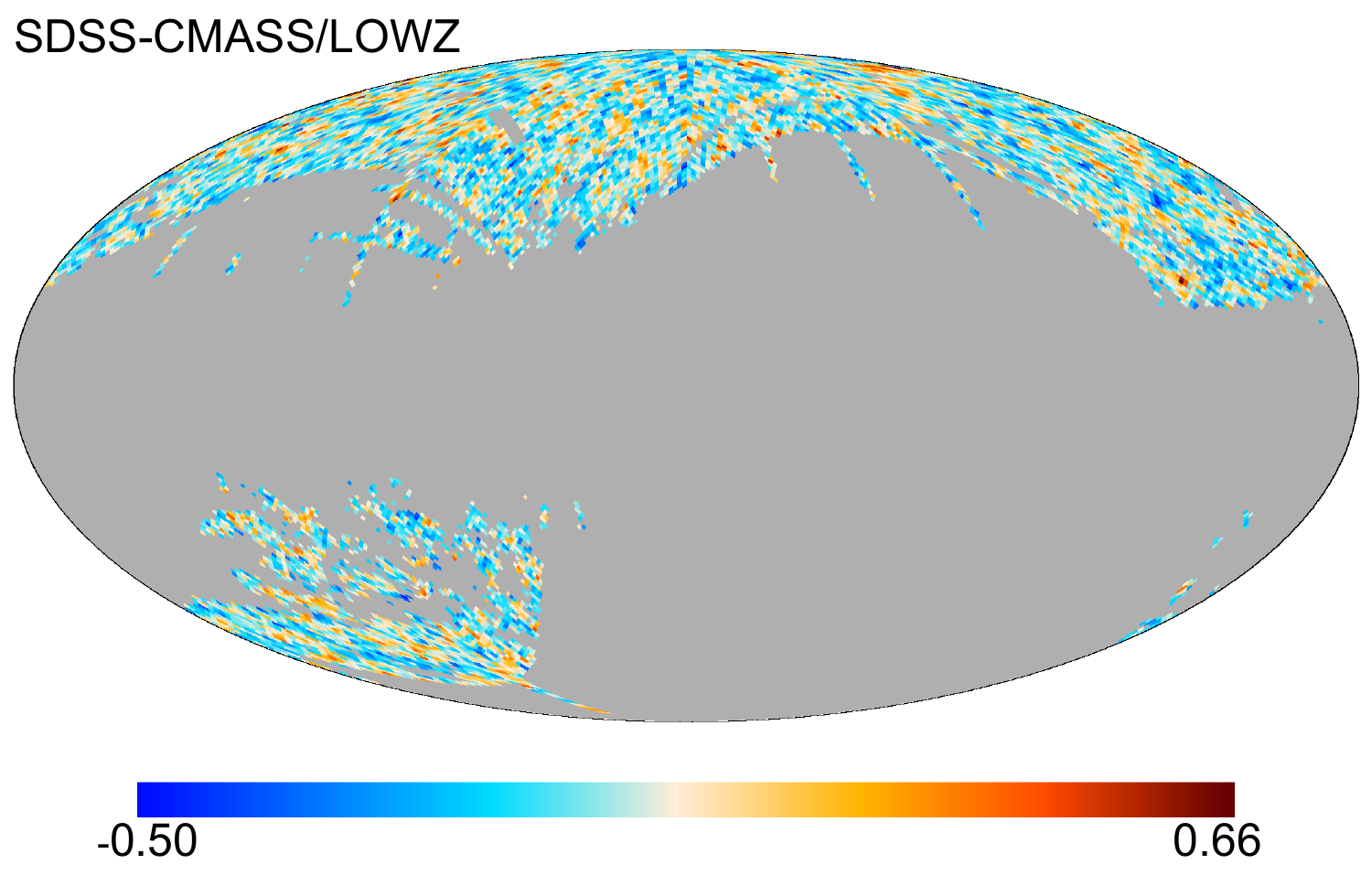}
\includegraphics[width=0.495\textwidth]{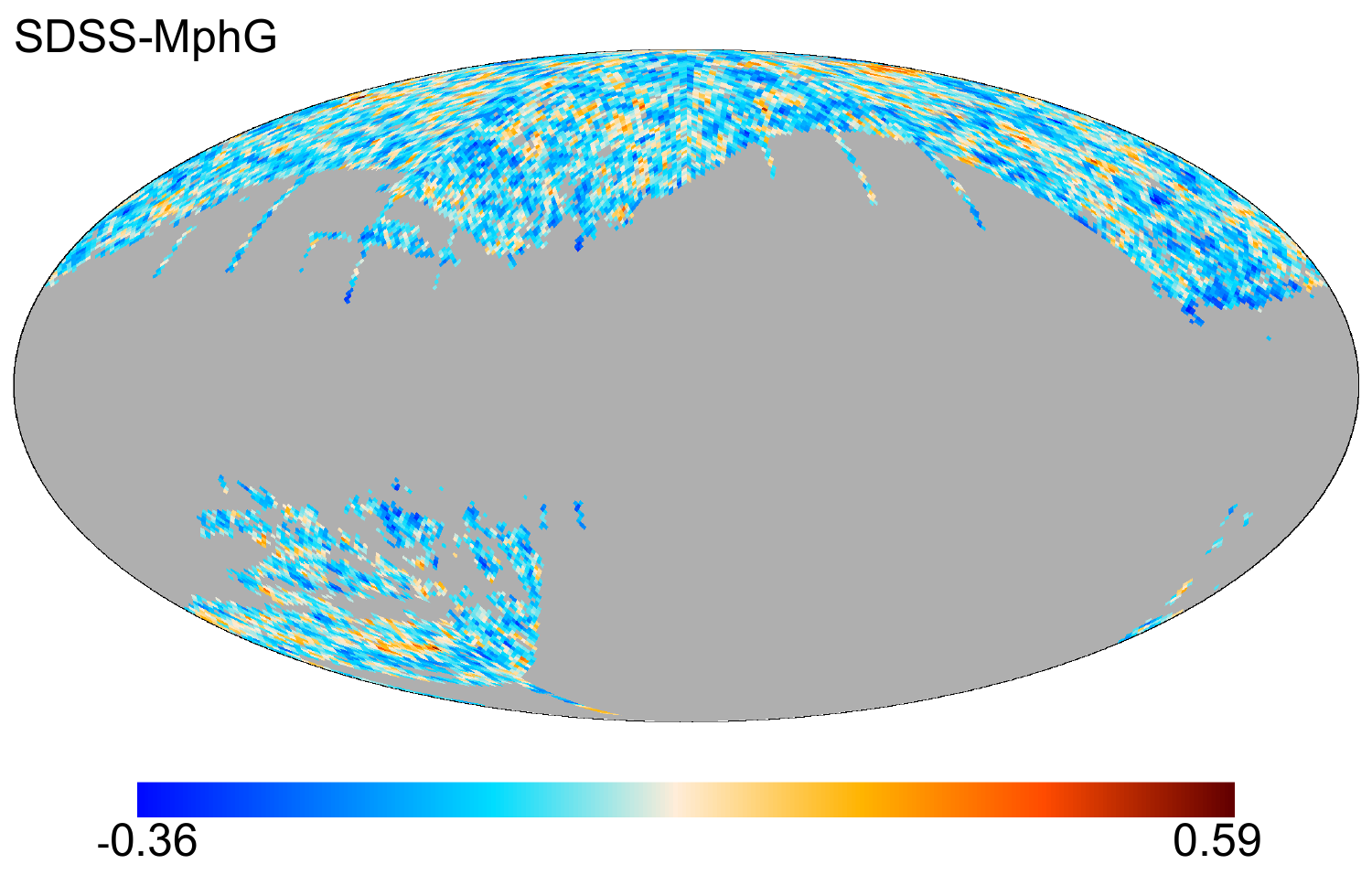}
\includegraphics[width=0.495\textwidth]{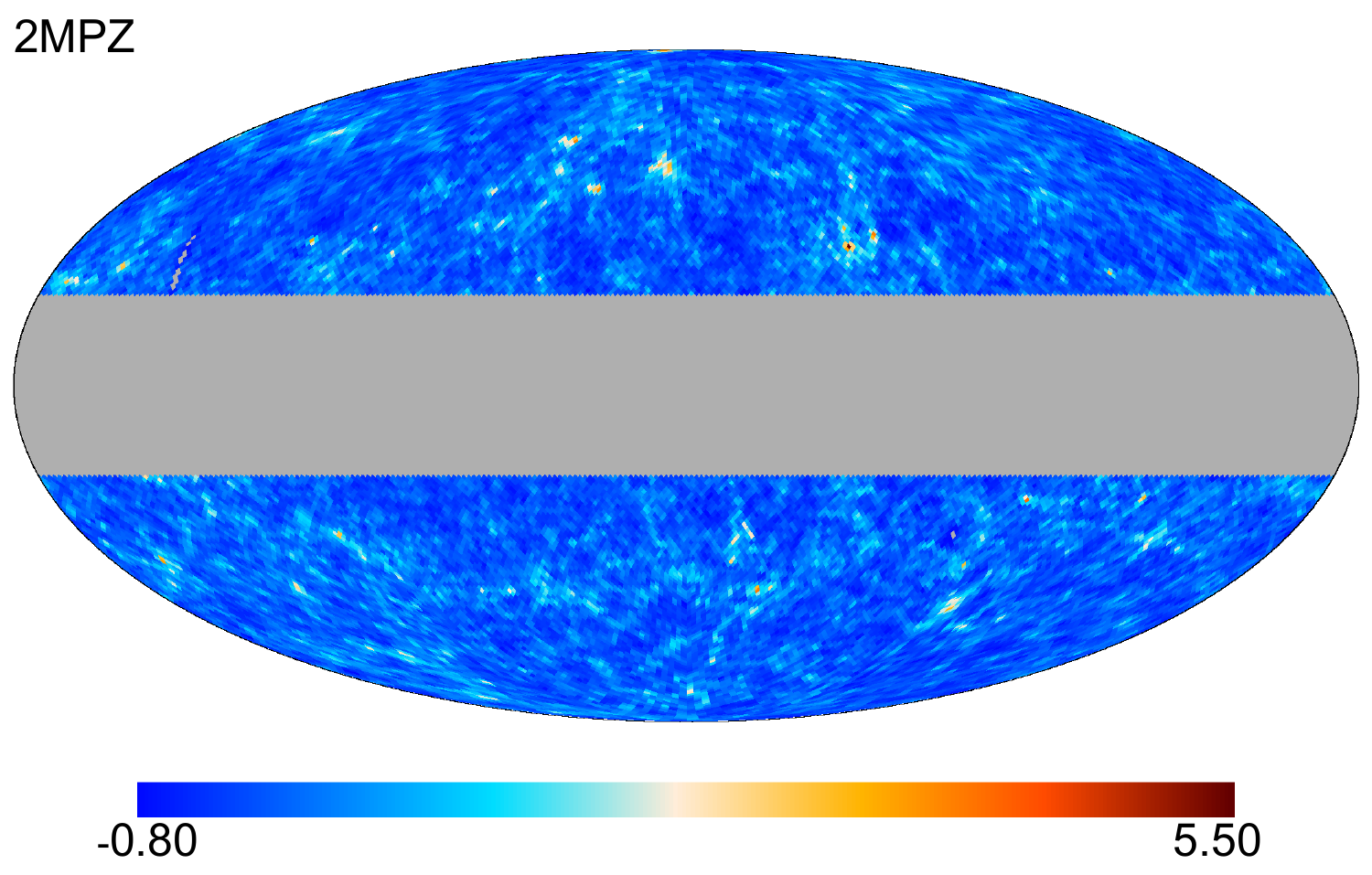}
\includegraphics[width=0.495\textwidth]{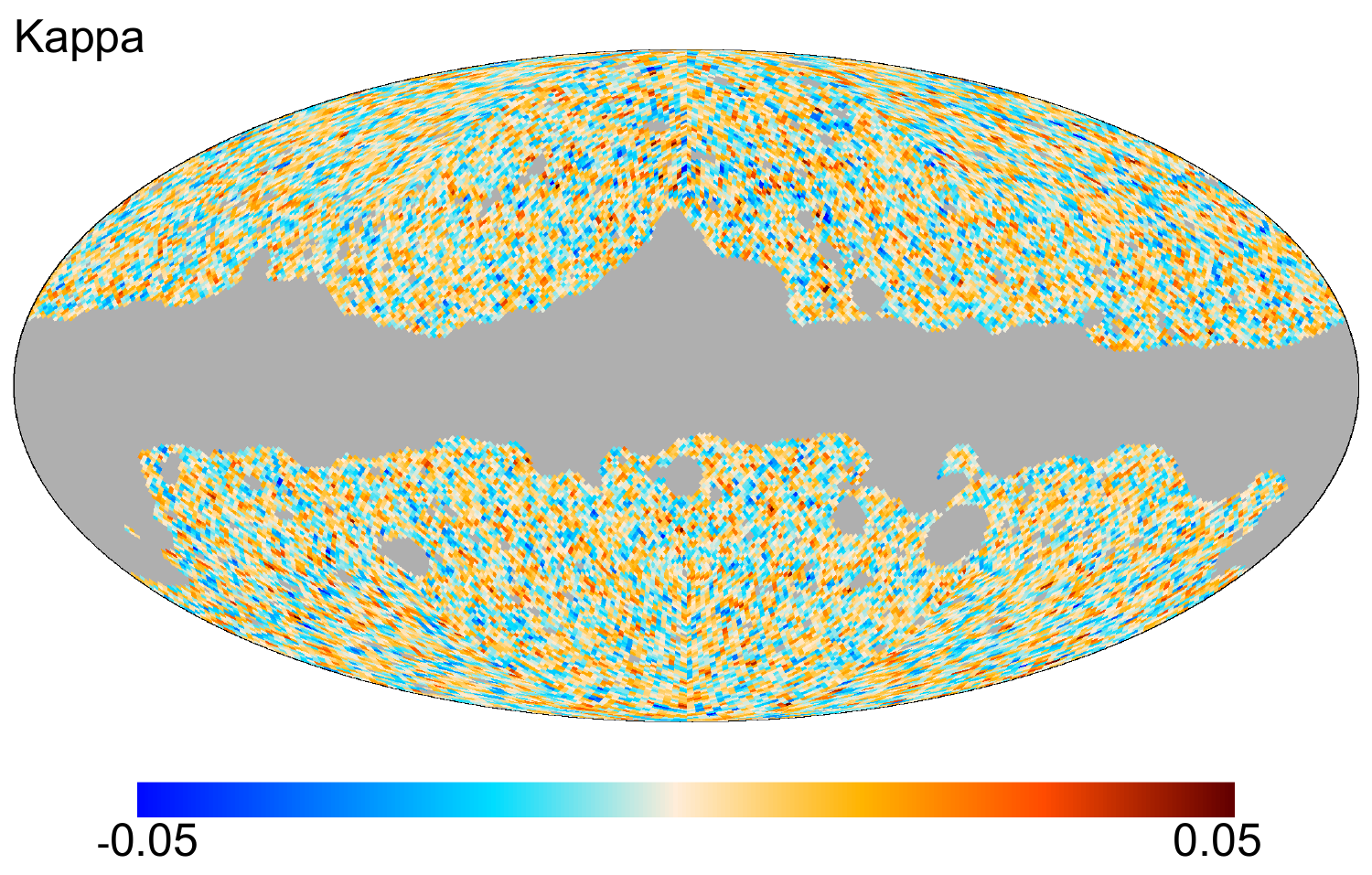}
\caption{Density contrast maps obtained from the galaxy catalogues at
$N_\mathrm{side}$ = 64. From left to right and from top to bottom: \nvss; \wagn; \wg; \lrg; \mg; and \tmpz. The \Planck\ lensing convergence map (\kap) is given in the fourth row. For visualization purposes, all these maps are Wiener-filtered versions of the original data. Maps are in dimensionless units here.}
\label{fig:surveys_maps}
\end{figure*}

For a better visualization, Wiener-filtered versions of the all-sky density projection of the external catalogues, as well as the \Planck\ \kap\ map, are shown in Fig.~\ref{fig:surveys_maps}. These are constructed from the theoretical power spectra obtained as described in Sect.~\ref{sec:data_sims}.
In Fig.~\ref{fig:surveys_cls}, we show the angular auto- and the cross-power angular spectra for all the LSS tracers: dashed lines and points correspond to the theoretical model and the data measurements, respectively (red for auto-spectra, and blue for cross-spectra); and grey areas represent the $1\,\sigma$ sampling uncertainties due to cosmic variance. 
All of these spectra have been corrected for the mask coupling following the {\tt MASTER} approach~\citep{Hivon2002}.
Notice that the \Planck\ lensing convergence map only contains information for multipoles $\ell > 8$~\citep[see][for details]{planck2014-a17}. The two maps based on the  \wise\ catalogues (\wagn\ and \wg) exhibit some extra signal at the largest scales. We identify this with some systematic effect present in these catalogues and, therefore, as a baseline, we only consider multipoles $\ell > 9$ for these two surveys. This cut implies only a minor loss of the ISW signal, while permitting a more robust determination of it. The rest of the auto-spectra are in reasonably good agreement with the theoretical predictions. Notice that any mismatch on the auto-spectra could suffer, not only from systematic effects, but from an inaccurate description of the statistical properties of the catalogues. In this sense, cross-spectra are, in principle, less affected by systematics (at least, among catalogues from different experiments), and, therefore, are more useful for identifying possible problems in the adequacy of the galaxy redshift distribution and galaxy biases. 
We emphasise that, in this sense, the \Planck\ \kap\ map could in principle be a more robust LSS probe, since it does not suffer from these kinds of uncertainty and, therefore, its correlation with the rest of the surveys is very useful for highlighting potential issues related to the catalogue characterization. In this sense, from Fig.~\ref{fig:surveys_maps}, it seems that the measured cross-correlation of the lensing potential with the galaxy catalogues is very good, indicating that, within the current uncertainties, the description of the surveys is accurate.
The three maps (\kap, \wagn, and \wg) shown in~Fig.~\ref{fig:surveys_maps} do not include the cut multipoles.

Besides the galaxy surveys described above, we also use superstructure catalogues to study the ISW effect through the stacking of the CMB anisotropies on the positions of clusters and voids. We concentrate on the supercluster and void catalogue of~\cite {Granett2008b}, obtained from SDSS (\gr08), since, as shown in~\cite{planck2013-p14}, its reported strong signal would be a challenge for the standard $\Lambda$CDM cosmology if it is solely caused by the ISW effect.

Below we provide a description of all these LSS tracers. For those catalogues already used in our previous publication (\nvss, \lrg, \mg, and \gr08) only a summary is provided; more detailed description can be found in~\cite{planck2013-p14}.
\begin{table*}[tb]
\begingroup
\newdimen\tblskip \tblskip=5pt
\caption{Main characteristics of the galaxy catalogues used as tracers of the gravitational potential. From left to right, the columns indicate the number of galaxies per steradian, the fraction of the sky covered by each survey, the galaxy redshift distribution ($dn/dz$), the galaxy bias, and the mean redshift. Whereas for the \nvss\ and the \mg\ catalogues there are analytical expressions of the galaxy redshift distribution, for the other tracers there are only numerical estimations.
\label{tab:surveys}}
\nointerlineskip
\vskip -3mm
\footnotesize
\setbox\tablebox=\vbox{
   \newdimen\digitwidth 
   \setbox0=\hbox{\rm 0} 
   \digitwidth=\wd0 
   \catcode`*=\active 
   \def*{\kern\digitwidth}
   \newdimen\signwidth 
   \setbox0=\hbox{+} 
   \signwidth=\wd0 
   \catcode`!=\active 
   \def!{\kern\signwidth}
\halign{\hfil#\hfil\tabskip=0.8cm& \hfil#\hfil\tabskip=0.5cm&
 \hfil#\hfil\tabskip=0.5cm& \hfil#\hfil\tabskip=0.5cm& \hfil#\hfil\tabskip=0.5cm&  \hfil#\hfil\tabskip=0.cm\cr 
\noalign{\doubleline}
 \noalign{\vskip -2pt}
Galaxy catalogue & $\bar{n}$& $f_{\rm sky}$ & $dn/dz$ & bias & $\bar{z}$\cr 
\noalign{\vskip 3pt\hrule\vskip 5pt}
\nvss&
$1.584\times10^5$&0.73&$\propto \left(z/0.33\right)^{0.33}e^{-0.37\left(z/0.33\right)}$&$0.90\left(1+0.54\left(1+z\right)^2\right)$& 1.22\cr
\wagn&$1.552\times10^5$&0.45&numerical&$0.48\left(1+0.54\left(1+z\right)^2\right)$&1.03\cr
\wg&$1.187\times10^7$&0.45&numerical&$0.79\left(1+z\right)$&0.38\cr
\lrg&$5.558\times10^5$&0.22&numerical&2.03&0.45\cr
\mg&$9.680\times10^6$& 0.22&$\propto \left(z/0.16\right)^{1.5}e^{\left(-z/0.34\right)^{2.30}}$&1.20&0.32\cr
\tmpz&$8.328\times10^4$&0.66&numerical&1.35&0.09\cr
\noalign{\vskip 5pt\hrule\vskip 3pt}}}
\endPlancktable                    
\endgroup
\end{table*}

\subsubsection{The \nvss\ radio-galaxies catalogue}
\label{sub:nvss}

The luminous AGN are very powerful radio sources, which can be seen also at high redshifts. These sources are able to trace the cosmic density field for both the redshift evolution and the spatial distribution. Therefore they can probe the spatial distribution of large-scale potential wells that contribute to the ISW effect during the dark energy era.

The sources we use in this paper are the same samples we used in \citet{planck2013-p14}, i.e., the NRAO Vary-Large-Array (VLA) Sky Survey~\citep{Condon1998}. This NVSS survey was conducted by using VLA at $1.4\,$GHz, and covers up to an equatorial latitude of $b_{\rm E}=-40\deg$, with an average noise level of $0.45$\,mJy\,beam$^{-1}$. There are roughly $1.4\times10^6$ sources above a flux threshold of $2.5$\,mJy. Figure~\ref{fig:surveys_maps} (top-left panel) shows the all-sky density projection for the NVSS galaxies, where the grey area indicates regions not observed or disregarded by the surveys. Figure~\ref{fig:surveys_cls} includes a subplot to show the angular power spectra (blue points) of the NVSS survey.

For the galaxy bias, we use the Gaussian bias evolution model of~\citet{Xia2011}, i.e., the bias of the survey is given by a mass-weighted average,
\begin{linenomath*}
\begin{equation}
b(z) = \frac{\int_{M_{\rm min}}^\infty \mathrm{d}M \ b(M,z) \
n(M,z)}{\int_{M_{\rm min}}^\infty \mathrm{d}M \ n(M,z)} \ ,
\label{eq:biasxia}
\end{equation}
\end{linenomath*}
where $n(M,z)$ is the halo mass function for which we adopt the Sheth-Tormen~\citep{Sheth1999} form and $b(M,z)$ is the bias of halos with comoving mass $M$. This bias (as a function of redshift) can be approximated as a second-order polynomial, as given in Table~\ref{tab:surveys}. In addition, the redshift distribution is parametrized by:
\begin{linenomath*}
\begin{equation}
\frac{\mathrm{d}n}{\mathrm{d}z} =
n_0 \left( \frac{z}{z_0} \right)^\alpha e^{-\alpha z / z_0} \ ,
\label{eqn:bias}
\end{equation}
\end{linenomath*}
where $z_0=0.33$ and $\alpha=0.37$, and $n_0$ is a constant to normalize the distribution to unity. The function is given by the red line in Fig.~\ref{fig:surveys_dndz}. We refer the interested readers to our previous paper \citet{planck2013-p14} for more details of the possible systematic effects for the samples.

\subsubsection{The \textit{Sloan Digital Sky Survey} catalogues}
\label{sub:sdss}

We use two subsamples of \sdss: the \sdss\ luminous galaxy samples (\lrg); and the main photometric \sdss\ galaxy sample (\mg).
The redshift distributions of the two samples are shown in blue and green lines in Fig.~\ref{fig:surveys_dndz}. The sky coverage for the two subsamples are shown in the subplots of Fig.~\ref{fig:surveys_maps}, and the angular power spectra in the subplots of Fig.~\ref{fig:surveys_cls}.

\subsubsection*{\lrg}
\noindent
We use the photometric luminous galaxy (LG) catalogue from the Baryonic Oscillation Spectroscopic Survey (BOSS) of the SDSS III.
The data used consist of two subsamples: CMASS; and LOWZ. In this paper we will use a combination of them, i.e., \lrg\ in our data analysis.

The CMASS sample has roughly constant stellar mass, and is mostly contained in the redshift range $z=0.4$--$0.7$, with a galaxy number density close to
$110\,{\rm deg}^{-2}$. With the colour selection criteria, it is a catalogue of about one million sources, in an area of $10500\,{\rm deg}^2$.  Photometric redshifts of this sample are calibrated using a selection of about 100{,}000 BOSS spectra as a training sample for the photometric catalogue.

The photometric LOWZ sample selectes luminous, highly biased, mostly red galaxies, placed at an average redshift of $\bar{z} \approx 0.3$ and below the redshifts of the CMASS sample ($z<0.4$). 
With a total number of sources of roughly 600000, the number density of galaxies in the southern part of the footprint is higher than in the northern one (by more than 3\,\%). Both SDSS-CMASS and SDSS-LOWZ samples are further corrected for any scaling introduced by possible systematics like stars, mask value, seeing, sky emission, air mass and dust extinction, since the high star density tends to ``blind'' galaxy detection algorithms. The algorithm followed to correct for systematics is described in \citet{Hernandez2014}.

The LOWZ and CMASS galaxy samples come from two different colour and magnitude selections on \sdss\ photometry, and they effectively probe two different redshift ranges. If studied individually, the expected sensitivity of CMASS is larger than for LOWZ, resulting in the total signal-to-noise obtained when adding the contribution of each survey separately being very similar (to 10--15\,\%) to the result from a combination of the two surveys into one single galaxy sample. Most of this difference comes from assuming an effectively constant bias for the joint survey. 
Therefore, the improvement is quite low, particularly taking into account that this is a survey that, at the end of the day, provides very little of the signal-to-noise in the total 4$\,\sigma$ detection (see Sect.~\ref{sec:xcorr}). On the other hand, the effective combination simplifies the overall analysis. 
Regarding the mask, we also remark that, since it is determined by systematic effects such as the star density, airmass, or Galactic extinction, and is built independently of the colour and magnitude cuts applied on the galaxy sample from \sdss, we ended up having the same mask for both the LOWZ and CMASS samples (even if one could argue for more or less conservative masks).

\subsubsection*{\mg}
\noindent 
These are the photometrically selected galaxies from the SDSS-DR8 catalogue, which covers a total sky area of $14\,555\,{\rm deg}^2$ \citep{Aihara2011}. The total number of objects labelled as galaxies in this data release is 208 million. However, for correcting extinction and restricting redshift ranges, our final sample consists of about 42 million, with redshifts distributed around a median value of $\approx 0.35$. We use the analytical function
\begin{linenomath*}
\begin{equation}
\frac{dn}{dz}=\frac{\beta}{\Gamma\left(\frac{m+1}{\beta}\right)}\frac{z^m}{z_0^{m+1}}e^{-(z/z_0)^\beta}~,
\label{eq:dndz_xia}
\end{equation}
\end{linenomath*}
with parameters $m=1.5$, $\beta=2.3$, and
$z_0=0.34$, for the number density distribution, and the constant galaxy bias $b=1.2$ by fitting the $\Lambda$CDM prediction to the observed auto-correlation function of the galaxies. As for the LOWZ/CMASS LRG samples, this galaxy sample was also corrected for systematics following the approach of \citet{Hernandez2014}.
\begin{figure*}
\begin{flushright}
\includegraphics[width=0.1380\textwidth]{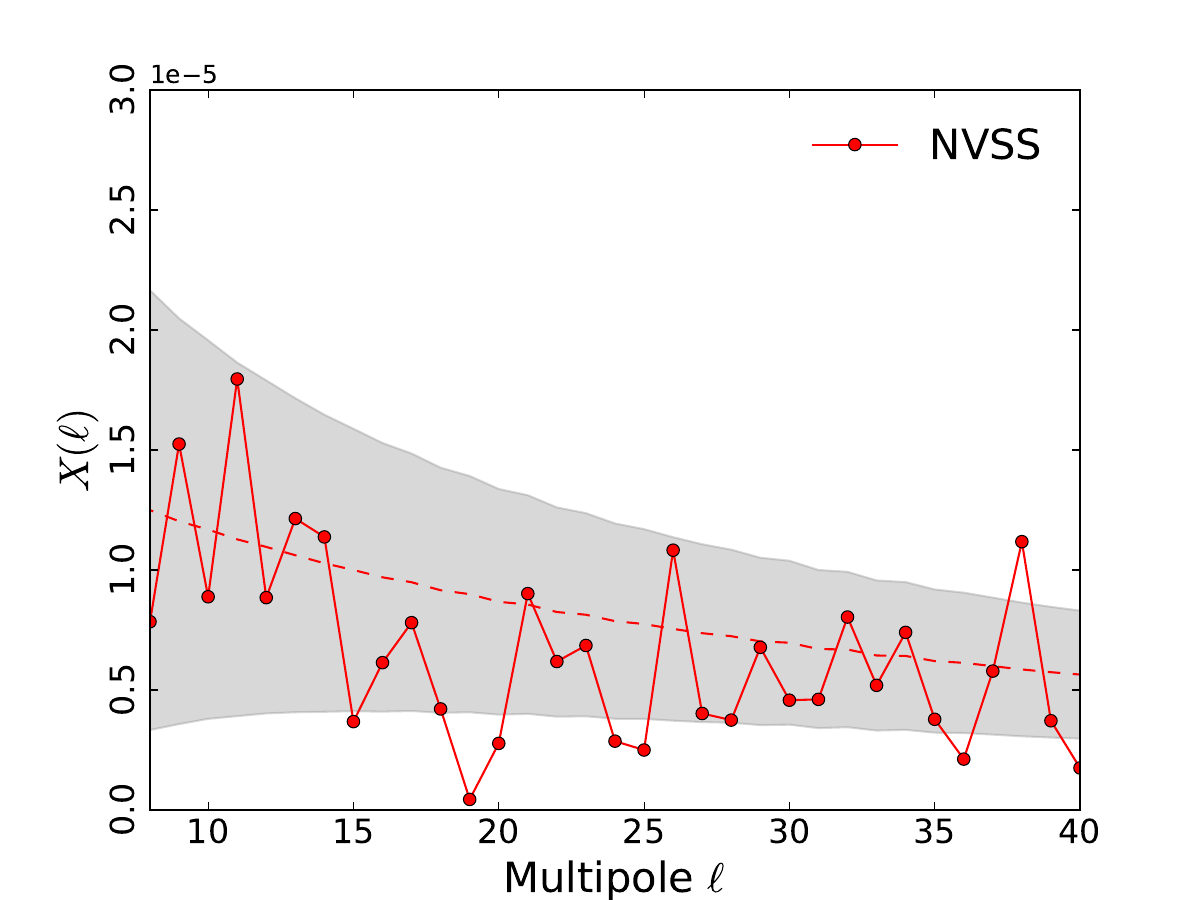}
\includegraphics[width=0.1380\textwidth]{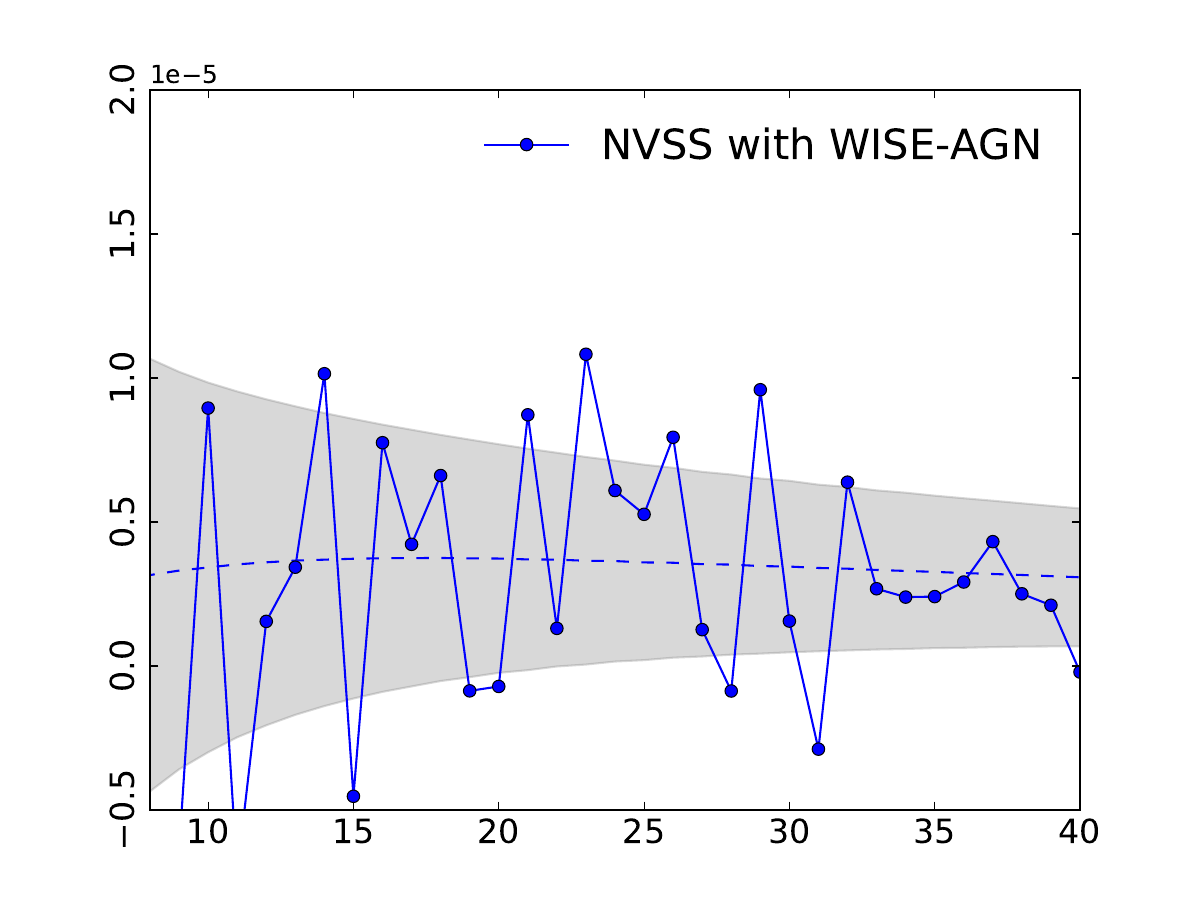}
\includegraphics[width=0.1380\textwidth]{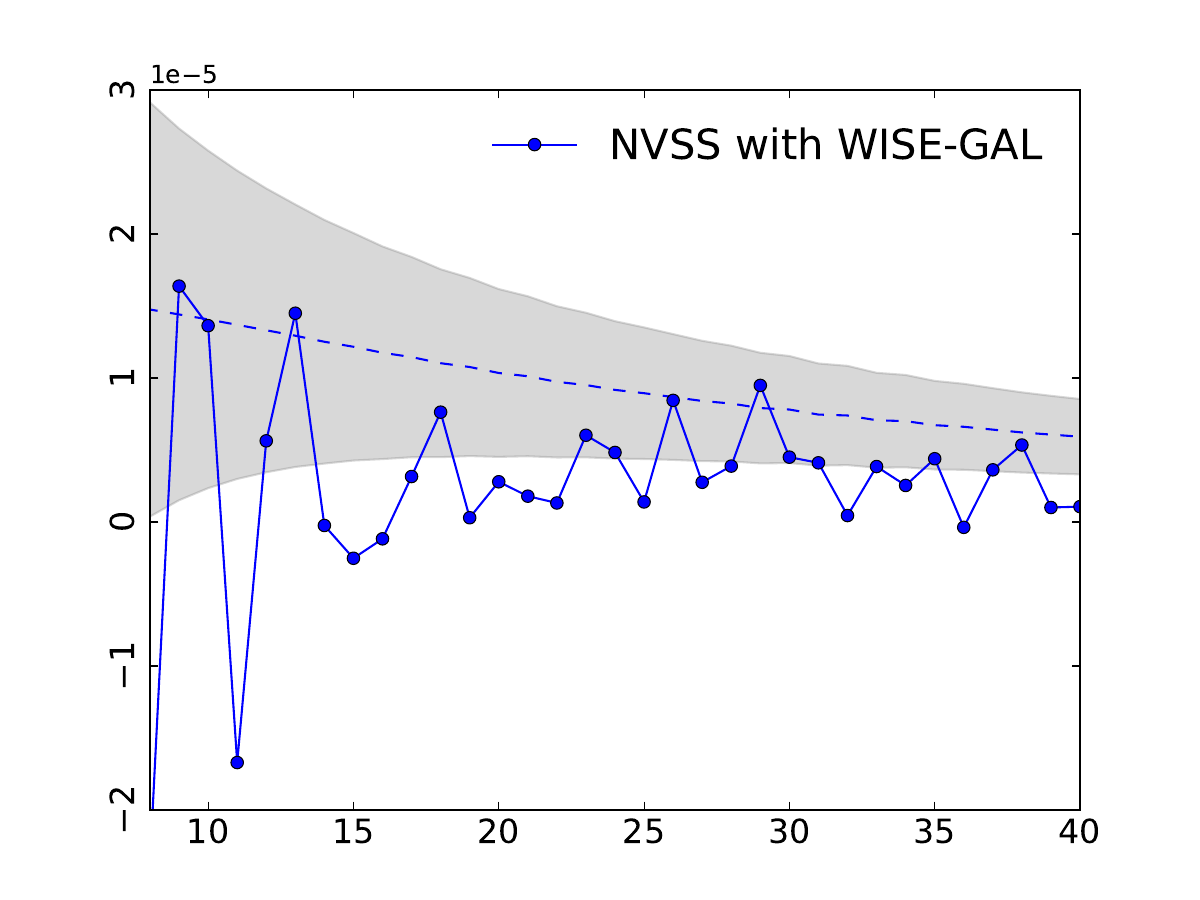}
\includegraphics[width=0.1380\textwidth]{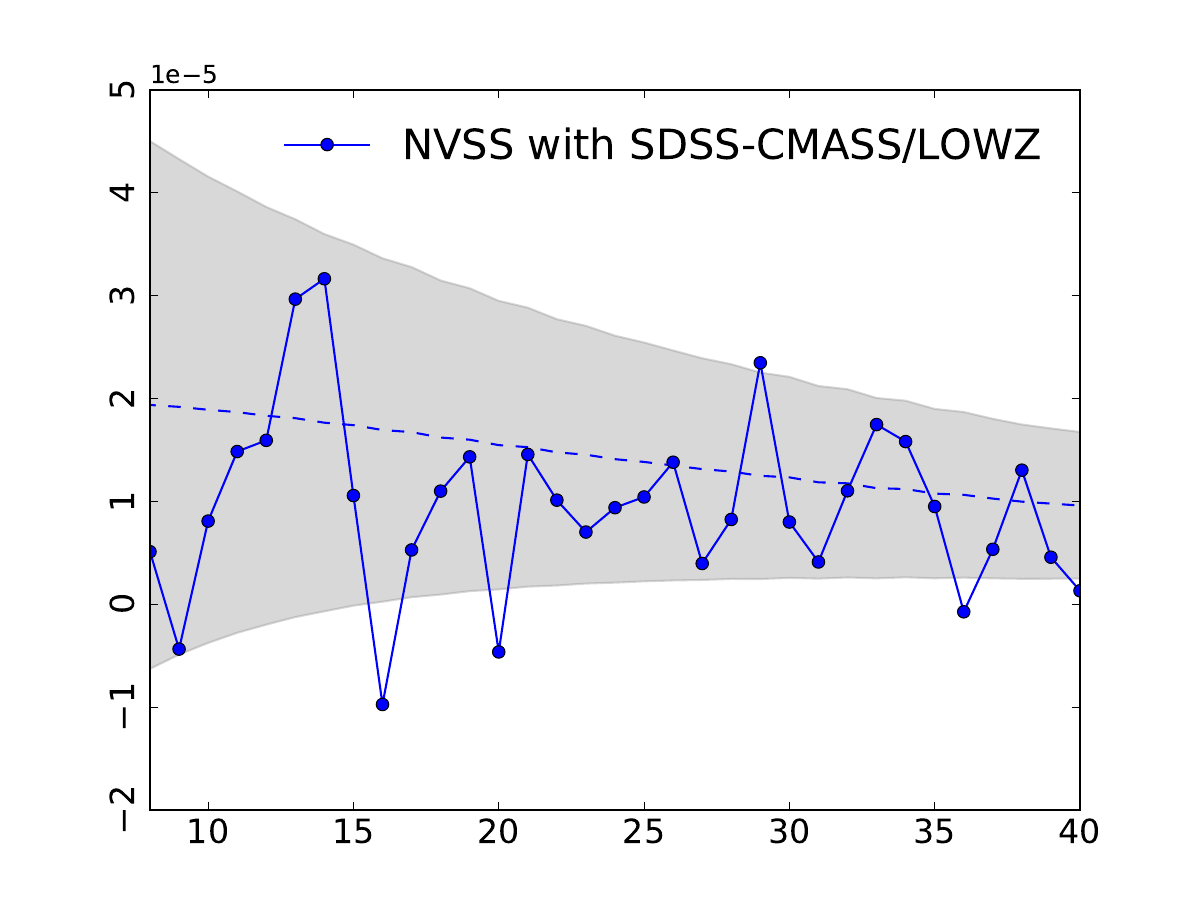}
\includegraphics[width=0.1380\textwidth]{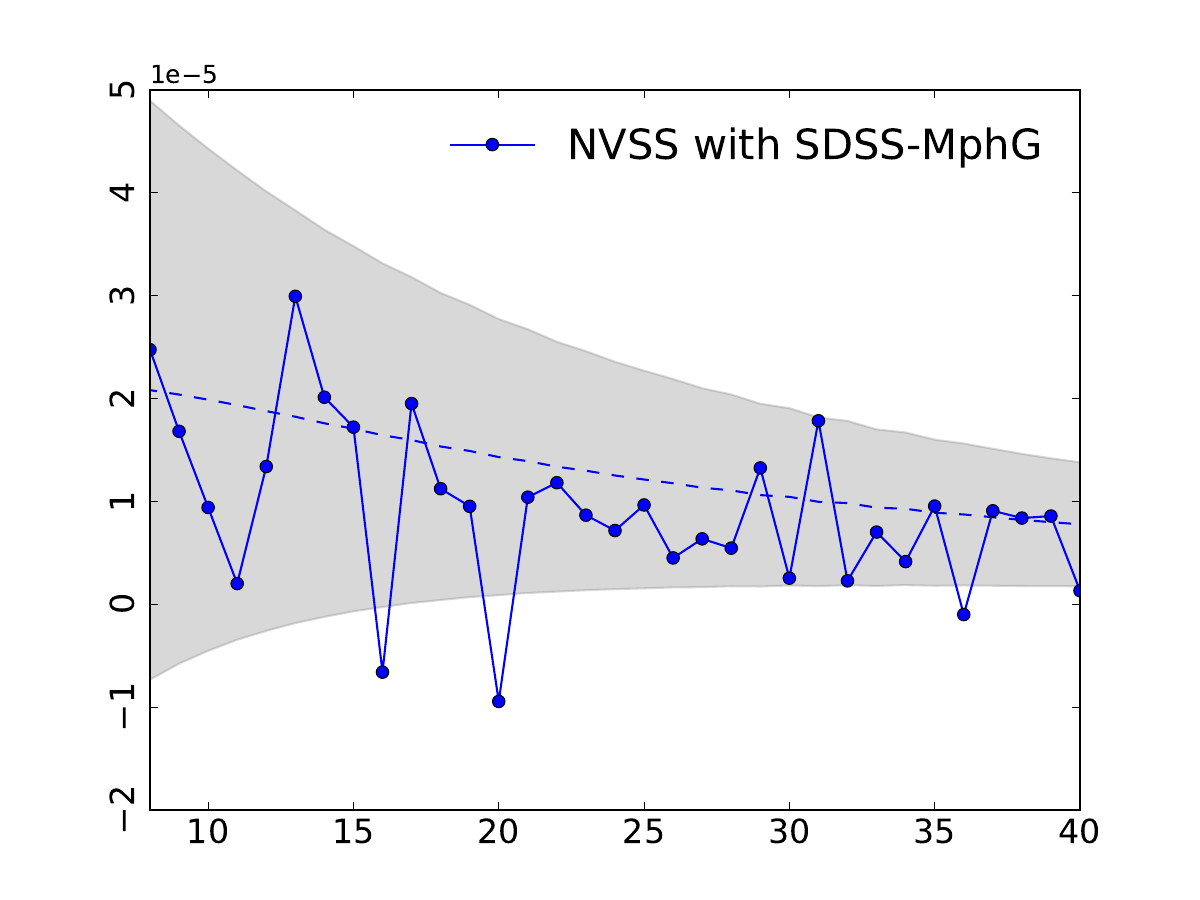}
\includegraphics[width=0.1380\textwidth]{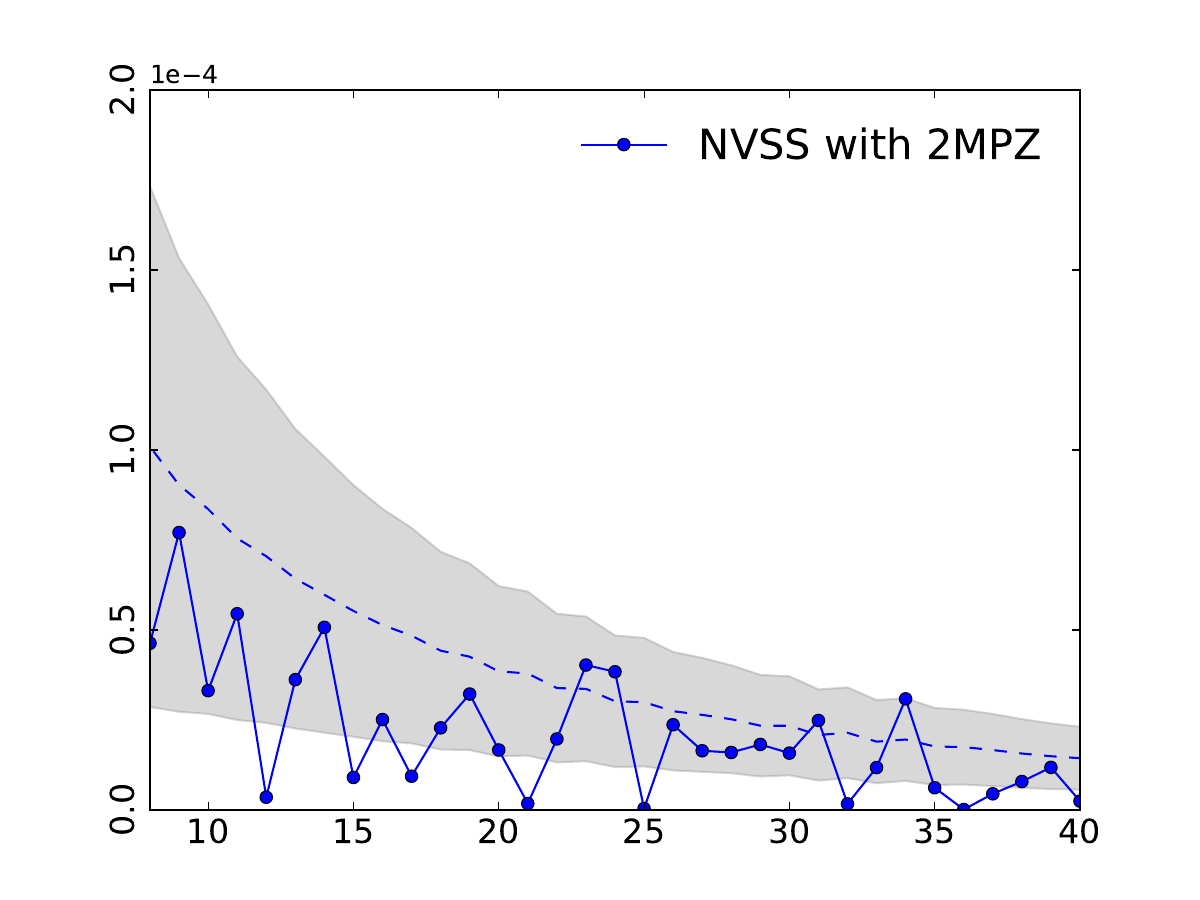}
\includegraphics[width=0.1380\textwidth]{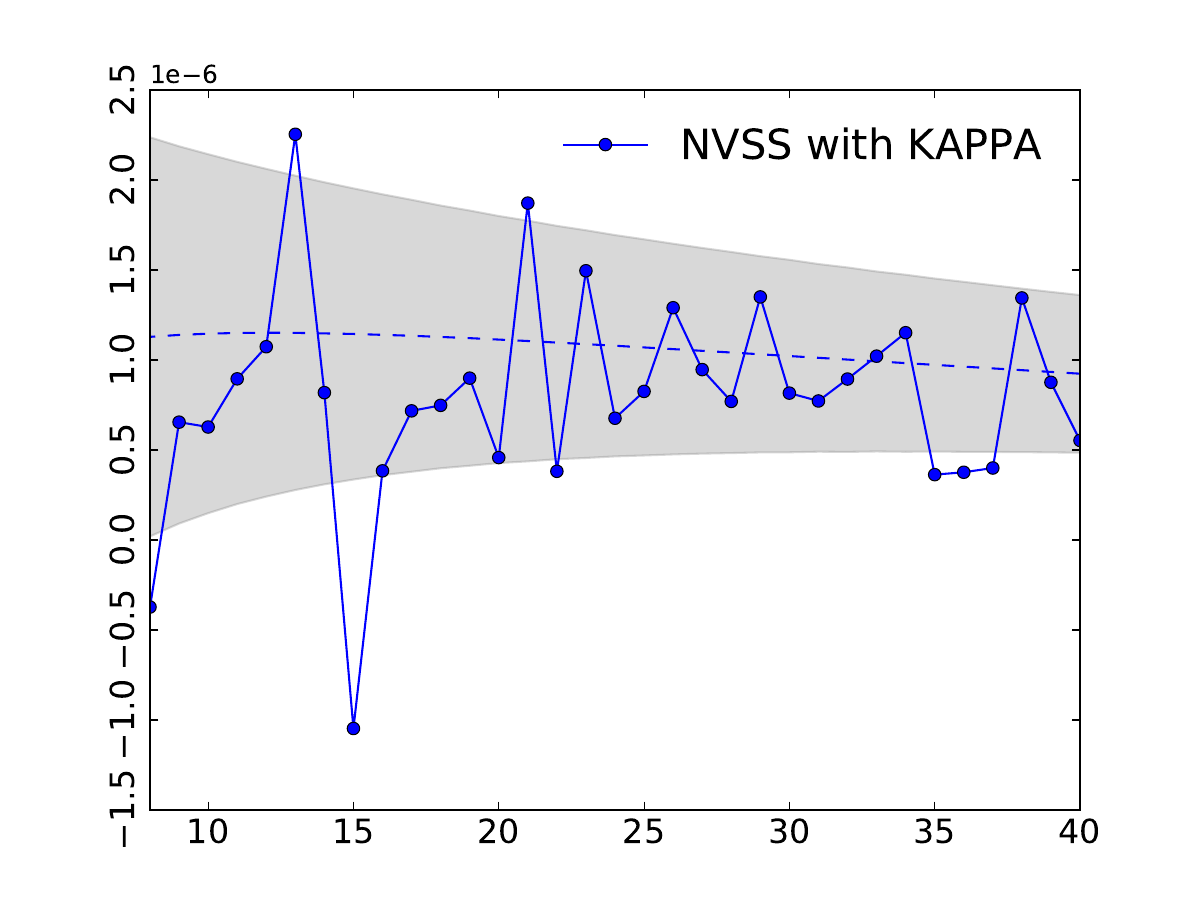}\\
\includegraphics[width=0.1380\textwidth]{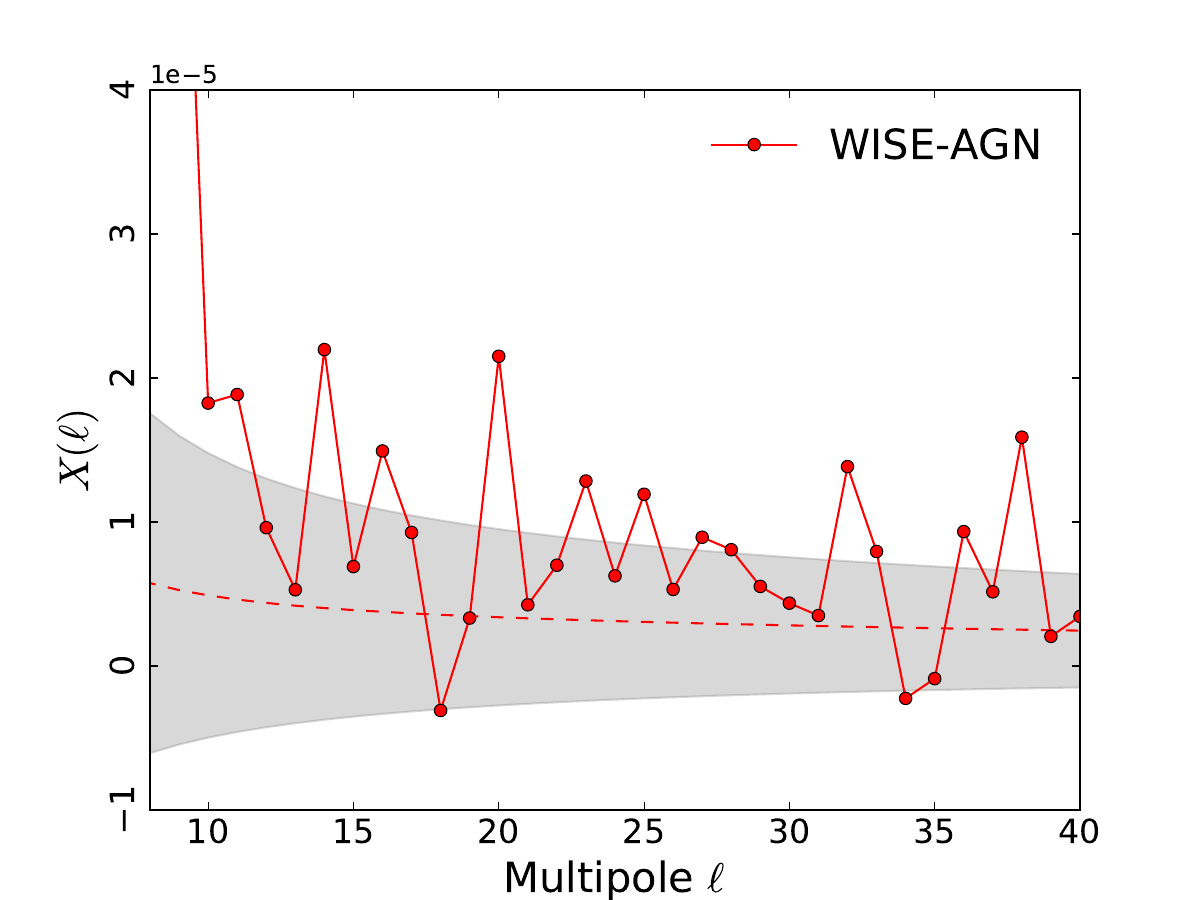}
\includegraphics[width=0.1380\textwidth]{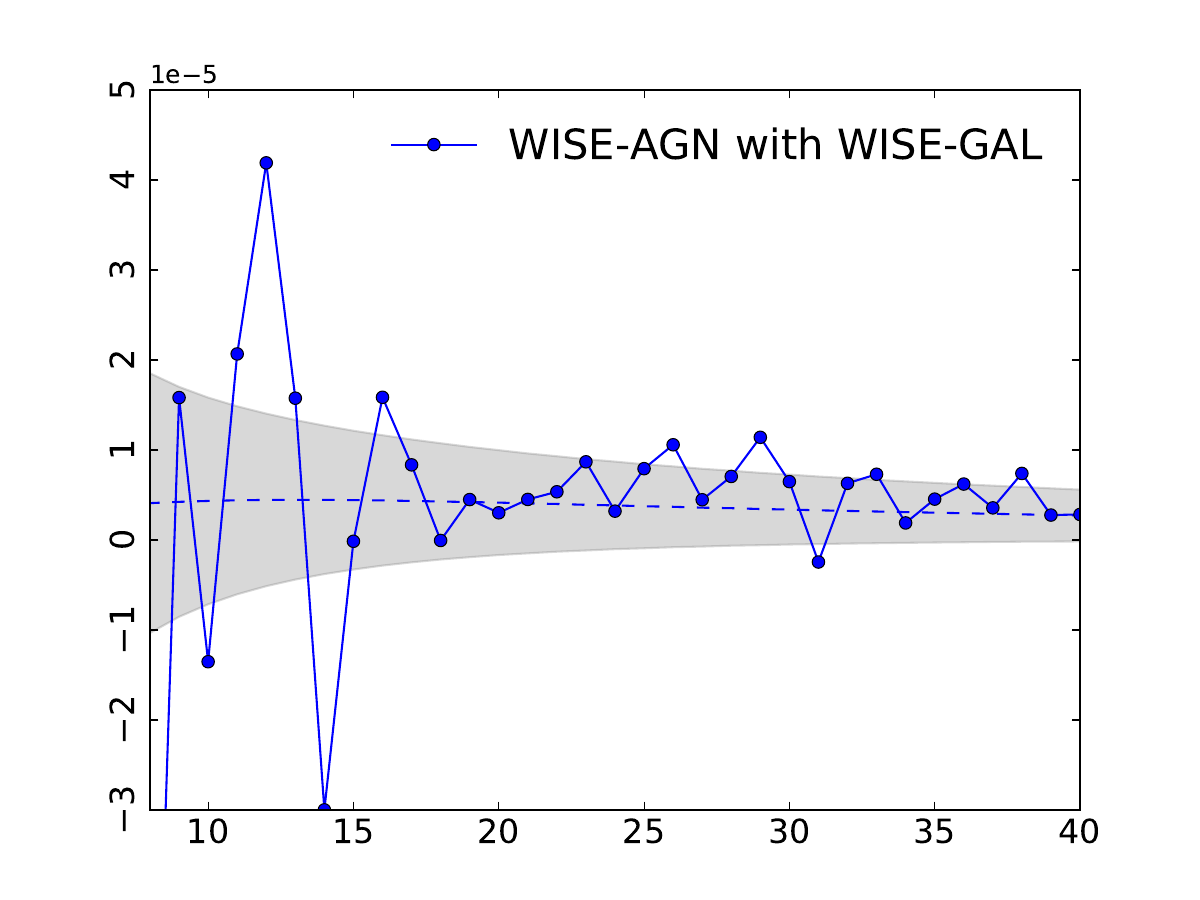}
\includegraphics[width=0.1380\textwidth]{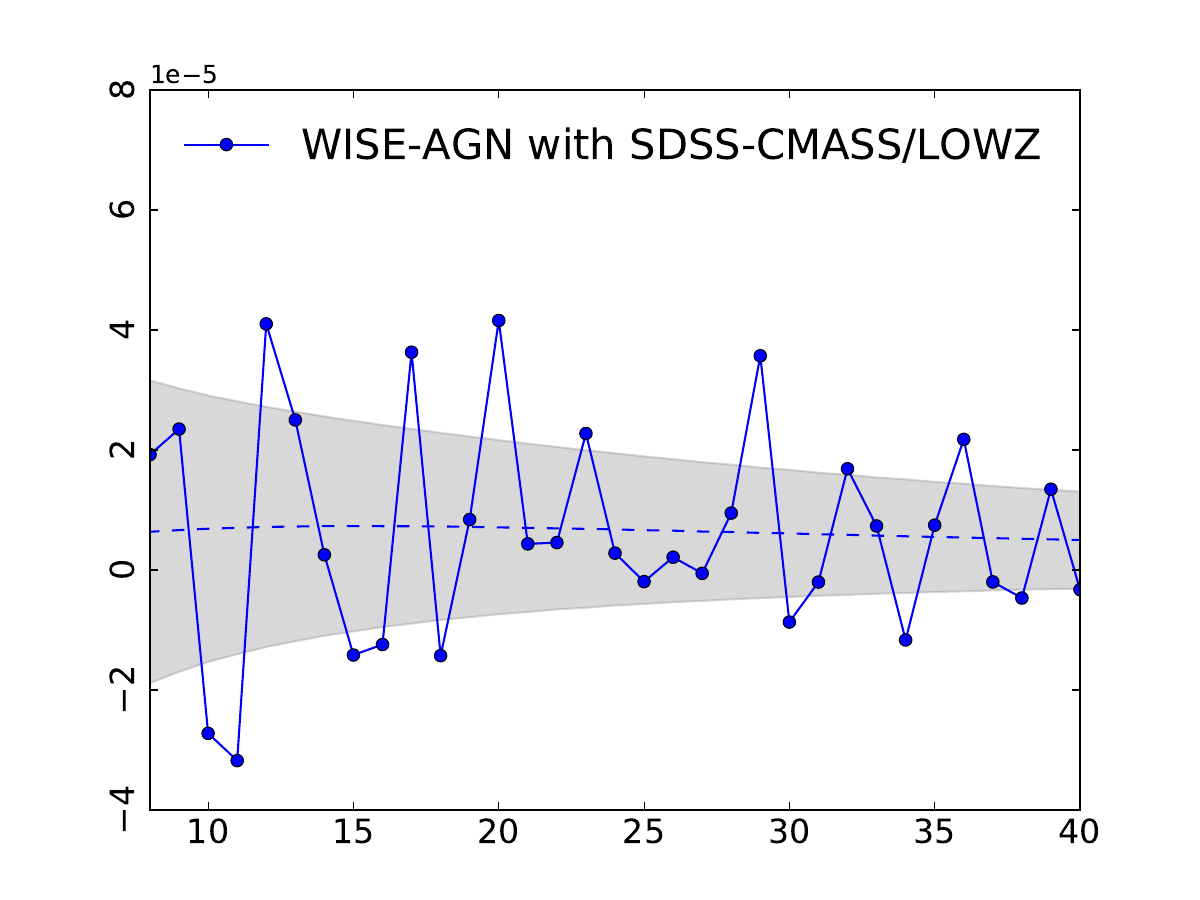}
\includegraphics[width=0.1380\textwidth]{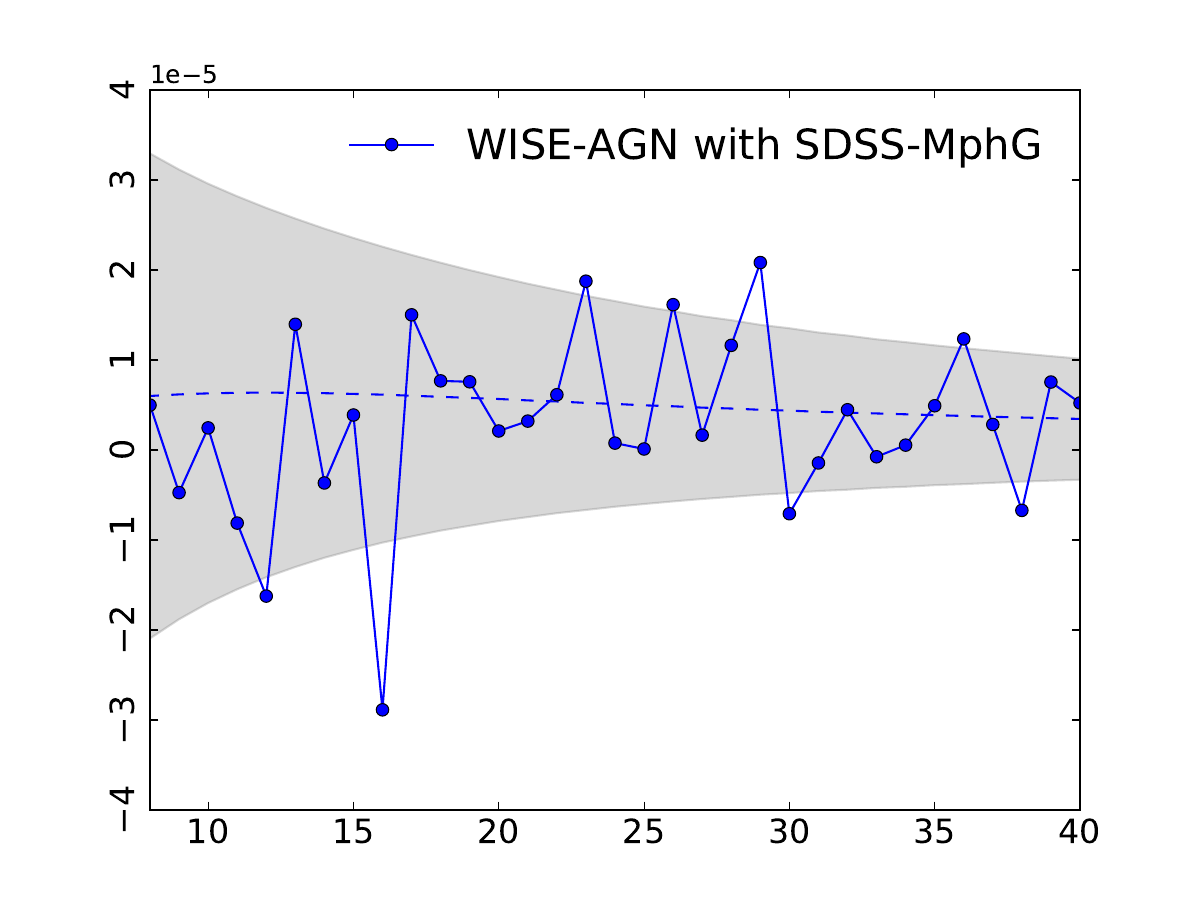}
\includegraphics[width=0.1380\textwidth]{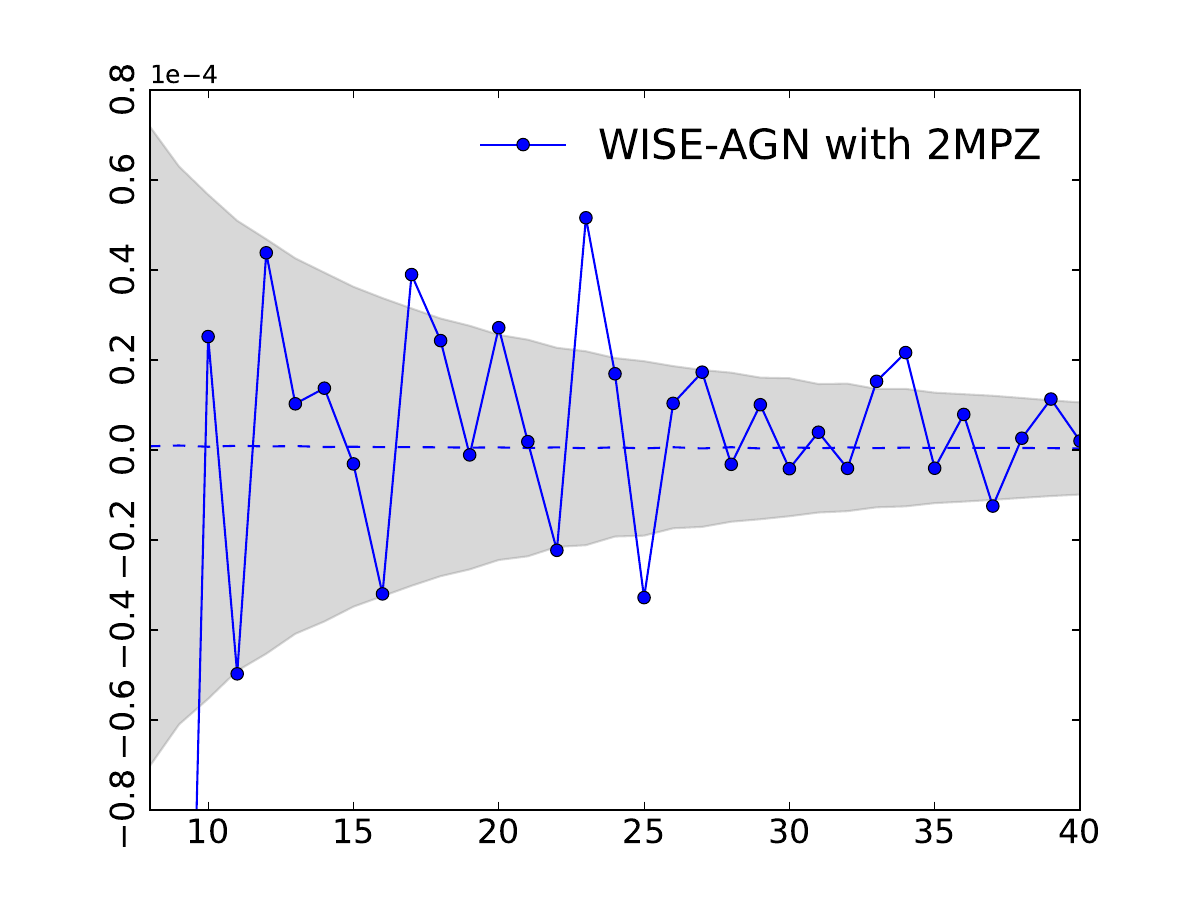}
\includegraphics[width=0.1380\textwidth]{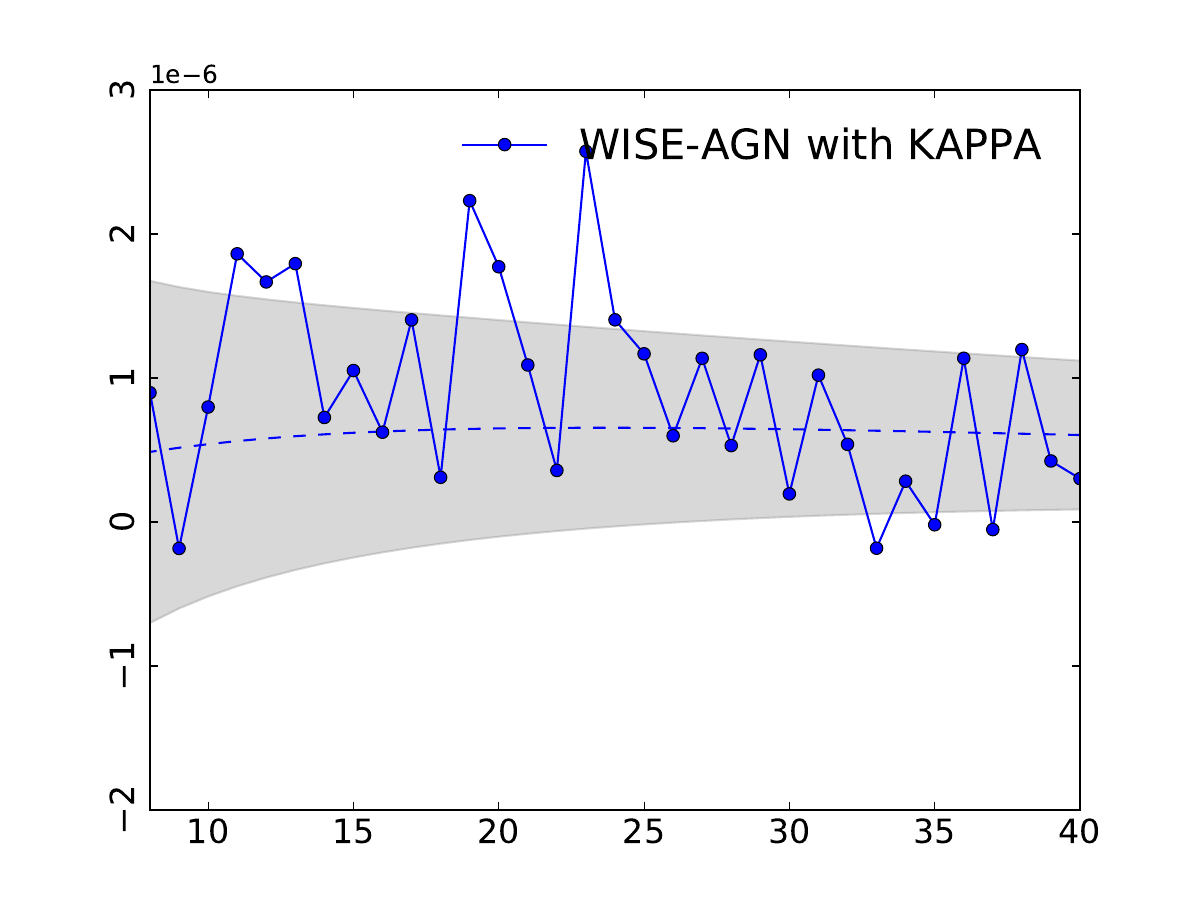}\\
\includegraphics[width=0.1380\textwidth]{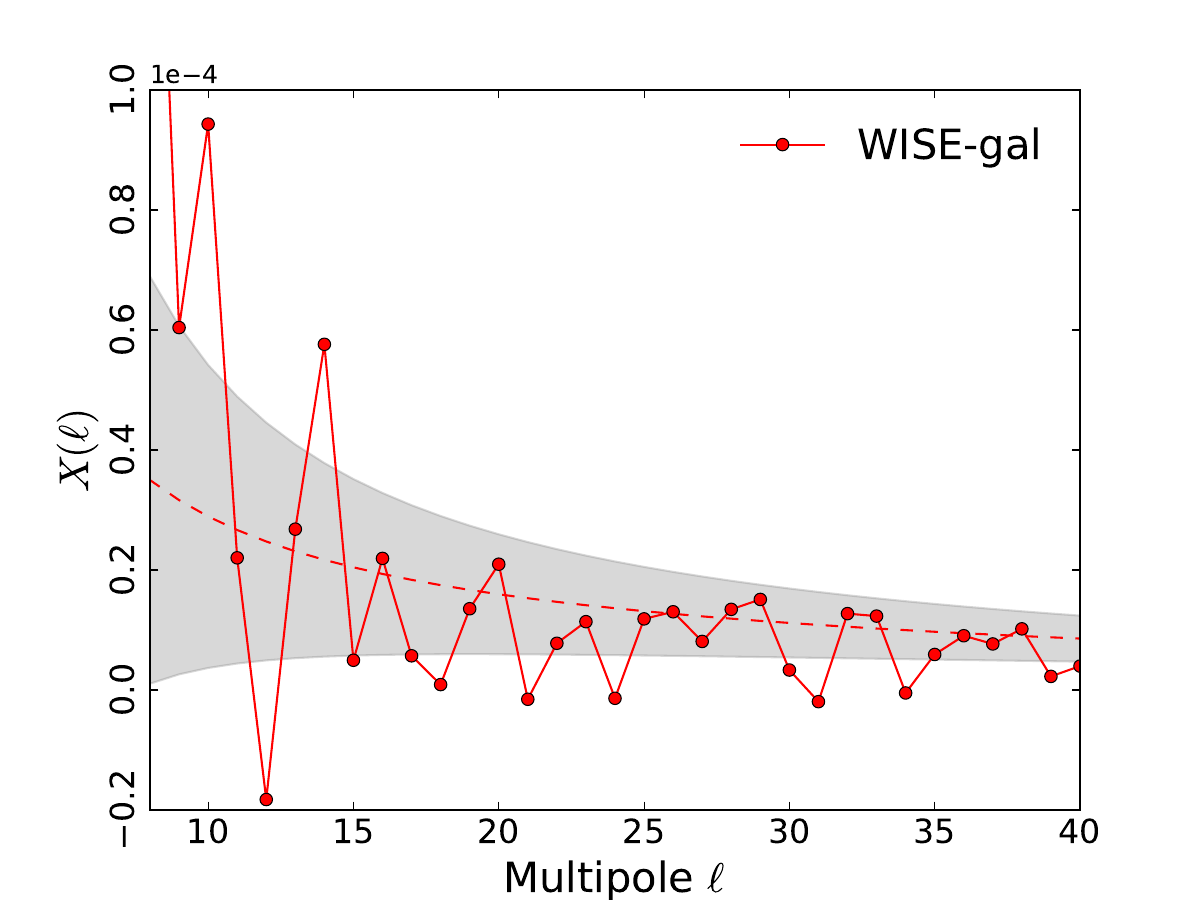}
\includegraphics[width=0.1380\textwidth]{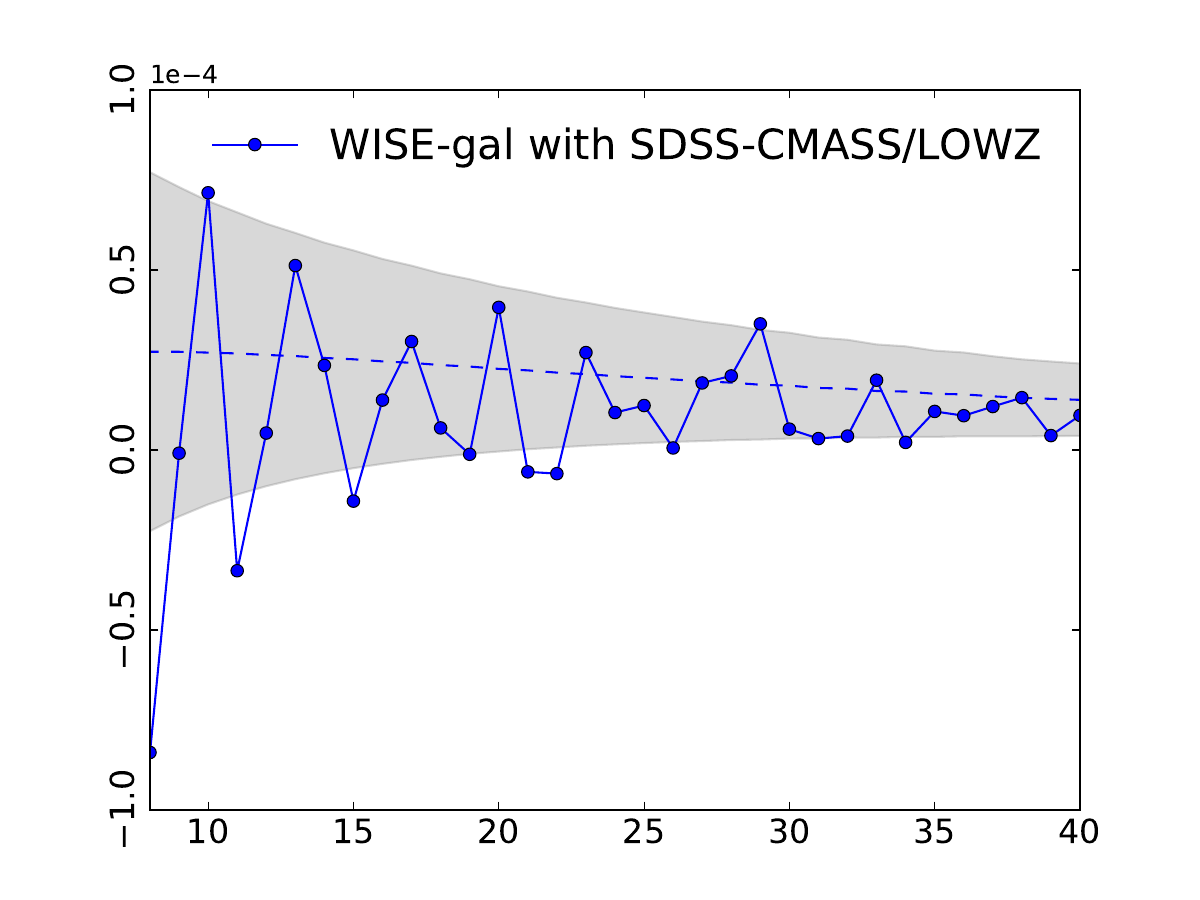}
\includegraphics[width=0.1380\textwidth]{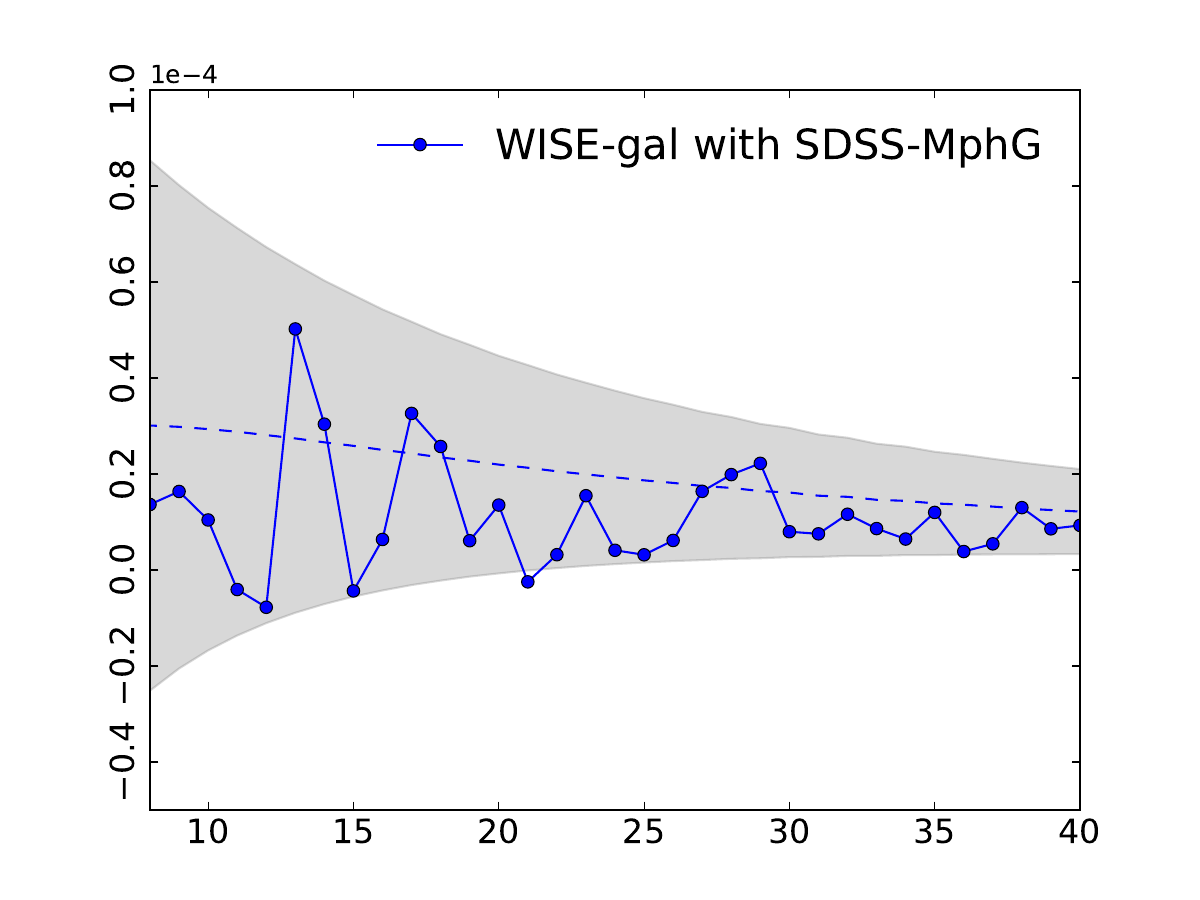}
\includegraphics[width=0.1380\textwidth]{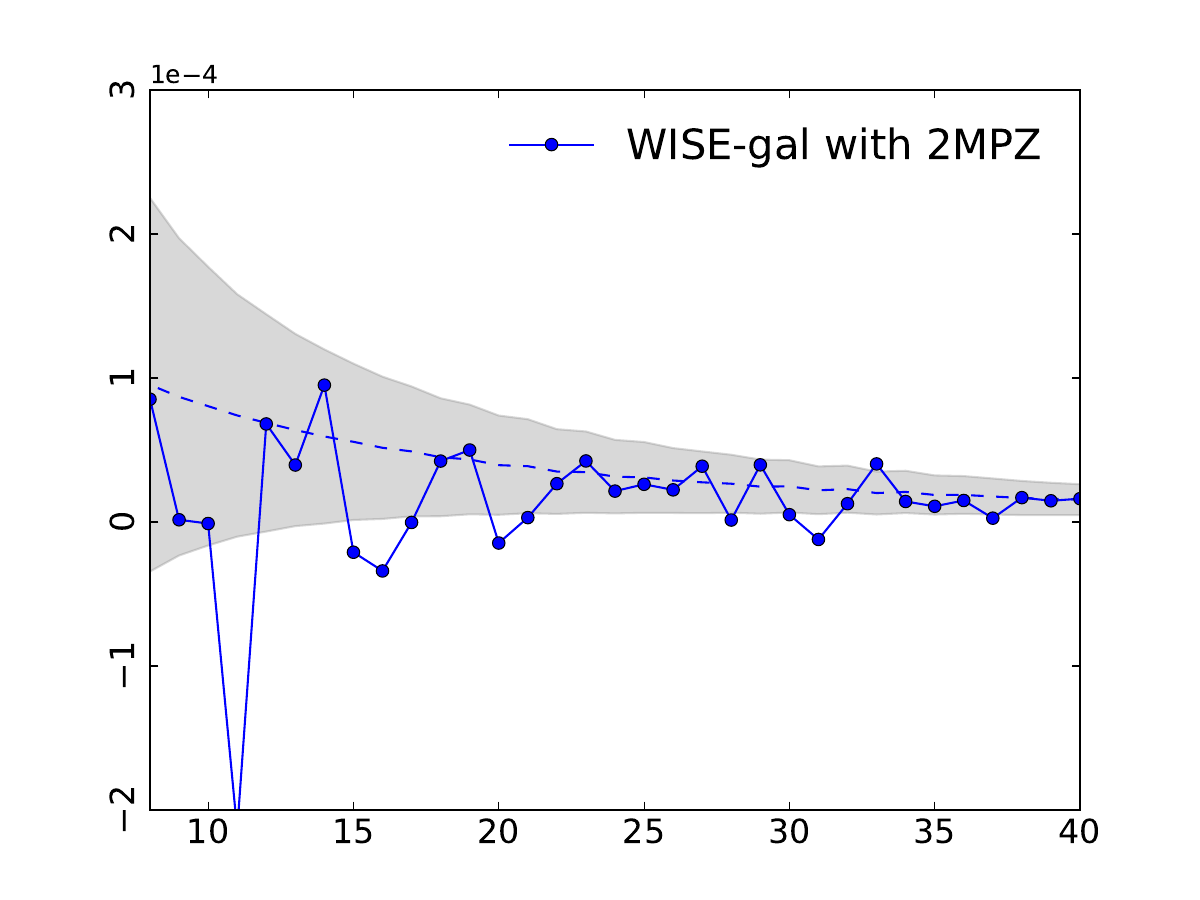}
\includegraphics[width=0.1380\textwidth]{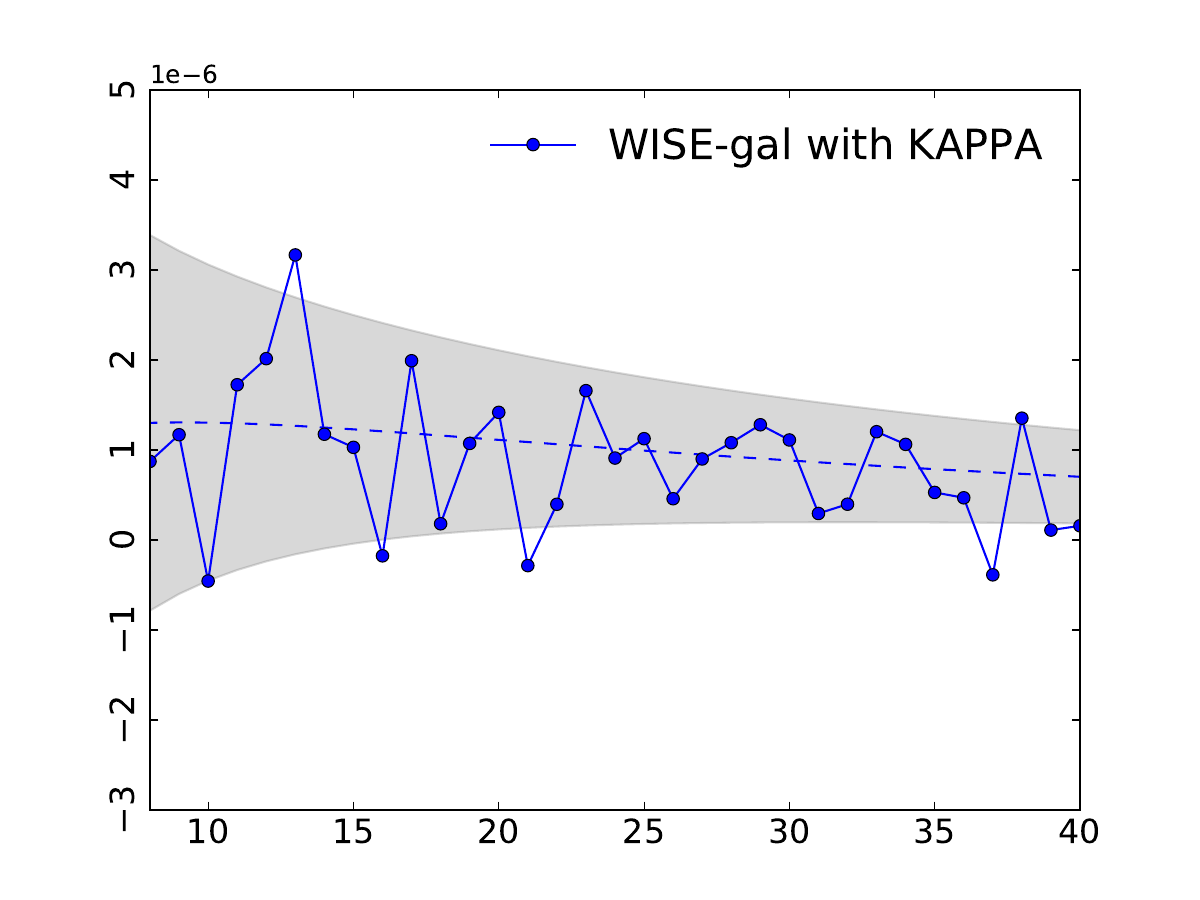}\\
\includegraphics[width=0.1380\textwidth]{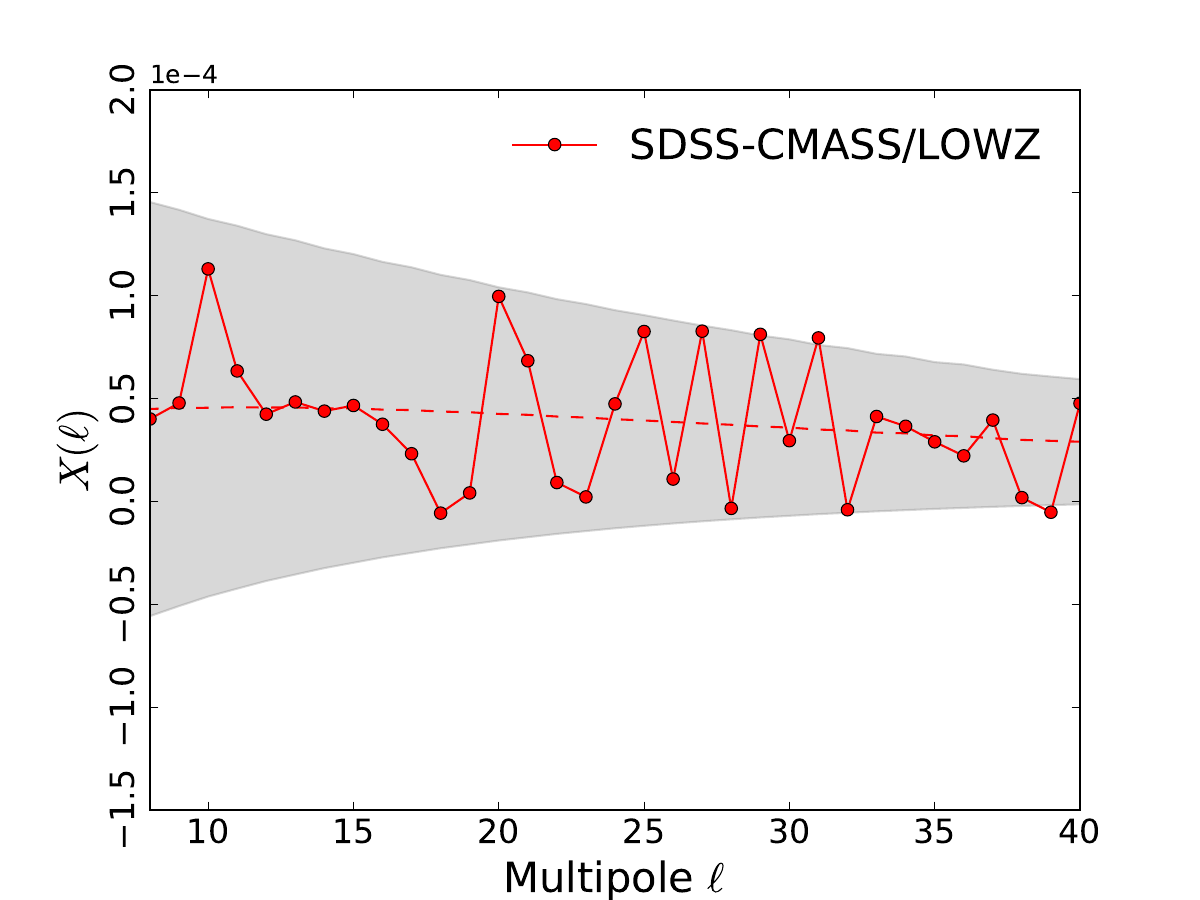}
\includegraphics[width=0.1380\textwidth]{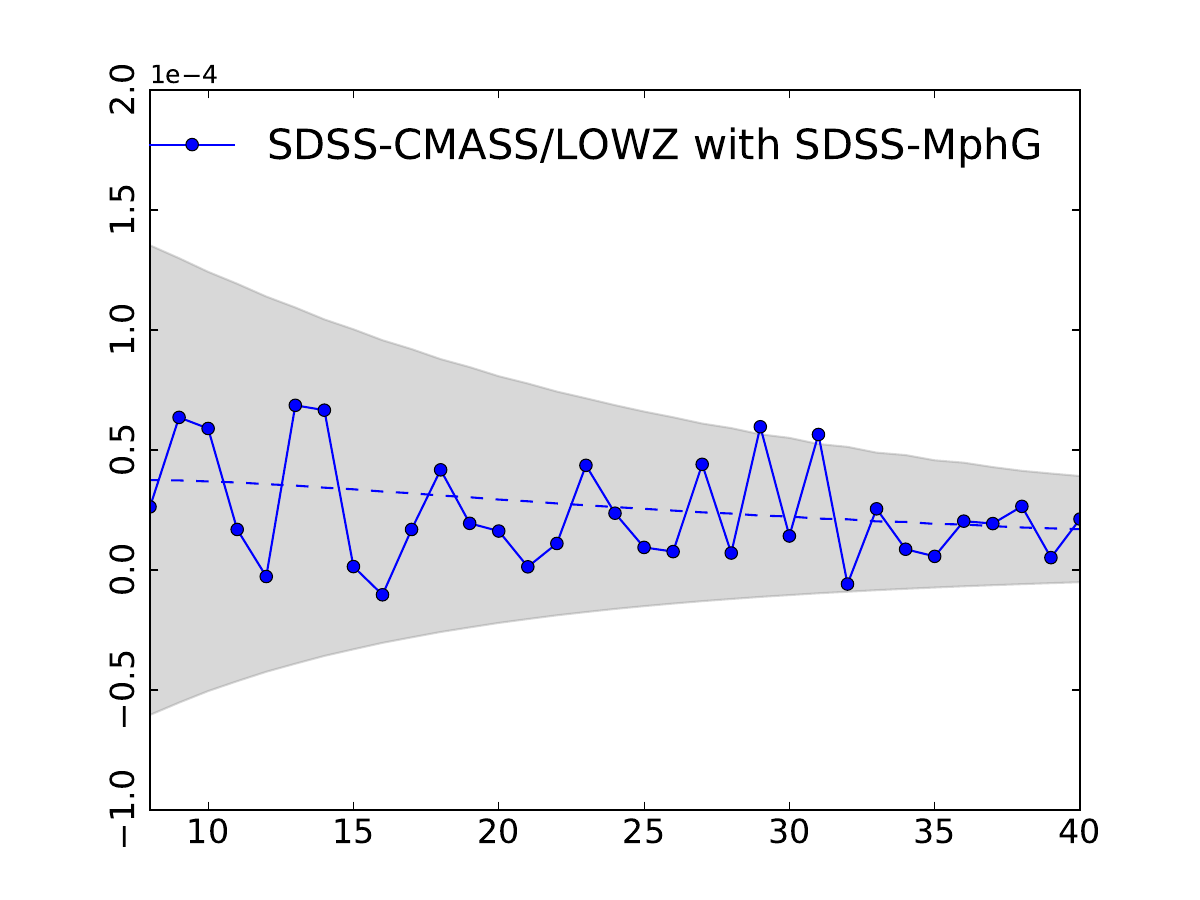}
\includegraphics[width=0.1380\textwidth]{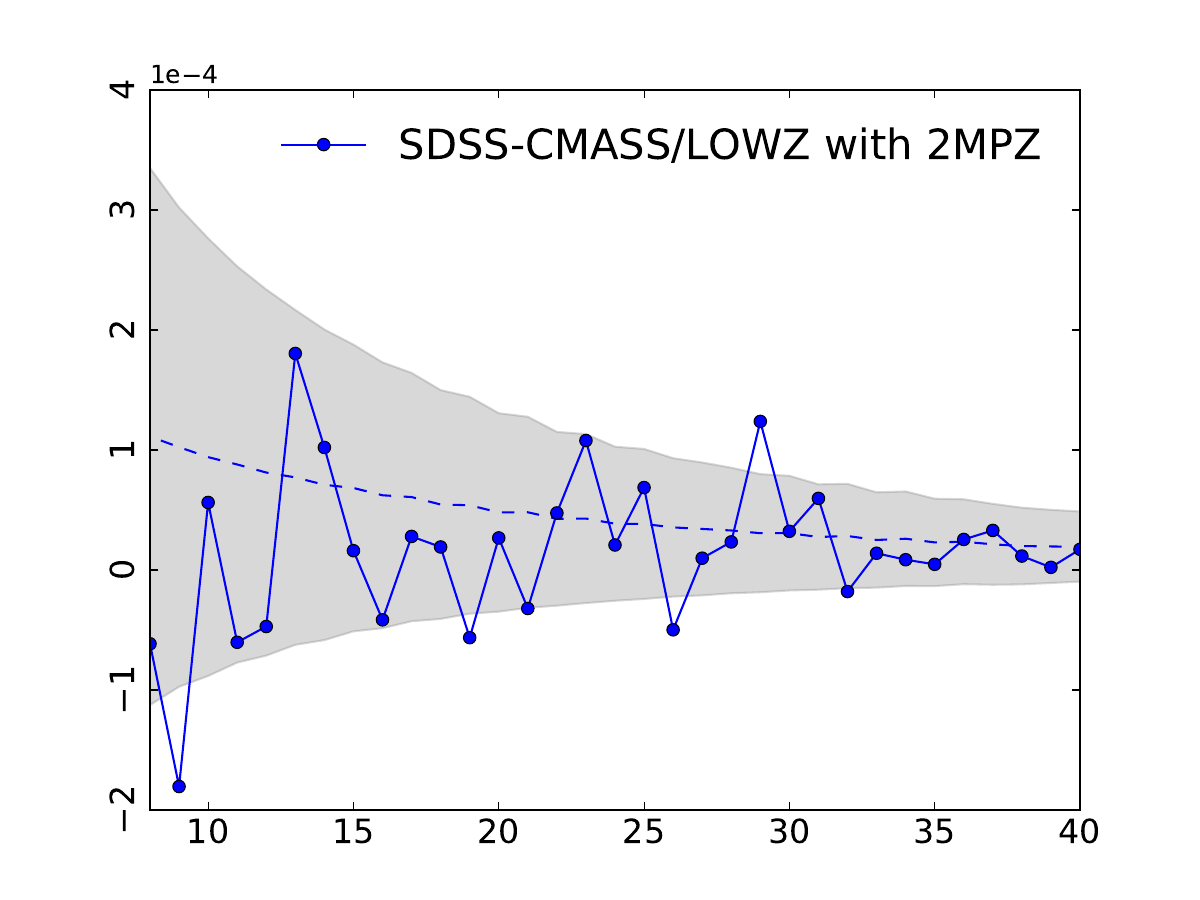}
\includegraphics[width=0.1380\textwidth]{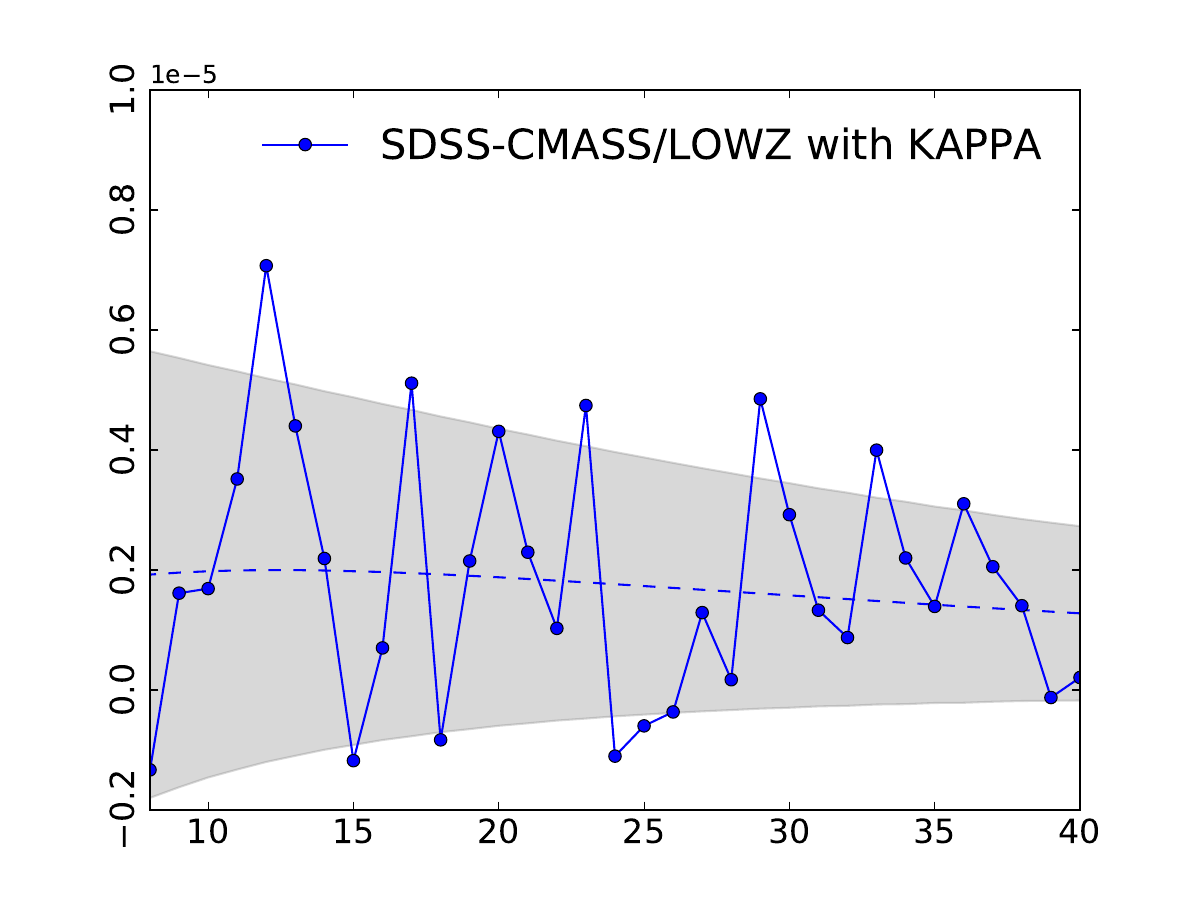}\\
\includegraphics[width=0.1380\textwidth]{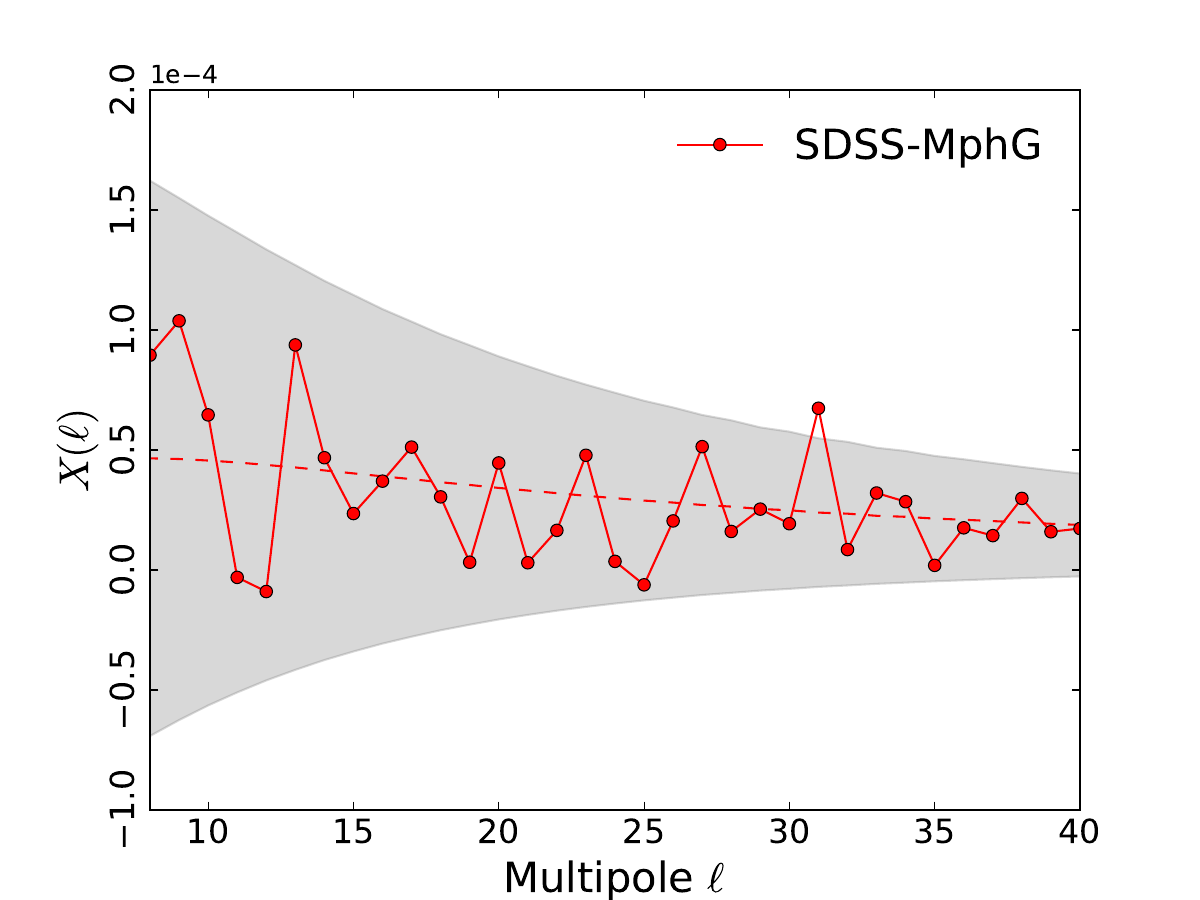}
\includegraphics[width=0.1380\textwidth]{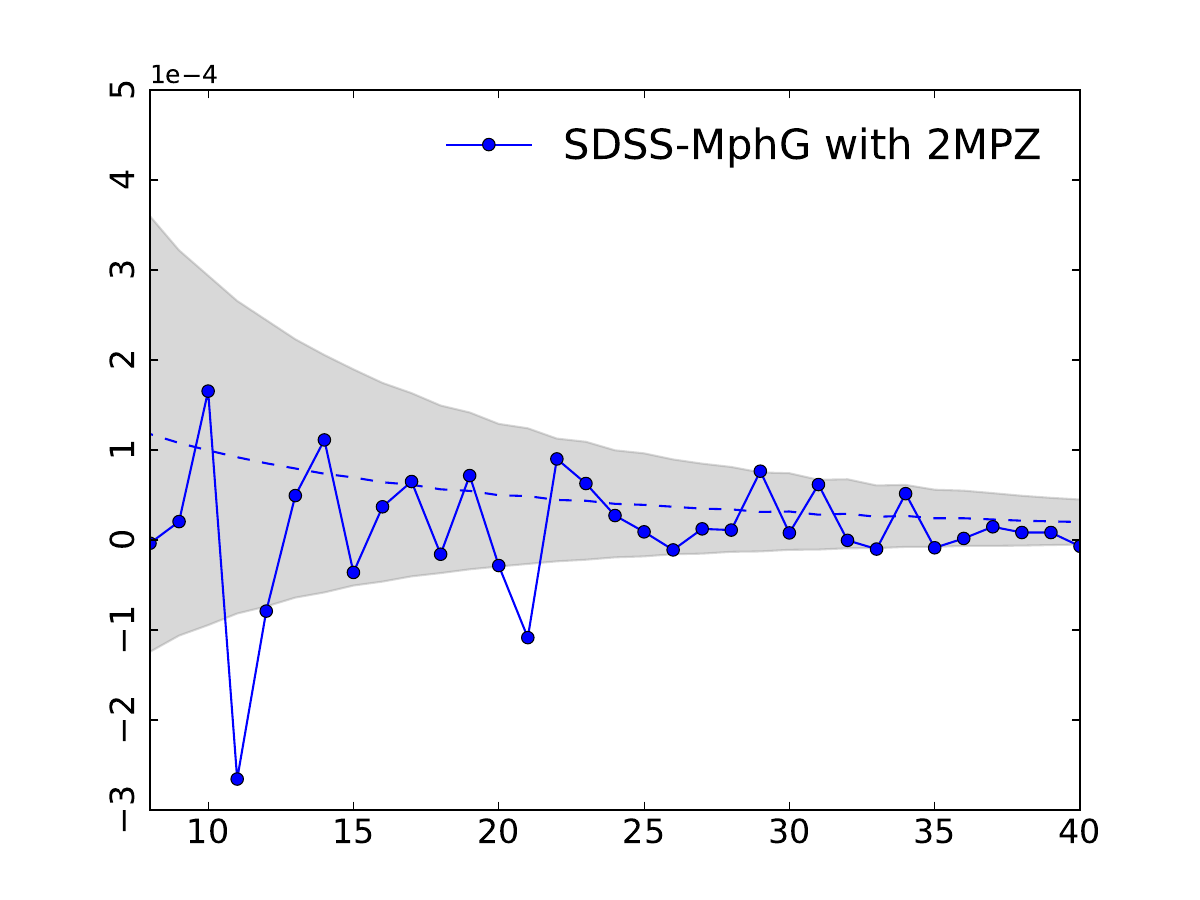}
\includegraphics[width=0.1380\textwidth]{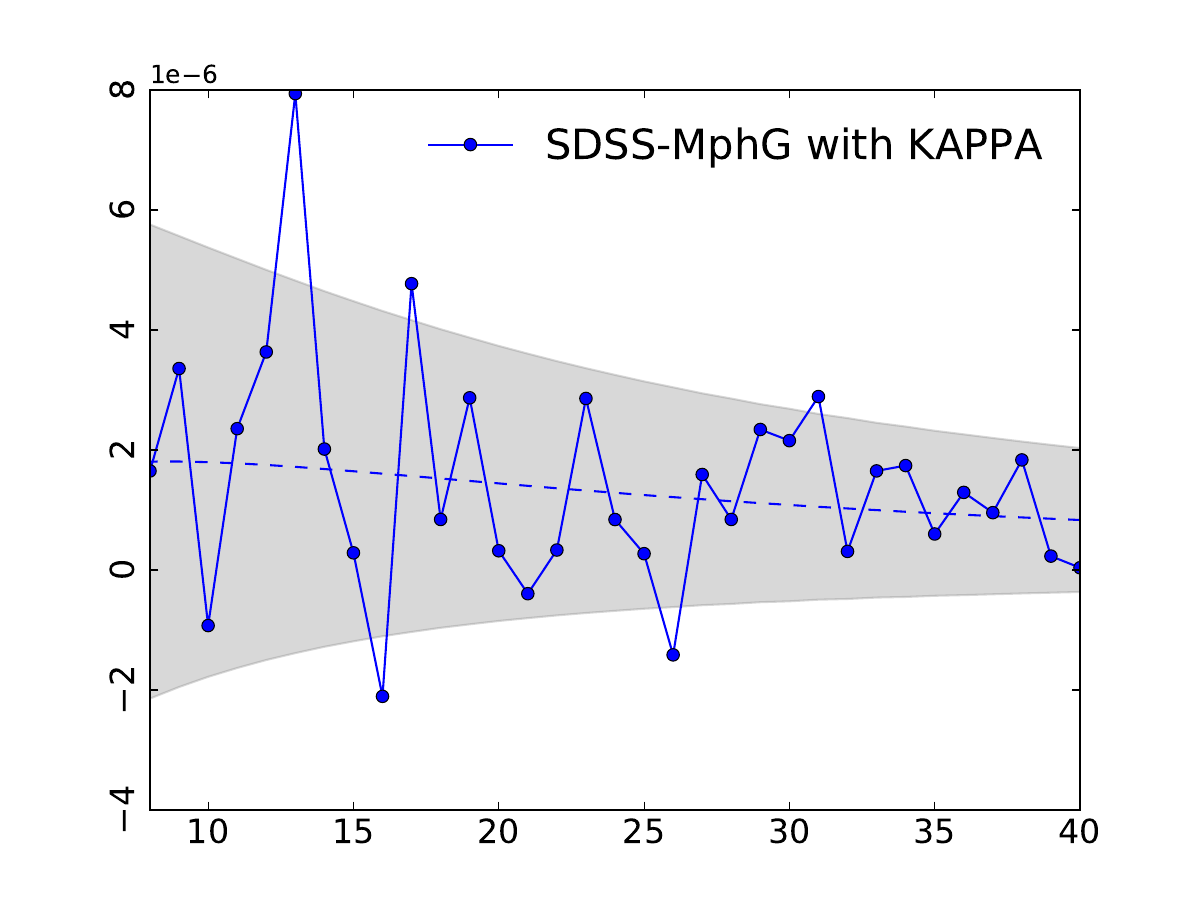}\\
\includegraphics[width=0.1380\textwidth]{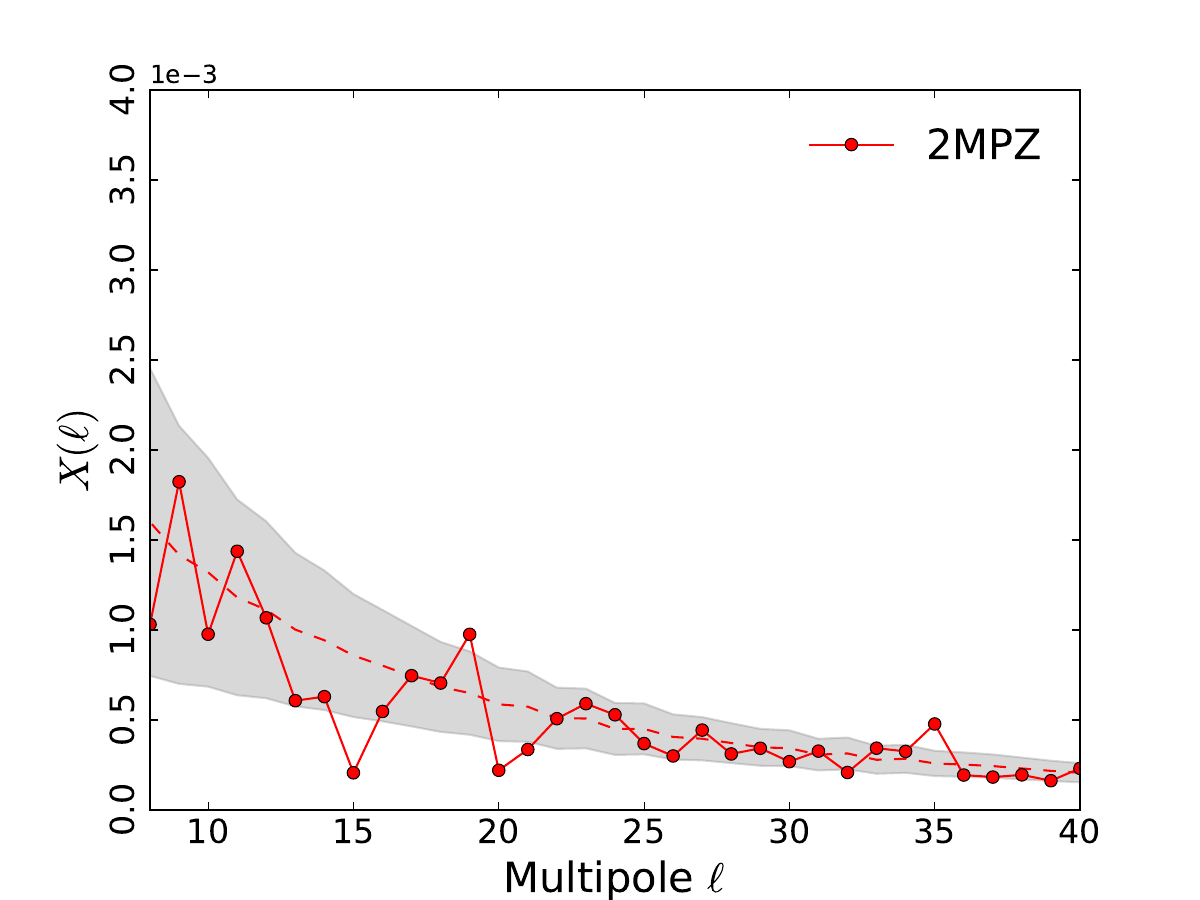}
\includegraphics[width=0.1380\textwidth]{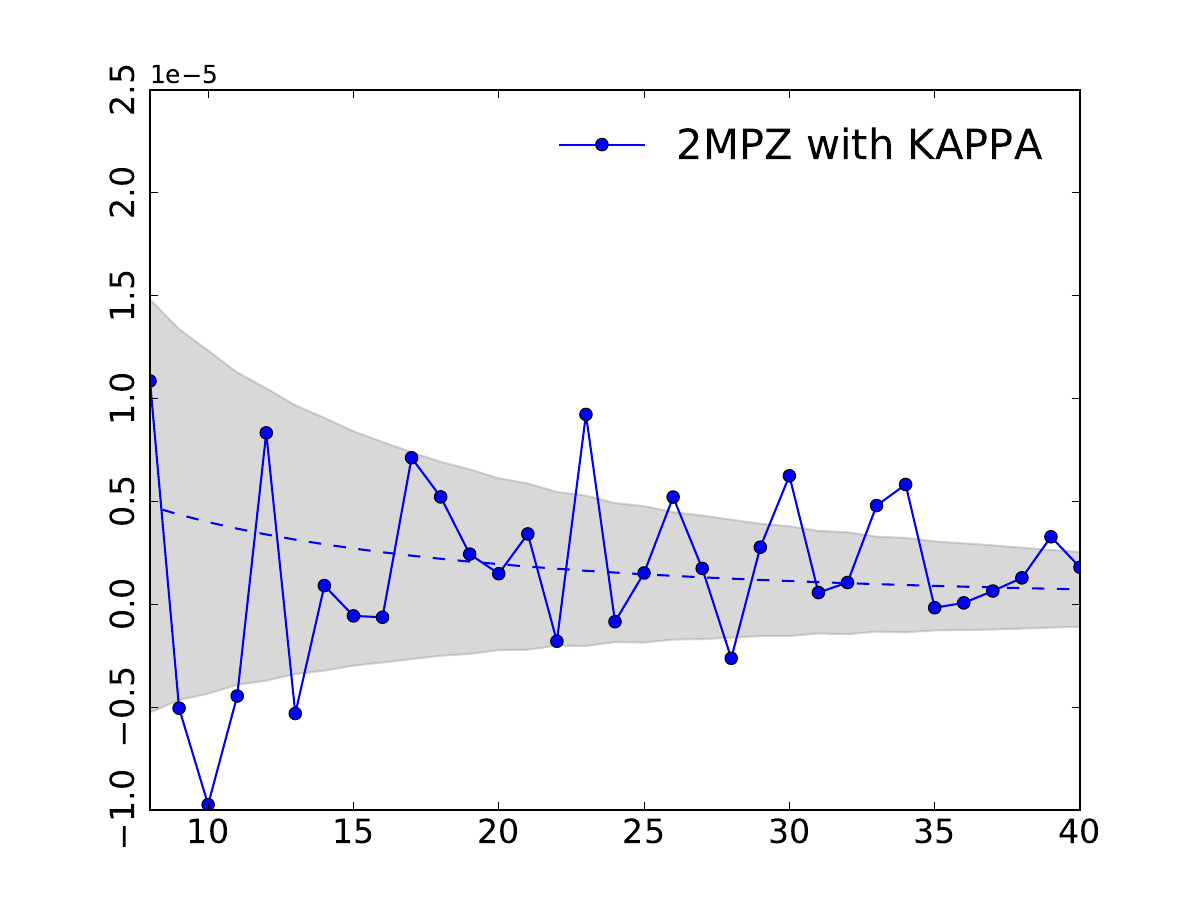}\\
\includegraphics[width=0.1380\textwidth]{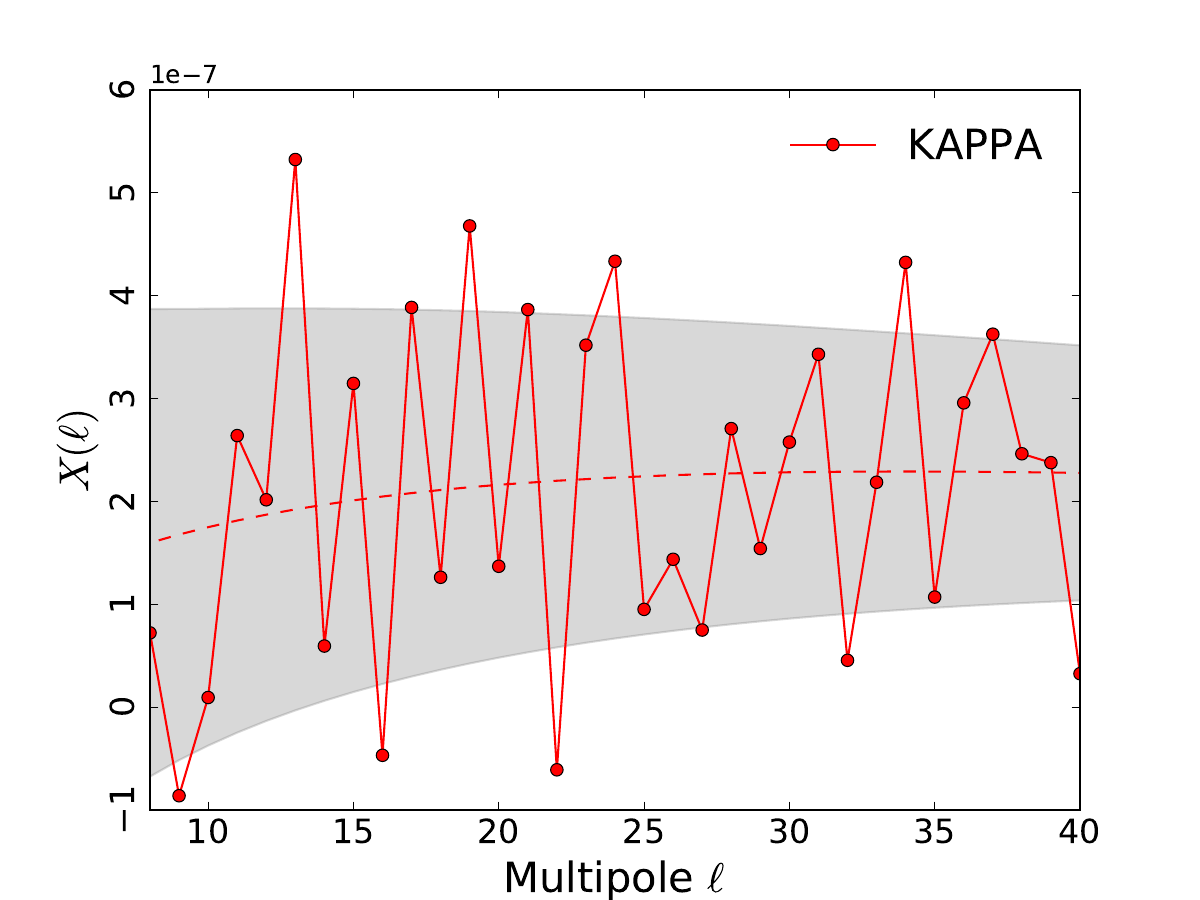}\\
\end{flushright}
\caption{Angular power spectra from the maps in Fig.~\ref{fig:surveys_maps} used to study the ISW effect through the CMB-LSS cross-correlation. From top to bottom: \nvss; \lrg; \mg; \wg; \wagn; \tmpz; and \kap. The observed spectra are the points (red for auto-spectra and blue for the cross-spectra), while the theoretical models are represented by the dashed lines (the grey areas correspond to the sampling variance). 
}\label{fig:surveys_cls}
\end{figure*}

\subsubsection{The \textit{Wide-Field Infrared Survey Explorer} extragalactic catalogues}
\label{sub:wise}

\noindent We next describe the use of the extragalactic sources detected by the textit{Wide-Field Infrared Survey Explorer}  \citep[\wise, see ][]{Wright2010} in our ISW studies. \wise\  scanned the full sky at 3.4, 4.6, 12 and 22\,$\mu$m (which constitute bands W1 to W4). These observations give a deeper view of the infrared sky than previous surveys like 2MASS or IRAS, and provide an extensive extragalactic catalogue of more than 500 million sources \citep[see][]{Wright2010}. The four W1 to W4 bands are sensitive to either UV radiation reproduced by dust grains in star forming galaxies or to infrared emission from stars, either in our Galaxy or in extragalactic sources. The W1 band turns out to be the deepest, sampling the deep Universe by detecting massive galaxies up to $z \approx 1$ and with a median redshift of $0.3$ \citep[][]{Yan2013}

In the context of ISW studies, our approach is very similar to that of \citet{Ferraro2014}. We focus our efforts on two different sets of extragalactic sources: star-forming galaxies; and AGN. This requires a careful separation of the stars in the catalogue. Since the sky scanning of the \wise\ satellite is not homogeneous, a magnitude cut of W1$<16.6$ is imposed on all sources at high galactic latitude, since for this cut \citet{Ferraro2014} found a uniform sample. Given that stray light from the Moon may cause faint detections and other spurious effects in the data, we discard all sources with a flat {\tt moon\_lev} $> 4$, while also dropping all sources suspected of being artefacts ({\tt cc\_flags}$\neq 0$).

Following the colour cuts given in \citet{Yan2013} and \citet{Ferraro2014}, we impose the cut W1-W2$>0$ to isolate the galaxies from stars, while the stricter conditions W1$-$W2$>0.85$ and W2$<$15.0 should separate the AGN from star forming galaxies, although the former constitute a very small fraction of the latter. With these cuts, we obtain about 140 million galaxies and 1.4 million AGN in the entire sky. 

The presence of systematics causes clear excess in the auto-power spectra of these two \wise -based surveys on the largest scales. This poses a problem, since, in order to predict the level of cross-correlation with the CMB maps generated by the ISW component, we need first to characterise the bias of each tracer. For this purpose we follow exactly the same approach as in \citet{Ferraro2014}: we use the cross-correlation of these two galaxy surveys with lensing convergence maps from \Planck\ in order to estimate the bias. This requires adopting some models for the redshift distribution of \wise\ galaxies and AGN, and in particular we use those given in \citet{Yan2013}, which were obtained after cross-matching \wise\ and \sdss\ data. We also adopt some redshift dependences for the bias that are identical to those used in \citet{Ferraro2014}. For \wg\ our fiducial model has this very simple redshift dependence:
\begin{linenomath*}
\begin{equation}
b^\mathrm{\wg}\left(z\right) = b_0^\mathrm{\wg} \left(1+z\right).
\label{eq:bwisegals}
\end{equation}
\end{linenomath*}
For the \wagn\ the suggested redshift dependence is quadratic:
\begin{linenomath*}
\begin{equation}
b^\mathrm{\wagn}\left(z\right) = b_0^\mathrm{\wagn} [0.53 + 0.289 (1+z)^2].
\label{eq:bwiseagn}
\end{equation}
\end{linenomath*}
After cross-correlating with \Planck\ lensing convergence maps in the multipole range $\ell=10 - 400$, we find the following values for the fiducial parameters: $b_0^\mathrm{\wg} \simeq 0.79$; and $b_0^\mathrm{\wagn} \simeq 0.90 $.

\subsubsection{The 2MASS Photometric Redshift catalogue}
\label{sub:2mpz}

\noindent The 2MASS Photometric Redshift catalogue \citep[hereafter 2MPZ][]{Bilicki2014} is a combination of the 2MASS XSC \citep[][]{Jarrett2000}, \wise, and SuperCOSMOS \citep[][]{Hambly2001} surveys, yielding an all sky extragalactic source catalogue with a typical uncertainty in redshift of $\sigma_z = 0.016$. This is achieved by employing an artificial neural network approach \citep[the ANNz algorithm, see][]{Collister2004} on the above quoted surveys, and after training it with the 2MRS \citep[][]{Huchra2012}, SDSS\footnote{\url{http://www.sdss.org}}, 6dFGS \citep[]{Jones2009}, 2dFGRS\footnote{\url{http://www2.aao.gov.au/\~TDFgg/}}, and ZCAT \citep[]{Huchra1995} spectroscopic surveys. 

The resulting catalogue contains almost one million sources with a median redshift of 0.09. Out of those sources, more than three hundred thousand contain spectroscopic redshifts. In Fig.~\ref{fig:surveys_dndz} we display the histogram of the photometric redshift distribution of these sources, with a high redshift tail extending up to $z \approx 0.3$. When using {\tt MASTER} \citep[][]{Hivon2002} to compute its angular power spectrum, we find that its effective linear bias for large scales ($\ell<70$) is close to unity ($b\simeq 1.35$). However, at small scales there is clear evidence for non linear power.

\subsubsection{\Planck\ lensing map}
\label{sub:lens}

\noindent The clustered matter forming the cosmic web modifies the 2D projected distribution of the CMB anisotropies on the sky, via the weak gravitational lensing effect. This distortion breaks the isotropy of the intrinsic CMB fluctuations, introducing correlations among multipoles. Optimal inversion methods~\citep{Hu2002b,Okamoto2003} allows us to recover the projected density field ($\phi$), which is proportional to the gravitational field ($\Phi$). 

As part of its official products release, \Planck\ provides a map of the estimated lensing field~\citep{planck2013-p12,planck2014-a17}, which can be used to probe the ISW effect, through its cross-correlation with the CMB map, in the same manner as is done with external galaxy catalogues. The lensing signal measured from this map is detected at about $40\,\sigma$, using the full-mission temperature and polarization data to construct the lensing map estimation.

In fact, what \Planck\ releases is the lensing convergence map ($\kappa$), which has a whiter angular power spectrum than the raw lensing potential ($\phi$): $\kappa_{\ell m} = \phi_{\ell m} \ell(\ell+1)/2$. This lensing map was obtained from the \smica\ CMB solution, and it is shown (Wiener-filtered) in Fig.~\ref{fig:surveys_maps}, bottom row. It covers 67\,\% of the sky, and has a multipole range $8 \leq \ell \leq 2048$. 
The convergence map traces the matter distribution through a wide redshift range (see Fig.~\ref{fig:surveys_dndz}).

\subsubsection{The \textit{Sloan Digital Sky Survey} superstructures}
\label{sub:granett}

\noindent We use here the catalogue of superstructures\footnote{\url{http://ifa.hawaii.edu/cosmowave/supervoids/}} from \citet{Granett2008a}, also used in \cite{planck2013-p14}. This sample consists of 50 superclusters and 50 supervoids identified from the Luminous Red Galaxies (LRGs) in the SDSS (sixth data release, DR6, \citealt{Adelman2008}), which covers an area of $7500\,{\rm deg}^2$ on the sky. These authors used publicly available algorithms, based on the Voronoi tessellation, to find 2836 superclusters \citep[using VOBOZ, VOronoi BOund Zones,][]{Neyrinck2005} and 631 supervoids \citep[using ZOBOV, ZOnes Bordering On Voidness,][]{Neyrinck2008} above a $2\,\sigma$ significance level (defined as the probability of obtaining, in a uniform Poissonian point sample, the same density contrasts as those of clusters and voids).

The 50 superclusters and 50 supervoids published in the \citet{Granett2008a} catalogue correspond to density contrasts of about $3\,\sigma$ and $3.3\,\sigma$, respectively. They span a redshift range of $0.4<z<0.75$, with a median of around 0.5, and inhabit a volume of about 1.6\,Gpc$^3$. These superstructures can potentially produce measurable ISW signals, as suggested in \cite{Granett2008a, Granett2008b}. For each structure, the catalogue provides: the position on the sky of its centre; the mean and maximum angular distance between the galaxies in the structure and its centre; the physical volume; and three different measures of the density contrast (calculated from all its Voronoi cells, from only its over- or under-dense cells, and from only its most over- or under-dense cell).

\subsection{Simulations}
\label{sec:simulations}
\label{sec:data_sims}

We have performed a set of 11000 correlated simulations of the CMB and different LSS tracers, which are used to study the CMB-LSS cross-correlation up to $\ell \approx 190$. In fact, the maps are produced directly at \nside = 64, since we have checked that these maps already capture all the information required for our analyses. From the CMB side, each simulation consists of two independent signals: an ISW map; and the rest of CMB anisotropies in a \tp\ map. Notice that, since no polarization information at the largest scales is provided in this release, this signal is not generated. At these large scales, instrumental noise is negligible and, therefore, we have not included it in the simulations.
From the galaxy-surveys side, we simulate galaxy density maps for \nvss, \lrg, \mg, \wagn, and \wg.  Although it is not used for studying the CMB-LSS cross-correlation, we also generate a \tmpz\ map that helps to assess the quality of the ISW recovery from this photometric catalogue (see Sect.~\ref{sec:recov}).
Shot noise is added to the simulated galaxy density maps, accordingly to the corresponding  mean number of galaxies per pixel given in Table~\ref{tab:surveys}.
Finally, we also produce a coherent lensing convergence (\kap) map. Noise is also added to this map, following the uncertainty level estimated in the \Planck\ lensing paper~\citep{planck2014-a17}.
 
The maps are simulated by assuming that the perturbations are Gaussian. This is a good approximation for the \kap\ map, as well as for all the distributed galaxy density maps. In particular, for the galaxy catalogues, although they follow a Poisson distribution, the mean number of galaxies per pixel is large enough (about $40$, for the worst case, \tmpz). Therefore, all the required information to perform the coherent simulations is given by all the angular auto- and cross- power spectra~\cite[see, for instance,][for details]{Barreiro2008}. In particular, given two surveys $a$ and $b$, the theoretical cross-power spectra between the surveys read as:
\begin{linenomath*}
\begin{equation}
C^{ab}_\ell = 4 \pi \int \frac{\mathrm{d}k}{k} \ \Delta^2(k) I^a_\ell(k)
I^b_\ell(k) \ ,
\end{equation}
\end{linenomath*}
where $\Delta^2(k)$ is the matter power spectrum per logarithmic interval, and $I_{\ell}^a(k)$ is a transfer function represented by the redshift integral:
\begin{linenomath*}
\begin{equation}
I^a_{\ell}(k) = \int_0^\infty \mathrm{d}z \ W_a(z,k) j_\ell(kr(z)) \ .
\end{equation} 
\end{linenomath*}
Here $r(z)$ is the comoving distance as a function of the redshift, and $j_{\ell}$ are the spherical Bessel functions, which project the window function $W_a (z,k)$ into each multipole ${\ell}$ of the power spectrum. In the case of a galaxy survey, the window function is independent of $k$ and is given by
\begin{linenomath*}
\begin{equation}
W_a(z) = b_a(z) D_+(z) \frac{\mathrm{d}n_a}{\mathrm{d}z} \ .
\end{equation}
\end{linenomath*}
This depends on the galaxy redshift distribution and the bias function $b_a(z)$ of each survey. The growth factor $D_+(z)$ in this expression takes into account the linear evolution of the matter perturbations. 

For lensing, the efficiency window function $W_\kappa(\chi)$, which relates the density perturbations $\delta$ to the weak lensing convergence \kap\ in a line of sight integration,
\begin{linenomath*}
\begin{equation}
\kappa = \int\mathrm{d}\chi\:W_\kappa(\chi)\delta,
\end{equation}
\end{linenomath*}
is given by
\begin{linenomath*}
\begin{equation}
W_\kappa(\chi) = \frac{3\Omega_m}{2\chi_H^2}\frac{D_+}{a}\frac{\chi}{\chi_\mathrm{CMB}}\left(\chi_\mathrm{CMB}-\chi\right),
\end{equation}
\end{linenomath*}
with the comoving distance $\chi_\mathrm{CMB}$ to the surface of last scattering, which is approximately $15~\mathrm{Gpc}$ in the $\Lambda$CDM cosmology considered here and $\chi_H=c/H_0$ is again the Hubble distance. In order to compare the lensing efficiency function $W_\kappa(\chi)$ to the redshift distributions of other LSS surveys, we convert it by analogy into a dimensionless function by multiplying it by the inverse Hubble parameter:
\begin{linenomath*}
\begin{equation}
W_\kappa(z)\mathrm{d}z = W_\kappa(\chi)\mathrm{d}
\chi\,\rightarrow\,W_\kappa(z) = W_\kappa(\chi)\frac{\mathrm{d}\chi}{\mathrm{d}z} = W_\kappa(\chi)\frac{c}{H(z)}.
\end{equation}
 \end{linenomath*}

For the ISW effect, the window function involves the evolution of the potential with redshift:
\begin{linenomath*}
\begin{equation}
W_\mathrm{ISW}(z,k) = -3 \Omega_\mathrm{m} \left( \frac{H_0}{ck} \right)^2
\frac{\mathrm{d}}{\mathrm{d}z} \left[ (1+z) D_+(z) \right] \ .
\end{equation}
\end{linenomath*}
This depends on $k$ due to the Poisson equation relating the matter and the potential. If the Universe is matter-dominated, then the function $(1 + z)D_+(z)$ is constant and the ISW effect vanishes. 
\\
 
All the angular power spectra used in the present paper have been calculated using a modified version of the {\tt CAMB}\footnote{\url{http://camb.info}} code. The fiducial $\Lambda$CDM cosmological model assumed is: $\Omega_\mathrm{b} h^2 = 0.0222$, $\Omega_\mathrm{c} h^2 = 0.119$, $\Omega_\nu h^2 = 0$, $\Omega_K = 0$, $n_\mathrm{s} = 0.9615$, $A_\mathrm{s} = 2.1740\times10^{-9}$, $\tau=0.077$, and $h=H_0/100\,\, \mathrm{km\,s}^{-1}\mathrm{Mpc}^{-1}$, fully compatible with the \Planck\ fiducial model~\citep{planck2014-a15}.

%
\section{CMB correlation with tracers of the gravitational potential}
\label{sec:xcorr}
\begin{figure*}
\begin{center}
\includegraphics[width=0.325\textwidth]{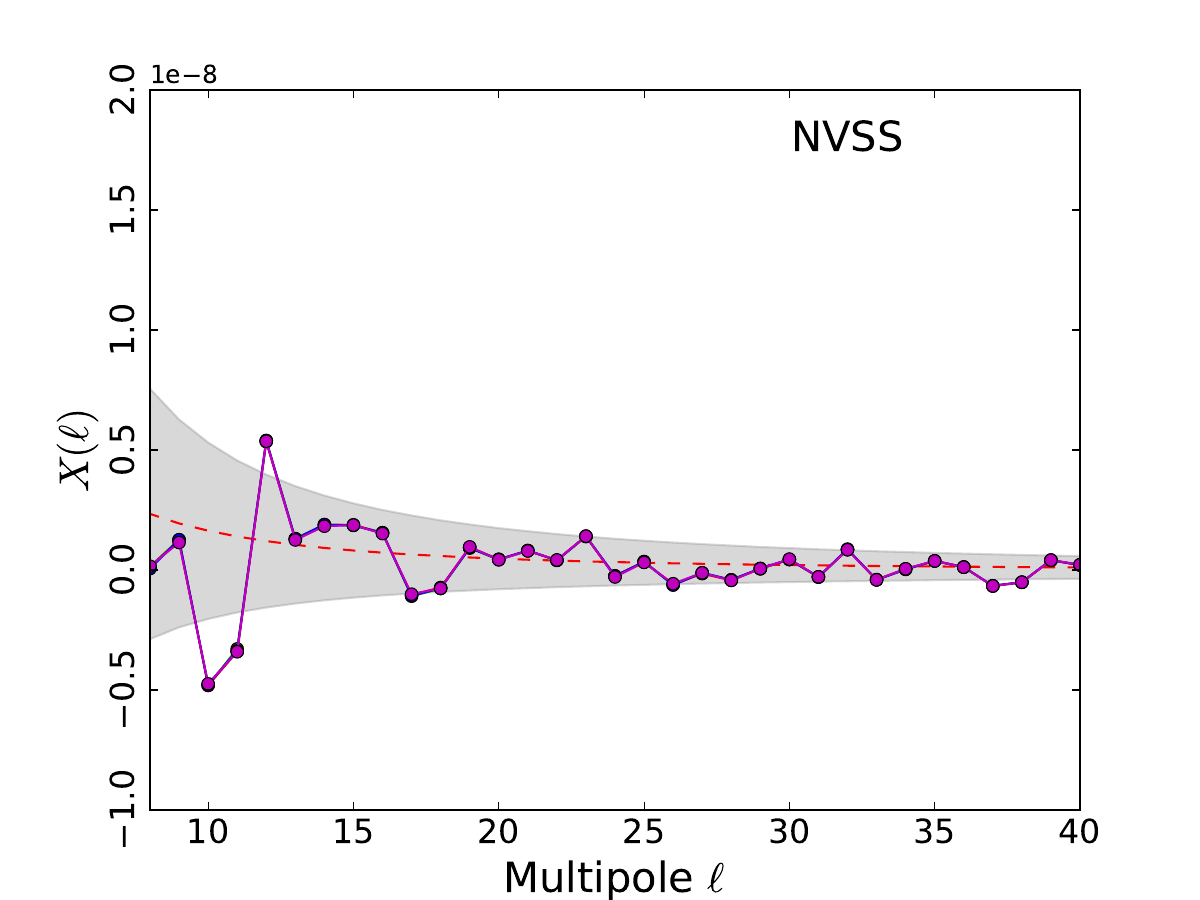}
\includegraphics[width=0.325\textwidth]{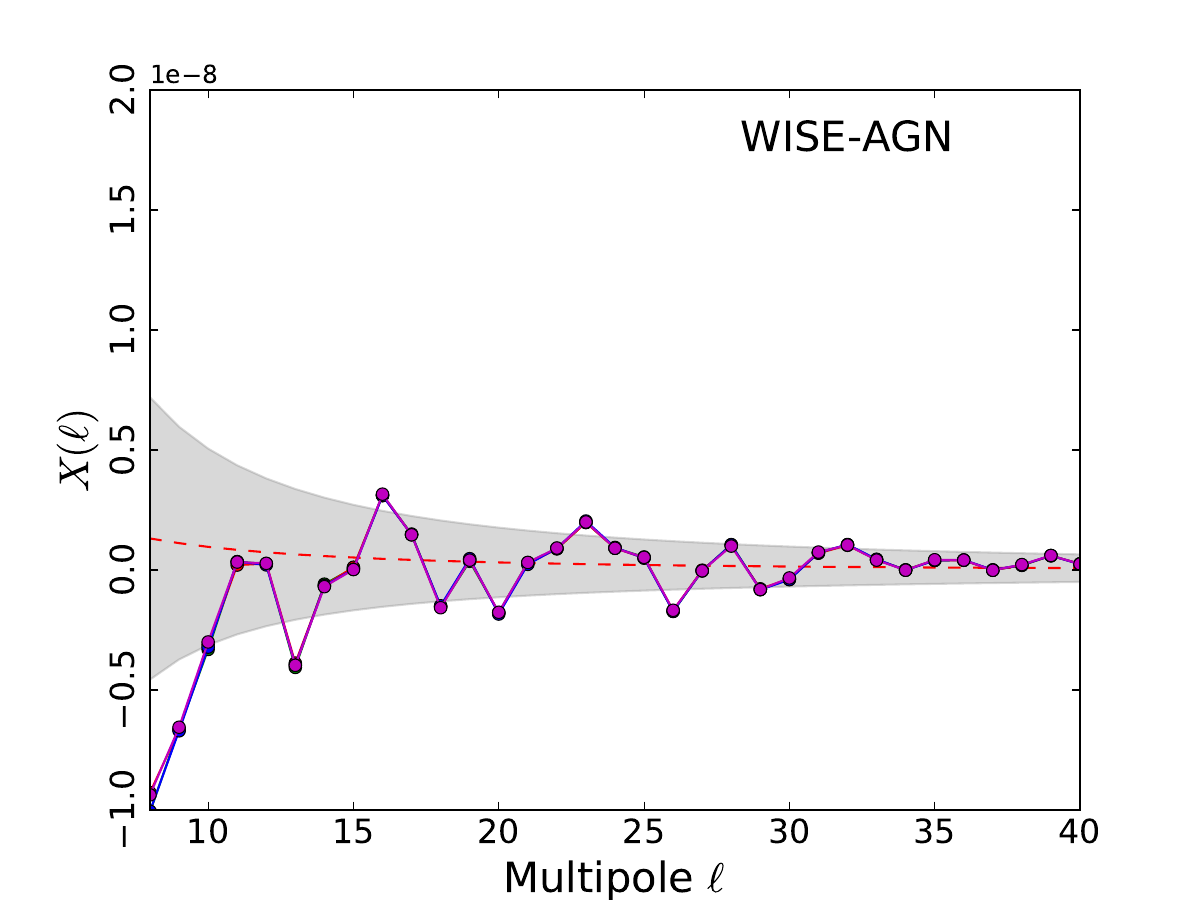}
\includegraphics[width=0.325\textwidth]{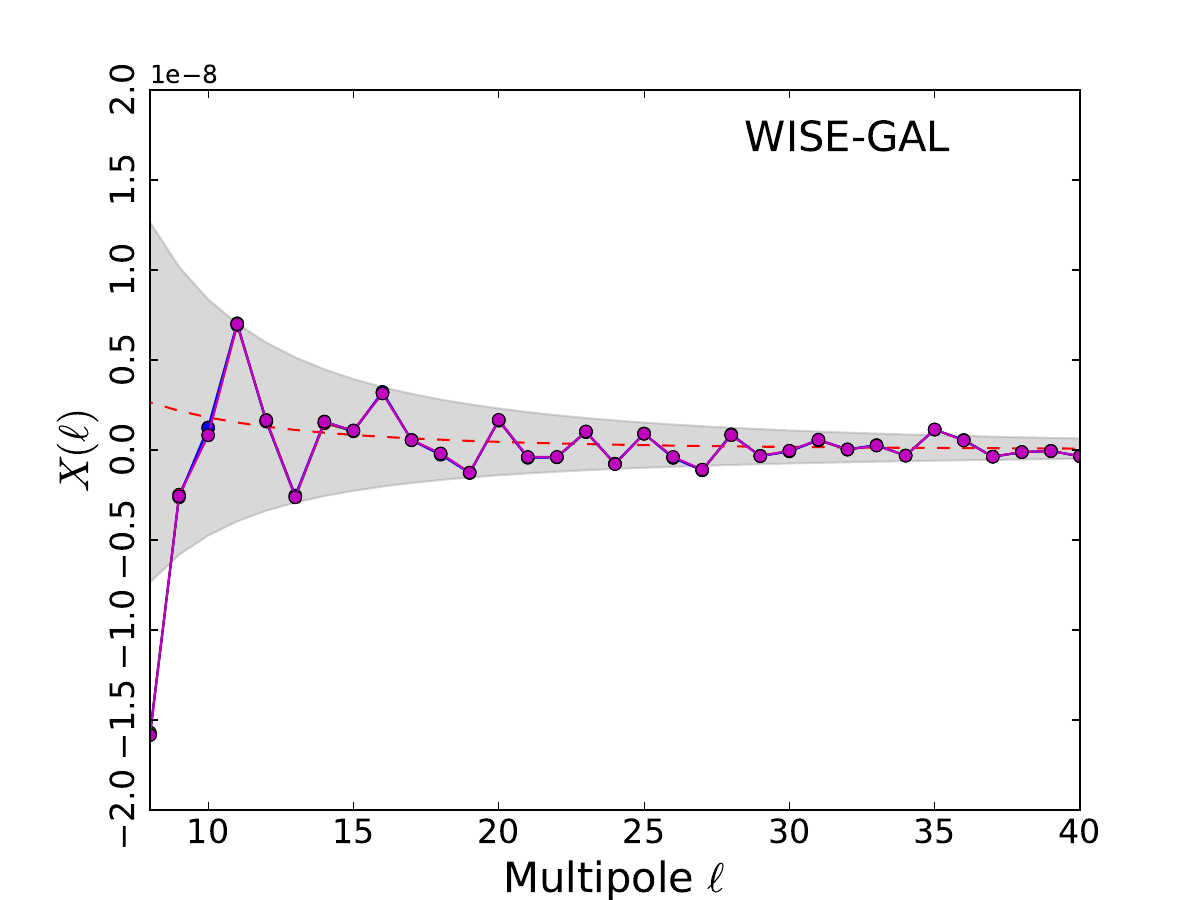}\\
\includegraphics[width=0.325\textwidth]{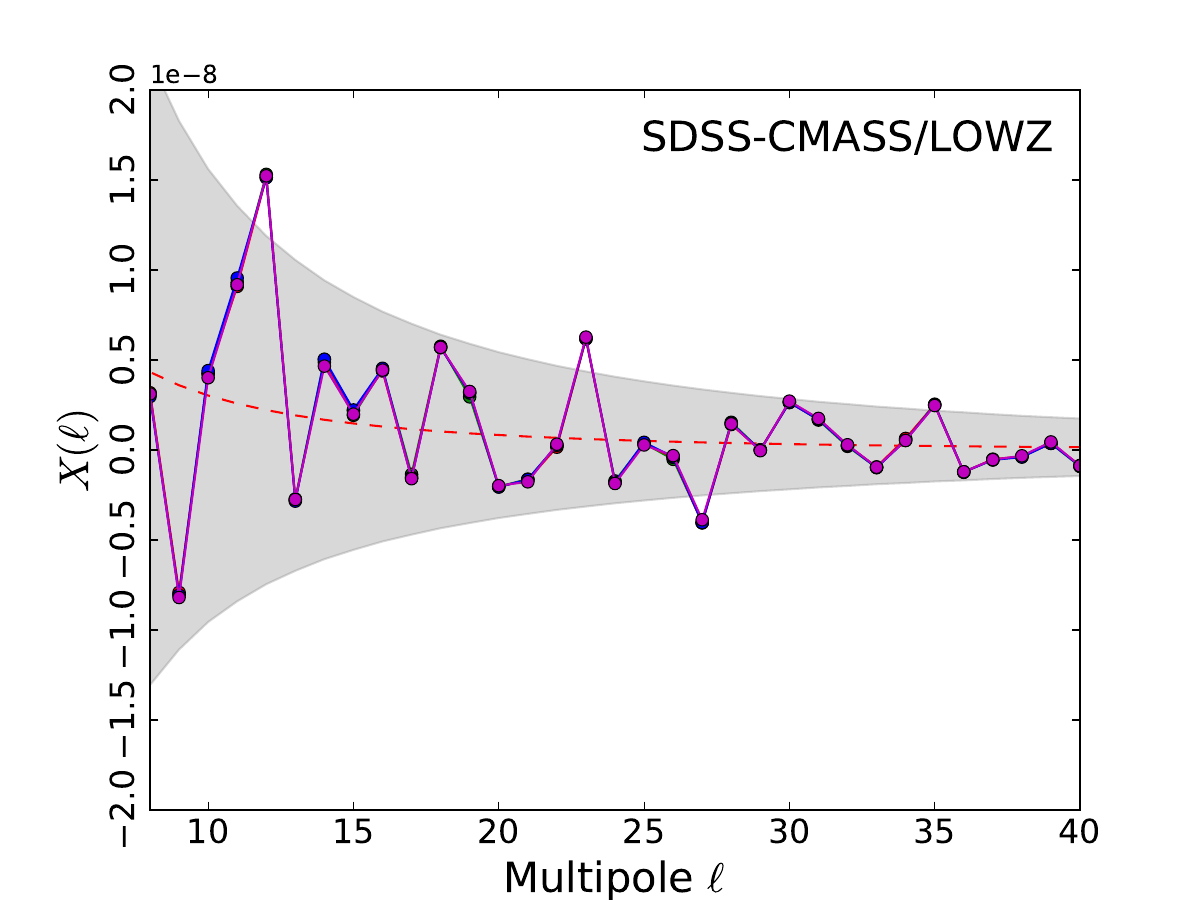}
\includegraphics[width=0.325\textwidth]{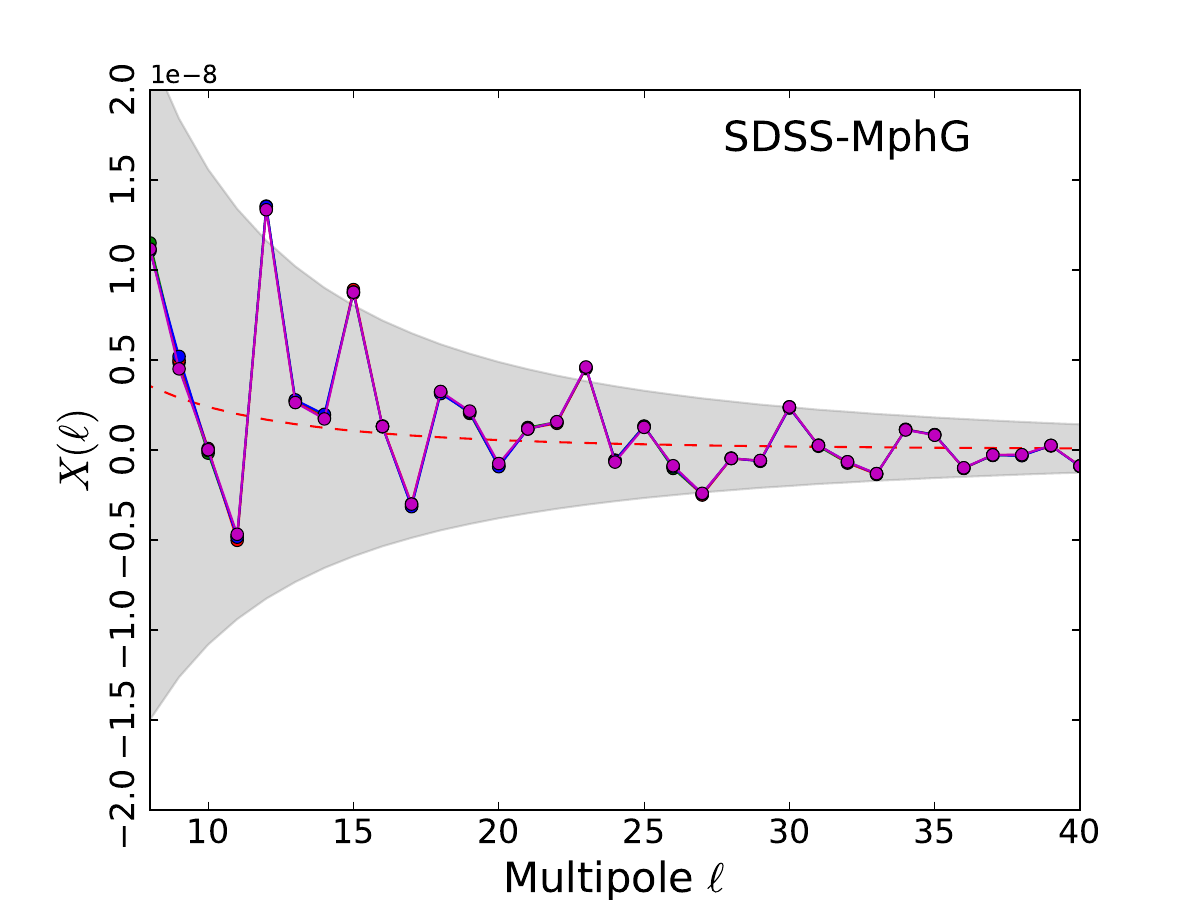}
\includegraphics[width=0.325\textwidth]{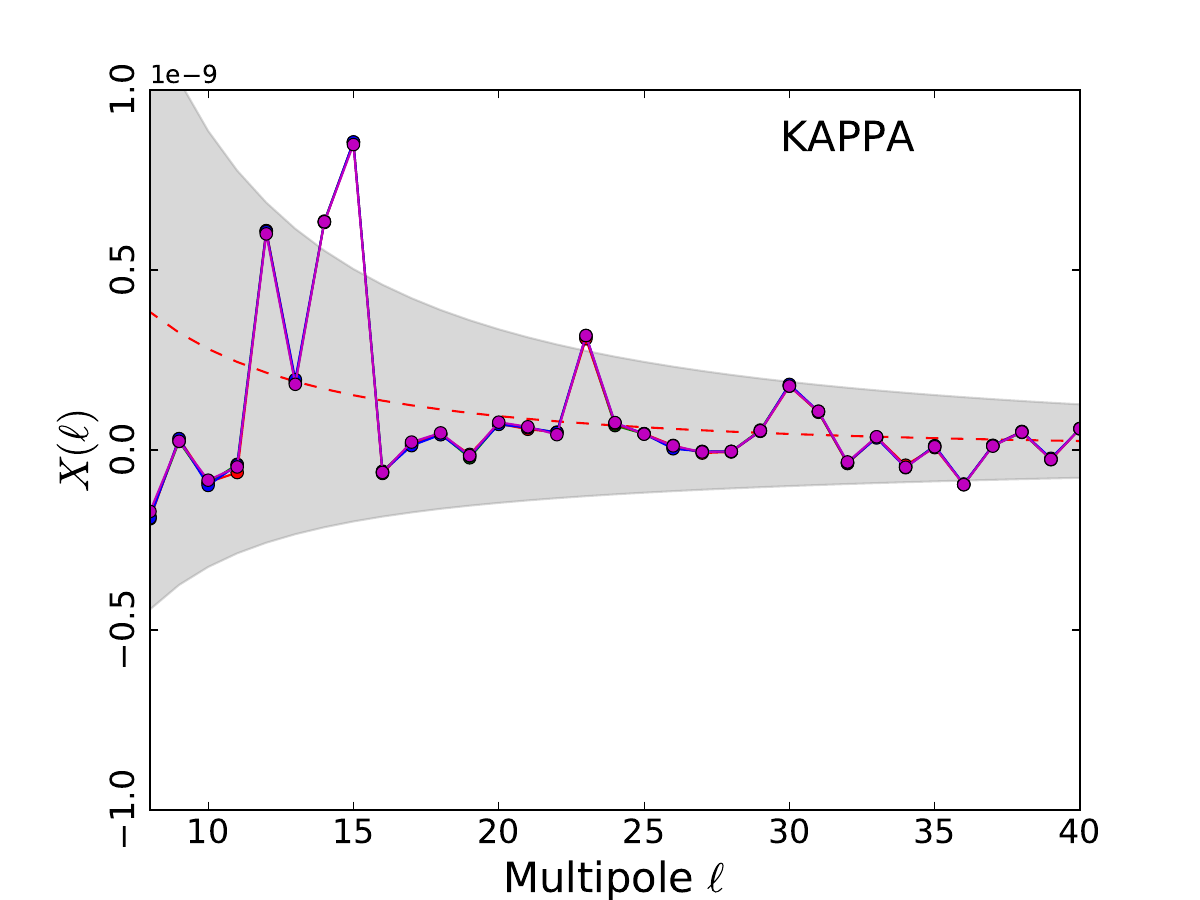}
\end{center}
\caption{\label{fig:xl_data} Measured ISW-LSS cross-spectra (CAPS). From left to right, snd top to bottom, the panels show the cross-correlation of the four CMB maps with \nvss, \wagn, \wg, \lrg, \mg, and \kap. Grey areas represent $\pm 1\,\sigma$ uncertainities derived from simulations. Spectra derived from the  \Planck\ CMB maps are virtually the same.}
\end{figure*}
\begin{figure*}
\begin{center}
\includegraphics[width=\textwidth]{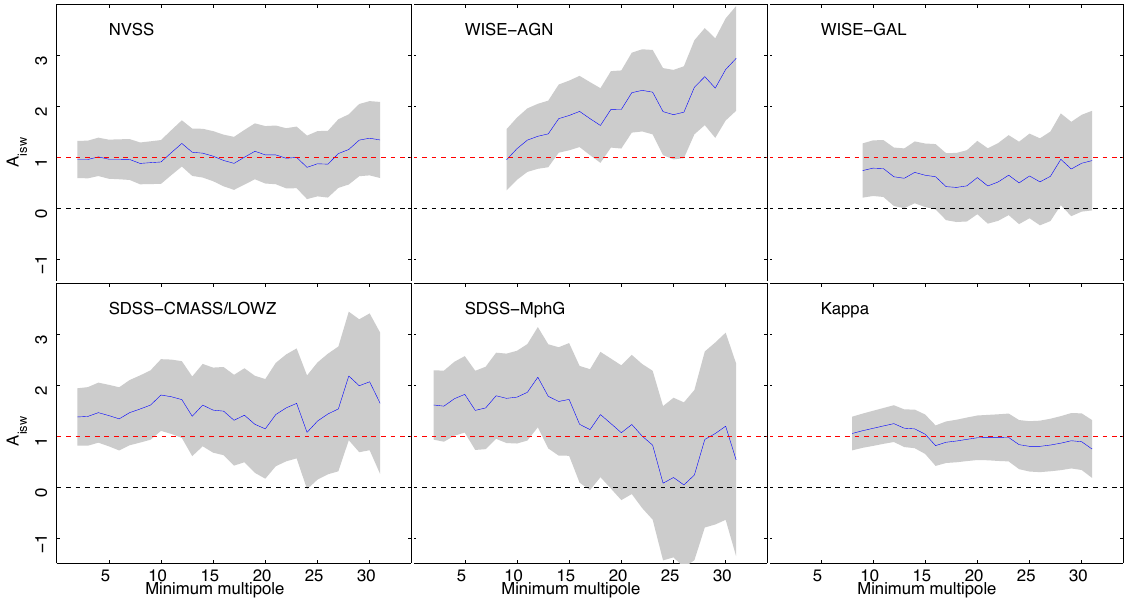}
\end{center}
\caption{\label{fig:a_dep} Dependence of the estimated $A_{\rm ISW}$ amplitude for the different surveys as a function of the $\ell_{\rm min}$ considered in the amplitude estimation.}
\end{figure*}

The CMB cross-correlation with LSS tracers of the matter distribution is the most classical approach to study the IWS effect; it offers the possibility of extracting these secondary anisotropies, otherwise swamped by the primordial CMB anisotropies.
The seminal work by~\cite{Crittenden1996} proposed to use a galaxy catalogue as the LSS tracer, and the first positive detection following this approach was done by~\cite{Boughn2004} using \wmap\ data and radio and X-ray tracers. As discussed in Sect.~\ref{sec:intro}, several other works quickly came after this, confirming the detection of the ISW effect with additional galaxy tracers.

In our previous study~\citep{planck2013-p14}, we performed, for the first time, the detection of the ISW effect using only CMB data, by cross-correlating the \Planck\ CMB map with the \Planck\ lensing potential map, which is naturally used as a tracer of the matter distribution. This cross-correlation is nothing but an estimator of the ISW-lensing bispectrum induced on the Gaussian CMB anisotropies by the deflection caused by the lensing effect~\citep[see, e.g.,][]{Lewis2011}.

This cross-correlation has been studied using different tools: the cross-correlation function (CCF); the covariance of wavelet coefficients (Wcov); and the angular cross-power spectrum (ACPS). These three estimators were first compared in~\cite{Vielva2006}, and were also applied to study the cross-correlation of the \Planck\ CMB data with several surveys in~\cite{planck2013-p14}.

The ACPS is the most natural tool for studying the cross-correlation, since the entire cross-correlation signal is fully included in it. In theory, the CCF and the Wcov estimators are also optimal, as long as they are evaluated at a sufficient number of angles/scales. In fact, for a given case, the CFF and the Wcov, using a relatively small number of evaluations, can achieve a significant fraction of the total signal-to-noise. The clear advantage of ACPS is that, under certain conditions, it provides statistics with uncorrelated elements in cases that full-sky maps can be used. Even for incomplete sky signals, an angular pseudo-spectrum, obtained through a {\tt MASTER} approach~\citep[e.g.,][]{Hivon2002,Hinshaw2003}, provides a very good approximation to our problem.

We showed in~\cite{planck2013-p14} that the three estimators (CCF, Wcov, and ACPS) render a similar detection level, although  ACPS obtained tighter limits than the other two approaches (especially as compared to the CCF). For that reason, in this release we only use the ACPS estimator. The measured ACPS between \sevem\ and the LSS tracers considered in this analysis (\nvss, \wagn, \wg, \lrg, \mg, and \kap) are shown in Fig.~\ref{fig:xl_data}.

\subsection{Methodology}

We aim to study the ISW cross-correlation by estimating the best-fit amplitude of the ACPS for a given fiducial model (the one mentioned in Sect.~\ref{sec:data_sims}). This approach allows us to check the compatibility of the data with the ISW effect, and provides an estimate of the signal-to-noise ratio of the measured signal. This method is complementary to an alternative approach, in which the signal is compared to the null hypothesis of no correlation between the CMB and the LSS. Using a Bayesian hypothesis test, we showed in~\cite{planck2013-p14} clear evidence in favour of the alternative hypothesis as compared to the null one. 

Let us denote the expected ACPS of two maps ($x$ and $y$) by $\xi_\ell^{xy}$, where $\ell$ represents a given multipole, and we assume that the two signals are given in terms of a fluctuation field (i.e., with zero mean and dimensionless units). 

In our particular case, $x$ can be seen as the CMB signal, and $y$ represents for one or more surveys. In other words, we can pursue the estimation of the ISW amplitude by a single correlation of the CMB with a given survey, or with several surveys jointly. In this latter case, $\xi_\ell^{xy}$ is a vector of $\ell_\mathrm{max}$ components, where the first $\ell_{\mathrm{max}_1}$ components correspond to the CMB cross-correlation with the first survey, the next $\ell_{\mathrm{max}_2}$ components correspond to the correlation with the second survey, and so on. Obviously, when $x \equiv y$, $\xi_\ell^{xy}$ represents an auto-correlation. 

The full description of $\xi_\ell^{xy}$ and its covariance $\tens{C}_{\xi^{xy}}$, in terms of the theoretical model and the specific sky coverage, is given in~\cite{planck2013-p14}. Since the ACPS is a very fast estimator, in particular for the \nside = 64 resolution maps, it is also possible to determine these quantities from the coherent simulations described in Sect.~\ref{sec:data_sims}: we use 10\,000 out of the 11\,000 performed simulations to estimate both the expected signal ($\xi_\ell^{xy}$) and its covariance ($\tens{C}_{\xi^{xy}}$). This is the approach followed in the past \Planck\ release, and it is also the one adopted in this work.

Denoting the observed cross-correlation by $\hat{\xi}_\ell^{xy}$, a simple $\chi^2$ can be formed to estimate the amplitude $A$, such that
$A\times\xi_\ell^{xy}$ is the best-fit solution to $\hat{\xi}_\ell^{xy}$:
\begin{linenomath*}
\begin{equation}
\chi^2\left(A\right) = \left[\hat{\xi}_\ell^{xy} - A\times\xi_\ell^{xy} \right]^{\rm T}
 \tens{C}_{\xi^{xy}}^{-1} \left[\hat{\xi}_\ell^{xy} - A\times\xi_\ell^{xy}\right],
\label{eq:xcorr_fit}
\end{equation}
\end{linenomath*}
where $\tens{C}_{\xi^{xy}}$ is the covariance matrix (of dimension
$\ell_\mathrm{max} \times \ell_\mathrm{max}$) of the expected cross-correlation $\xi_\ell^{xy}$, i.e., $\tens{C}_{{\xi^{xy}}_{i,j}} \equiv \left\langle \xi_{\ell_i}^{xy} \xi_{\ell_j}^{xy} \right\rangle$.
It is straightforward to show that the best-fit amplitude $A$, its error, and the significance are given by
\begin{linenomath*}
\begin{eqnarray}
\label{eq:fit}
A&=&\left[\hat{\xi}_\ell^{xy}\right]^T \tens{C}_{\xi^{xy}}^{-1} \xi_\ell^{xy}
 \left[\left[\xi_\ell^{xy}\right]^T
 \tens{C}_{\xi^{xy}}^{-1} \xi_\ell^{xy} \right]^{-1},  \\
 \sigma_A&=&\left[\left[\xi_\ell^{xy}\right]^T
 \tens{C}_{\xi^{xy}}^{-1} \xi_\ell^{xy} \right]^{-1/2}, \nonumber \\
 A/ \sigma_A&=&\left[\hat{\xi}_\ell^{xy}\right]^T \tens{C}_{\xi^{xy}}^{-1} \xi_\ell^{xy}
 \left[\left[\xi_\ell^{xy}\right]^T
 \tens{C}_{\xi^{xy}}^{-1} \xi_\ell^{xy} \right]^{-1/2}  \nonumber.
\end{eqnarray}
\end{linenomath*}
%

\subsection{Cross-correlation results}
\label{sec:xcorr_results}
\begin{table*}[tb]
\begingroup
\newdimen\tblskip \tblskip=5pt
\caption{ISW amplitudes $A$, errors $\sigma_A$, and significance levels S/N = $A/\,\sigma_A$ of the CMB-LSS cross-correlation (survey-by-survey and for different combinations). These values are reported for the four \Planck\ CMB maps: \cruler; \nilc; \sevem; and \smica. The last column gives the expected S/N within the fiducial $\Lambda$CDM model. \label{tab:s2n_data}}
\nointerlineskip
\vskip -3mm
\footnotesize
\setbox\tablebox=\vbox{
   \newdimen\digitwidth 
   \setbox0=\hbox{\rm 0} 
   \digitwidth=\wd0 
   \catcode`*=\active 
   \def*{\kern\digitwidth}
   \newdimen\signwidth 
   \setbox0=\hbox{+} 
   \signwidth=\wd0 
   \catcode`!=\active 
   \def!{\kern\signwidth}
\halign{#\hfil\tabskip=0.3cm& \hfil#\hfil\tabskip=0.3cm&
 \hfil#\hfil\tabskip=0.3cm& \hfil#\hfil\tabskip=0.3cm&
 \hfil#\hfil\tabskip=0.3cm& \hfil#\hfil\tabskip=0.3cm&
 \hfil#\hfil\tabskip=0.3cm& \hfil#\hfil\tabskip=0.3cm&
 \hfil#\hfil\tabskip=0.3cm& \hfil#\hfil\tabskip=0.cm\cr 
\noalign{\doubleline}
 \noalign{\vskip -2pt}
LSS data& \cruler&& \nilc\hfil&& \sevem&& \smica\hfil&& Expected\cr 
\noalign{\vskip 2pt\hrule\vskip 3pt} 
& $A \pm \sigma_A$ & S/N & $A \pm \sigma_A$ & S/N & $A \pm \sigma_A$ & S/N & $A \pm \sigma_A$ & S/N &S/N\cr
\noalign{\vskip 3pt\hrule\vskip 5pt}
\nvss            & $0.95\pm0.36$&2.61& $0.94\pm0.36$&2.59& $0.95\pm0.36$&2.62& $0.95\pm0.36$&2.61&2.78\cr 
\noalign{\vskip 3pt\hrule\vskip 5pt}
\wagn\ ($\ell_{\rm min} \geq 9$)
& $0.95\pm0.60$&1.58& $0.96\pm0.60$&1.59& $0.95\pm0.60$&1.58& $1.00\pm0.60$&1.66&1.67\cr 
\noalign{\vskip 3pt\hrule\vskip 5pt}
\wg\  ($\ell_{\rm min} \geq 9$)
& $0.73\pm0.53$&1.37& $0.72\pm0.53$&1.35& $0.74\pm0.53$&1.38& $0.77\pm0.53$&1.44&1.89\cr 
\noalign{\vskip 3pt\hrule\vskip 5pt}
\lrg             & $1.37\pm0.56$&2.42& $1.36\pm0.56$&2.40& $1.37\pm0.56$&2.43& $1.37\pm0.56$&2.44&1.79\cr 
\noalign{\vskip 3pt\hrule\vskip 5pt}
\mg              & $1.60\pm0.68$&2.34& $1.59\pm0.68$&2.34& $1.61\pm0.68$&2.36& $1.62\pm0.68$&2.38&1.47\cr 
\noalign{\vskip 3pt\hrule\vskip 5pt}
\kap\ ($\ell_{\rm min} \geq 8$)           
& $1.04\pm0.33$&3.15& $1.04\pm0.33$&3.16& $1.05\pm0.33$&3.17& $1.06\pm0.33$&3.20&3.03\cr 
\noalign{\vskip 3pt\hrule\vskip 5pt}
\nvss\ and \kap  & $1.04\pm0.28$&3.79& $1.04\pm0.28$&3.78& $1.05\pm0.28$&3.81& $1.05\pm0.28$&3.81&3.57\cr 
\noalign{\vskip 3pt\hrule\vskip 5pt}
\wise\             & $0.84\pm0.45$&1.88& $0.84\pm0.45$&1.88& $0.84\pm0.45$&1.88& $0.88\pm0.45$&1.97&2.22\cr 
\noalign{\vskip 3pt\hrule\vskip 5pt}
\sdss\             & $1.49\pm0.55$&2.73& $1.48\pm0.55$&2.70& $1.50\pm0.55$&2.74& $1.50\pm0.55$&2.74&1.82\cr 
\noalign{\vskip 3pt\hrule\vskip 5pt}
\nvss\ and \wise\ and \sdss\  & $0.89\pm0.31$&2.87& $0.89\pm0.31$&2.87& $0.89\pm0.31$&2.87& $0.90\pm0.31$&2.90&3.22\cr 
\noalign{\vskip 3pt\hrule\vskip 5pt}
All              & $1.00\pm0.25$&4.00& $0.99\pm0.25$&3.96& $1.00\pm0.25$&4.00& $1.00\pm0.25$&4.00&4.00\cr   
\noalign{\vskip 5pt\hrule\vskip 3pt}}}
\endPlancktablewide                    
\endgroup
\end{table*}

The fundamental CMB-LSS cross-correlation results are summarized in Table~\ref{tab:s2n_data}, where we report the estimated ISW amplitude ($A$), its error ($\sigma_A$), and the detection level $A/\,\sigma_A$, derived for the surveys described in~\ref{sec:data_lss}, applying Eqs.~(\ref{eq:fit}). Results obtained from the four \Planck\ CMB maps (\cruler, \nilc, \sevem, and \smica) are given, showing perfect agreement among them, indicating a robust recovery of the largest CMB anisotropies.

In addition to estimating the ISW amplitude by fitting individual surveys, we also consider several combinations: \nvss\ and \kap; the two \sdss\ surveys (\lrg\ and \mg); the two \wise\ catalogues (\wagn\ and \wg); the five external tracers (\nvss, and \wise\ and \sdss\ surveys); and the six surveys together. As expected, the lowest error is achieved by combining all the surveys, taking into account all their mutual correlations. For the fiducial $\Lambda$CDM model a total signal-to-noise of $4\,\sigma$ is predicted, and, that is the actual value estimated from the data.

The highest contribution comes from the \Planck\ convergence lensing map (\kap), which provides a detection level of $3.2\,\sigma$, followed by \nvss\ that allows us to detect the ISW effect at $2.6\,\sigma$. In fact, the combination of these two LSS tracers almost provides the full detection achieved with the six surveys, $3.8\,\sigma$.

The ISW effect characterized from the \sdss\ catalogues has a signal-to-noise level of around $2.4\,\sigma$ for each survey, and $2.7\,\sigma$ when they are considered jointly. The \wise\ surveys provide the lowest signal-to-noise: $1.6\,\sigma$ for \wagn; $1.4\,\sigma$ for \wg; and $1.9\,\sigma$ for the combination of both. The signal-to-noise achieved by the combination of the five external tracers is $2.9\,\sigma$. All these detection levels refer to \sevem, although the levels achieved from the analysis of the other \Planck\ CMB maps are virtually the same.

All the estimated amplitudes are compatible with unity, within the corresponding $1\sigma$ level. In fact, the value of the ISW amplitude is quite stable, independent of the lowest multipole ($\ell_{\rm min}$) considered in the amplitude estimation. This is graphically represented in Fig.~\ref{fig:a_dep}, where the best-fit amplitude $A$ (solid-blue lines) and the $1\sigma$ error (grey areas) are shown. It is remarkable that the estimated amplitude is very constant as a function of $\ell_{\rm min}$, and compatible with unity. The case in which there is a mild incompatibility is for \wagn, where, for $\ell_{\rm min} \gtrsim 18$, the departure from unity is in tension at about $1\sigma$. This could be the indication of some systematics or contamination, still present in this catalogue. Looking at Fig.~\ref{fig:surveys_cls}, we see that the \wagn\ auto-spectra presents some deviations with respect to the fiducial model, not only at the lowest multipoles, which, as already mentioned, are removed from the analyses, but, in general, over the whole $\ell$ range, showing some extra power. Interestingly, there is no systematic discrepancies between the cross-spectra of \wagn\ and the rest of the catalogues. The same is observed in the cross-correlation of CMB with \wagn\ in Fig.~\ref{fig:xl_data}. This could indicate that the \wagn\ catalogue could present some contamination, which does not correlate neither with other surveys nor with the CMB, but introduces some bias on the ISW amplitude estimation.

\begin{figure*}
\begin{flushright}
\includegraphics[width=0.195\textwidth]{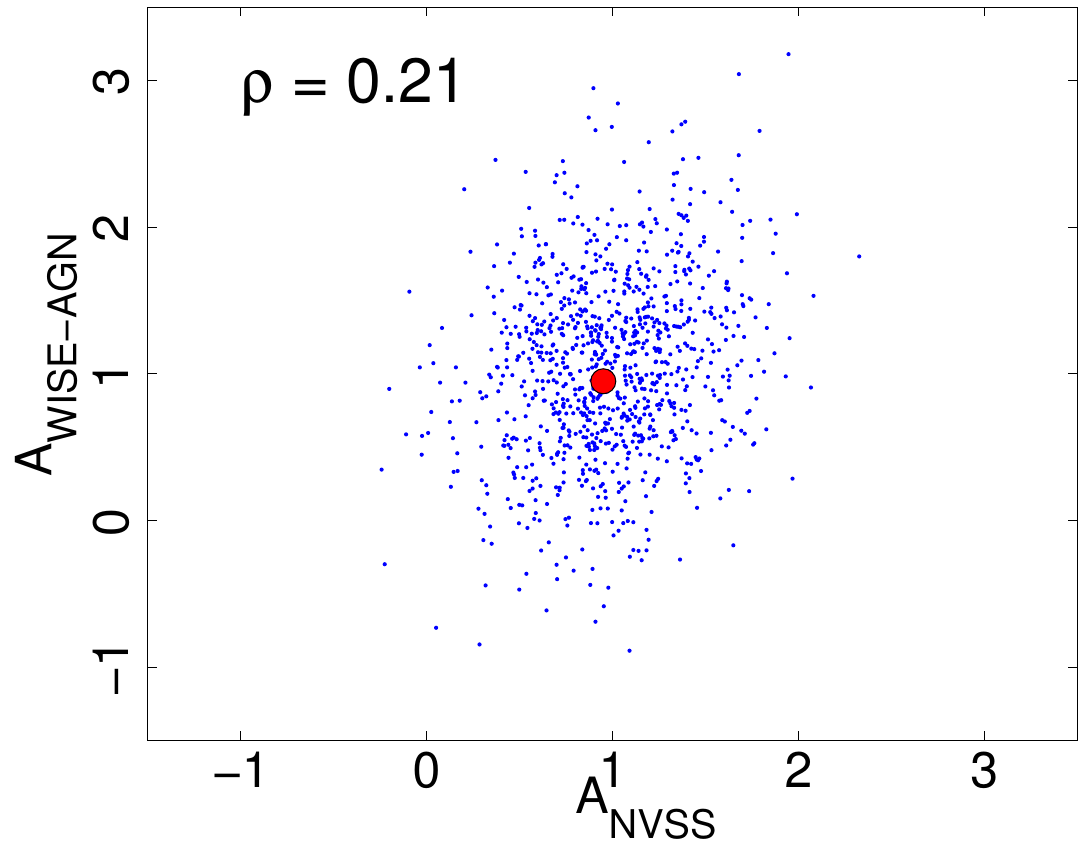}
\includegraphics[width=0.195\textwidth]{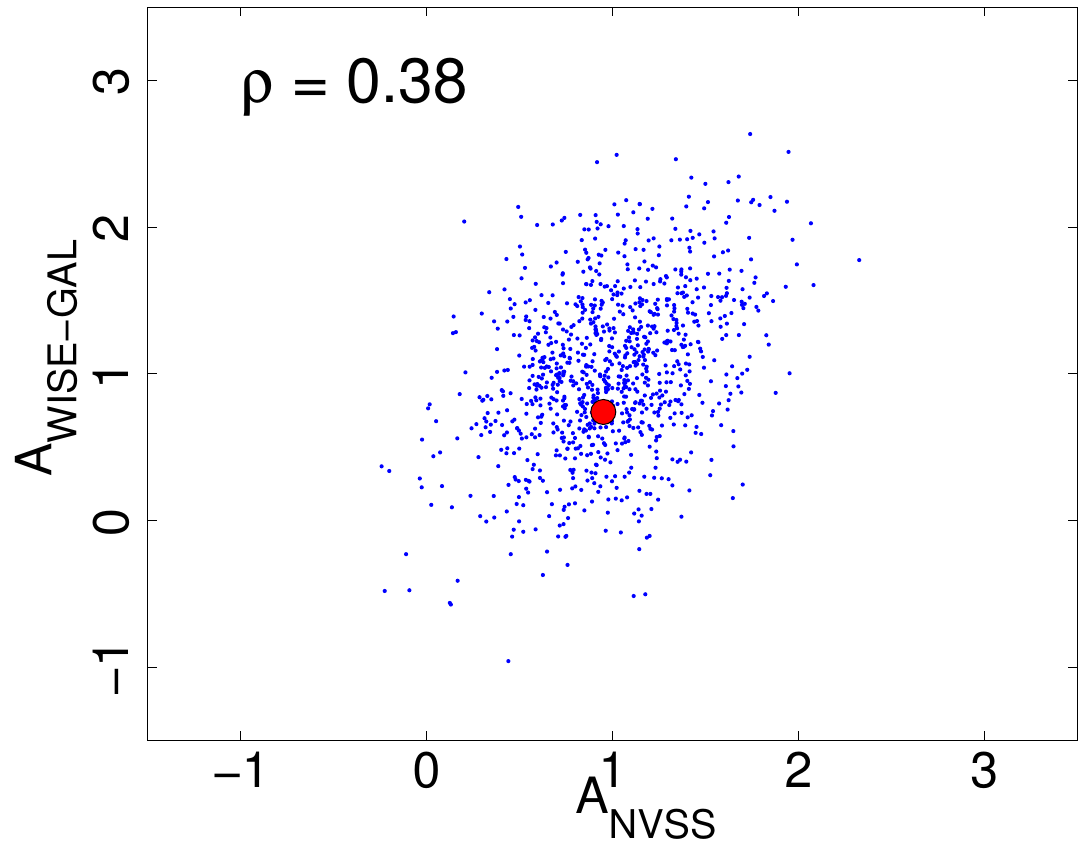}
\includegraphics[width=0.195\textwidth]{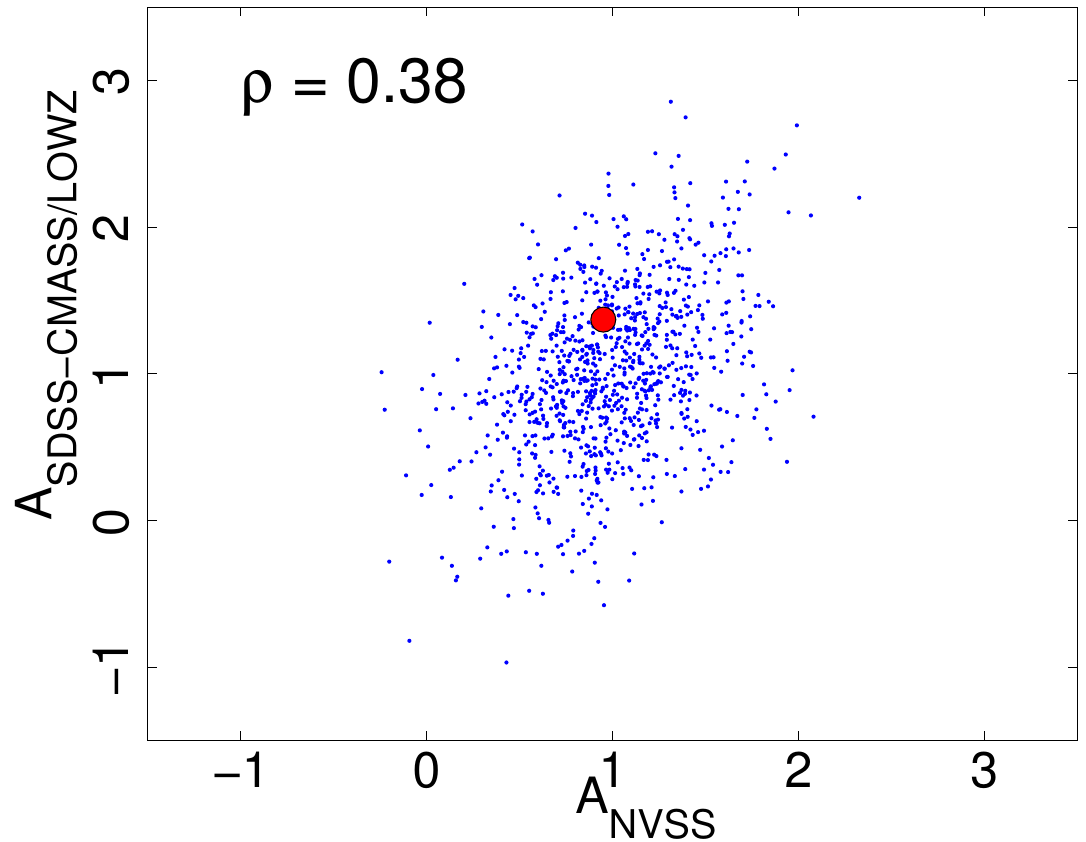}
\includegraphics[width=0.195\textwidth]{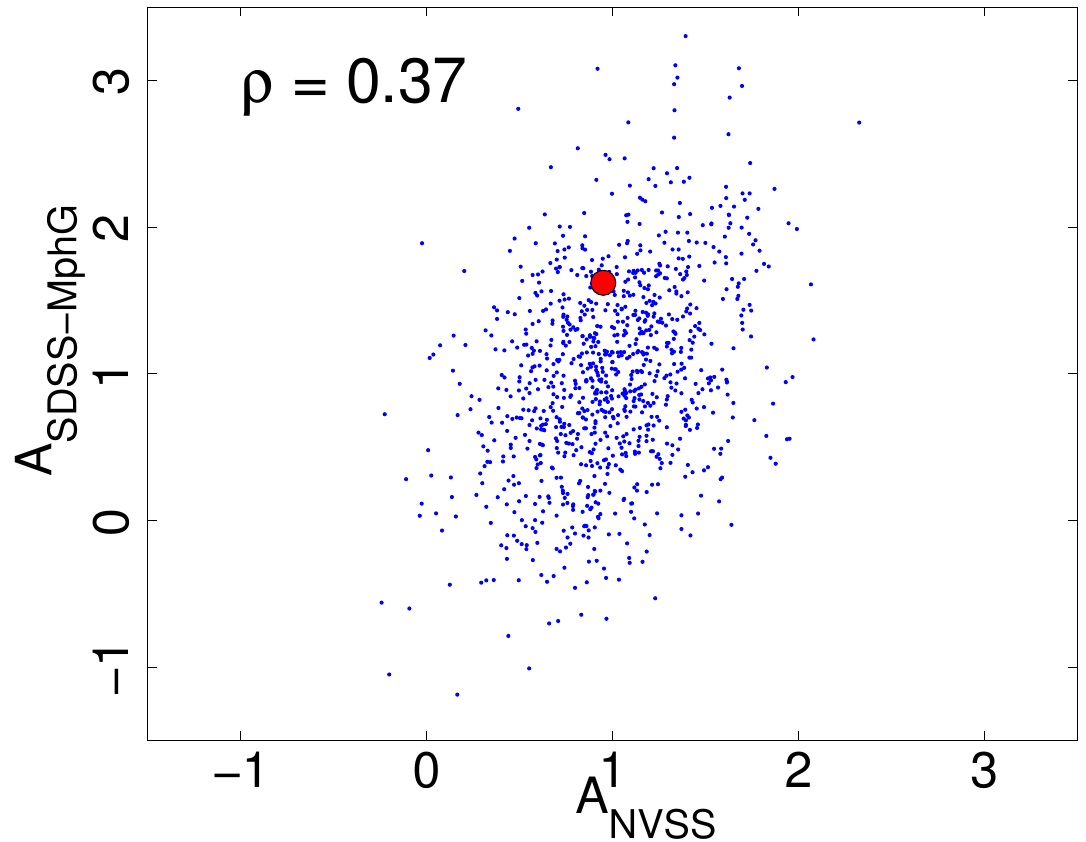}
\includegraphics[width=0.195\textwidth]{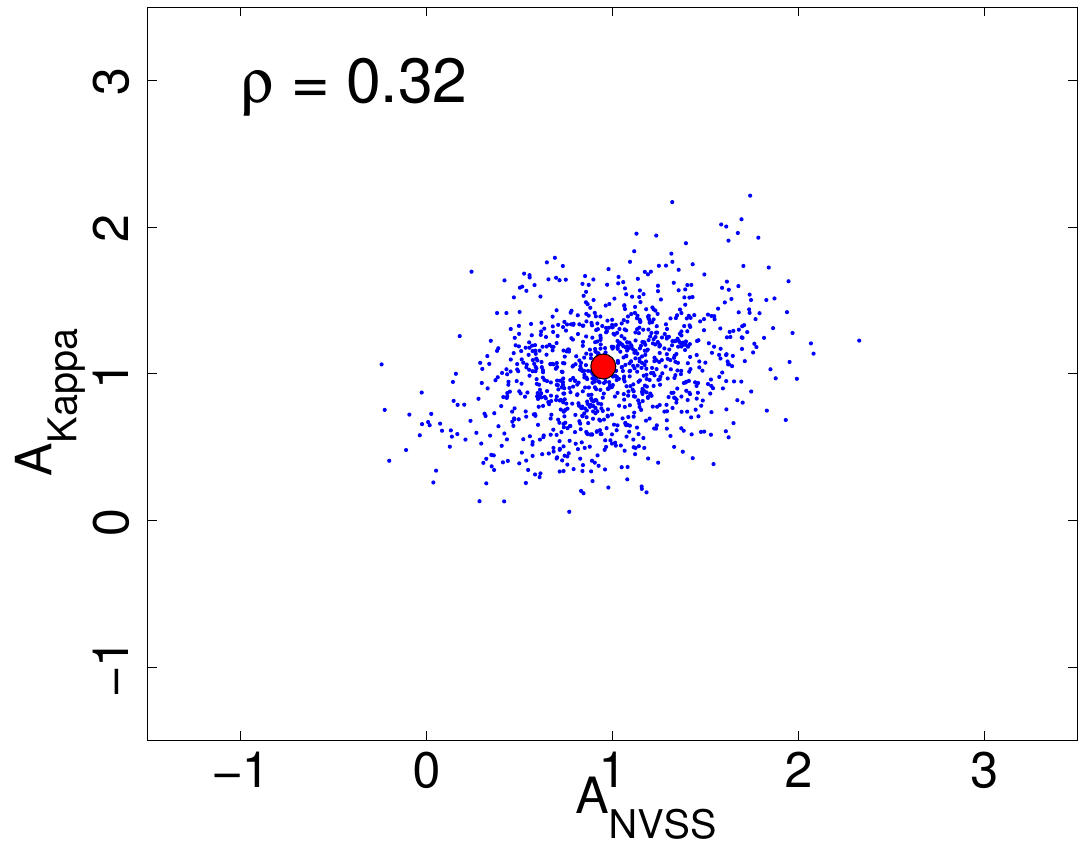} \\
\includegraphics[width=0.195\textwidth]{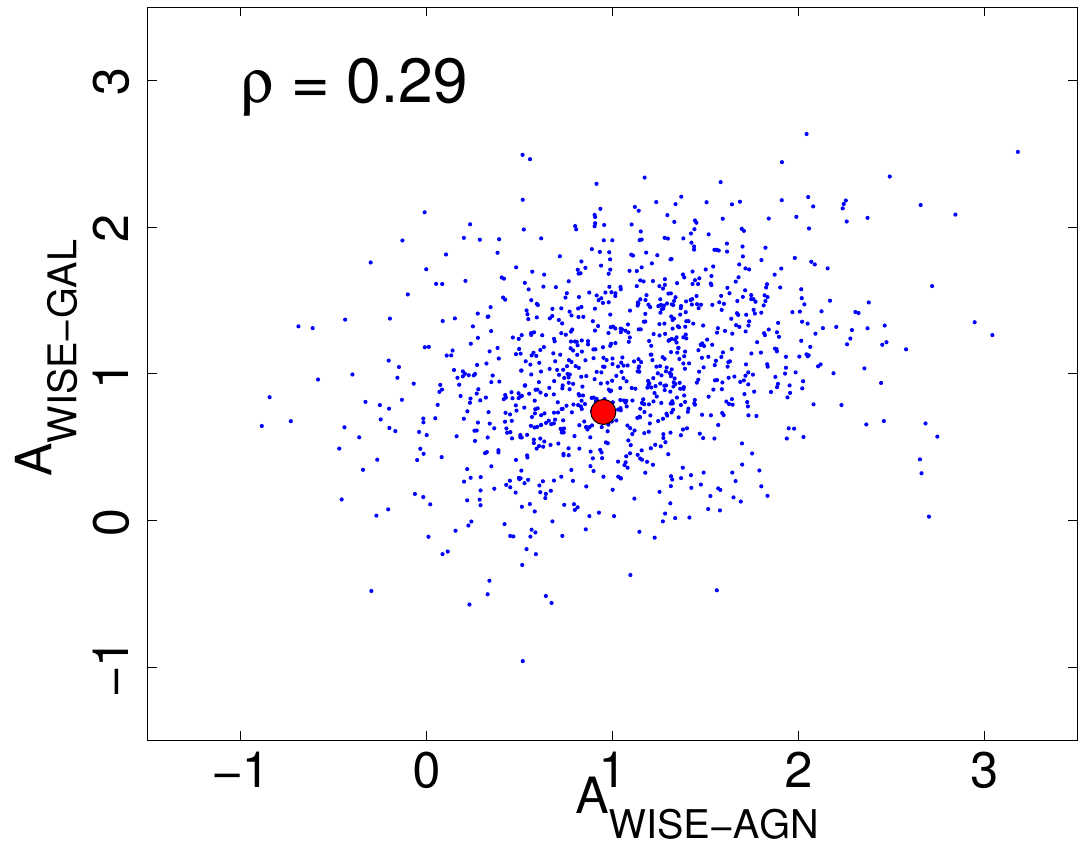}
\includegraphics[width=0.195\textwidth]{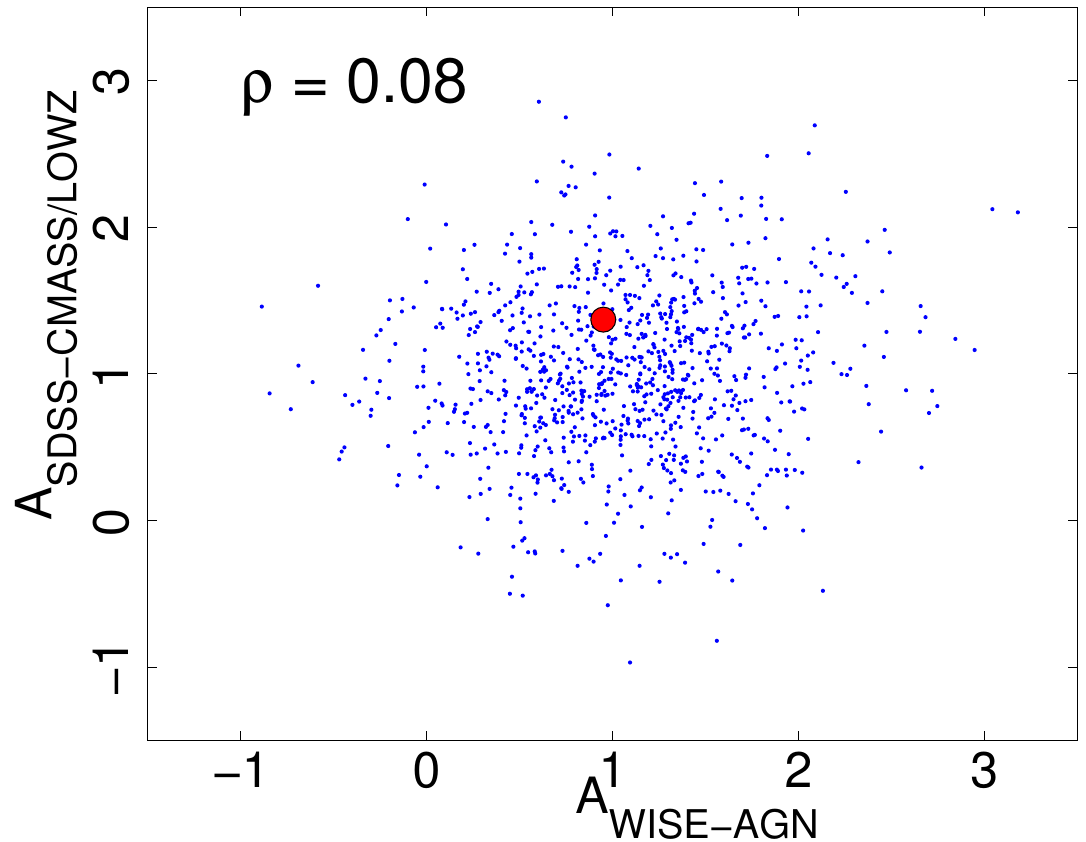}
\includegraphics[width=0.195\textwidth]{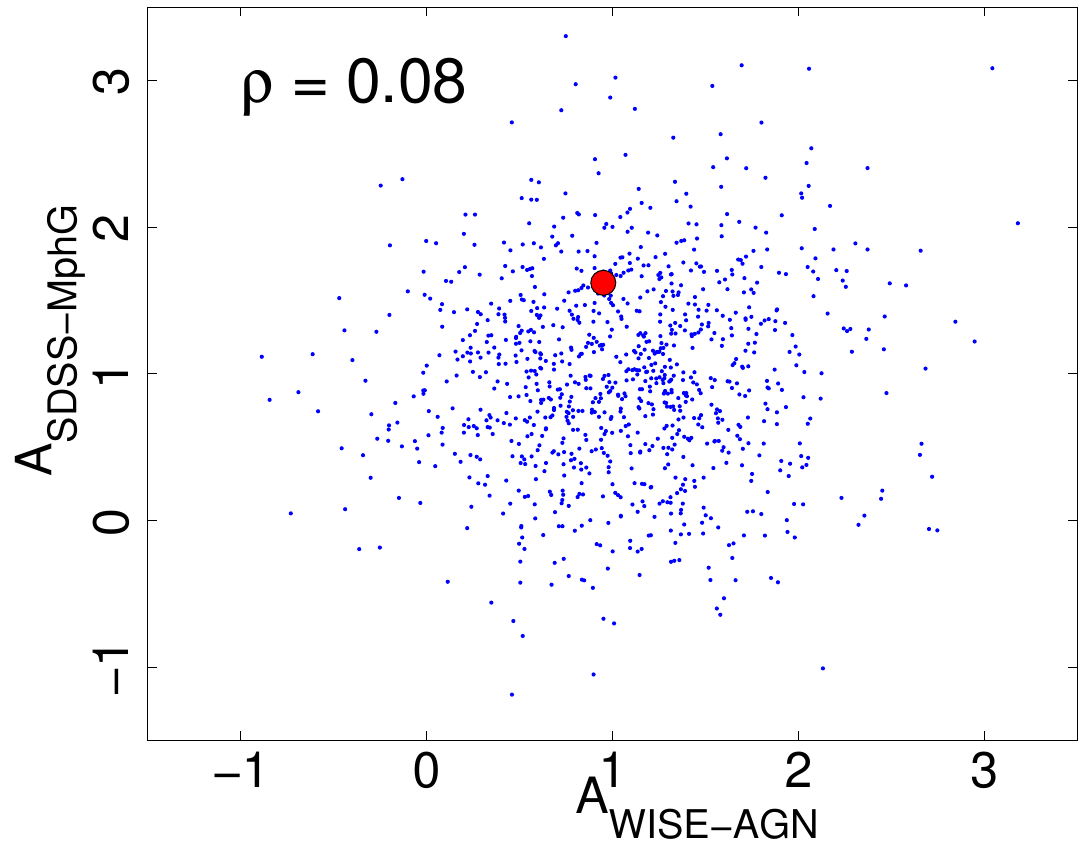}
\includegraphics[width=0.195\textwidth]{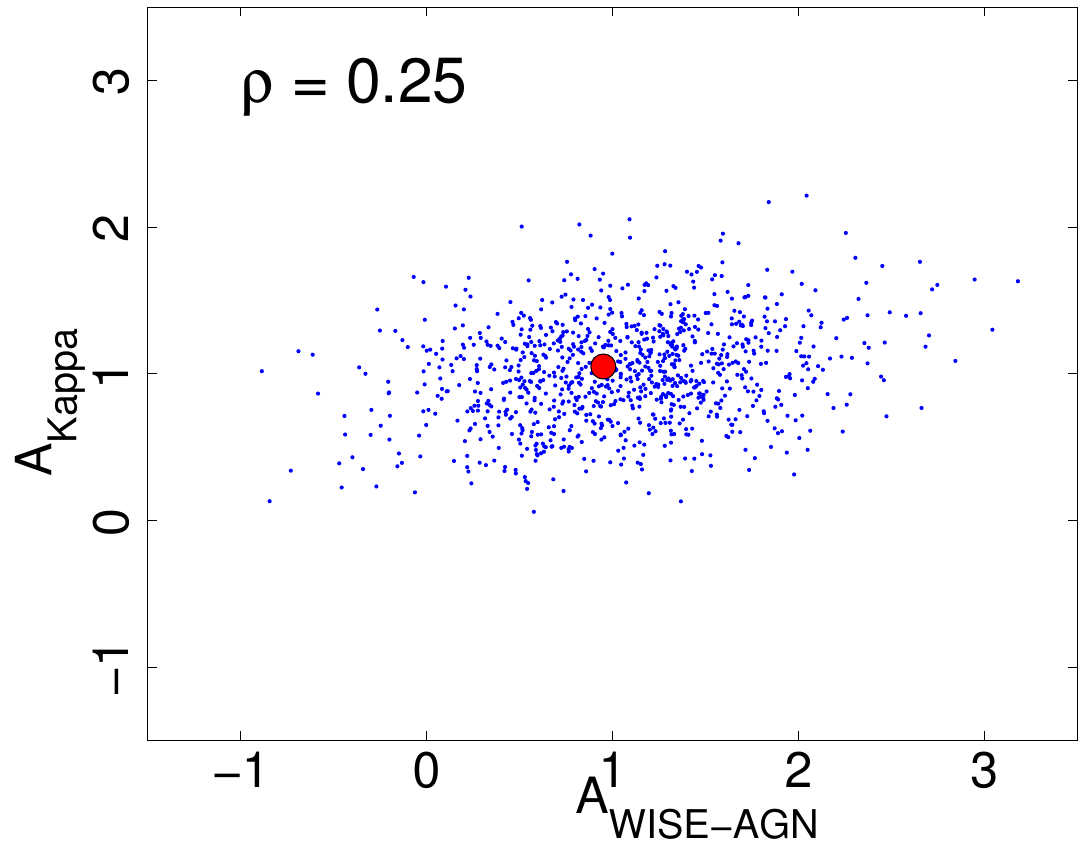}\\
\includegraphics[width=0.195\textwidth]{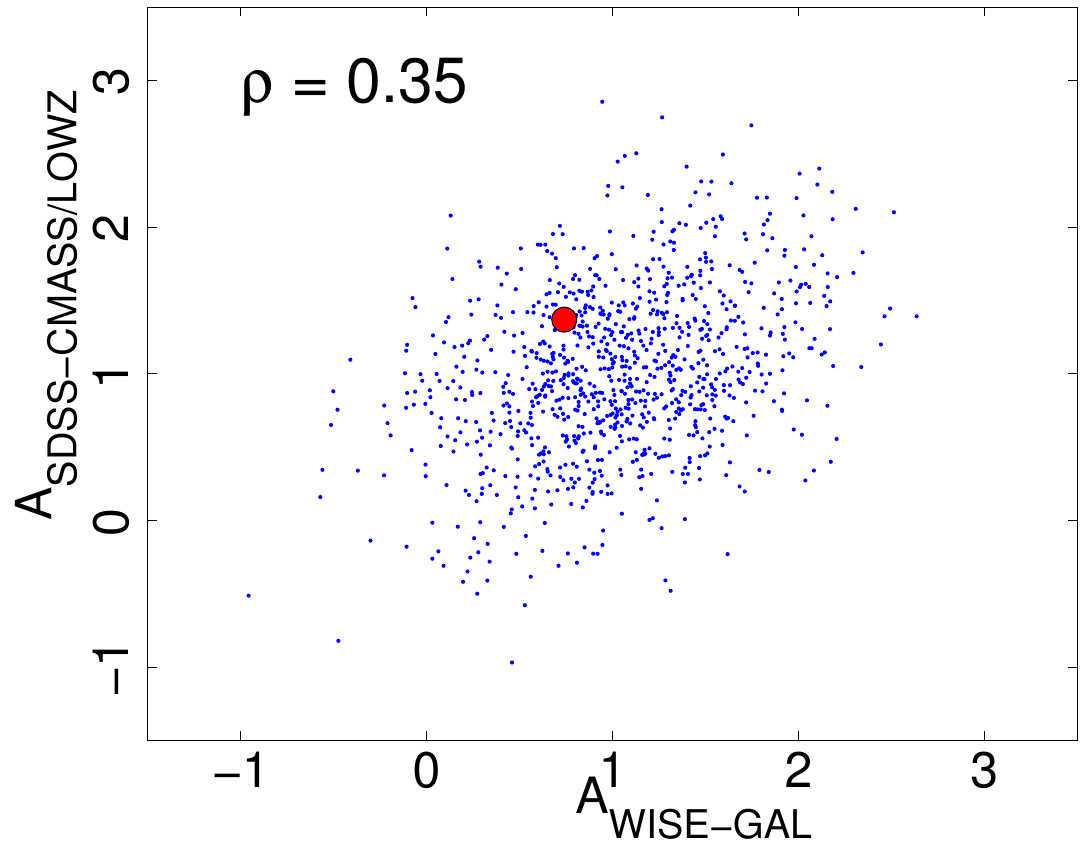} 
\includegraphics[width=0.195\textwidth]{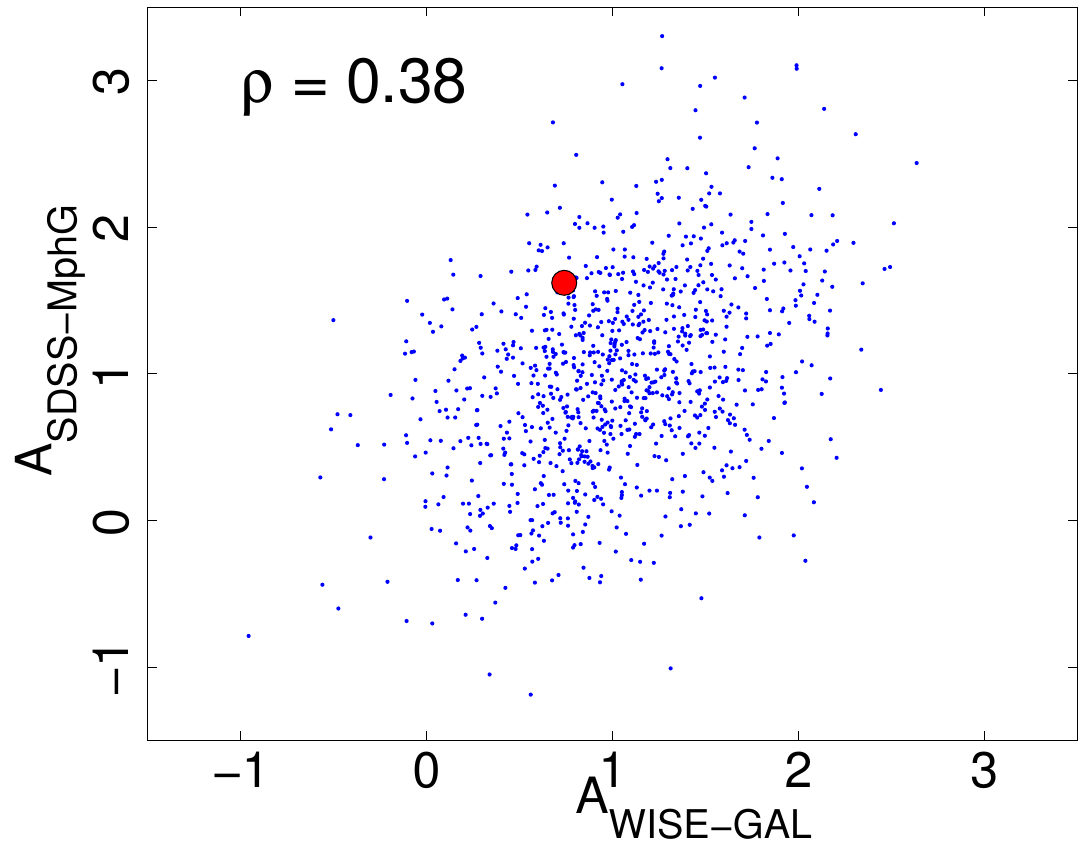}
\includegraphics[width=0.195\textwidth]{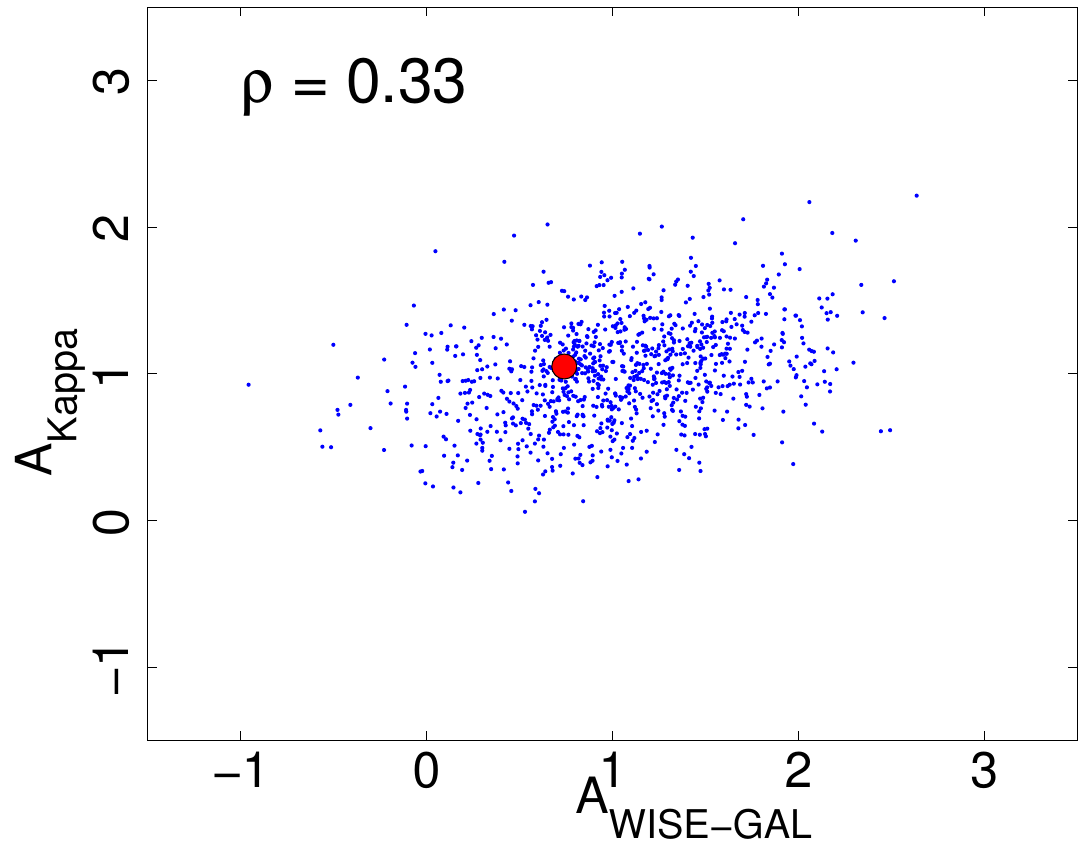}\\
\includegraphics[width=0.195\textwidth]{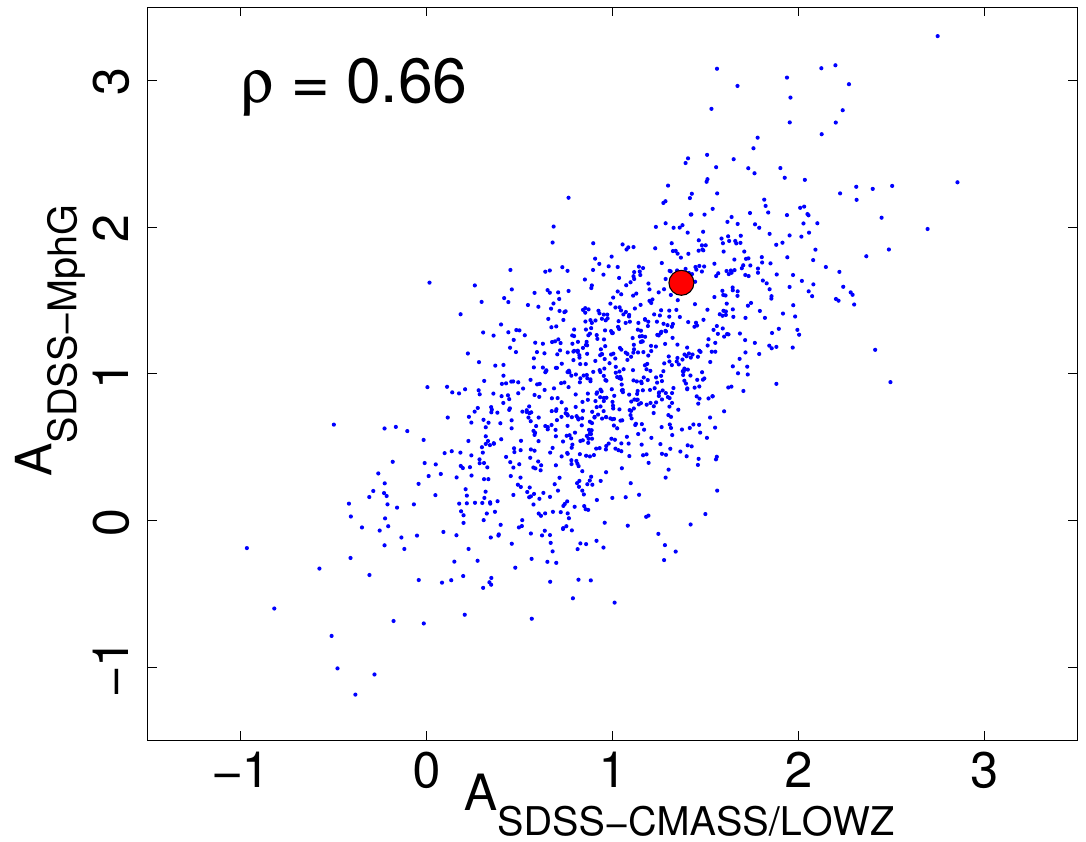}
\includegraphics[width=0.195\textwidth]{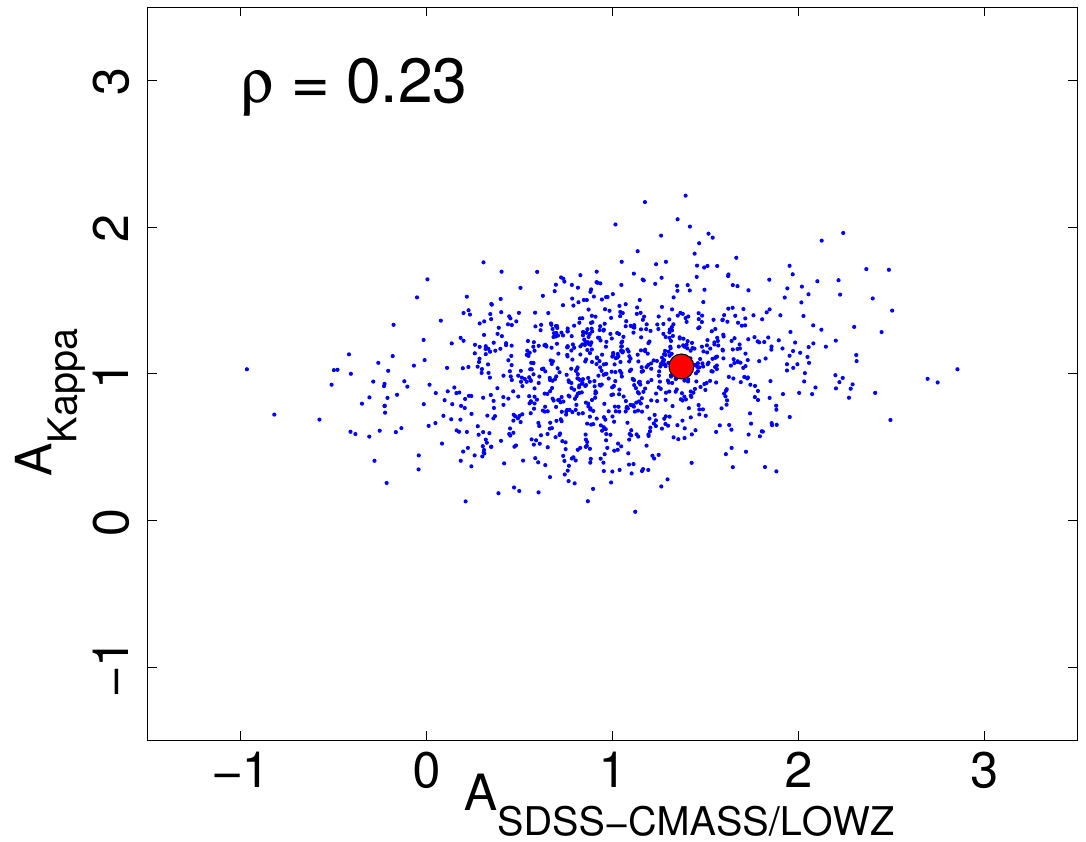}\\
\includegraphics[width=0.195\textwidth]{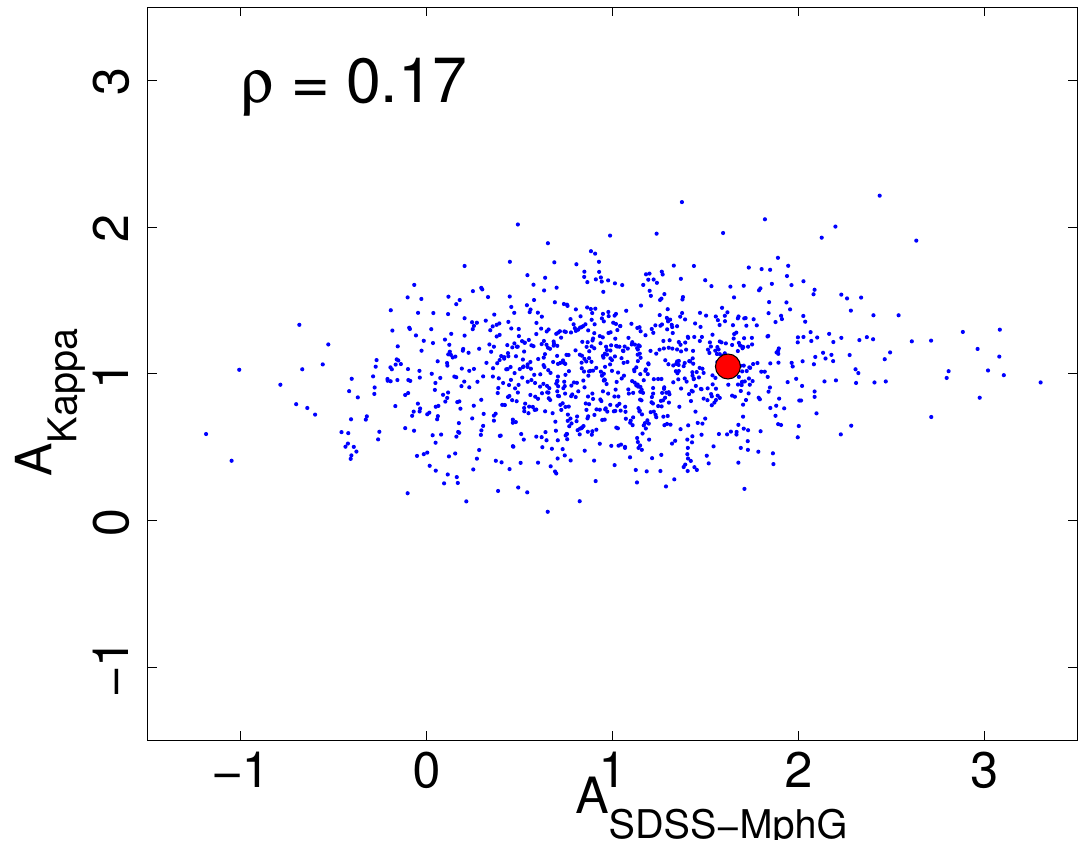}
\end{flushright}
\caption{\label{fig:a_rho} Correlation among the estimated $A_{\rm ISW}$ amplitudes for the different surveys. The small blue dots are the amplitudes estimated from the simulations described in Sect.~\ref{sec:data_sims}, whereas the large red dot stands for the amplitudes estimated from the data. For each pair, the correlation coefficient is indicated.}
\end{figure*}

Another interesting aspect is shown in Fig.~\ref{fig:a_rho}, where the correlation between ISW amplitudes is given, for all the possible survey-survey combinations. Using our coherent simulations, we have studied the correlation coefficient ($\rho$) among the estimated $A$ amplitudes from each of the six surveys. On each panel of this figure, we show a scatter plot obtained from 1000 simulations, confronting the ISW estimation for two surveys. We also plot (red circle) the value corresponding to the data. Finally, the correlation coefficient is also given.  


As previously mentioned, the CMB-\kap\ cross-correlation is one of the most robust results, since it represents a detection of the ISW effect, fully obtained from \Planck\ data, which simplifies consideration of possible sources of systematics present in galaxy catalogues. This correlation is nothing but an estimation of the ISW-lensing bispectrum~\citep{Lewis2011} induced by the lensing effect suffered from the CMB photons, as they pass through the gravitational potential.

This bispectrum represents a bias when determining some primordial bispectrum shapes, and one needs to account properly for it.
Within the \Planck\ collaboration, this alternative way of measuring the ISW-lensing correlation is carried out in~\cite{planck2014-a19}. In fact, a similar cross-correlation to the one described here is also performed in the \Planck\ lensing paper~\citep{planck2014-a17}, as a check to establish the reliability of the lensing map reconstruction. In the following section, we summarize these alternative ISW-lensing estimations performed within the present 2015 release.

\subsection{Results on the ISW-lensing bispectrum}
\label{sec:iswl}

\noindent The \Planck\ 2013 results~\citep{planck2013-p14,planck2013-p12,planck2013-p09a} showed for the first time evidence of the lensing-ISW CMB bispectrum by using the \Planck\ 2013 
temperature-only data release. The lensing-ISW non-Gaussian signal is an independent and direct probe of the influence of dark energy on the evolution of structure in the Universe, which only relies on CMB data.

The lensing potential $\phi$ and the CMB temperature $T$ are correlated, since it is the same gravitational matter distribution at redshifts less than about 2 that leads to both the gravitational lensing of the CMB and the ISW effect. Moreover, since the gravitational lensing leads to changes in the small-scale power of the CMB, and the ISW effect affects the large-scale CMB temperature, we obtain a non-zero lensing-ISW bispectrum of a predominantly squeezed shape, correlating one large scale with two much smaller scales \citep[see, e.g.,][]{Goldberg1999, Seljak1999, Hu2000, Hu2002b, Verde2002, 
Giovi2003, Okamoto2003, Giovi2005a, Lewis2006, Serra2008, Mangilli2009a, Hanson2009a, Hanson2010, Smith2011, Lewis2011}. 

The 2015 \Planck\ release offers us the possibility of including polarization in the estimation of the ISW-lensing bispectrum. As shown in \cite{Cooray2006}, the direct ISW-lensing correlation in $E$ polarization due to re-scattering of the temperature quadrupole generated by the ISW effect is negligible. However, as explained in \cite{Lewis2011}, there is an important correlation between the lensing potential and the large-scale $E$ polarization generated by scattering at reionization. Because the lensing potential is highly correlated with the ISW signal, this does in the end also lead to a non-zero ISW-lensing bispectrum in polarization. Although the current high-pass filtering of the polarization data reduces this cross-correlation somewhat, it is in principle still detectable.
Explicit expressions for the ISW-lensing bispectrum template can be found
in \cite{planck2014-a19}.

In this section we summarize the ISW-lensing estimations performed in three different papers of the present \Planck\ 2015 release, and we comment on their comparison. First, as explained in the previous subsection, we have implemented an estimator (see Eq.~\ref{eq:fit}) of the ISW-lensing bispectrum in terms of the CMB and lensing cross-correlation~\citep{Lewis2011}. An independent implementation of the same estimator can be found in the \Planck\ lensing paper~\citep{planck2014-a17}. However, whereas the latter uses the FFP8 simulations (which include the actual non-Gaussian signal induced by the lensing of the CMB anisotropies), the implementation performed in this paper uses Gaussian simulations that form part of the set of coherent CMB and LSS tracer maps. Despite this difference, both implementations yield very similar results, the ISW paper estimator gives $A=1.06\pm0.33$, whereas the lensing paper estimator finds $A=0.90\pm0.28$ (both for \smica).

The \Planck\ primordial non-Gaussianity paper~\citep{planck2014-a19} studies the ISW-lensing signal primarily to determine the bias this induces on the different primordial bispectrum shapes. However, it also gives results for the actual amplitude of the ISW-lensing signal. Three different estimators have been considered in this paper: the KSW estimator~\citep{Komatsu2003}; the modal estimator~\citep{Fergusson2010}, and the binned bispectrum estimator \citep{Bucher2010}. Whereas the two former methods are only implemented to work with temperature data, the binned bispectrum estimator is also able to include polarization. All these estimators use the FFP8 simulations to characterize the expected signal and the uncertainties. The binned bispectrum estimator finds an amplitude of the ISW-lensing bispectrum of $A=0.82\pm0.27$ for the \smica\ map using both temperature and polarization. The values obtained using temperature alone with the KSW, modal and binned bispectrum estimators are, $A=0.79\pm0.28$, $A=0.72\pm0.26$,
and $A=0.59\pm0.33$, respectively.

We performed a study on a set of 100 FFP8 simulations that have passed through
the \smica\ component separation pipeline to investigate any biases in
the estimators and their correlations. For reasons explained above, the two
implementations of the estimator based on the lensing reconstruction are not
exactly the same. There are also differences between the three bispectrum
estimators; the KSW estimator implements the ISW-lensing template exactly
(since it is separable), while the modal and binned estimators use 
approximations. Unlike all other templates studied in \cite{planck2014-a19},
the ISW-lensing template is difficult to bin and the correlation between
the exact and binned template is relatively low. Another difference is that
the bispectrum estimators use $\ell_\mathrm{min}=40$ in polarization, while
the lensing reconstruction estimators use $\ell_\mathrm{min}=8$. For all 
these reasons we do not expect the correlation between the different 
estimators to be perfect, which leads to slight differences in the results.
The result of the study is that all the bispectrum estimators agree on
the average value, which is slightly low at around 0.85. The other estimators find
higher values. Since these same simulations are used to determine uncertainties on the final result, all the error bars have been divided by the average
that each estimator finds. Regarding the correlations, we find that the
KSW and modal estimator are correlated at about 95\,\%, while their correlation
with the binned estimator is about 80\,\%. The two lensing reconstruction
estimators are also correlated at about 80\,\%, while the correlation between
the two types of estimator classes is about 60--70\,\%.

Despite these differences, we conclude that all results are 
consistent with the expected value for the ISW-lensing bispectrum amplitude
$A=1$, and that the absence of any ISW-lensing signal ($A=0$) is excluded at 
the level of about $3\sigma$.

\subsection{Derived cosmological constraints on dark energy}
\label{sec:param} 
We have explored the possibility of constraining some cosmological parameters through the ISW detection reported in~\ref{sec:xcorr_results}. In principle, the ISW effect depends on the full parameter set of a dark energy (or curvature) cosmology, but the weak overall significance of the signal makes it necessary to restrict parameter measurements to a single or at most a pair of parameters, while the remaining parameters need to be constrained from other observations. A  exhaustive study on dark energy constraints can be found in the \Planck\ dark energy and modify gravity paper~\citep{planck2014-a16}.

We have assumed a Gaussian shape for the ISW likelihood ${\cal{L}}(\Theta)$, where $\Theta$ stands for a general set of cosmological parameters:
\begin{linenomath*}
\begin{equation}
- 2 \ln[ {\cal{L}}(\Theta)] = \chi^2(\Theta)-\chi^2_{\rm
min}.
\label{likelihood}
\end{equation}
\end{linenomath*}
Here the corresponding quadratic $\chi^2(\Theta)$-functional is given by
\begin{linenomath*}
\begin{equation}
\label{chi2}
\chi^2(\Theta) = \left[ C_{\ell}^{TG,{\rm obs}}-C_{\ell}^{TG}(\Theta) \right]
{\cal C}^{-1}_{\ell \ell'} \left[C_{\ell'}^{TG,{\rm obs}} -
C_{\ell'}^{TG}(\Theta) \right].
\end{equation}
\end{linenomath*}
The covariance matrix $\cal{C}_{\ell \ell'}$ describes the Gaussian variation of the measured spectrum $C_\ell^{TG,obs}$ around the theoretical expectation $C_\ell^{TG}$, and is estimated from the 10\,000 simulations described in Sec.~\ref{sec:data_sims} and used in the cross-correlation analyses in the previous section:
\begin{linenomath*}
\begin{equation}
\label{cova}
{\cal C}_{\ell \ell'}\left(\Theta_0\right) =  
\sum_{i=1}^{N} 
\frac{
\Delta(C_{\ell \,i})
\Delta(C_{\ell \,j})
}{N},
\end{equation}
\end{linenomath*}
where $\Delta\left(C_{\ell \,i}\right) = C_{\ell \,i}^{TG}(\Theta_0)-\bar{C}_{\ell}^{TG}(\Theta_0)$, $C_{\ell \,i}^{TG}$ are the estimates for every single realization $i$, and  $\bar{C}_{\ell}^{TG}$ is their theoretical value. 
Non-zero off-diagonal entries describe correlations between different multipoles due to broken homogeneity, which is mainly caused by masking of emission from the Milky Way.

We expect that the covariance matrix does not change strongly with the cosmological model and therefore, that the fiducial model $\Theta_0$ given in Sec.~\ref{sec:data_sims} provides a suitable uncertainty characterization for all considered cases. The Gaussian likelihood adopted above is the common choice for this effect~\citep[see e.g.,][]{Nolta2004,Vielva2006,Ho2008}. In our case, this likelihood is used to explore the conditional probability of a given cosmological parameter (e.g., $\Omega_\Lambda$),
keeping constant the remaining cosmology. In this case, it is trivial to prove that the estimator is unbiased.

\begin{figure}
\centering
\includegraphics[width=0.495\textwidth]{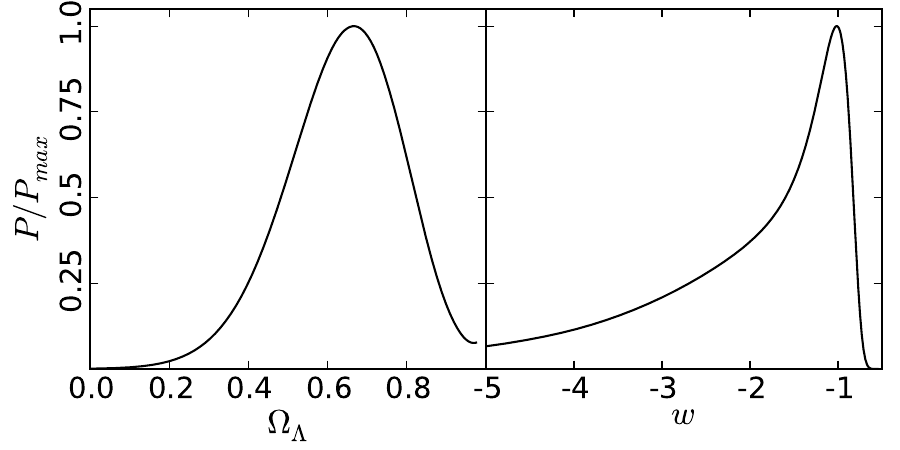}
\caption{\label{fig:like_omegal_w}Conditional probabilities on $\Omega_\Lambda$ (left panel) and $w$ (right panel) derived from the ISW likelihood, based on the CMB-\nvss\ and the CMB-\kap\ cross-correlations.}
\end{figure}

For simplicity, the data used for the ISW likelihood is the joint cross-correlation of the \Planck\ CMB map with the \nvss\ and the \kap\ tracers, which already captures 95\,\% of the total detection of the ISW effect (see Table~\ref{tab:s2n_data}). First, we have determined the conditional probability~\citep[where the rest of the cosmological parameters are fixed to the \Planck\ fiducial model,][]{planck2014-a15} for $\Omega_\Lambda$, obtaining the best-fit for $\Omega_\Lambda=0.67$ and $0.49<\Omega_\Lambda<0.78$ at 68\,\% CL. In particular, we find $\Omega_\Lambda>0$ at more than $3\,\sigma$.  Second, we have estimated the conditional probability on the equation of state parameter of the dark energy, obtaining the best-fit for $w=-1.01$ and $-4.45<w<-1.07$ at 68\,\% CL. These conditional probabilities are shown in Fig.~\ref{fig:like_omegal_w}.

\section{Stacking of CMB temperature and polarization data}
\label{sec:stack}
As an alternative approach to the detection of the ISW signal, we can focus on the objects expected to yield the strongest effect, namely the largest (tens to hundreds of Mpc) voids and clusters in the Universe. In order to measure the effect produced by individual structures, one can stack patches of the CMB anisotropy map centred at the locations of superstructures on the sky. Such a stacking technique allows us to detect and characterize a signal that, otherwise, would be undetectable due to the weakness of the ISW effect compared to the primordial CMB anisotropies.

Following this approach,  \citep[][hereafter GR08]{Granett2008a} found a potentially significant ISW signal by studying 100 superstructures identified in the SDSS DR6 LRG catalogue. The presence of this signal has since been confirmed and more precisely studied with the latest CMB data \citep{planck2013-p14}. However, the statistical significance of this detection is still debated, as well as its supposed ISW nature \citep{Hernandez2013a,Ilic2013,planck2013-p14} and the compatibility of its high amplitude with $\Lambda\mathrm{CDM}$ predictions of the ISW effect from such structures \citep{Granett2008a, Hernandez2013a, Cai2014, Hotchkiss2014}. Moreover, more recent catalogues of superstructures have since been used for similar studies \citep{planck2013-p14,Kovacs2015}, but none of them has yielded a signal with the same level of significance as the GR08 catalogue.

A crucial point in stacking studies is to determine what fraction of the signal detected using this method is either due to the ISW effect of the observed structures, or random and fortuitous anisotropies of the primordial CMB, or a mixture or both. In the present section, we attempt to address this question for the results obtained with the GR08 catalogue, sine it is to date the only result to apparently show a significant discrepancy with respect to $\Lambda\mathrm{CDM}$ expectations. The main novelty of the present analysis compared to previous works in the literature will be the use of a variety of statistical tests that rely on the latest polarization data from \Planck. Indeed, the CMB polarization map should prove to be a valuable asset for our purposes; any ISW signal found in temperature is expected to have no counterpart in CMB polarization, whereas we expect that a primordial CMB signal will be correlated at some level with the CMB polarization. Therefore, and despite the lack of the largest scales (see Sect.~\ref{sec:data_cmb}) in the polarization data, it can be used as a discriminant to separate genuine ISW detections from false positives due to random primordial anisotropies. 

In practice, our objective here will be to answer the following questions. Can polarization data help us to prove the ISW nature of the GR08 signal? Or disprove it -- i.e., show that it is actually caused (partially or entirely) by the primordial part of the CMB? We should keep in mind that the answers to these two questions could very well be negative, if the discriminating power of polarization data proves to be insufficient for stacking studies. In addition, the validity of the GR08 catalogue as an LSS tracer is also addressed by stacking patches from the \Planck\ lensing map.


\begin{figure}
\begin{center}
\includegraphics[width=0.495\textwidth]{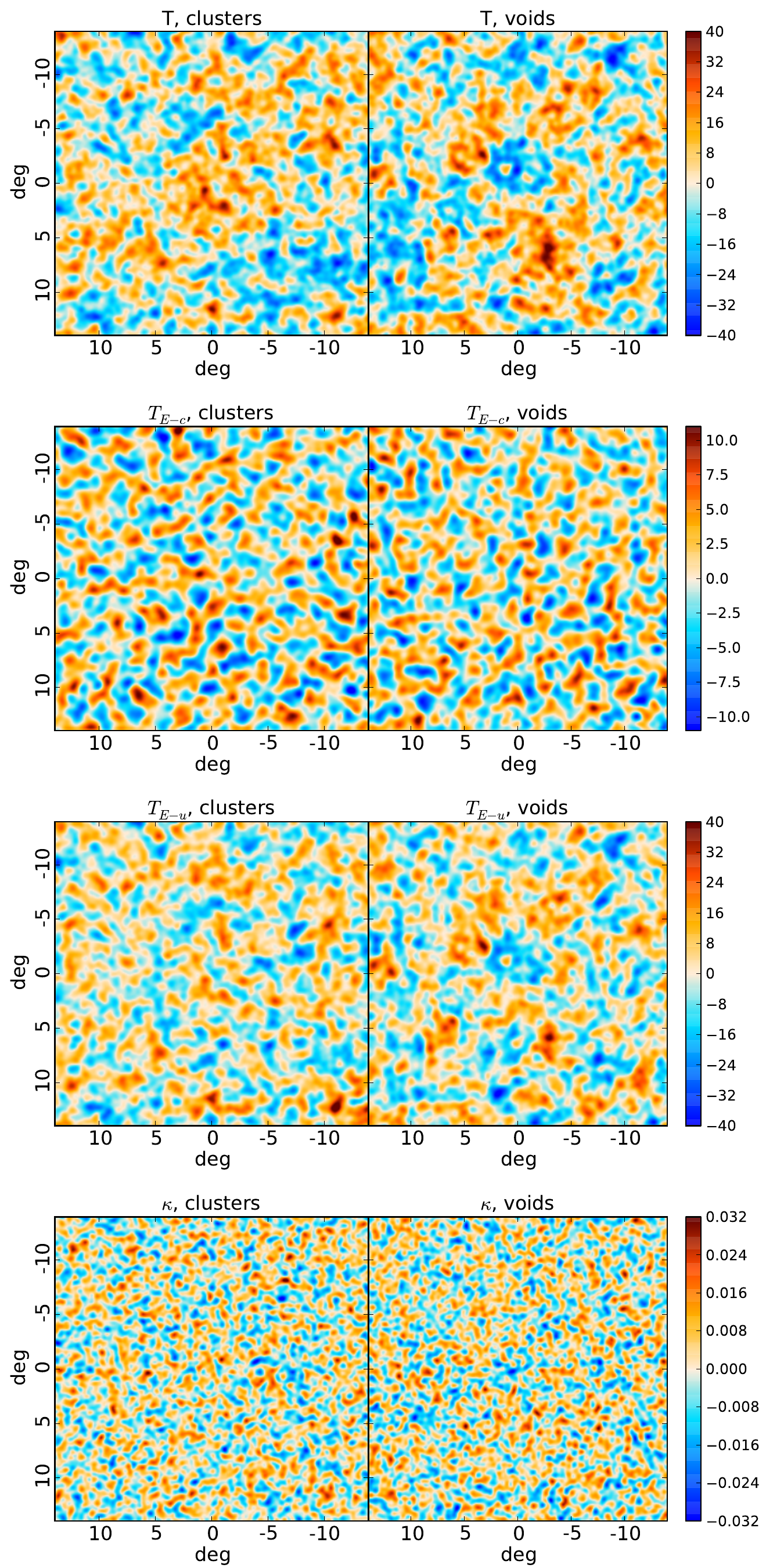}
\caption{Stacked patches of the scalar components from the \sevem\ solution, at the supercluster (first column) and supervoid (second column) positions from GR08. From top to bottom: $T$, \tce, \tue and $\kappa$ components. Temperature maps are given in $\mu \mathrm{K}$ units.}
\label{fig:patches_stacking_isw_T} 
\end{center}  
\end{figure}

\begin{figure}
\begin{center}
\includegraphics[width=0.495\textwidth]{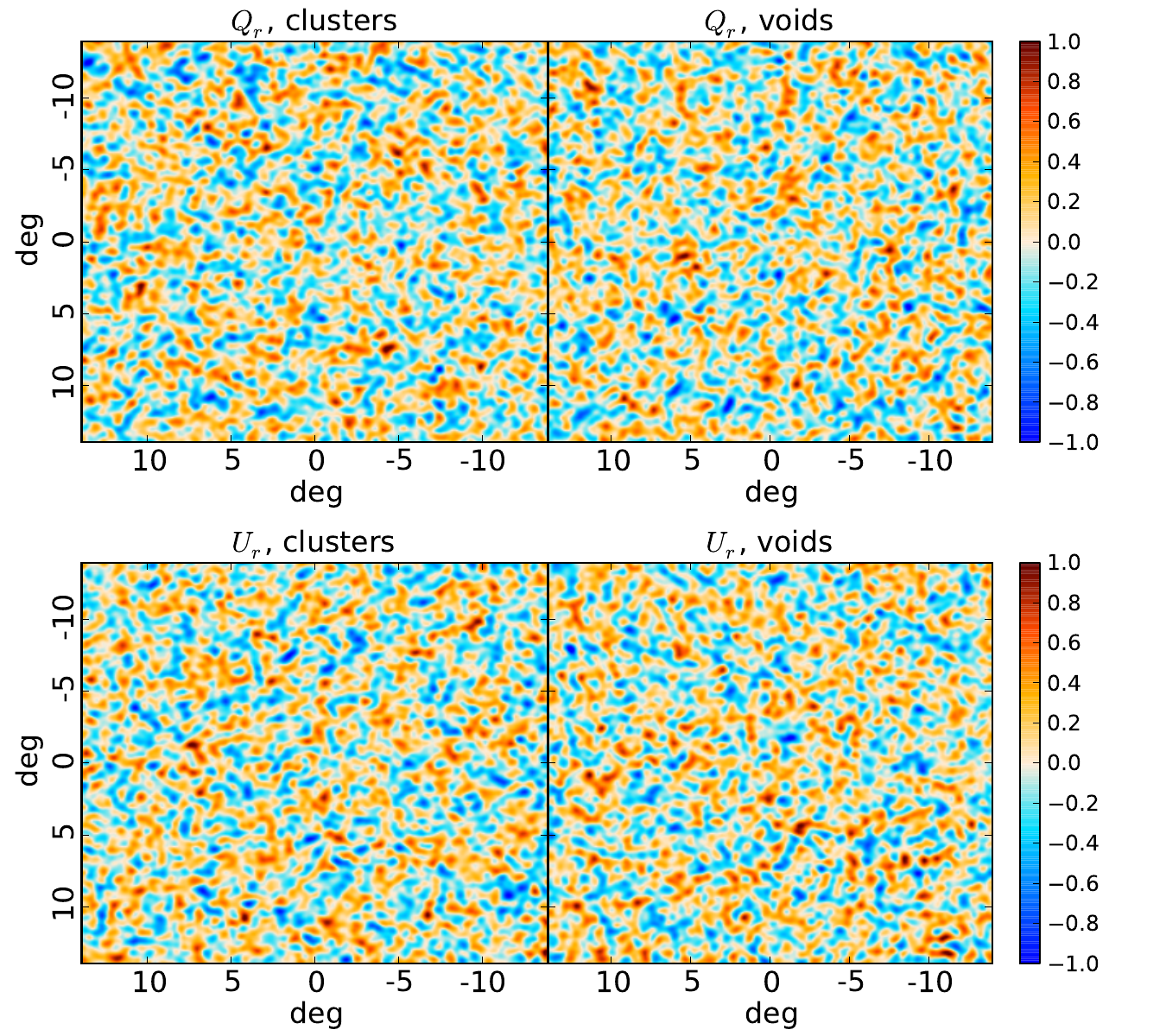}
\caption{Stacked patches of the $Q_r$ (top) and $U_r$ (bottom) components from the \sevem\ solution in $\mu \mathrm{K}$, at the supercluster (first column) and supervoid (second column) positions from GR08.}
\label{fig:patches_stacking_isw_pol} 
\end{center}  
\end{figure}

\subsection{Stacking methodology in polarization}

The main procedure for  stacking of CMB patches in the ISW context has been detailed in \citet{planck2013-p14}. However, the process for stacking patches of polarization data is not as straightforward as for scalar signals like the CMB temperature or the \ep\- mode polarization; indeed, the $Q$ and $U$ tensorial components are referred to a local frame, and patches at different locations cannot be directly stacked together. Here instead, we employ a configuration of the Stokes parameters that allows for superposition; more precisely, we use the following locally defined rotation of the Stokes parameters:
\begin{linenomath*}
\begin{equation} \begin{array}{ccc}
Q_\mathrm{r}\left(\boldsymbol{\theta}\right) & = & -Q\left(\boldsymbol{\theta}\right)\cos{(2\phi)}-U\left(\boldsymbol{\theta}\right)\sin{(2\phi)}  \\
U_\mathrm{r}\left(\boldsymbol{\theta}\right) & = & Q\left(\boldsymbol{\theta}\right)\sin{(2\phi)}-U\left(\boldsymbol{\theta}\right)\cos{(2\phi)}  \\
\end{array},
\label{eq:def_qr}
\end{equation}
\end{linenomath*}
where $\boldsymbol{\theta} = \theta \left( \cos{\phi}, \sin{\phi}\right)$ and $\phi$ is the angle defined by the line that connects the location considered at the centre of the reference system and a position at an angular distance $\theta$ from the centre. This definition, first proposed by \citet{Kamionkowski1997}, decomposes the linear polarization into a radial ($Q_\mathrm{r} > 0$) and a tangential ($Q_\mathrm{r} < 0$) contribution around the reference positions. \citet{Komatsu2011} provided a recipe to compute the theoretical $T$, $Q_\mathrm{r}$ and $U_\mathrm{r}$ angular profiles from stacked patches centred on temperature peaks, making explicit its dependence on the correlations of the CMB primordial anisotropies.

In practice, we also remove the monopole and dipole from the temperature maps outside the mask, before computing the $Q_\mathrm{r}$ signal around each location of the GR08 structures. Similarly to the work done in \citet{planck2013-p14}, we then compute two types of profiles from each $T$, $Q_\mathrm{r}$ and $U_\mathrm{r}$ patch. On the one hand, the radial angular profile is obtained as the mean of the pixels in rings of fixed width. We choose $150$ different angular scales from $0^{\circ}$ and $15^{\circ}$, with a width of $\delta \theta = 0\pdeg5$ for each ring. On the other hand, the value of the photometry profile at a given scale is defined as the difference between the average signal within the disk of radius $\theta$ and the surrounding ring of equal area (i.e., between radius $\theta$ and $\sqrt{2}\theta$). In this case, $150$ angular scales are also taken into account, defining aperture sizes between $0^{\circ}$ and $15^{\circ}/\sqrt{2} \simeq 10\pdeg6$. The final step is to compute the average of all profiles (radial or photometric) for all of the selected locations.

Complementary to the temperature analysis, the stacked profiles are also computed for the \ep\-correlated (\tce) and the \ep\-uncorrelated (\tue) temperature maps (see Sect.~\ref{sec:data_cmb}), as well as for the \Planck\ lensing map, where we search for a counterpart to the anomalous temperature signal. The stacked images for every map and set of structures considered in this section are shown in Figs. \ref{fig:patches_stacking_isw_T} and \ref{fig:patches_stacking_isw_pol}.


\begin{figure}
\begin{center}
\includegraphics[width=0.495\textwidth]{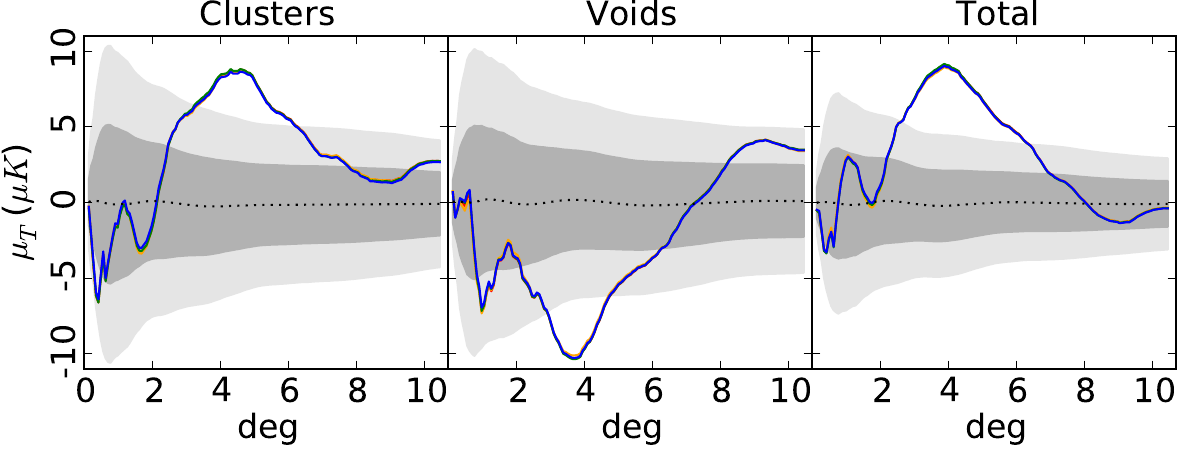}
\caption{Photometry profiles of the stacked temperature patches at the supercluster (first panel) or supervoid (second panel) positions from the GR08 catalogue. The third panel shows the difference between cluster and void profiles. Coloured lines correspond to the different component separation methods: \cruler\ (red); \nilc\ (orange); \sevem\ (green); and \smica\ (blue). Notice that the four lines are almost exactly on top of each other. The dotted black lines correspond to the mean values of the null profiles, i.e., computed at the same locations as the real superstructures, but in $1000$ FFP8 simulations processed through the \sevem\ pipeline. The shaded regions show the $\pm 1\sigma$ and $\pm 2\sigma$ uncertainties. Similar levels are obtained for the different component separation methods.}
\label{fig:stacking_granett_T} 
\end{center}  
\end{figure}

\subsection{Temperature analysis}

In order to confirm the result presented in \citet{planck2013-p14}, we carry out the stacking of temperature patches at the locations of the GR08 structures. For the whole analysis, we use \healpix\ maps at $N_{\mathrm{side}} = 512$ with a filter of $\mathrm{FWHM} = 20\ \mathrm{arcmin}$. The mean radial and photometry profiles are computed from each set of $50$ superclusters and supervoids, respectively. Simultaneously, we perform the same analysis on $1000$ FFP8 simulations of CMB temperature, and derive the statistical properties of the resulting profiles (mean and standard deviation at all scales). Finally, we determine if the profiles measured on real CMB data present any significant deviation from those derived from the simulations. Since we expect the simulated maps to have no correlation with the actual large-scale structures of the Universe, this procedure corresponds to carrying out a null hypothesis test.
 
As we show in Fig. \ref{fig:stacking_granett_T}, we observe the peculiar shape for the profiles already detected in \citet{planck2013-p14} using the CMB temperature maps supplied by the different component separation methods, i.e., an excess of temperature signal at scales around $5^\circ$ in the photometry profiles computed on the supercluster positions, and a deficit at scales around $4^\circ$ in the corresponding supervoid locations. The deviation is even more evident if the total photometry profiles are computed as the difference between the profiles from clusters and voids, as is shown in the third panel of Fig. \ref{fig:stacking_granett_T}. 

\begin{figure}
\begin{center}
\includegraphics[width=0.495\textwidth]{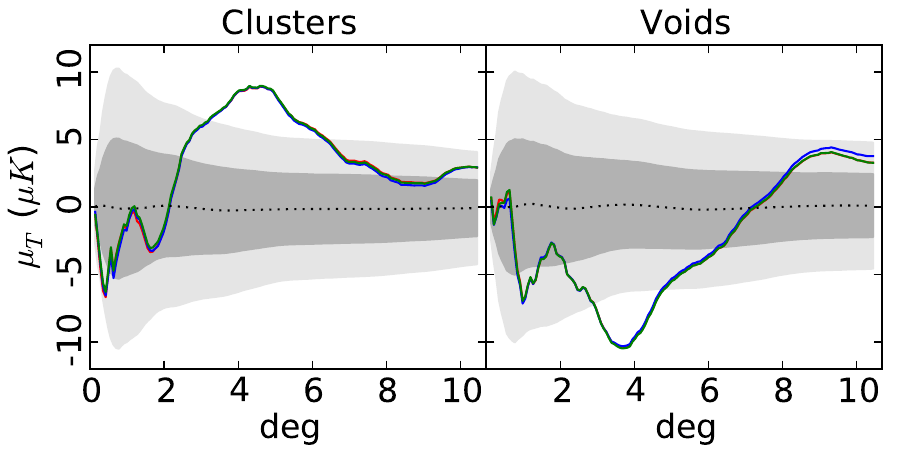}
\caption{Mean photometry profiles of the stacked temperature patches at the supercluster (first panel) and supervoid (second panel) positions of the GR08 catalogue. The CMB data used are the $100\ \mathrm{GHz}$ (red), $143\ \mathrm{GHz}$ (blue), and $217\ \mathrm{GHz}$ (green) cleaned maps supplied by \sevem. The dotted black line and shaded regions show the mean, $\pm 1\sigma$ and $\pm 2\sigma$ uncertainties of the null profiles, i.e., computed at the same locations as the real superstructures but in $1000$ FFP8 simulations processed through the $143\, \mathrm{GHz}$ \sevem\ pipeline.}
\label{fig:stacking_granett_T_sevem_freq} 
\end{center}  
\end{figure}

A multi-frequency analysis on \sevem\ maps is also performed to check if these deviations are monochromatic or, conversely, show a specific frequency dependence. In Fig. \ref{fig:stacking_granett_T_sevem_freq}, we show the mean temperature profiles computed in the $100\ \mathrm{GHz}$, $143\ \mathrm{GHz}$, and $217\ \mathrm{GHz}$ maps. The error bars are estimated as the dispersion of the mean profiles computed at the GR08 positions in $1000$ FFP8 simulations processed through the corresponding \sevem\ pipeline. As we show in Fig. \ref{fig:stacking_granett_T_sevem_freq}, the temperature signal is frequency independent, as already checked in \citet{planck2013-p14}. The uncertainties plotted in the panel correspond to the $143\ \mathrm{GHz}$ case only, since the level of the corresponding uncertainties for the other frequencies is similar.

In summary, the analyses performed in this section confirm that, as expected, the \Planck\ 2015 temperature data also exhibit an anomalous signal that can be associated with the GR08 catalogue.


\subsection{Polarization analysis}

One of the most attractive developments for this \Planck\ release is the possibility of exploring the counterpart in polarization of these temperature anomalies. Our motivation here is the following: if the $4\,\sigma$ signal measured in temperature is dominated by the primordial part of the CMB, it is reasonable to assume that it will have some form of counterpart in polarization (although it might not be detectable within the total polarization signal). Conversely, if these temperature deviations are created as a result of the presence of large clusters and voids, then no correlated signal is expected in polarization. 

In the following, we take several approaches in order to make the most of the potentially discriminating power of the polarization data, focusing on the study of the GR08 results. It should be noted, however, that the strength of these tests could be diminished by the high-pass filtering of the \Planck\ 2015 polarization data release.

\subsubsection{$Q_\mathrm{r}$/$U_\mathrm{r}$ profile significance estimation}

\noindent Since the $Q_\mathrm{r}$ signal is proportional to the correlation between temperature and \ep\-mode polarization~\citep[e.g.,][]{Komatsu2011}, it represents a valuable observable for studying a potential polarization counterpart to the previously observed temperature signal. It should be noted that, a priori, no signal is expected in the $U_\mathrm{r}$ map since it would depend on $TB$ correlations, which are null in the standard model.

\begin{figure}
\begin{center}
\includegraphics[width=0.495\textwidth]{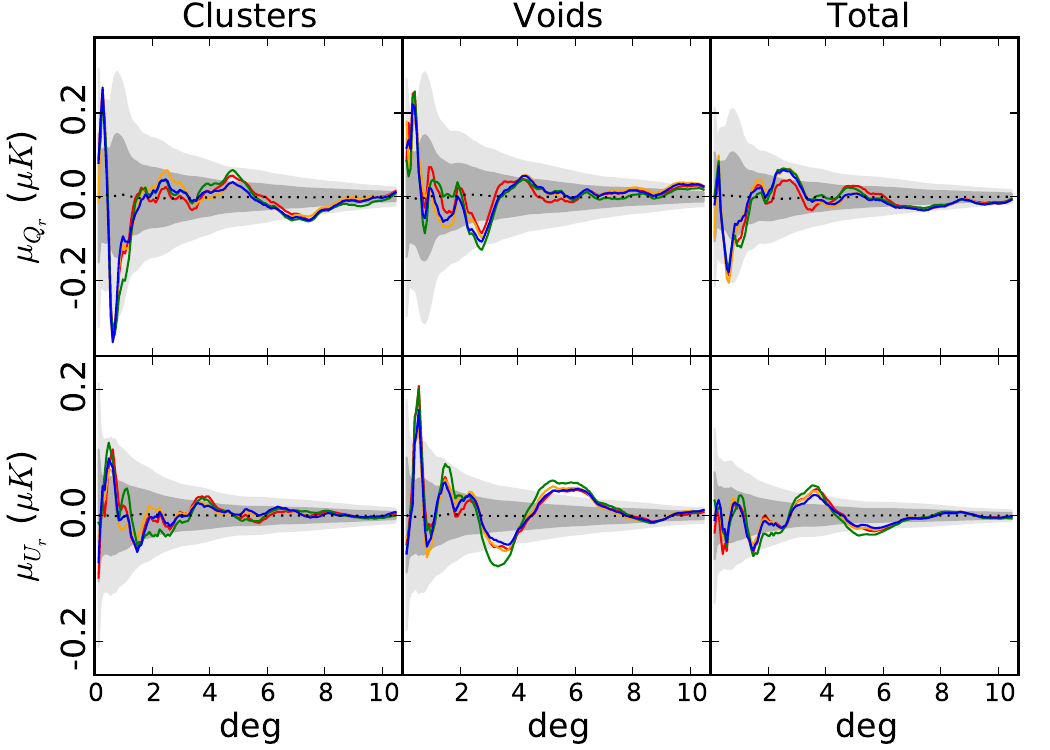}
\caption{Mean photometry profiles of the $Q_\mathrm{r}$ (first row) and $U_\mathrm{r}$ (second row) components stacked at the $50$ supercluster (first column) and $50$ supervoid (second column) positions of GR08. The CMB maps used come from \cruler\ (red); \nilc\ (orange); \sevem\ (green); and \smica\ (blue). The third column shows the difference between the supercluster and the supervoid profiles. The dotted black lines show the mean value of our estimator computed with $1000$ FFP8 simulations processed through the \sevem\ pipeline at the same locations as the real superstructures. Shaded regions show the $\pm 1\,\sigma$ and $\pm 2\,\sigma$ uncertainties of these null profiles; those computed for the rest of component separation methods reach a similar level.}
\label{fig:stack_H0rnd} 
\end{center}  
\end{figure}

The aperture photometry profiles are shown in Fig. \ref{fig:stack_H0rnd} and overall do not present any significant signal at large angular scales (greater than $1^\circ$). A notable exception comes from the $U_\mathrm{r}$ photometry profile for the voids, which does show two significant excesses around $3^\circ$ and $6^\circ$. However, since no $TB$ correlations are expected either for the primordial CMB or for the ISW effect, these features are most likely caused either by a fortuitous signal, and/or systematics in the polarization map that remain to be characterized, which is not in the scope of this paper.
The deviations seen at angular scales below $1^\circ$, especially in the supercluster case, are somewhat reminiscent of the expected primordial $Q_\mathrm{r}$ peaks, which appear due to the dynamics of the photon flows around over-dense and under-dense regions at the last-scattering surface~\citep[i.e., hot and cold spots in the primordial CMB, see, for instance][for a description of these dynamics]{Hu1997,Komatsu2011,planck2014-a18}. However, this similarity is most likely fortuitous since the shape of the temperature profiles does not bear much similarity to that obtained from the stacking of the extrema of the primordial CMB. In addition, the GR08 positions do not appear to correspond with the positions of CMB extrema. The total polarization photometry profiles, computed as the differences between clusters and voids, are shown in the third column of Fig.\ref{fig:stack_H0rnd}. 

We should note, however, that the significance and interpretation of these results are complicated by the use of high-pass filtering in the polarization data, which could mitigate the signal, in principle, at all scales in the profiles (since they include contributions from a large range of multipoles). 

\subsubsection{Covariance analysis}

\noindent We have also tried a different, more general approach to the problem by focusing on the following question: what should we expect in the $Q_\mathrm{r}$/$U_\mathrm{r}$ stacking signal, if the GR08 temperature signal originates purely from primordial anisotropies?

To answer this question, we perform a set of 100\,000 simulations of CMB $T$, $Q$, and $U$ maps, using the \Planck\ best-fit cosmological model as input. For each one of these sets of maps, we derive the $T$, $Q_\mathrm{r}$, and $U_\mathrm{r}$ stacked images corresponding to the 50 sky positions of the 50 voids of the GR08 catalogue (in order to keep the same, potentially relevant, configuration of positions in the sky). For these images, we derive the radial and photometry profiles, and end up with a collection of 2 (temperature and photometry) $\times$ 3 ($T$, $Q_\mathrm{r}$, $U_\mathrm{r}$) $\times$ 100\,000 profiles. More precisely, for each one of the 100\,000 sets of CMB $T$, $Q$ ,and $U$ maps, we can construct the corresponding vector in which we put end to end the three radial profiles (of the $T$, $Q_\mathrm{r}$, $U_\mathrm{r}$ stacked images) and the three photometry profiles. Using these 100\,000 vectors, we construct a covariance matrix $M_{ij}$, which contain the covariance between any combination of angular scales of any of the $T$/$Q_\mathrm{r}$/$U_\mathrm{r}$ profiles. We also derive the correlation matrix $N_{ij}$, defined as $N_{ij} = M_{ij}/\sqrt{M_{ii} M_{jj}}$.

In the resulting matrices, we look for the existence of significant correlations between a temperature signal, with features similar to the GR08 one (i.e., peaking around a scale of $4^\circ$), and a polarization signal at any scale. The idea here is that in the simulated maps and associated stacked images that we use here, we can be certain that any stacked signal that arises in temperature is fortuitous and due to primordial anisotropies. Starting from this point of view, the covariance analysis allows us to obtain a general picture of how a primordial stacked signal in temperature is correlated to its (potential) polarization counterpart. This provides us with valuable insight when trying to test the hypothesis that the GR08 signal is purely (or partially) primordial.

After performing this analysis, the covariance/correlation matrices obtained show that the temperature photometry at around $4^\circ$ is indeed correlated with a polarization signal, both in the $Q_\mathrm{r}$ radial and photometry profiles (with a maximum correlation around $4^\circ$ for both). The existence of these correlations is quite robust, thanks to the large number of simulations, and confirms that if a significant, primordial CMB signal appears in temperature, it will have a counterpart in polarization. However, it shows that for a GR08-like signal only due to primordial CMB, the biggest correlation factors with polarization are below about $15\,\%$. Therefore a $3\,\sigma$ signal in temperature would only translate on average into $0.5\,\sigma$ signal in $Q_\mathrm{r}$, making it effectively impossible to detect among the rest of the polarization signal.


\subsubsection{\tce\ and \tue\ maps}

\noindent An alternative to the use of the $Q_\mathrm{r}$ and $U_\mathrm{r}$ components is to perform the stacking at the locations of the GR08 structures, but using the \ep\-correlated and \ep\-uncorrelated temperature maps of the CMB described earlier (see Sect.~\ref{sec:data_cmb}). If we were to find that most of the GR08 signal is contained in the stacked image associated with the E-correlated temperature map, this would be a strong argument towards a primordial nature of this signal. Conversely, if it is found mostly in the \ep\-uncorrelated temperature stacked image, it would give credence to the ISW signal hypothesis.

\begin{figure}
\begin{center}
\includegraphics[width=0.495\textwidth]{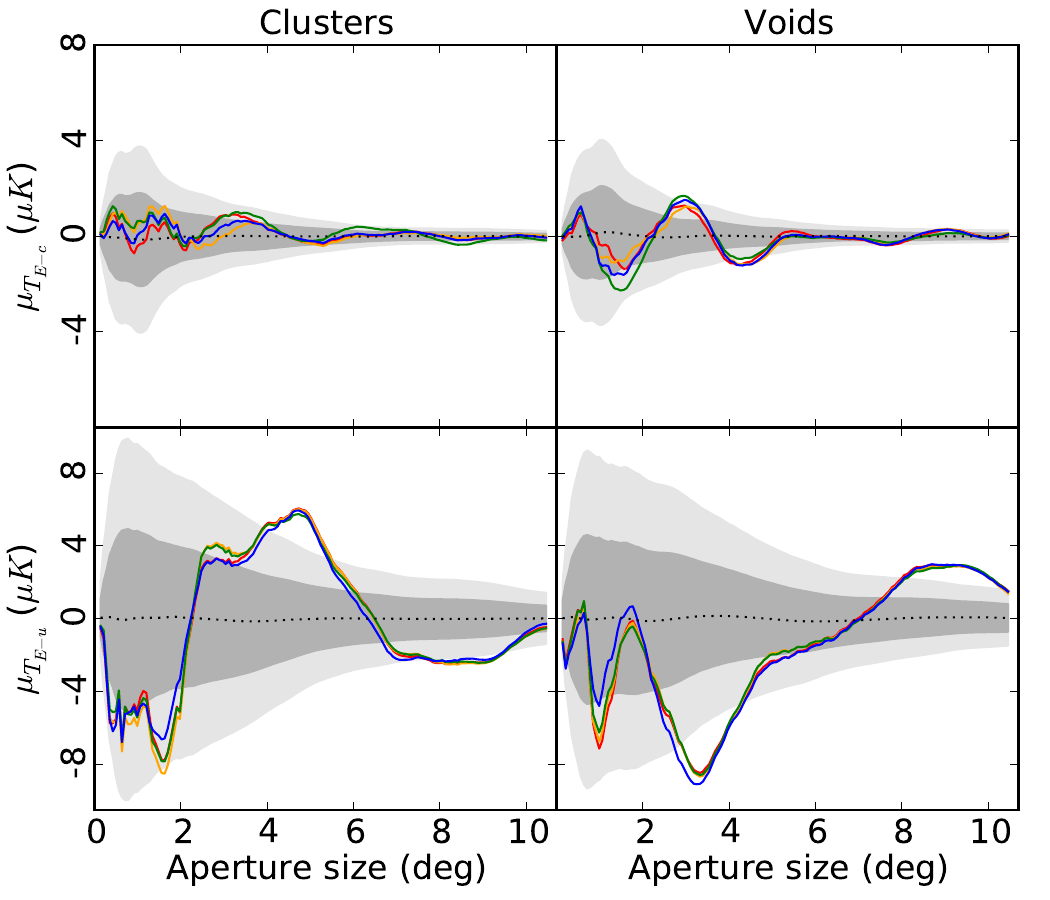}
\caption{Mean photometry profiles of the images for the $50$ superclusters (first column) and the $50$ supervoid (second column) positions of GR08, when stacking in the \tce\ (first row) and \tue\ (second row) maps computed from the CMB maps supplied by \cruler\ (red); \nilc\ (orange); \sevem\ (green); and \smica\ (blue). The dotted black line represents the null hypothesis computed as the mean value of the photometry profile at $50$ random positions in $1000$ FFP8 simulations processed through the \sevem\ pipeline, according to the noise properties of the CMB data at the GR08 superstructure locations. The shaded regions show the $\pm 1\,\sigma$ and $\pm 2\,\sigma$ uncertainties of these profiles, computed as the dispersion of the mean photometry profiles of the simulations. The corresponding error bars for the different component separation methods reach a similar level.}
\label{fig:stacking_corr_uncorr_sevem} 
\end{center}  
\end{figure}

In Fig. \ref{fig:stacking_corr_uncorr_sevem}, we show the mean photometry profiles computed from the two aforementioned maps. It appears quite clearly that for both the superclusters and the supervoids, most of the signal originally observed in temperature is contained in the E-uncorrelated map, thus apparently strengthening the hypothesis of the ISW nature of the signal, however we have to bear in mind that, by construction, the \tue\ map contains more power than the \tce\ map.
We remark that, although the polarization data have been high-pass filtered, most of the relevant scales responsible for the anomalous temperature signal could be still present in the analysed maps, since we have checked that the photometry profiles on high-pass filtered temperature data are very similar to those plotted in Fig.~\ref{fig:stacking_granett_T}.
On the other hand, the \ep\-correlated part of the signal appears to sit within the expected values from simulations.


\subsection{\Planck\ lensing convergence map}

\noindent Another physical observable explored in the present analysis is the \Planck\ lensing convergence map. As explained in Section \ref{sec:data_lss}, the \kap\ map is proportional to the gravitational field and is therefore expected to be correlated with the distribution of the large-scale structures, as well as the individual objects that generate the ISW signal in the CMB. Although complicate projection (and possibly cancellation) effects are expected to be involved here, we can expect that the stacking of the lensing map at the locations of the GR08 structures will give a significant signal with respect to the null hypothesis. On the other hand, an absence of signal could indicate a problem with the structures and it could put into question the method and data used to identify them in the SDSS and therefore even question their existence, or at the very least their reported properties (sizes, redshifts, etc). The legitimacy of such questions is reinforced by recent studies \citep[see][]{Kovacs2015} that failed to detect some of the GR08 structures in newer SDSS data that cover the same survey volume.

\begin{figure}
\begin{center}
\includegraphics[width=0.495\textwidth]{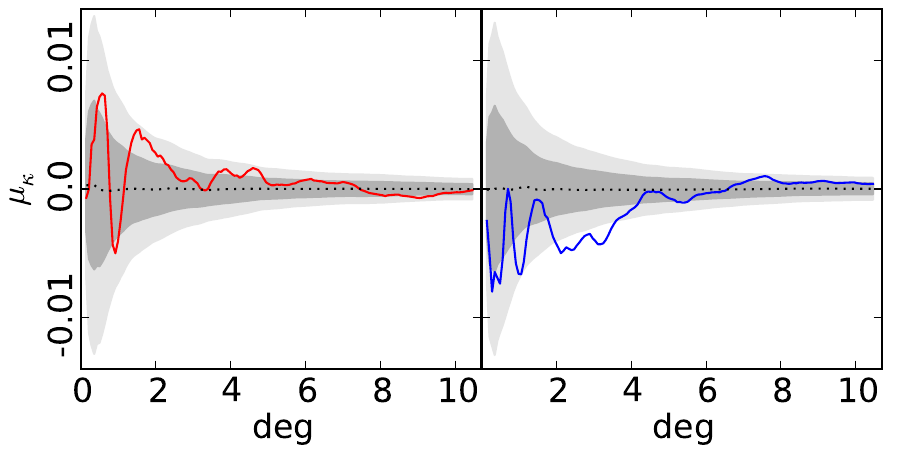}
\caption{Mean photometry profile of the stacked images from the \Planck\ convergence lensing map, at the locations of the $50$ supercluster (left panel) and the $50$ supervoid (right panel) from the GR08 catalogue. The dotted black line represents the null hypothesis computed as the mean value of the photometry profiles at the corresponding \gr\ positions in $1000$ simulations according to the lensing model. The shaded regions show the $\pm 1\,\sigma$ and $\pm 2\,\sigma$ uncertainties of these profiles, computed as the dispersion of the mean photometry profiles of the simulations.}
\label{fig:stacking_kappa} 
\end{center}  
\end{figure}

The photometry profiles computed from the \kap\ map at the positions of the GR08 structures are shown in Fig. \ref{fig:stacking_kappa}. The error bars are estimated with simulations generated according to the lensing model. We should be cautious here when drawing any conclusions, since the lensing map is known to be very noisy. However, it should be noted that the photometry profiles for the clusters and voids show some relatively significant features, with opposite signs, as expected if these profiles result from the averaging of several gravitational wells and hills, respectively. The significance of these features reaches as high as $3\,\sigma$  for the void profiles, although it is hard to pinpoint any typical scale for either of the two cases. However, the fact that the profiles tend to be positive in the case of clusters and negative for voids could be pointing in favour of the ISW interpretation of the temperature signal observed.

To shed light on this question beyond these qualitative statements, a $\chi^2$ estimator is used to evaluate the compatibility of a binned version of these profiles with the null hypothesis, taking into account the correlation between different scales. In the case of superclusters, the p-value reaches a value of $50\,\%$, which could represent an evidence that the observed positive trend is not significant (and without any further evidence in favour of the ISW interpretation). In the case of supervoids, the same probability is around $2\,\%$, such that the observed feature could be a hint of a significant signal with respect to the null hypothesis from the GR08 structures.


\subsection{Summary of stacking}

A stacking analysis of the GR08 positions reveals a temperature signature, as in \citet{planck2013-p14},  which according to the literature compares poorly with the ISW predictions. The major deviation appears at about $4.5^{\circ}$ for clusters and $3.5^{\circ}$ for voids. An analysis of different cleaned frequency maps that \sevem\ provides shows that the photometry profiles are not frequency dependent. This is both compatible with a pure CMB component or an ISW signal, and effectively rules out, for instance a hypothetical foreground contribution.

The use of polarization as a discriminant is the main novelty of this analysis with respect to the previous \Planck\ results. However, the large-scale information of \Planck\ 2015 polarization is suppressed with a high-pass filter, and the conclusions derived from these data should be therefore taken with some caution. The $Q_\mathrm{r}$ photometry profile is revealed to be mostly compatible with the expected signal from random positions. The absence of a counterpart in polarization is expected for a contribution caused by a secondary anisotropy. On the other hand, a theoretical covariance analysis shows that a primary temperature anisotropy does have a counterpart in polarization, but at such a weak level that it would be difficult to detect.

The analysis of the \ep\-correlated and \ep\-uncorrelated temperature maps at the GR08 locations supplies a complementary, and supposedly cleaner way to access the potential polarization counterpart. We found that the largest part of the temperature excess appears in the \ep\-uncorrelated component, and is comparable to the signal recovered from a high-pass filtered version of the total temperature map. Moreover, the stacking of the \ep\-correlated maps seems compatible with the contribution of random positions. Although we cannot conclude that the excess is not present in the primordial contribution, we do assert that it is compatible with a contribution caused by a secondary anisotropy, and therefore with an ISW signal. To summarize, we have found some hints that seem to point towards an ISW interpretation of the stacked signal observed in temperature at the position of the GR08 superstructures. However, our analysis of the current high-pass filtered CMB polarization maps cannot yet completely confirm nor invalidate an ISW origin. These analyses have to be further explored with the next \Planck\ data release, where the CMB polarization is expected to be recovered at all angular scales.

As an additional test, we performed a similar stacking of the same positions in the \kap\ map. It revealed a significant negative signal in the photometry profile associated to the voids positions. This most likely confirms the presence of supervoids (more precisely gravitational potentials) at these positions, which helps clearing some doubts about the detection and use of such structures, even if the overall amplitude remains unexplained.

\section{ISW map recovery}

\label{sec:recov}
As we did in~\cite{planck2013-p14}, we apply the linear covariance-based (LCB) filter first introduced by~\cite{Barreiro2008}, and recently extended in~\cite{Manzotti2014} and~\cite{Bonavera2016} to deal with several LSS tracers jointly.

An alternative approach to estimate the ISW temperature map that does not require the input of any CMB data, but only LSS data, is discussed in Appendix~\ref{sec:appen}, following a simplified version of the method proposed by~\cite{Kitaura2010} and~\cite{Jasche2010}, who estimated the 3D gravitational potential of a galaxy network given in terms of a redshift catalogue.

\label{sec:lcb}

The LCB filter is able to combine all the information encoded in the CMB and LSS data about the ISW effect, in order to recover an actual map of this weak signal. In particular, the LCB filter \citep{Barreiro2008,Barreiro2013} was originally developed to recover the ISW map by combining CMB intensity data and one LSS tracer. The method has now been extended to deal with any number of LSS surveys \citep{Manzotti2014}, as well as to include polarization information \citep{Bonavera2016} .

\begin{figure*}
\centering
\includegraphics[width=0.495\textwidth]{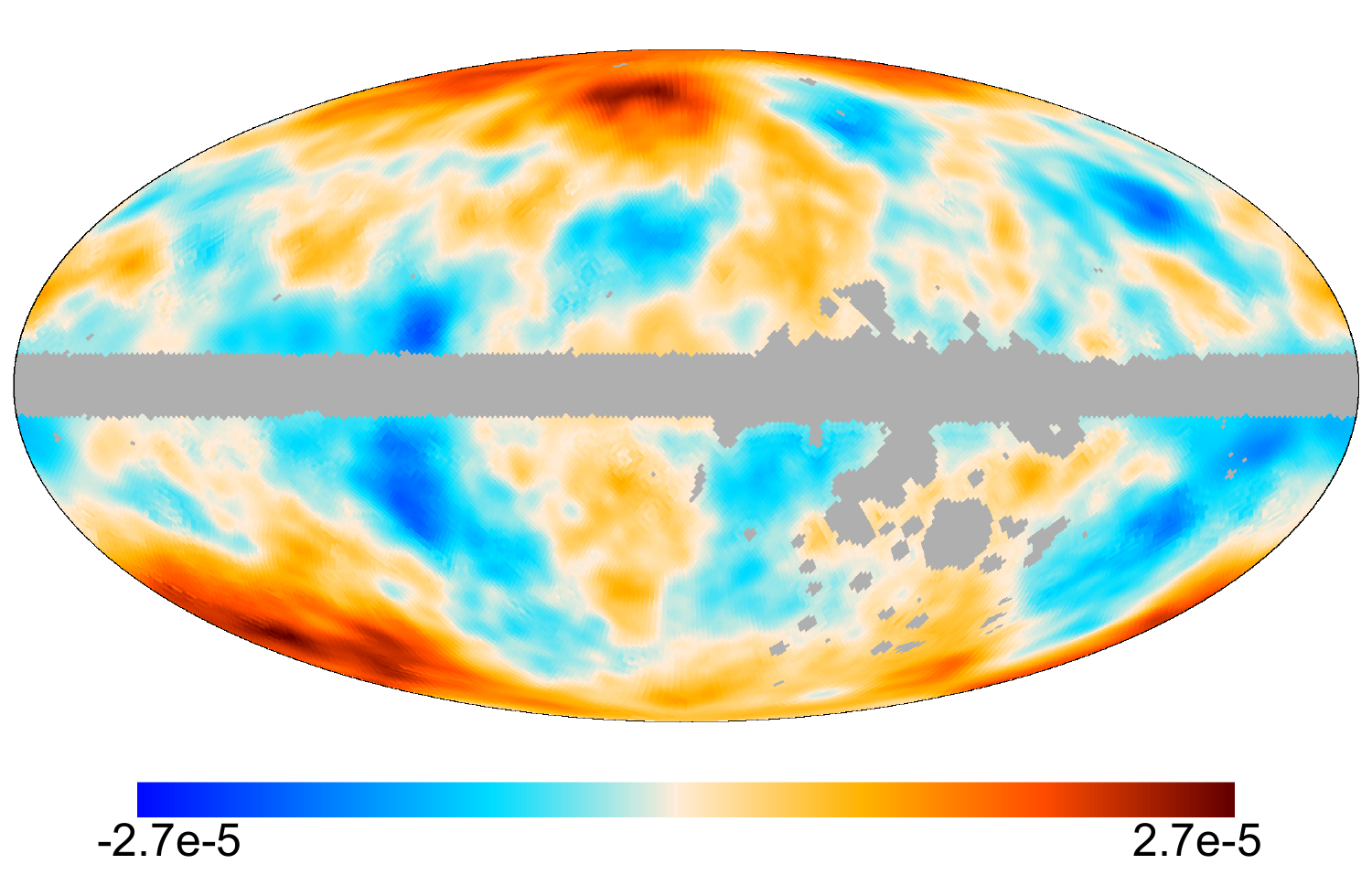}
\includegraphics[width=0.495\textwidth]{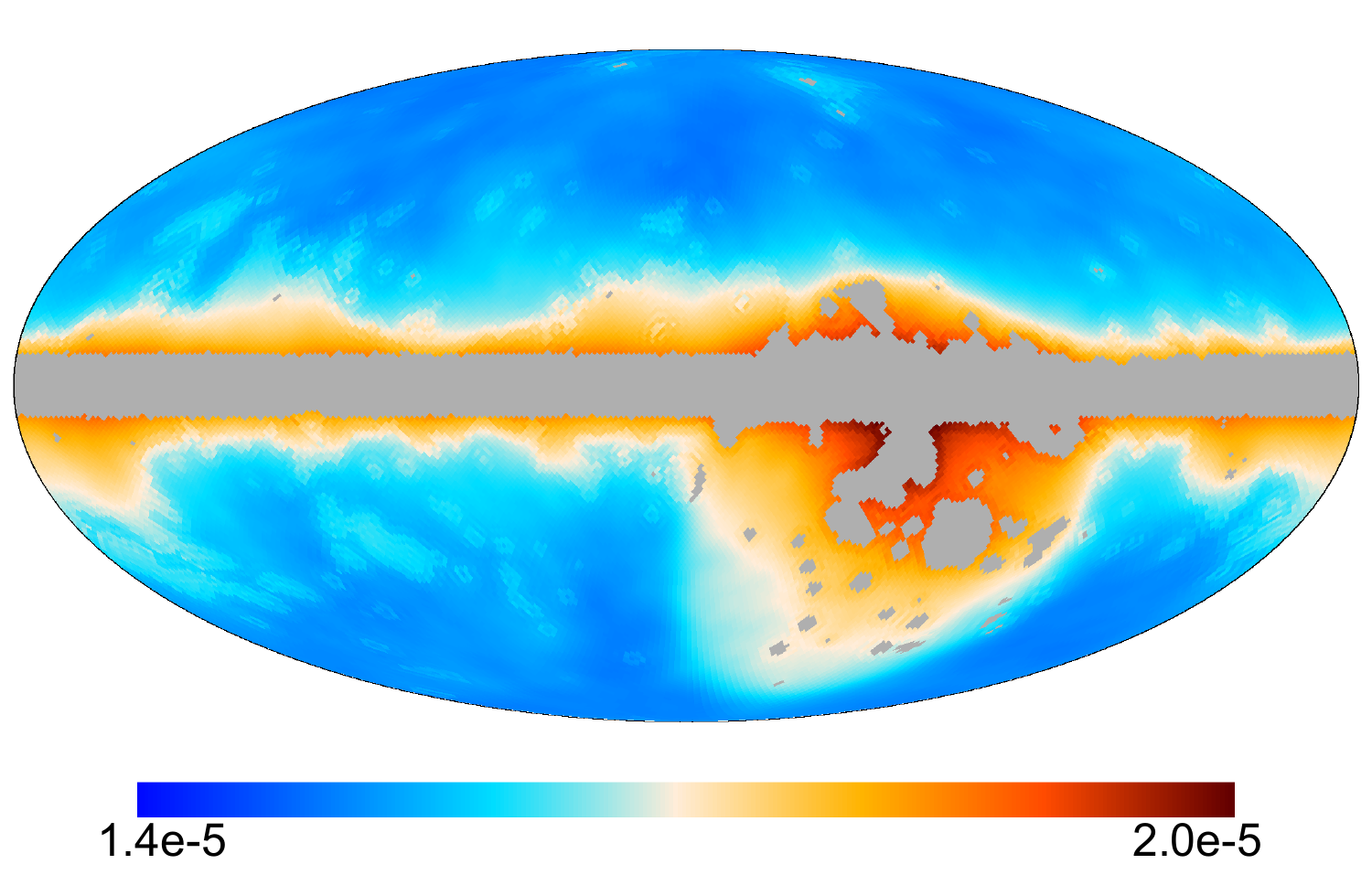}\\
\includegraphics[width=0.495\textwidth]{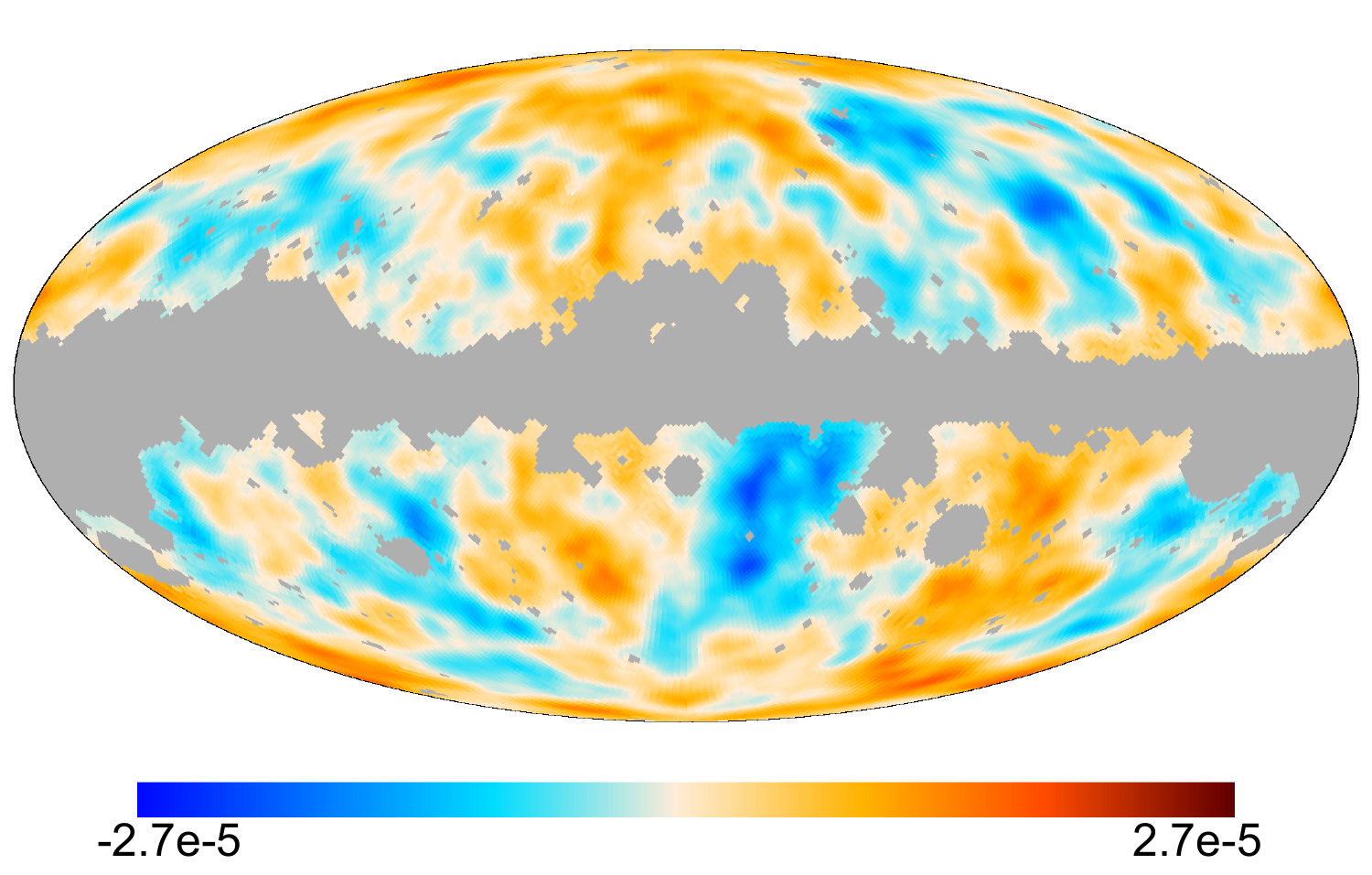}
\includegraphics[width=0.495\textwidth]{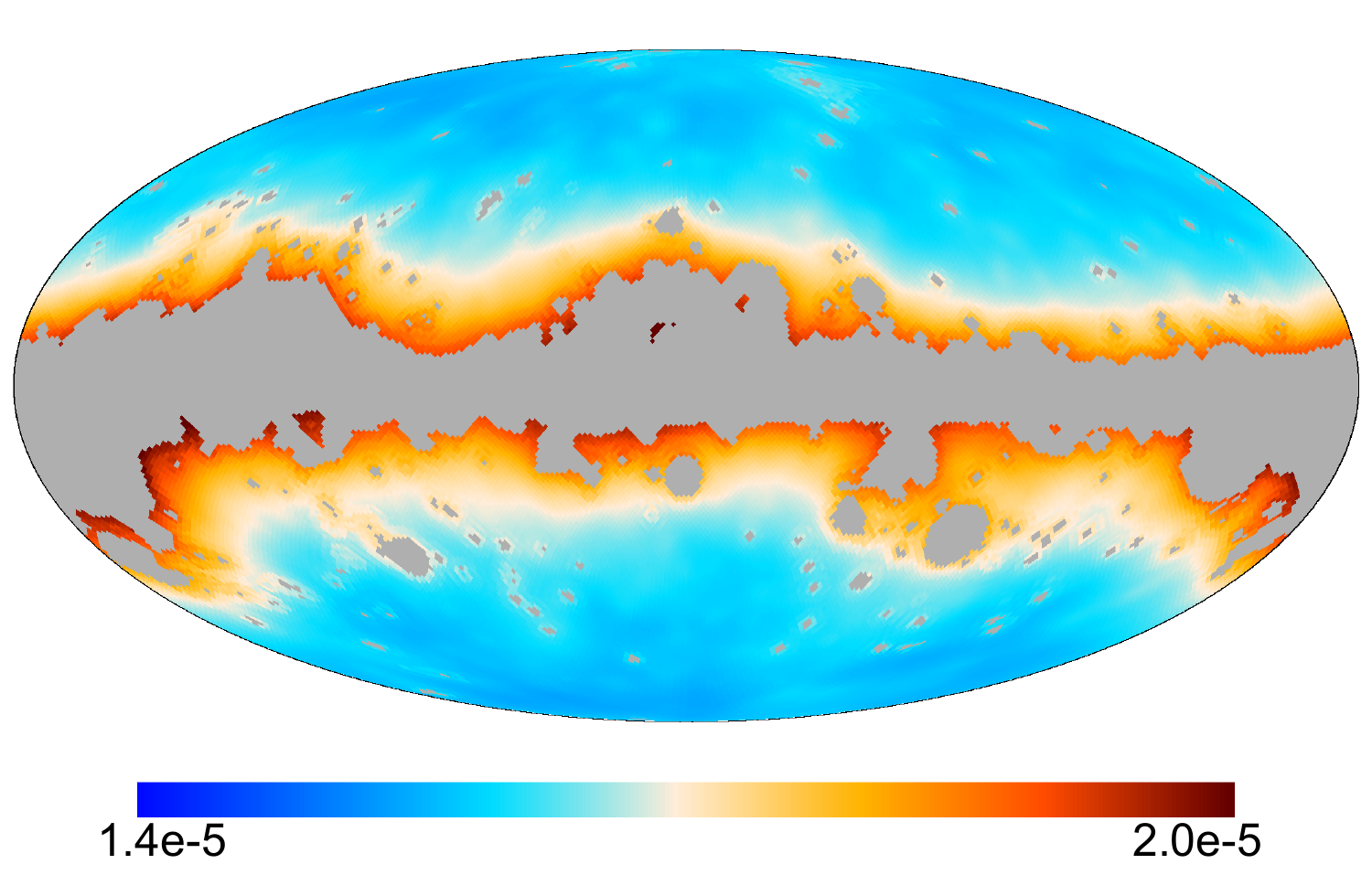}\\
\includegraphics[width=0.495\textwidth]{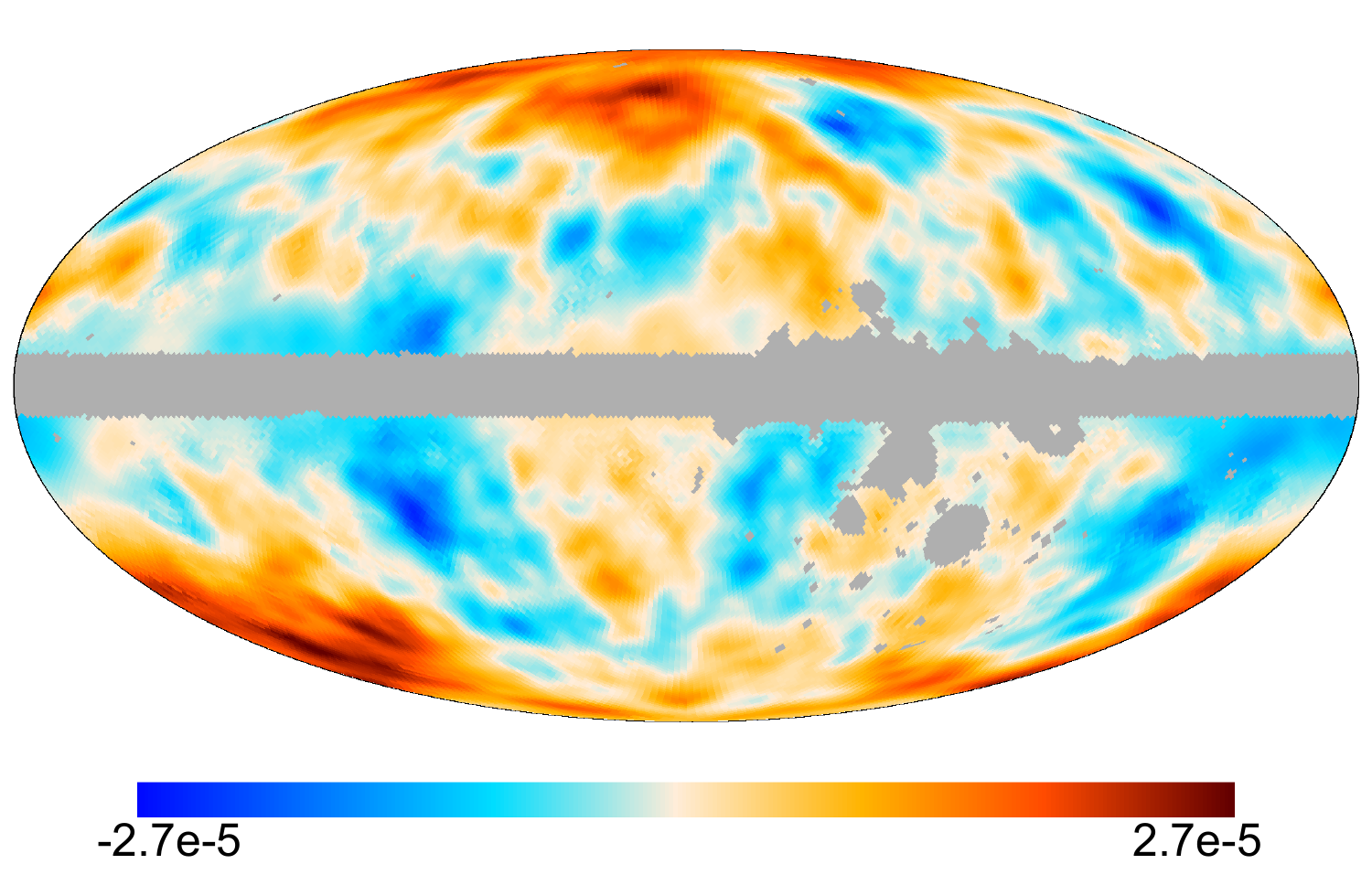}
\includegraphics[width=0.495\textwidth]{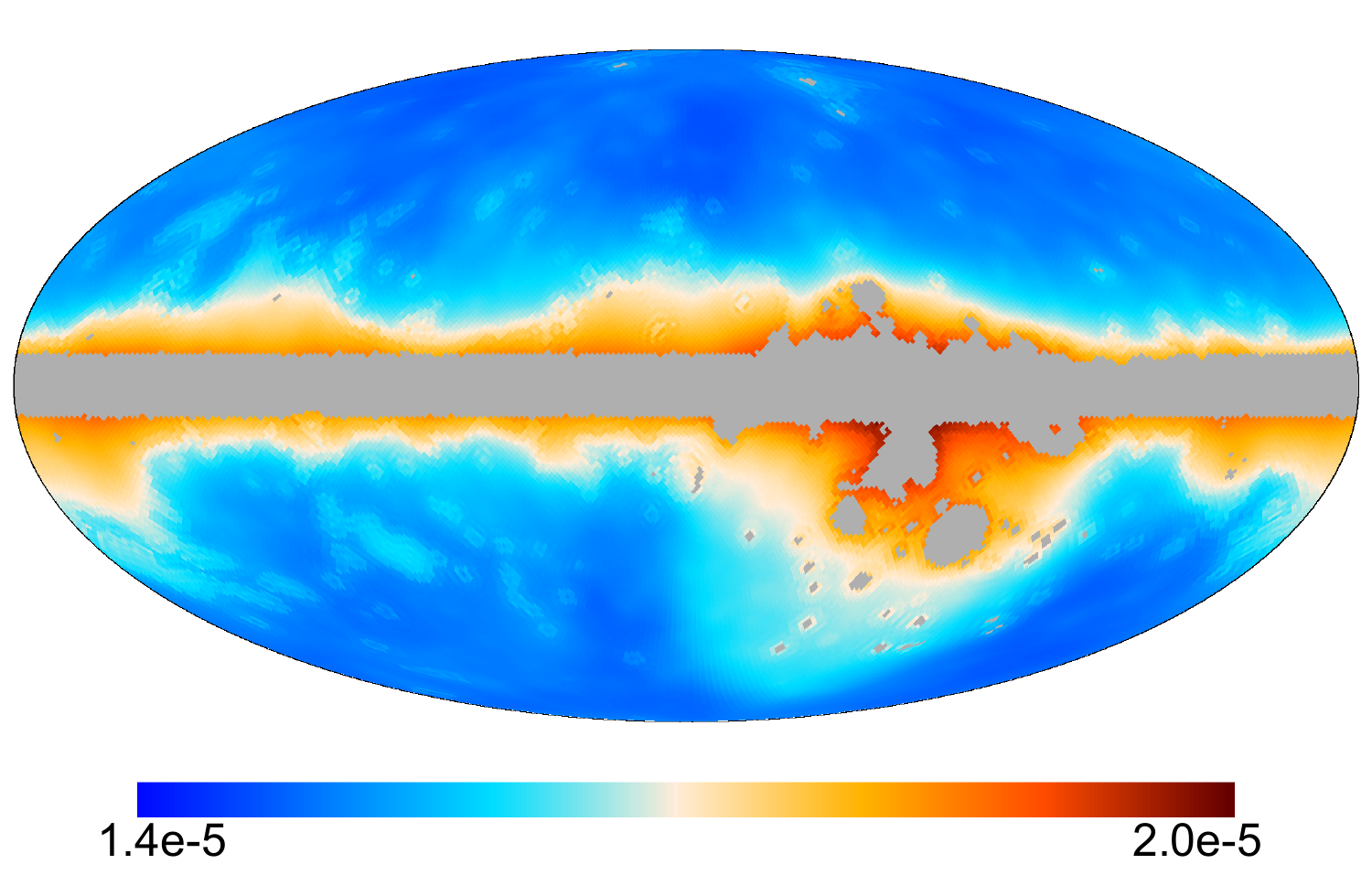}\\
\caption[]{Map of the recovered ISW anisotropies (left column) and the corresponding estimated error per pixel  (right column) obtained from the combination of the \Planck\ \sevem\ CMB map with: the \nvss\ survey (top); the \Planck\ lensing map (middle); and both tracers jointly (bottom). The units here are in Kelvin.} 
\label{fig:lcb_nvss_kappa}
\end{figure*}

\subsection{Methodology}

\noindent We briefly describe here the formalism of the extended method that will be used in this paper. In order to construct the filter for $n$ given surveys, the covariance matrix $\tens{C}(\ell)$ between ISW and LSS data is assumed to be known. We note that, at each multipole, $\tens{C}$ is a square matrix of order $t=n+1$. To simplify the notation, the matrix is written such that the first $n$ elements (whose harmonic coefficients are given by $g_{j=1,n}(\ell,m)$) refer to the auto- and cross-spectra involving only the $n$ LSS tracers, while the $n+1$ element contains the auto- and cross-spectra that include the ISW effect ($d(\ell,m)$ being the harmonic coefficients of the CMB intensity map). Through a Cholesky decomposition of the covariance matrix, we construct the matrix $\tens{L}$ satisfying $\tens{C}(\ell)=\tens{L}(\ell)\tens{L}^\mathrm{T}(\ell)$. 
The estimated ISW map $\hat{s}(\ell,m)$ at each harmonic mode is then given by:
\begin{linenomath*}
\begin{eqnarray}
\label{eq:rec_n}
\begin{split}
\hat{s}(\ell,m) = \sum\limits_{i=1}^{n} \left[ \tens{L}_{it}
\left(\sum\limits_{j=1}^{n}\left(\tens{L}^{-1}\right)_{ij}g_j(\ell,m)\right)\right]  + \frac{\tens{L}_{tt}^2}{\tens{L}_{tt}^2+C_{\ell}^n}\\
\left\lbrace  d(\ell,m)-\sum\limits_{i=1}^{n}\left[ \tens{L}_{it} \left( \sum\limits_{j=1}^{n}\left(\tens{L}^{-1}\right)_{ij}g_j(\ell,m)\right)\right] \right\rbrace 
\end{split}
\end{eqnarray}
\end{linenomath*}
where $C_{\ell}^n$ corresponds to the power spectrum of the CMB signal, without including the ISW contribution. To simplify the notation, we have dropped the dependence of the Cholesky matrix $\tens{L}$ on $\ell$. We note also that although, in principle, the inversion of $\tens{L}$ should be performed only on the $n \times n$ submatrix, it is equivalent to use of the full matrix, since it is triangular.

The realistic case of incomplete sky coverage or the presence of Poissonian noise in the surveys can be accommodated in the previous equation. In particular, the contribution of the Poissonian noise is simply added to the auto-spectrum of the corresponding survey. For those data with partial sky coverage, their corresponding cross- and auto-spectra are replaced in the filter by its {\it masked} version, i.e., correlations among different multipoles in the power spectra are integrated with the {\tt MASTER} algorithm.

\subsection{Results}

\begin{figure*}
\centering
\includegraphics[width=0.495\textwidth]{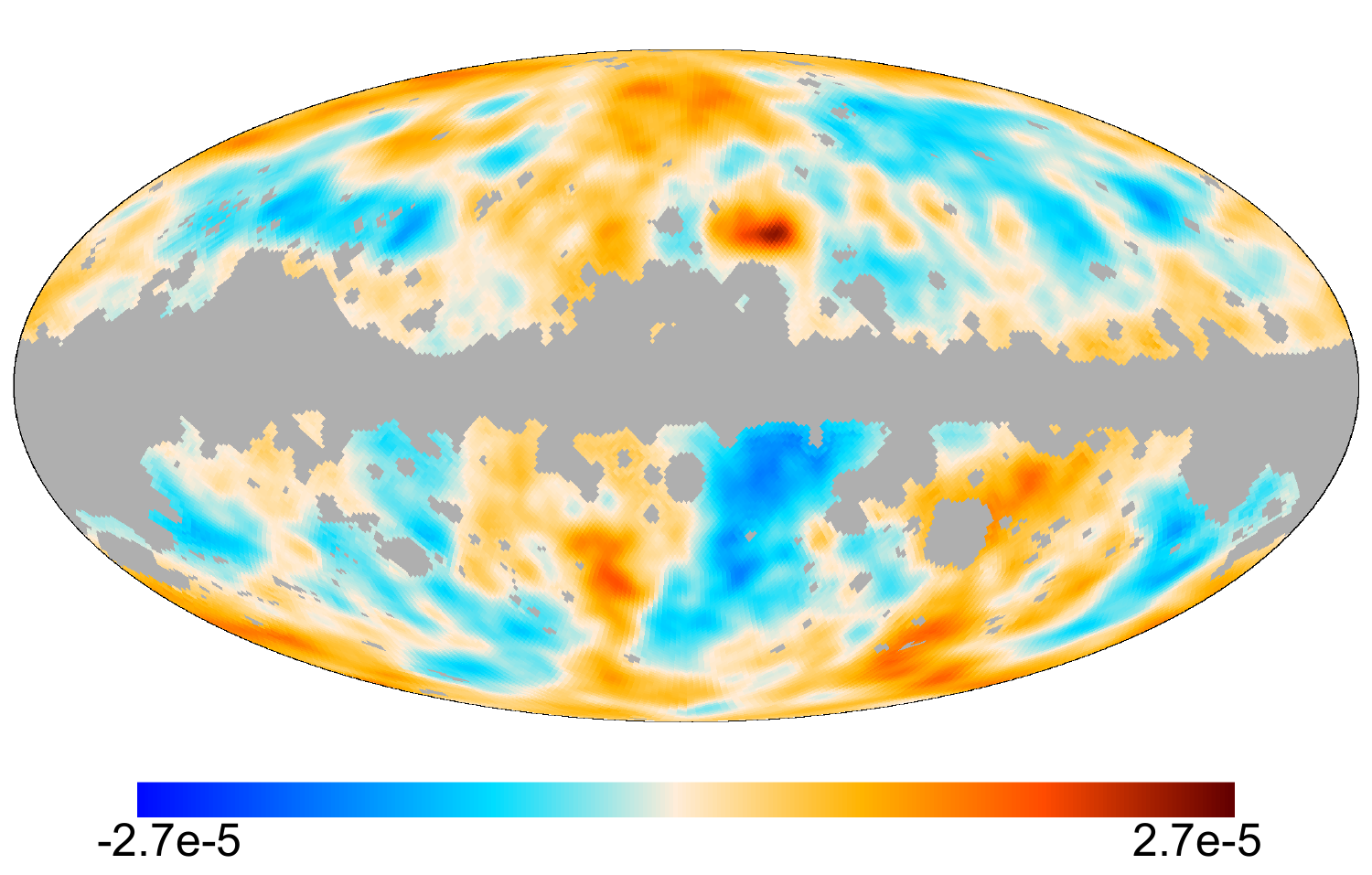}
\includegraphics[width=0.495\textwidth]{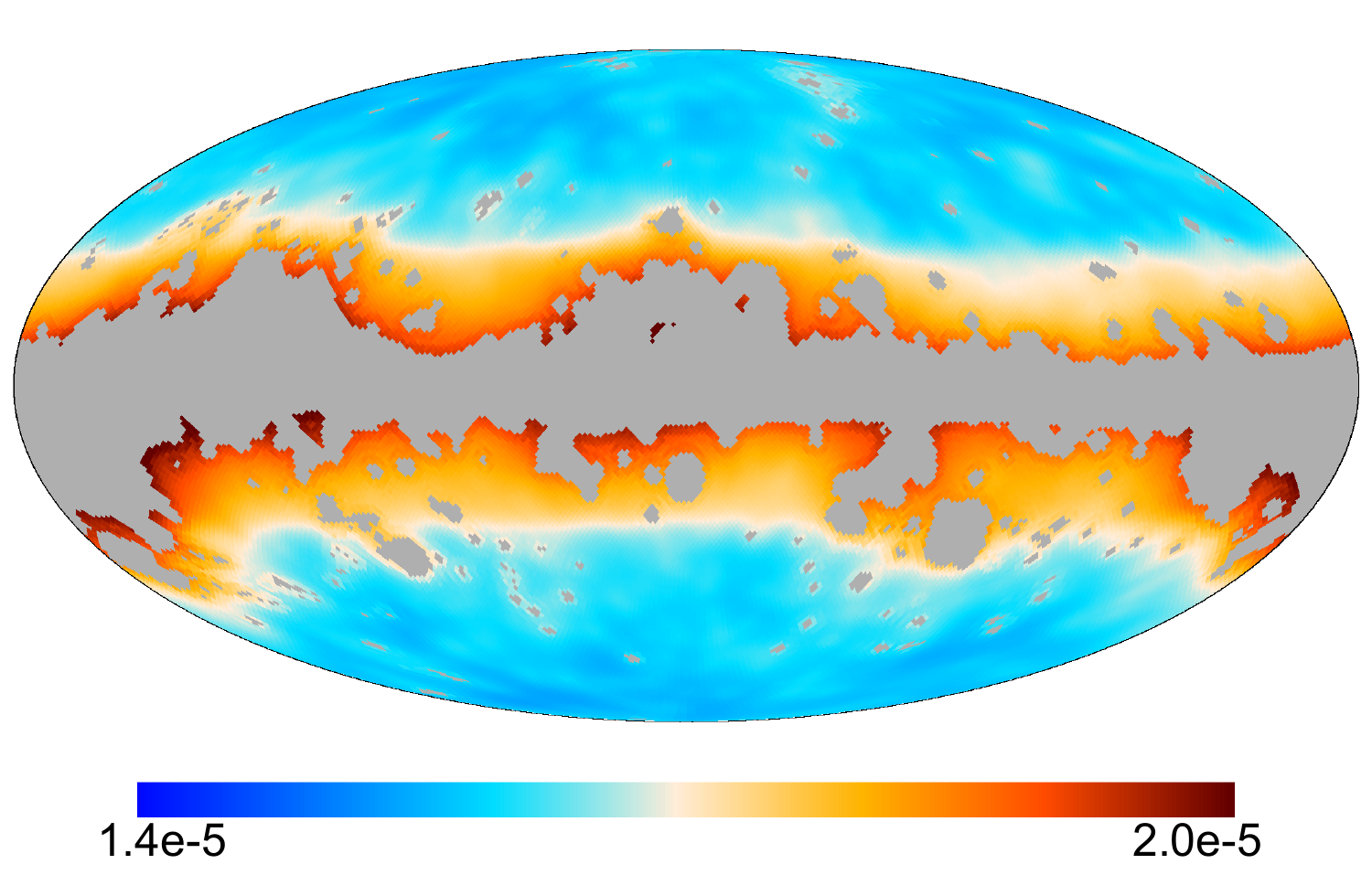}\\
\includegraphics[width=0.495\textwidth]{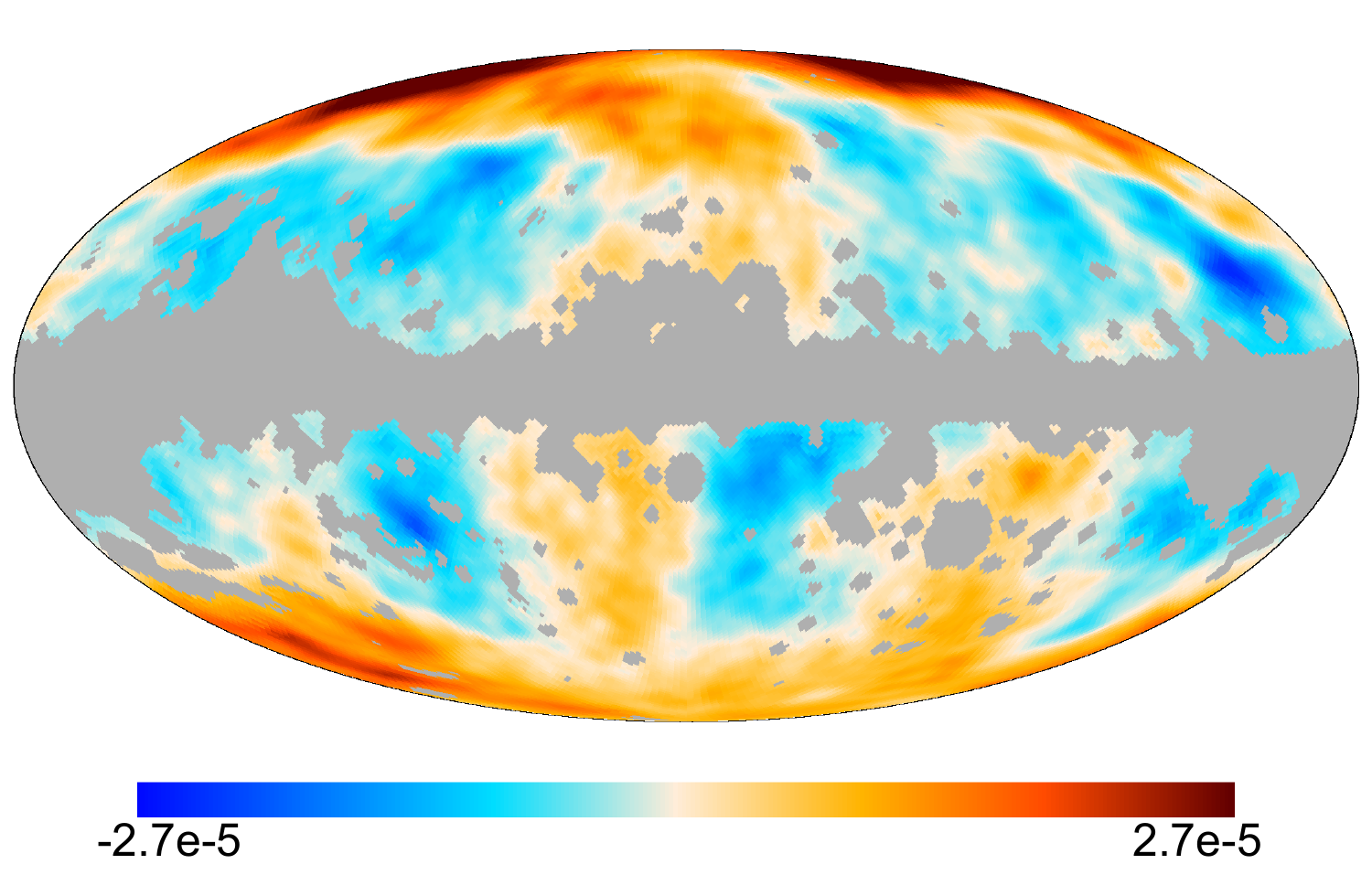}
\includegraphics[width=0.495\textwidth]{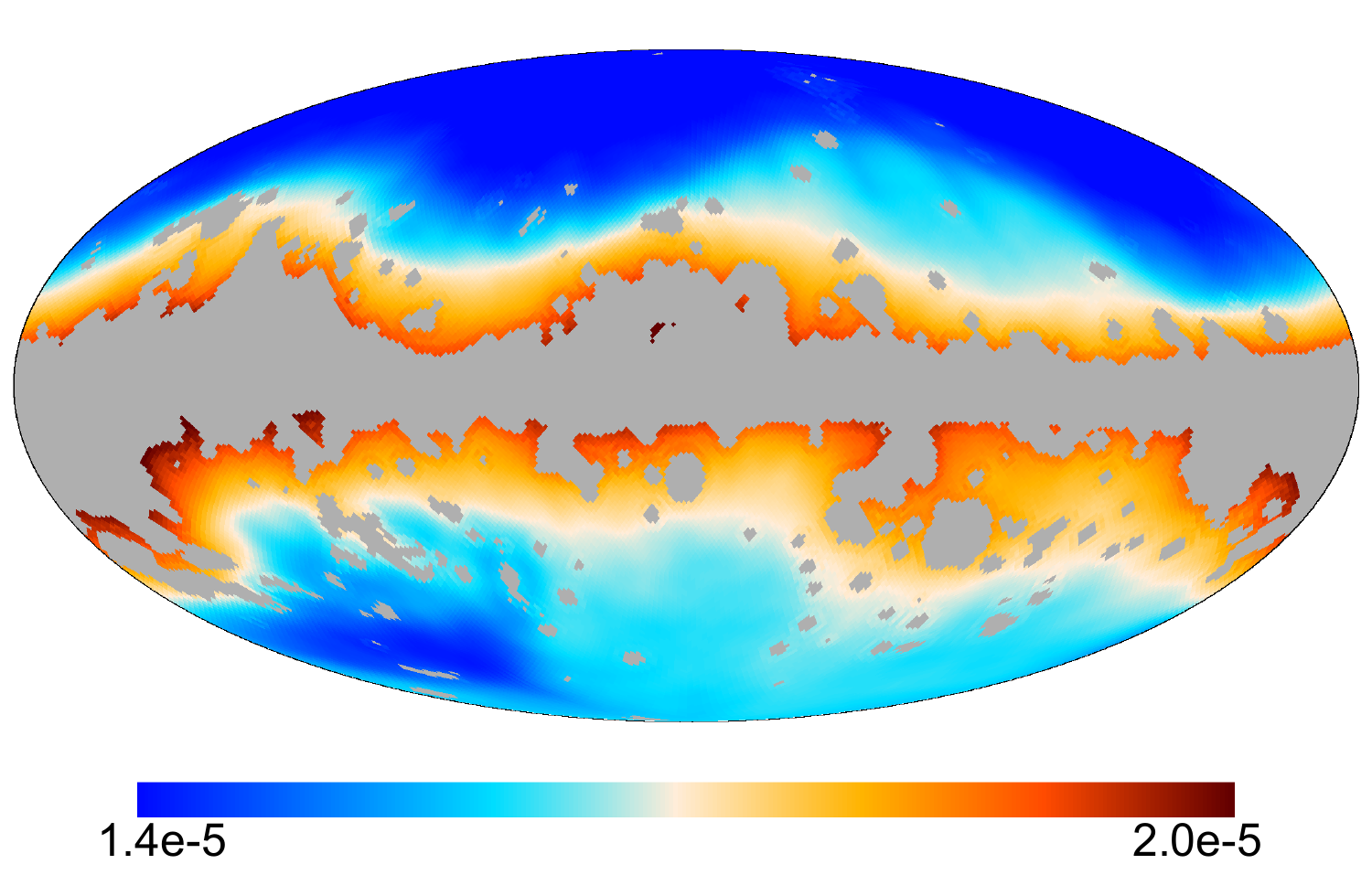}\\
\caption[]{Map of the recovered ISW anisotropies (left column) and the corresponding estimated uncertainty per pixel (right column) from the combination of the \Planck\ \sevem\ CMB map with: the two \wise\ surveys (top); and the two \sdss\ tracers (bottom). The units here are Kelvin.} 
\label{fig:lcb_wise_sdss}
\end{figure*}

\begin{figure*}
\centering
\includegraphics[width=0.495\textwidth]{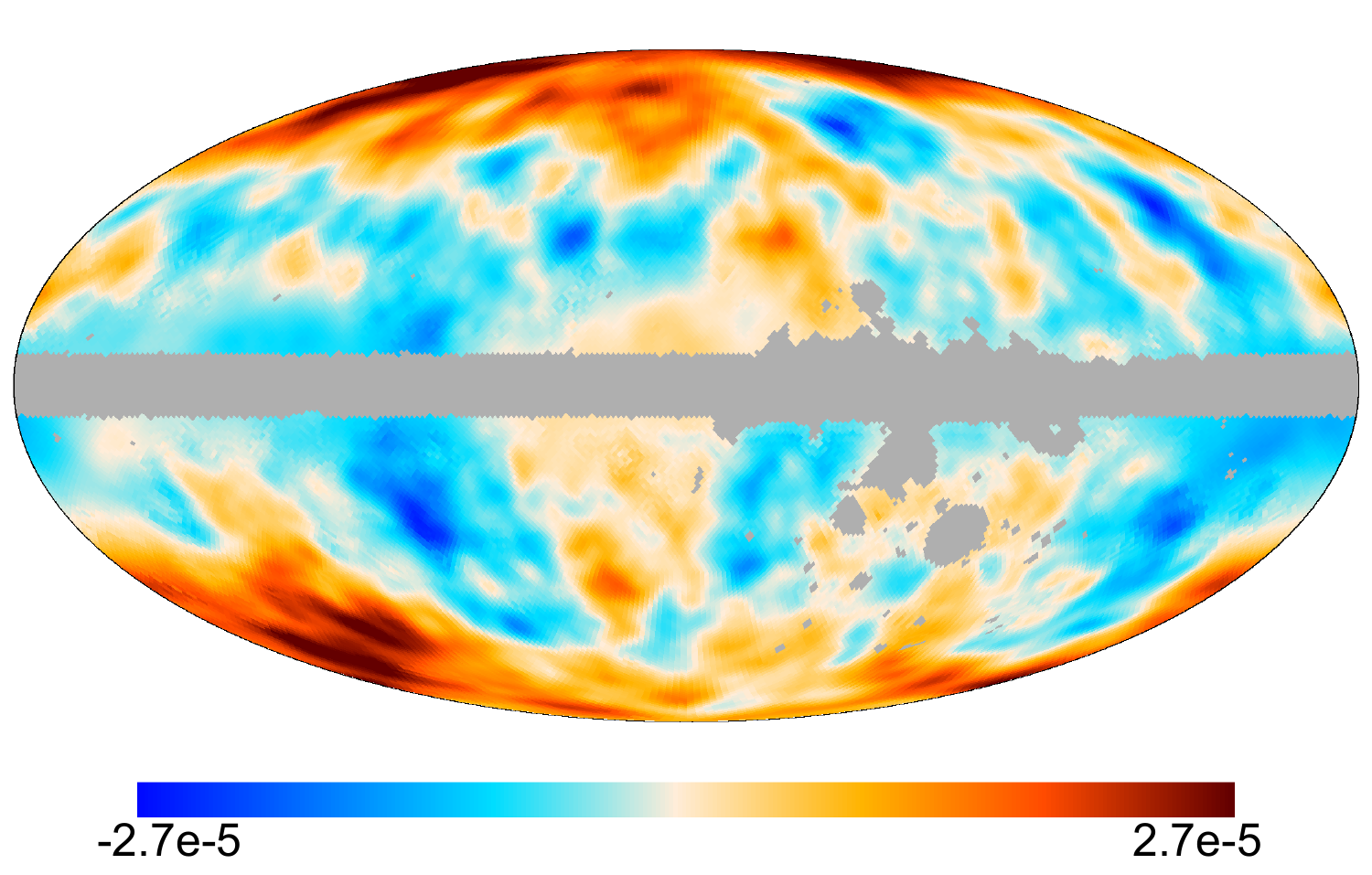}
\includegraphics[width=0.495\textwidth]{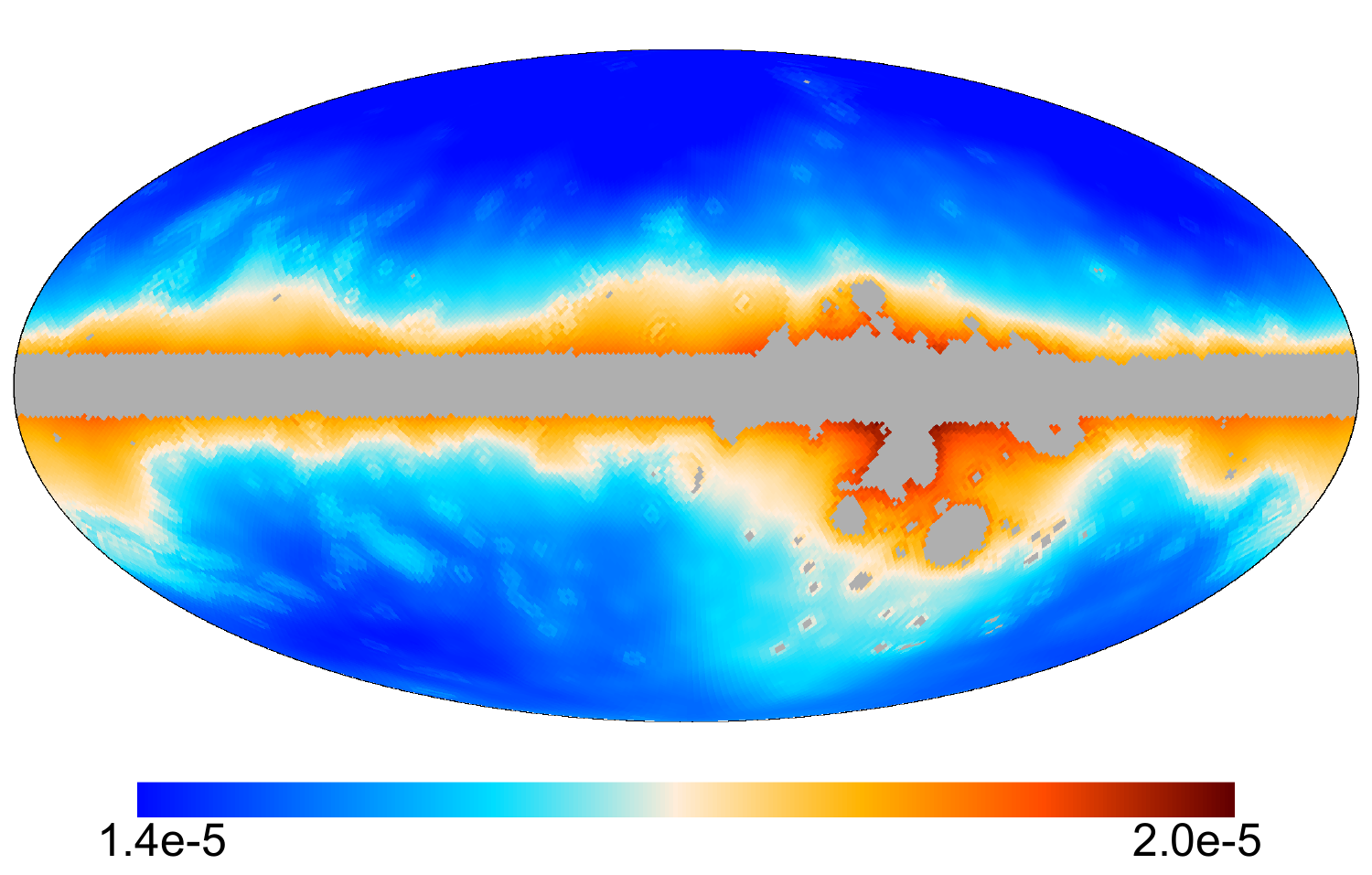}\\
\includegraphics[width=0.495\textwidth]{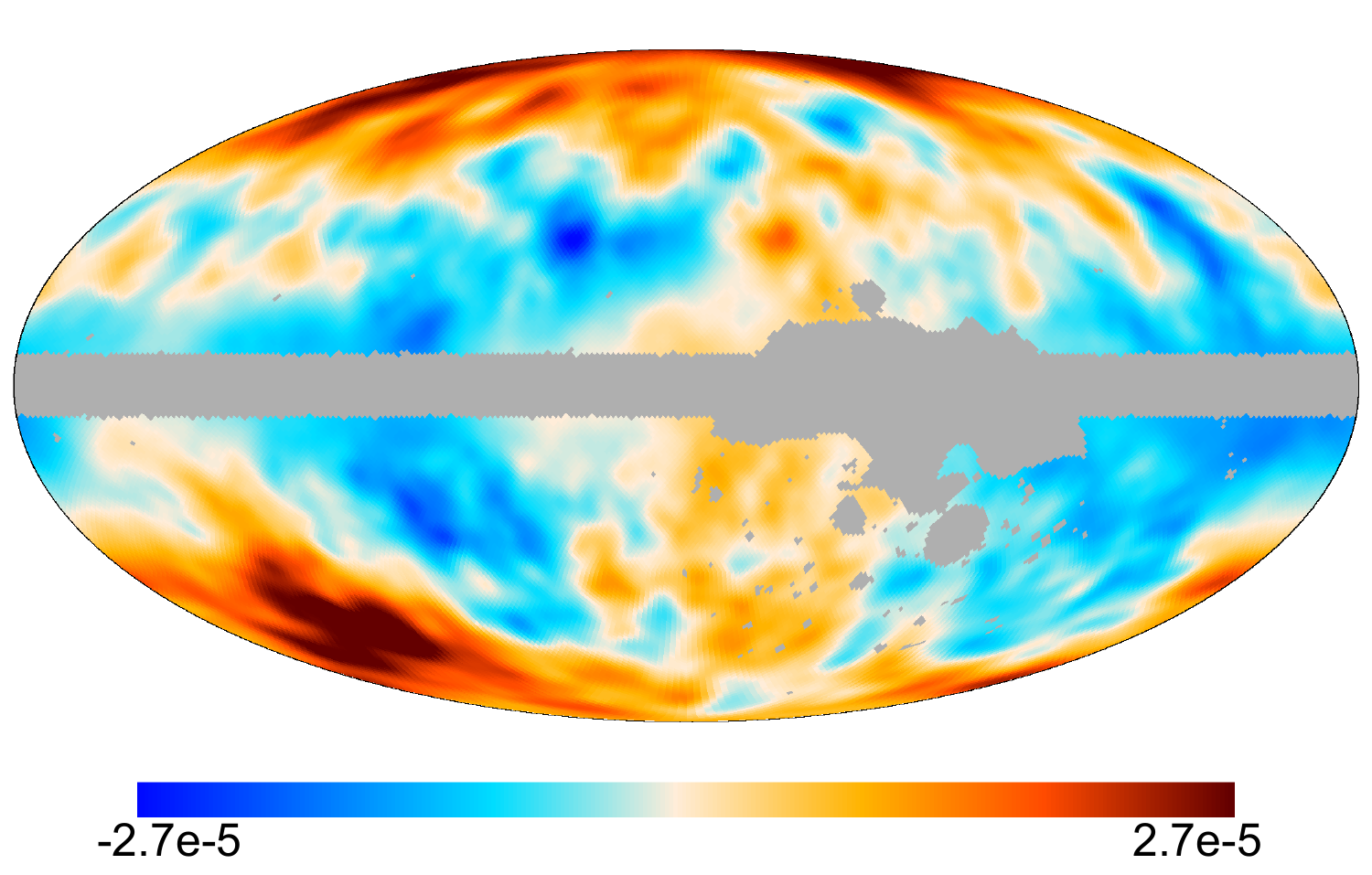}
\includegraphics[width=0.495\textwidth]{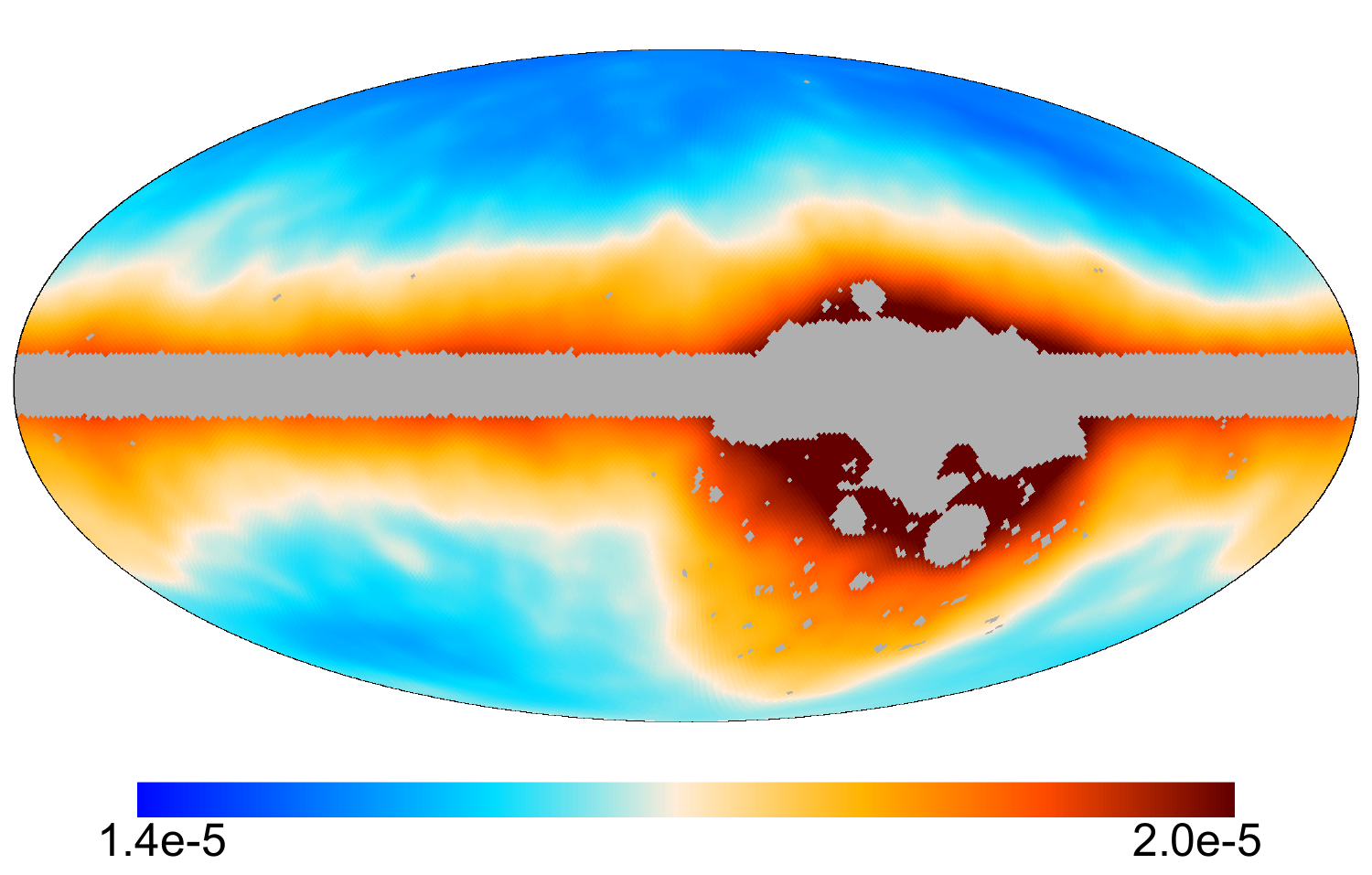}\\
\includegraphics[width=0.495\textwidth]{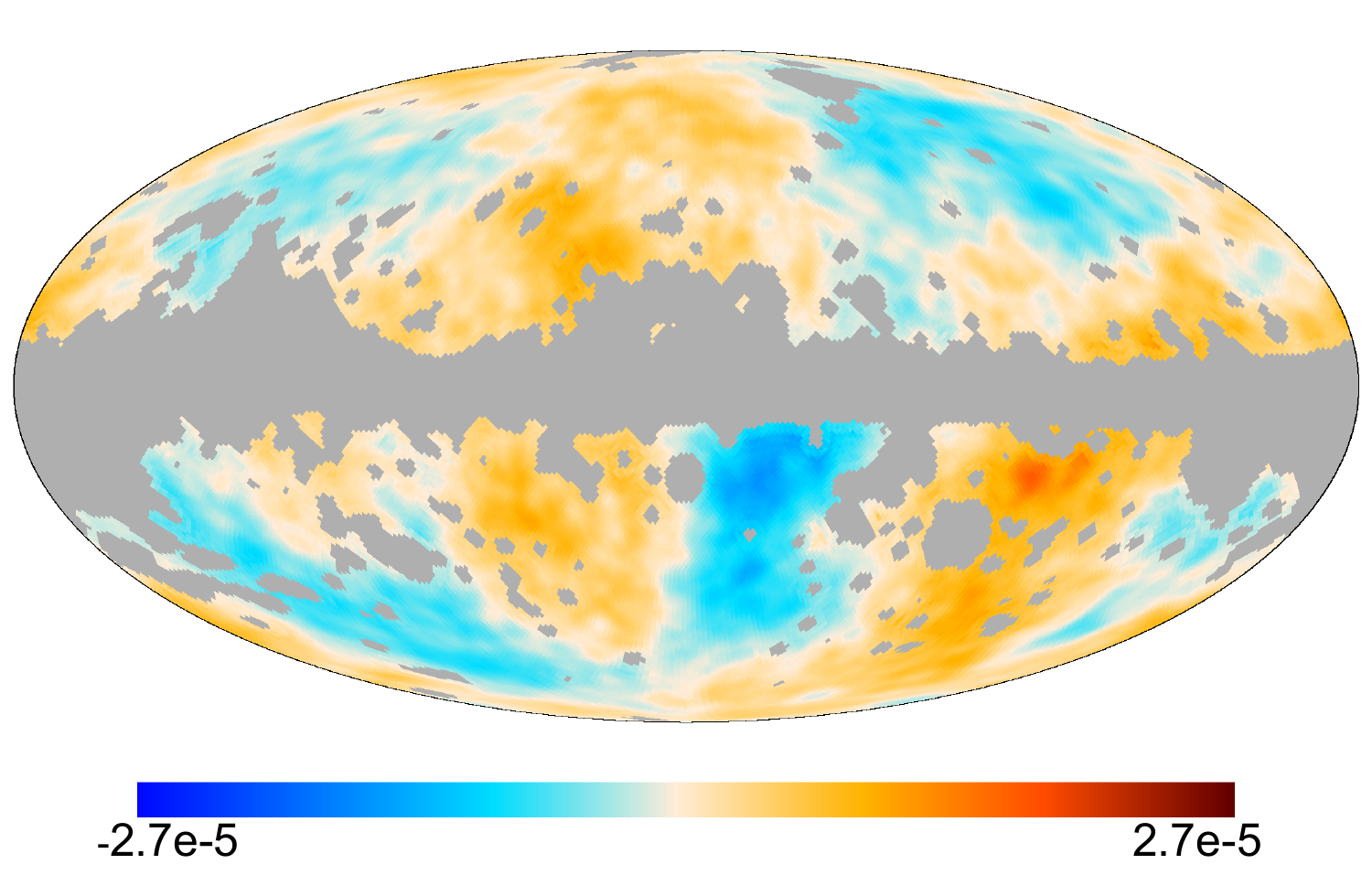}\\
\caption[]{Maps of the recovered ISW anisotropies (left column) and the corresponding estimated uncertainty per pixel (right column) from the combination of the \Planck\ \sevem\ CMB map and all the surveys (\nvss, \wagn, \wg, \lrg, \mg, and \kap; top), and only considering the information from these LSS tracer surveys (middle). The bottom panel gives the difference between both reconstructions, with the CMB intensity mask applied. The units here are Kelvin.} 
\label{fig:lcb_all}
\end{figure*}

\begin{figure}
\centering
\includegraphics[width=0.495\textwidth]{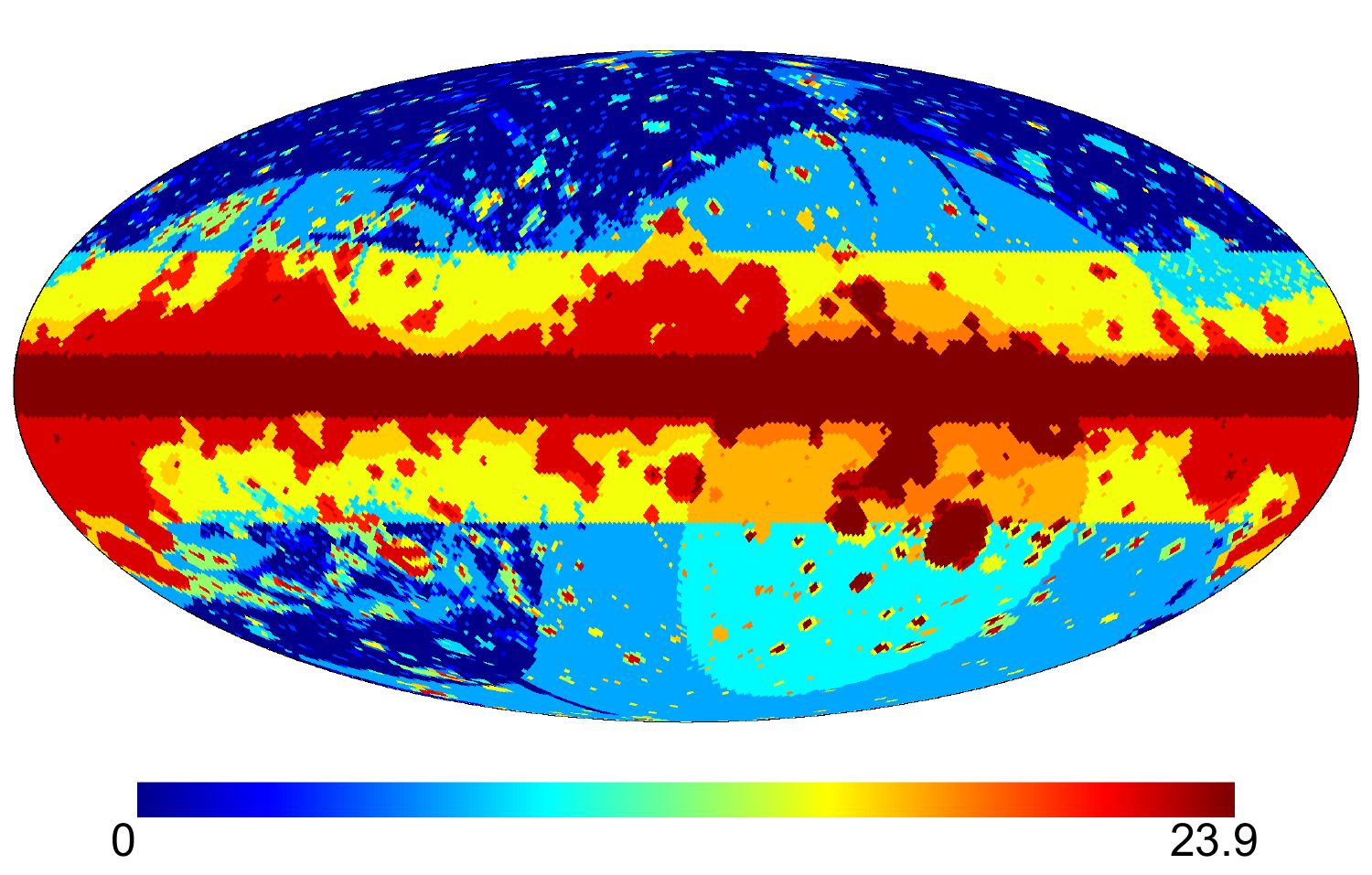}
\caption[]{Regions defined by the different intersections of the masks considered for the recovery of the ISW signal, using the LCB filter. To produce this figure, for each mask we construct a map with a constant value given by $\sqrt{k_{m}}$ in the excluded pixels and zero otherwise. These maps are then added together, producing the pattern seen in the figure. In particular, we choose $ k_m = \{ 2,4,8,16,32,64 \} $ corresponding to the masks used for \kap, \nvss, \lrg, \mg, CMB, and \wise, respectively (the square root function is introduced to allow for a better visualization).
} 
\label{fig:mask_regions}
\end{figure}

\noindent We have applied the LCB filter to the \Planck\ CMB temperature map and to different combinations of the surveys described in Sect.~\ref{sec:data_lss}. Before applying the filter, the different data sets have been masked using an apodized version of the masks shown in Sect.~\ref{sec:data}. The apodization of the masks is performed to reduce the spurious correlations introduced in the harmonic domain due to incomplete sky coverage. To construct the covariance matrix, we have made use of the models described in Sect.~\ref{sec:data_sims}. The different auto- and cross-spectra are then transformed to their \emph{masked} versions with couplings computed by the {\tt MASTER} algorithm~\citep{Hivon2002}. As in previous sections, when using the \kap\ and \wise\ maps, a cut for the lowest multipoles is imposed, meaning that these surveys do not contribute to the recovered ISW signal for $\ell < 8$ (for \kap) and for $\ell < 9$ (for \wise). For the CMB intensity map, we consider the \Planck\ \sevem\ cleaned CMB map, but similar results are expected for the other component separation methods.

To study the contribution of each data set to the final ISW map, we have applied the LCB filter to a total of seven different combinations of maps. For each of these combinations, we  consider two different types of masks to study the quality of the recovered map, namely the intersection and the union mask. The intersection mask only excludes those pixels that are masked by all the data sets considered, since the method will reconstruct the ISW signal providing there is at least one data map available for a given position in the sky (although, as one would expect the reconstruction error would depend on the number of observations available at each pixel). Conversely, the union mask only keeps those pixels which are allowed by all the individual masks and, therefore, the reconstruction error will be more uniform in the region of the sky considered, since the same information is available for all pixels. We note that these masks represent two extreme cases and they are only used to study the quality of the reconstruction. To obtain the recovered CMB map, all data sets are used after applying their own individual masks. The different combinations used to recover the ISW map are given in Table~\ref{tab:mean_cor}, together with the sky fraction allowed by the corresponding intersection and union masks.

\begin{table}[tb]
\begingroup
\newdimen\tblskip \tblskip=5pt
\caption{Mean correlation between the input and reconstructed ISW maps for different combinations of data sets.\label{tab:mean_cor}}
\nointerlineskip
\vskip -3mm
\footnotesize
\setbox\tablebox=\vbox{
   \newdimen\digitwidth 
   \setbox0=\hbox{\rm 0} 
   \digitwidth=\wd0 
   \catcode`*=\active 
   \def*{\kern\digitwidth}
   \newdimen\signwidth 
   \setbox0=\hbox{+} 
   \signwidth=\wd0 
   \catcode`!=\active 
   \def!{\kern\signwidth}
\halign{#\hfil\tabskip=0.3cm& \hfil#\hfil\tabskip=0.3cm&
 \hfil#\hfil\tabskip=0.3cm& \hfil#\hfil\tabskip=0.3cm&
 \hfil#\hfil\tabskip=0pt\cr 
\noalign{\doubleline}
 \noalign{\vskip -2pt}
\omit& \multispan2 Intersection & \multispan2 Union \cr
\omit& \multispan2 mask & \multispan2 mask \cr
\noalign{\vskip 3pt\hrule\vskip 5pt} 
& f$_{\rm sky}$& $\bar{\rho}$& f$_{\rm sky}$& $\bar{\rho}$\cr
\noalign{\vskip 3pt\hrule\vskip 5pt}
CMB and \nvss& 0.84& 0.56& 0.56& 0.58\cr
\noalign{\vskip 3pt\hrule\vskip 5pt}
CMB and \kap& 0.73& 0.50& 0.61& 0.51\cr
\noalign{\vskip 3pt\hrule\vskip 5pt}
CMB, \nvss\ and \kap&  0.85& 0.58& 0.51& 0.61\cr
\noalign{\vskip 3pt\hrule\vskip 5pt}
CMB and \wise& 0.70& 0.48&  0.42&  0.49\cr
\noalign{\vskip 3pt\hrule\vskip 5pt}
CMB and \sdss&  0.69&  0.53& 0.19& 0.60\cr
\noalign{\vskip 3pt\hrule\vskip 5pt}
CMB and all surveys&  0.85& 0.60&  0.16&  0.67\cr
\noalign{\vskip 3pt\hrule\vskip 5pt}
All surveys&  0.83&  0.49& 0.17& 0.61\cr
\noalign{\vskip 5pt\hrule\vskip 3pt}}}
\endPlancktablewide                    
\endgroup
\end{table}

Fig.~\ref{fig:lcb_nvss_kappa} shows the reconstruction attained by combining the CMB with \nvss, with \kap\ and with both surveys simultaneously as well, as their corresponding uncertainties per pixel. The intersection mask has been applied in each case.
The errors are obtained as the average dispersion of the input minus the reconstructed ISW obtained from 10\,000 coherent simulations of the different data sets. The first two cases (CMB plus \nvss\ and CMB plus \kap) were already presented in the \Planck\ 2013 paper, finding very similar results to the ones presented here. As one would expect, using both tracers jointly with the CMB (bottom row) improves the results with respect to the cases where only one tracer is used, although the improvement obtained by adding the \kap\ map is only moderate. This is due, at least in part, to the low multipole cut imposed in this tracer, which implies that the lowest multipoles are recovered using only the CMB. The quality of the ISW reconstruction can be further quantified by calculating the correlation $\rho$ between the input $s$ and reconstructed $\hat{s}$ maps using simulations. Before calculating the correlation, the monopole and dipole are subtracted from the input and reconstructed maps outside the considered mask. Table~\ref{tab:mean_cor} gives the average correlation obtained over 10\,000 simulations (outside the union and intersection masks) estimated for each simulation as:
\begin{linenomath*}
\begin{eqnarray}
\label{eq:cor_lcb}
\rho&=&\frac{\sum_i \omega_i \left(s_i - \mu_{s} \right) \left( \hat{s}_i -\mu_{\hat{s}} \right)}{ \sigma_{s} \sigma_{\hat{s}}}; \\
\omega_i &=& \frac{1/\sigma^2_i}{\sum{1/\sigma^2_i}}
\nonumber.
\end{eqnarray}
\end{linenomath*}
where the sum runs over all the pixels allowed by the considered mask and the weights at each pixel $\omega_i$ have been estimated from the error map $\sigma_i$ shown in the right column of Fig.~\ref{fig:lcb_nvss_kappa}. The quantities $\sigma_{s}$ and $\sigma_{\hat{s}}$ are the dispersion of the input and reconstructed map for each simulation obtained with the same weights, while $\mu_{s}$ and $\mu_{\hat{s}}$ correspond to the weighted mean values of the same maps. For the union mask, an average correlation coefficient of 0.61 is found when \nvss, \kap\ and the CMB are combined, to be compared to the cases when only one tracer is used, i.e., 0.58 and 0.51 for \nvss\ and \kap, respectively.

Fig.~\ref{fig:lcb_wise_sdss} gives the ISW signal reconstructed from CMB and the \wise\ surveys (top) and from CMB and the \sdss\ surveys (bottom) as well as their corresponding uncertainties. A bright red area is seen in the northern Galactic region, just above the central part of the mask, which can be identified with systematics present in the \wise\ catalogues (see Fig.~\ref{fig:surveys_maps}). Due to the cut at low multipoles imposed in these surveys, the structure at the largest scales is suppressed in the reconstruction, which is reflected in a larger uncertainty. The correlation between input and reconstruction outside the union mask is 0.49, the lowest value found among all the considered cases (see Table~\ref{tab:mean_cor}). Regarding the reconstruction using CMB and the \sdss\ surveys (bottom), we find a large signal in the relatively small regions observed by these surveys. This is refelcted in a mean correlation between input and reconstruction of 0.60 in the region allowed by the common mask, showing that the \sdss\ provides a sensitive tracer of the ISW effect.

The top row of Fig.~\ref{fig:lcb_all}  shows the reconstructed ISW signal, as well as the estimated uncertainty, obtained from the CMB map together with the six mentioned surveys (\nvss, \kap, the two \wise\ and the two \sdss\ surveys).  As one would expect, by combining all the available information, we obtain the best ISW map with a reconstruction uncertainty of around 14\,$\mu$K and a mean correlation coefficient of 0.67 outside the union mask. This corresponds to a maximum S/N ratio greater than in certain regions of the sky. If the intersection mask is considered, the correlation coefficient is 0.60 obtained over 85\,\% of the sky. Finally, the last case (middle row) gives the ISW map reconstructed using only the six surveys, without including the CMB, which corresponds to the first term of Eq.~(\ref{eq:rec_n}). It is apparent that removing the CMB degrades the reconstruction, especially in those areas where less surveys are available, and this decreases the correlation coefficient to 0.60 (union mask). To show the contribution given by the CMB to the recovery of the ISW map, the difference between these two reconstructions is also shown in the bottom panel of the figure. The intensity CMB mask has been applied and the monopole and dipole removed outside this mask. We note that, as expected, the structure of this map mainly reflects that of the large scales of the intensity CMB data presented in Fig.~\ref{fig:cmb_data}.

It is interesting to point out that some common structures are visible among the different reconstructions, although the maps are not expected to look exactly the same, since each survey traces the ISW effect in a different way and, thus, each considered LSS tracer provides a partial reconstruction of the ISW signal.

As already mentioned, the structure of the error maps given in the right columns of Figs.~\ref{fig:lcb_nvss_kappa}, \ref{fig:lcb_wise_sdss}, and \ref{fig:lcb_all} reflects the different sky coverages of the surveys, showing the contribution of each data set to the final ISW reconstruction. This can be further explored by comparing these different structures with Fig.~\ref{fig:mask_regions}, which shows the intersection regions defined by the CMB and survey masks. Each colour corresponds to a region where the intersection of a different sets of masks occurs; the dark blue area is observed by all data sets whereas the dark red region gives those pixels that are not observed by any of the data sets.

\section{Conclusions}
\label{sec:discussion}   

We have presented a study of the ISW effect using the \Planck\ 2015 data release, which provides higher sensitivity temperature anisotropy maps with respect to the previous \Planck\ 2013 release, as well as CMB polarization data at angular scales below $5^\circ$. Compared to our past publication~\citep{planck2013-p14}, we have extended the analysis in the following ways.

First, we have included additional galaxy (\wg) and AGN (\wagn) catalogues from the WISE survey as LSS tracers to be correlated with the four \Planck\ CMB maps (\cruler, \nilc, \sevem, and \smica). These tracers, in combination with the \nvss\ radio catalogue, the photometric luminous galaxy (LG) catalogue from the Baryonic Oscillation Spectroscopic Survey (BOSS) of the SDSS III (\lrg), and the photometrically selected galaxies from the SDSS-DR8 catalogue (\mg), yield a detection of the ISW signal at $2.9\, \sigma$. This detection is dominated by the \nvss\ catalogue ($2.6\, \sigma)$, while the combination of the two \sdss\ catalogues provides a $2.7\, \sigma$ level, and the two \wise\ render a $1.9\, \sigma$ \stn\ ratio.

Second, we have also improved the characterization of the ISW effect through the ISW-lensing bispectrum, since the higher \stn\ ratio of the \Planck\ 2015 temperature data and the new polarization data allows us to improve the reconstruction of the \Planck\ lensing signal. In particular, we have increased the detection achieved in the previous release by about $20\%$, reaching rougly a $3\, \sigma$ detection. We have performed a new analysis in which the \Planck\ ISW-lensing is combined with the cross-correlation of the \Planck\ CMB with all the previously mentioned LSS tracers, obtaining a total detection of the ISW effect at the $4\, \sigma$ level. The four CMB maps provide similar detection levels for all the cross-correlation combinations.

Third, we have investigated the anomalous nature of the ISW signal detected through the stacking of the CMB anisotropies at the positions of known superstructures~\citep{Granett2008a}. We have confirmed that the aperture photometry profiles around the 50 supervoids and 50 superclusters of the GR08 catalogue exhibit a maximum amplitude of $-11\muK$ (at scales of around $3.5^\circ$) and $ 9\muK$ (at scales of around $4.5^\circ$), respectively. These amplitudes are much larger than expected in the context of the standard $\Lambda$CDM scenario. We have used the \Planck\ polarization data to further explore  the origin of this signal. We do not find evidence for a positive correlation of this signal in the polarization data, indicating  that the origin of the temperature signal is, indeed, compatible with a secondary anisotropy, as expected for the ISW effect. These aperture photometry results are extremely consistent for the four CMB polarization maps, as well as for the \sevem\ cleaned frequency maps at 100, 143, and 217 GHz. This greatly reduces the possibility that this signal is significantly affected by contamination from residual Galactic and extragalactic foregrounds. Similar conclusions are obtained through the analysis of the \ep -correlated and \ep -uncorrelated counterparts of the temperature signal; excess is found only in the latter, as expected for the ISW effect. Finally, we have also stacked patches of the \Planck\ lensing map at the locations of these superstructures, and find a positive correlation for both clusters and voids, which offers extra evidence in favour of the ISW hypothesis.

Finally, we have improved the recovery of the ISW fluctuations on the sky by using a generalization of the linear covariance-based filter. In particular, we have used the five galaxy catalogues mentioned above and the \Planck\ lensing convergence map to infer a map of the secondary anisotropies associated with the ISW effect caused by the LSS traced by these surveys. Using simulations, we have been able to provide an associated rms map with a mean value of $14\muK$ per pixel of about $1^\circ$. Our ISW reconstruction provides regions where the ISW fluctuations are recovered at more than $2\, \sigma$.
We have also explored an alternative approach to estimating a map of ISW anisotropies by attempting a direct inversion of the density field as traced by the 2MASS photometric redshift catalogue into its corresponding gravitational potential field. The typical rms of the ISW effect induced by these nearby structures is, as expected, very low ($\approx 0.6\,\muK$), and this is well below the level of the measured large angular CMB fluctuations. Nevertheless, the angular power spectrum of the ISW effect produced by these structures is accurately recovered for $\ell \lesssim 20$.

Therefore, the cross-correlation of the \Planck\ CMB maps with different tracers of the LSS confirms the detection of the ISW effect at the expected level for the $\Lambda$CDM model. The current detection level could be slightly improved, on the CMB side, by analysing the next \Planck\ release, which will include large-scale polarization data. In addition, the use of the future full polarization \Planck\ data could be very important to probe further the nature of the ISW stacked signal, in a more complete manner, since the current analysis (limited by the high-pass filetring) provides only a partial view of the problem.

\begin{acknowledgements}

The Planck Collaboration acknowledges the support of: ESA; CNES and CNRS/INSU-IN2P3-INP (France); ASI, CNR, and INAF (Italy); NASA and DoE (USA); STFC and UKSA (UK); CSIC, MINECO, JA, and RES (Spain); Tekes, AoF, and CSC (Finland); DLR and MPG (Germany); CSA (Canada); DTU Space (Denmark); SER/SSO (Switzerland); RCN (Norway); SFI (Ireland); FCT/MCTES (Portugal); ERC and PRACE (EU). A description of the Planck Collaboration and a list of its members, indicating which technical or scientific activities they have been involved in, can be found at \href{http://www.cosmos.esa.int/web/planck/planck-collaboration}{\texttt{http://www.cosmos.esa.int/web/planck/planck-collaboration}}. 
Some of the results in this paper have been derived using the {\tt HEALPix} package.
We acknowledge the computer resources, technical expertise and assistance
provided by the Spanish Supercomputing Network (RES) node at Universidad de
Cantabria, and the support provided by the Advanced Computing and e-Science team at IFCA. 
Part of this work was undertaken on the STFC COSMOS@DiRAC HPC Facilities at the University of Cambridge, funded by UK BIS NEI grants.
\end{acknowledgements}

\bibliographystyle{aat}
\bibliography{references_isw,Planck_bib}

\begin{appendix}
\section{Construction of an ISW map from 3D galaxy surveys}
\label{sec:appen}

In this Appendix we present an alternative approach to estimate the ISW temperature map that does not require the input of any CMB data, but only LSS data. We first outline the methodology for a generic survey, and then apply it to the \tmpz\ survey, which characteristics, in particular, the uncertainty associated to the galaxy photometric redshift estimates, are small enough to have a proper gravitational potential reconstruction. In addition, this survey traces the very local universe, whose contribution to the total ISW signal has been pointed out by some previous works~\citep[e.g.,][]{Francis2010b,Rassat2013} to be related to the large-scale \Planck\ anomalies.

The approach consists of using redshift information in galaxy catalogues to provide a full 3D gravitational potential reconstruction, which, under the assumption of a given cosmological framework, can be trivially converted into an ISW map estimate, for linearly evolving structures. We remark that in this case very high redshift precision is not required, since the gravitational potential sourcing the ISW effect is coming from large scales (at or above $100\,h^{-1}$\,Mpc typically), leaving room for redshift uncertainties at the level of $\Delta z = 0.01$--$0.03$. On such large scales, redshift space distortions can be safely ignored. 

\subsection{Methodology}

\noindent The procedure must invert a galaxy density field into a potential field in a given region of the Universe that is limited by the selection function of the survey and the sky mask. This is done by applying the Poisson equation in Fourier space, and expressing the gravitational potential in terms of the density contrast, namely
\begin{linenomath*}
\begin{equation}
-k^2 \Phi_{\vec{k}}= \frac{3}{2} H_0^2 \Omega_{\rm m} a^{-1} (z) \delta_{\vec{k}}.
\label{eq:Poisson1}
\end{equation}
\end{linenomath*}
In this equation, $\Phi_{\vec{k}}$ stands for the Fourier transform of the gravitational potential, $\Omega_{\rm m}$ denotes the total matter density parameter and $H_0$ corresponds to the Hubble constant. The factor $a^{-1}(z)=1+z$ corresponds to the inverse of the cosmological scale factor, and  $\delta_{\vec{k}}$ is the (time dependent) matter density contrast Fourier transform for the mode $\vec{k}$, as estimated from the galaxy density. The time dependence of the gravitational potentials is thus given by these last two quantities.
The use of the Poisson equation is justified, since we are considering scales that, despite being larger than typical density clustering lengths, are well inside the horizon. When handling the equation above, the presence of an effective volume mask (induced either by the sky mask and/or the survey selection function) may introduce biases in our gravitational potential estimates. In order to handle this, we choose to conduct our particular {\it Poissonian data augmentation}, consisting of the following steps

\begin{itemize}
\item We place the galaxies in a regular 3D grid in comoving coordinates. For that purpose we use the central value of the redshift assigned to each source. As will be shown below, we find that, when accounting for all other sources of uncertainty, the level of uncertainty associated with errors in the photometric redshifts is subdominant and thus can be neglected.
\item In those grid cells excluded by the sky mask, we introduce a number of mock galaxies that comes from a Poissonian realization with an average galaxy number density equal to the average number density of cells at the same distance but not excluded by the sky mask.
\item In all grid cells, we introduce another set of randomly, Poisson distributed mock galaxies, in order to make the selection function constant with respect to depth/redshift.
\end{itemize}

We assume that the selection function of the survey and the sky mask can be factorized separately, in such a way that the selection function depends exclusively on the depth/redshift. In practice, this may not be the case for regions with high extinction, but we assume that most of these regions should be discarded by the sky mask of the survey.

This procedure should provide a homogenized galaxy density field in the entire 3D grid, which can be inverted into a gravitational potential field. By conducting a set of simulations for each of the stages of our Poissonian data augmentation approach, it is possible to assess the dependence of the resulting gravitational potential field on each step. This potential reconstruction method is a very simplified version of more sophisticated approaches of inversion of observed galaxy surveys \citep[e.g.,][]{Kitaura2010,Jasche2010}.

Finally, the gravitational potential time derivative is obtained from the 3D gravitational potential after imposing that it follows  linear theory predictions of the reference cosmological model. In other words, we express the derivative of the gravitational potential field with respect to the radial comoving distance $\eta$ according to
\[
\frac{d\Phi_{\vec k}}{d\eta} = \frac{3}{2} H(z) H_{0}^2 \Omega_{\rm m}\,
 \biggl[ 1 + \frac{d\log D_{\delta}}{d\log (1+z)}\biggr] \frac{\delta_{\vec k}}{k^2}
 \]
 \begin{equation}
 \phantom{xxxxxxx} = \, -\Phi_{\vec k} \,H(z) a(z) \, \biggl[ 1 + \frac{d\log D_{\delta}}{d\log (1+z)} \biggr].
\label{eq:dphioverdt}
\end{equation}
In this equation, $H(z)$ denotes the Hubble parameter, $H_0$ its current value, and $D_{\delta}(z)$ the redshift dependent linear growth factor of the density perturbations. The final ISW map is obtained after integrating the $d\Phi_{\vec k}/d\eta$ 3D grid along the line of sight.

\subsection{Results}

\noindent Now we present the results of inverting the 2MPZ survey into a gravitational potential field whose time derivative is then projected along the line of sight. 
 
We place the galaxies of this survey in a 3D grid of 128$^3$ cells of $6\,h^{-1}$\,Mpc on a side, centred upon the observer. This means that the maximum redshift considered is $z_{\rm max}\simeq 0.13$, and that more than 85\,\% of the 2MPZ sources are actually placed inside the grid. This choice of $z_{\rm max}$ is motivated as a compromise between sampling a large cosmological volume and having a representative amount of galaxies tracing the potential wells; increasing $z_{\rm max}$ degrades potential reconstructions at large distances from the observer and provides little information about the ISW signal. 

Provided that the ISW effect is generated out to redshifts $z\approx 1$, our choice of $z_{\rm max}$ should contain a small fraction of the total ISW signal generated in our visible Universe. Nevertheless the contribution of this relatively nearby cosmological volume to the low multipole anisotropy power has been claimed not to be completely negligible, and it has been argued that it may be of relevance in the context of the CMB large-angle anomalies \citep[e.g.,][]{Francis2010b,Rassat2013}. 

\begin{figure}
\includegraphics[width=0.5\textwidth]{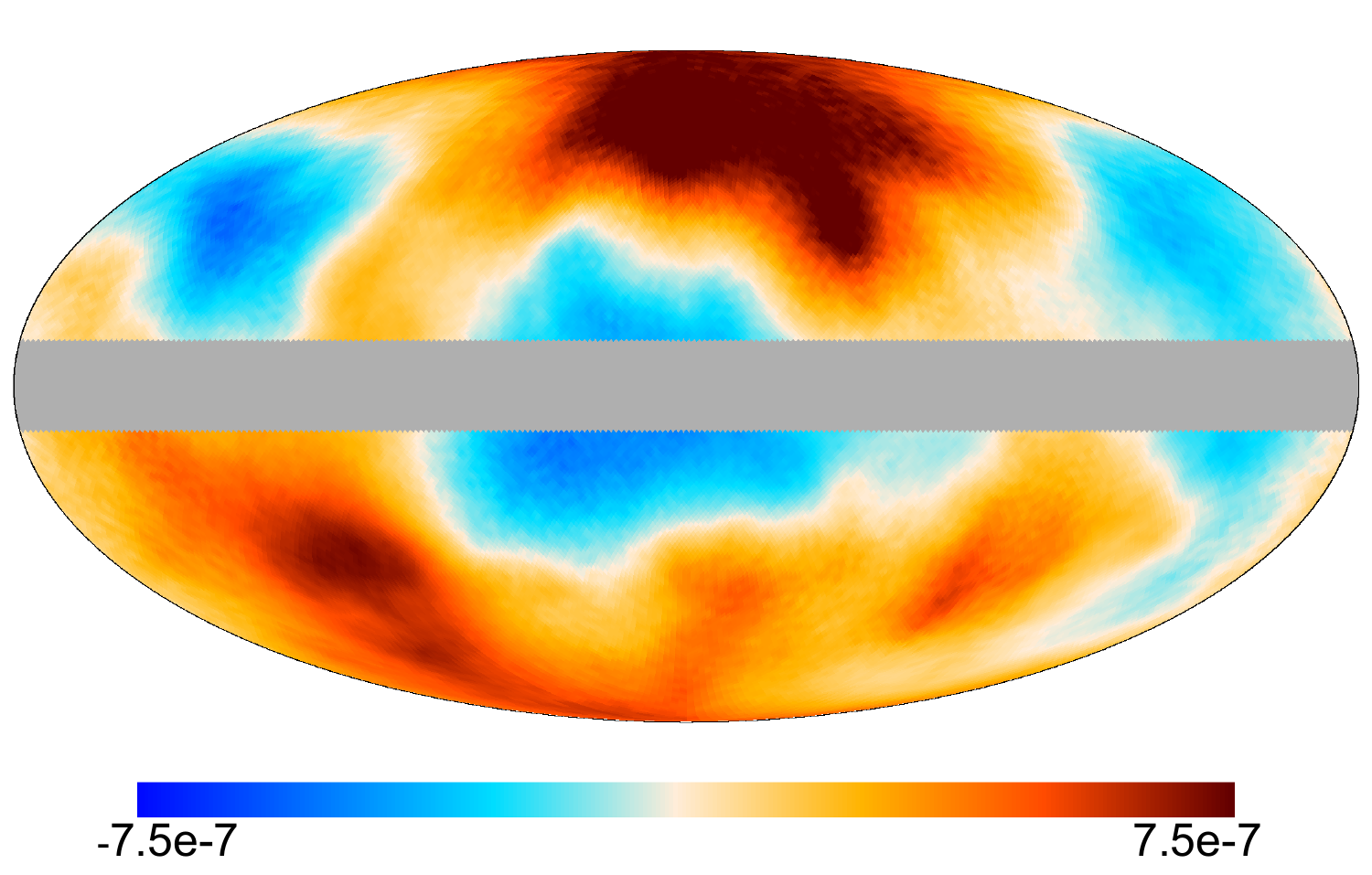}
\caption[fig:mapISW2mphz]{Map of the recovered ISW signal from the 2MPZ catalogue. Units here are Kelvin.} 
\label{fig:mapISW2mphz}
\end{figure}

When conducting the reconstruction, we impose a sky mask for all pixels with $|b_{\rm gal}| < 10\degr$ for which the Galaxy heavily impacts the selection function of the survey. The result of the inversion of the galaxy density field into the gravitational potential field and its time derivative is shown in Fig.~\ref{fig:mapISW2mphz}. The recovered ISW map resembles the large-scale structure of the projected density map (Fig.~\ref{fig:surveys_maps}). The positive structure of the ISW map traces, in the North Galactic hemisphere, the presence of well known superclusters like Ursa Major, Virgo, Centaurus, or Hydra, while in the southern hemisphere, at slightly negative Galactic latitudes, the most prominent negative spot corresponds to the Local Void. 

\begin{figure}
\includegraphics[width=9.cm]{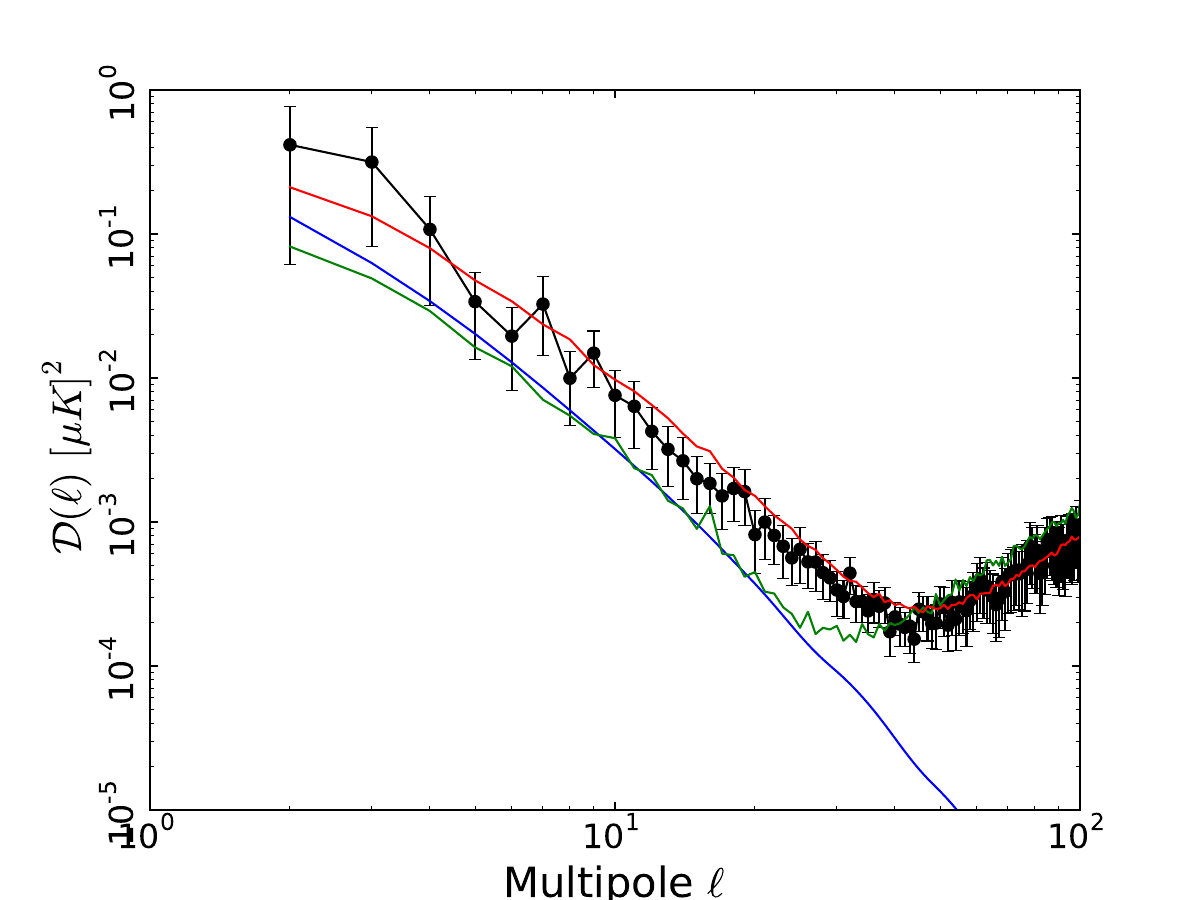}
\caption[fig:cls_rec]{Comparison of the ISW-recovered angular power spectrum from 2MPZ data (black solid circles) with theoretical expectations (blue solid line) and the average of ideal (green solid line) and realistic (red solid line) simulations of our density-to-potential inversion algorithm (see text for details). } 
\label{fig:cls_rec}
\end{figure}

We find, however, that the amplitude of the recovered ISW map is too small to contribute significantly to the total CMB map on the largest scales. In Fig.~\ref{fig:cls_rec} filled black circles display the angular power spectrum of the recovered ISW effect from 2MPZ.  The rms of this map is dominated by the quadrupole, whose amplitude is found to be $C_{\ell=2}^{\rm 2MPZ,\,ISW}=(0.44\pm 0.37 \,\mu \mathrm{K})^2$, driving the rms map to be at the level of only  $0.56\,\mu \mathrm{K}$. Thus our estimated quadrupole amplitude of the ISW map generated by 2MPZ seems to be in tension with the estimate of \citet[][]{Rassat2013}, since those authors quote a theoretical expectation for the ISW quadrupole of $(12 \pm 10\,\mu K)^2$, i.e., almost two orders of magnitude above our estimate.
 
We next compare the amplitude of the angular power spectrum of our recovered ISW map with theoretical expectations. For this purpose, we make use of a modified Boltzmann code that provides the ISW angular power spectrum for a generic galaxy sample that is probing the large-scale structure under the same selection function as the one estimated for 2MPZ. We remark that this estimate of the ISW amplitude is independent of the bias of the galaxy sample. Such a prediction is provided by the thick blue solid line in Fig.~\ref{fig:cls_rec}. We can see that the recovered ISW power spectrum is significantly higher than this expectation. In order to understand this, we run 100 Monte Carlo simulations of Gaussian density fields in the same 3D spatial grid used for the density-to-potential inversion of real data. These Gaussian simulations are obtained from a $\Lambda$CDM matter power spectrum corresponding to our fiducial cosmological model at $z=0$. In this set of MC {\it ideal} simulations we only impose the radial selection function of the 2MPZ survey at the time of conducting the line of sight integral of the time derivative of the gravitational potentials, but ignore all effects of radial selection function, photometric redshift errors, and shot noise, when producing the potential maps. The green solid line provides the average angular power spectrum obtained from this set of simulated maps. The agreement of this computation with the theoretical expectation is very good for multipoles $\ell < 20$; artefacts related to the projection of the finite grid cells on the sky introduce spurious power that becomes dominant on smaller scales. 
 
We also run a second set of MC {\it realistic} simulations which are based upon the same set of Gaussian simulations just described above, but after including the impact of the 2MPZ radial selection function, photometric redshift errors, and shot noise, as  was required for real data. The photometric redshift errors were simulated by adding a normal deviate of rms, $\sigma_z=0.015$ to the ''correct'' redshifts of the simulated galaxies. We note that in both sets of simulations the initial 3D Gaussian matter density field is identical, but in this case Poissonian augmentation was required: (1) in the sky-mask excluded areas; and (2) at large redshifts, in order to avoid radial galaxy density gradients associated with the radial selection function. The average angular power spectrum from this set of simulated ISW maps is displayed in Fig.~\ref{fig:cls_rec} by the red solid line. We find in this case much better agreement with the results from the real 2MPZ catalogue. From this set of simulations we obtain the uncertainties for the recovered 2MPZ ISW power spectrum multipoles.
   
Despite the uncertainties in the amplitude of the recovered ISW map, this analysis shows that it is highly implausible that, in a standard $\Lambda$CDM scenario, the ISW generated by the gravitational potentials hosting the 2MPZ galaxies can significantly modify the large-scale pattern of the CMB; the expected quadrupole of the ISW is about three orders of magnitude below the total CMB quadrupole amplitude.

\begin{figure}
\centering
\includegraphics[width=9.cm]{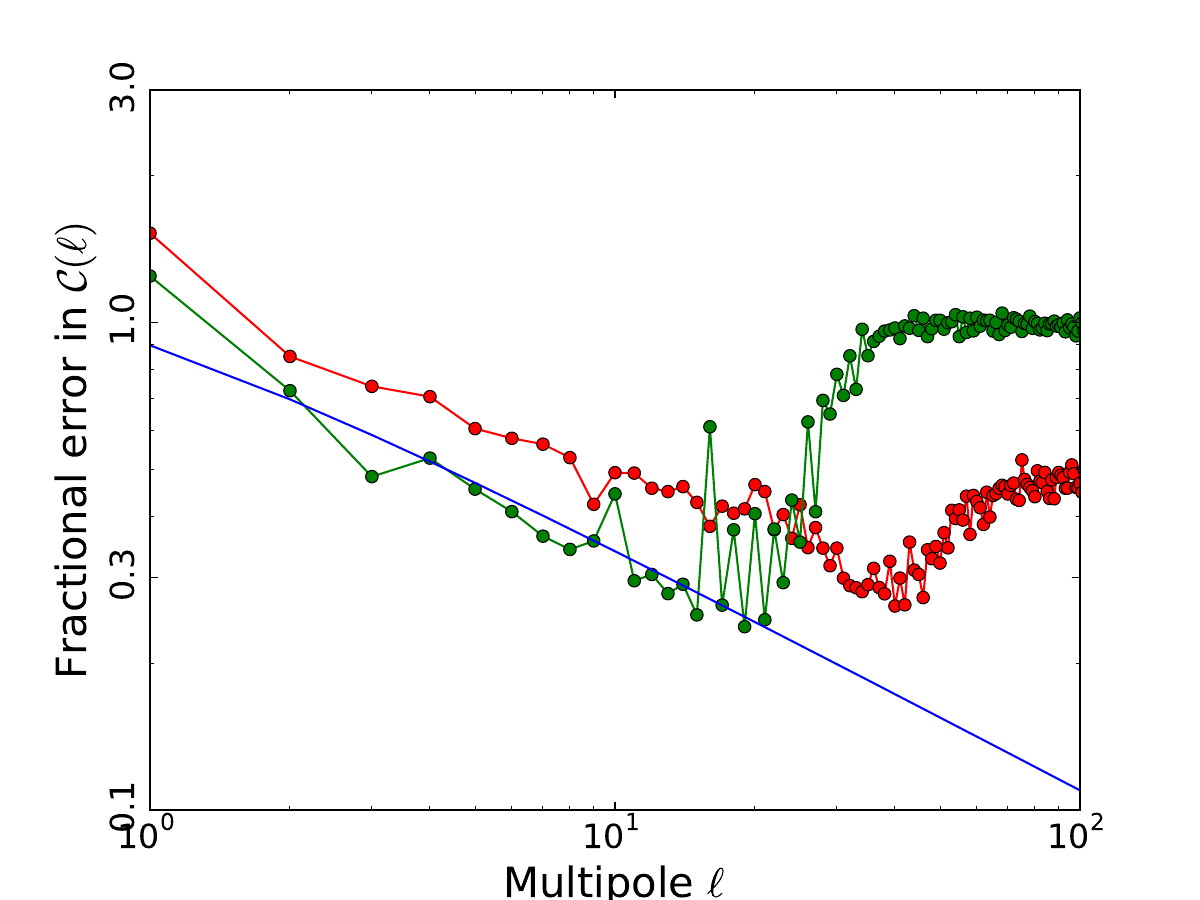}
\caption[fig:err_cls]{Relative uncertainty in the recovered ISW angular power spectrum multipoles. Red triangles and green circles refer to the {\it realistic} and {\it ideal} sets of MC simulations, respectively.} 
\label{fig:err_cls}
\end{figure}

These two sets of simulations should provide a fair description of the total error budget introduced by our approach. In Fig.~\ref{fig:err_cls}, filled red triangles display the ratio of the rms of the recovered angular power spectrum multipoles over their average value for the realistic set of simulations. For instance, for the recovered quadrupole this plot shows that the rms of the quadrupole amounts to roughly 90\,\% of its amplitude. We note that the quoted uncertainty in the recovered angular power spectrum multipoles includes the contribution from cosmic variance. For the sake of comparison, the solid blue line depicts this ratio for the case of a pure Gaussian field without any coupling between different multipoles and the same sky coverage as for 2MPZ and the Monte Carlo simulations; in the case this ratio obeys the simple form $\sqrt{2/f_{\rm sky}/(2l+1)}$.  This trend is closely followed by the output of the ideal MC simulations ignoring the impact of the radial selection function and the Poissonian augmentation (green filled circles). For multipoles below $\ell = 20$ errors in the recovered angular power spectrum multipoles are close to the Gaussian prediction, but on smaller scales errors associated with the line of sight integral become dominant.
\begin{figure}
\centering
\includegraphics[width=9.cm]{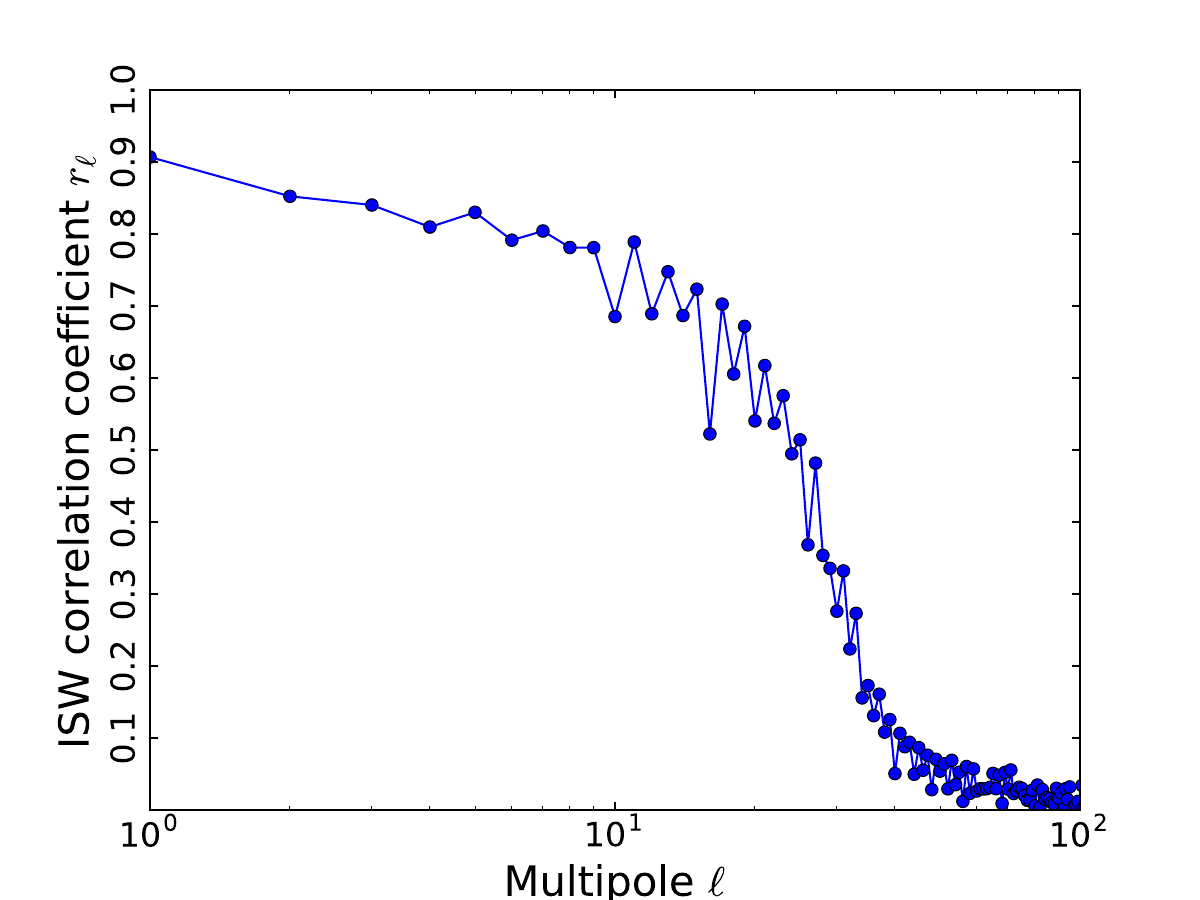}
\caption{\label{fig:corr_coeff_cls}Correlation coefficient between the ISW recovered maps in the realistic and ideal sets of simulations.}
\end{figure}

Finally, in order to provide an estimate of how the ISW maps recovered by our technique actually resemble the real, underlying ISW maps, Fig.~\ref{fig:corr_coeff_cls} displays the correlation coefficient between the ISW recovered map under the ideal and realistic set of simulations.  This correlation coefficient is defined as $r_{\ell} = \langle a_{\ell,m}^{\rm real} (a_{\ell,m}^{\rm ideal})^\ast\rangle / (C_{\ell}^{\rm real} C_{\ell}^{\rm ideal})^{1/2}$, that is, the ratio of the angular cross-spectrum of each pair of maps over the square root of the product of the auto spectra.  The correlation coefficient is, on average, about 0.8 in the multipole range $\ell =$ 2--10, and above 0.70 in $\ell =$ 2--20. On smaller angular scales there is little ISW power and spurious power erases any ISW information in the recovered maps. 

Finally, we performed a new realistic set of MC simulations for which the uncertainty in the radial distance to galaxies associated with photometric redshift errors is switched off. We found very little difference in the uncertainty in the recovered ISW angular power spectrum multipoles and in the correlation coefficients, suggesting that photo-$z$ errors at the level $\sigma_z=0.015$ are not a dominant source of uncertainty for our ISW reconstruction approach.

\end{appendix}
\raggedright

\end{document}